\pdfoutput=1
\documentclass[traditabstract]{aa}
\usepackage{graphicx}
\usepackage{txfonts}
\usepackage{natbib}
\bibpunct{(}{)}{;}{a}{}{,}

\begin{document}

\title 
{
A multi-scale, multi-wavelength source extraction method: \textsl{getsources}
}

%||||||||||||||||||||||||||||||||||||||||||||||||||||||||||||||||||||||||||||||||||||||||||||||||||||||||||||||||||||||||||||||||||

\author
{
A.~Men'shchikov\inst{1} 
\and Ph.~Andr\'e\inst{1} 
\and P.~Didelon\inst{1}
\and F.~Motte\inst{1}
\and M.~Hennemann\inst{1}
\and N.~Schneider\inst{1}
}

%||||||||||||||||||||||||||||||||||||||||||||||||||||||||||||||||||||||||||||||||||||||||||||||||||||||||||||||||||||||||||||||||||

\institute
{
Laboratoire AIM Paris--Saclay, CEA/DSM--CNRS--Universit{\'e} Paris Diderot, IRFU, Service d'Astrophysique, Centre d'Etudes de 
Saclay, Orme des Merisiers, 91191 Gif-sur-Yvette, France
}

\date{Received 09 Jan 2012 / Accepted 16 Apr 2012}

\offprints{Alexander Men'shchikov}
\mail{alexander.menshchikov@cea.fr}
\titlerunning{A multi-scale, multi-wavelength source extraction method}
\authorrunning{Men'shchikov et al.}

%||||||||||||||||||||||||||||||||||||||||||||||||||||||||||||||||||||||||||||||||||||||||||||||||||||||||||||||||||||||||||||||||||

\abstract{

We present a multi-scale, multi-wavelength source extraction algorithm called \textsl{getsources}. Although it has been designed
primarily for use in the far-infrared surveys of Galactic star-forming regions with \emph{Herschel}, the method can be applied to
many other astronomical images. Instead of the traditional approach of extracting sources in the observed images, the new method
analyzes fine spatial decompositions of original images across a wide range of scales and across all wavebands. It cleans those
single-scale images of noise and background, and constructs wavelength-independent single-scale detection images that preserve
information in both spatial and wavelength dimensions. Sources are detected in the combined detection images by following the
evolution of their segmentation masks across all spatial scales. Measurements of the source properties are done in the original
background-subtracted images at each wavelength; the background is estimated by interpolation under the source footprints and
overlapping sources are deblended in an iterative procedure. In addition to the main catalog of sources, various catalogs and images
are produced that aid scientific exploitation of the extraction results. We illustrate the performance of \textsl{getsources} on
\emph{Herschel} images by extracting sources in sub-fields of the \object{Aquila} and \object{Rosette} star-forming regions. The
source extraction code and validation images with a reference extraction catalog are freely available.

}


\maketitle

%||||||||||||||||||||||||||||||||||||||||||||||||||||||||||||||||||||||||||||||||||||||||||||||||||||||||||||||||||||||||||||||||||

\section{Introduction}
\label{introduction}

The \emph{Herschel} Space Observatory \citep{Pilbratt_etal2010} provides the best opportunity to study the earliest stages of star
formation. Prestellar cores and young (Class 0) protostars emit the bulk of their luminosities at wavelengths ~80--400\,{${\mu}$m},
which makes the \emph{Herschel} imaging instruments PACS \citep{Poglitsch_etal2010} and SPIRE \citep{Griffin_etal2010} with their 6
wavebands from 70 to 500\,{${\mu}$m} perfect for performing a census of these objects down to ~0.01--0.1\,{$M_{\sun}$} in the nearby
(distances $D \la$ 500 pc) molecular cloud complexes. In particular, the \emph{Herschel} Gould Belt survey \citep{Andre_etal2010}
aims at probing the link between diffuse cirrus-like structures and compact cores with the main goal to understand the physical
mechanisms of the formation of prestellar cores out of the diffuse medium, which is crucial for understanding the origin of stellar
masses. Furthermore, the \emph{Herschel} HOBYS survey \citep{Motte_etal2010} aims at performing a census of massive young stellar
objects, providing accurate bolometric luminosities and envelope masses for homogeneous and complete samples of the progenitors of
massive stars.

Preparing for these two \emph{Herschel} surveys, we had evaluated a few popular source extraction algorithms to check whether they
could be used in our surveys. The main problem was the absence of any \emph{multi-wavelength} extraction technique. None of the
methods was designed to handle multi-wavelength data, making it necessary to match the independent catalogs obtained at different
wavelengths using an association radius as a free parameter. This posed very serious problems for detecting and measuring sources in
the \emph{Herschel} images with angular resolutions differing by a factor of $\sim$\,\emph{7}. A direct consequence of the mismatch
in resolutions is that the degree to which sources in a region may be blended depends \emph{very} strongly on the waveband.

With such great differences in resolution, one cannot just match independent catalogs without introducing \emph{unknown} and
potentially very large errors in the association of sources detected across different wavebands and in their measured properties.
Studying star formation in the nearest clouds, one expects to resolve many crowded regions in the highest-resolution images at the
shortest wavelengths, but at the same time, one would progressively ``lose'' sources within much larger beams at the longer
wavelengths. In effect, the fluxes of such sources on the long-wavelength side of their spectral energy distributions (SEDs) would
have large and unknown errors when matching independent extraction catalogs, greatly reducing the extraction quality (detection
completeness and reliability, as well as the accuracy of the derived properties).

For large-scale projects, such as the \emph{Herschel} Gould Belt and HOBYS surveys, one needs a fully automated source extraction
method that can find as many sources as possible from the images in all bands, reliably distiguishing them from variable backgrounds
and noise, deblending them as accurately as possible in the crowded regions by preserving and utilizing all information obtained
from the higher-resolution images at shorter wavelengths. A very serious problem with some existing source extraction methods is
that they do not allow sources to overlap, thus deblending of crowded regions is impossible. One cannot expect from such methods
high levels of detection completeness or consistently accurate flux measurements in realistic conditions.

Note that most existing methods have been developed and oriented for use in different areas of astronomy, thus their performances
for a specific project must be carefully analyzed before an appropriate method can be chosen. The SPIRE SAG3 consortium has been
testing several popular source extraction algorithms using simulated skies of various degrees of complexity and those benchmarks
will be published elsewhere (Men'shchikov et al. in prep.). Below we introduce the reader to the basic concepts of those techniques,
to place our new method\footnote{For a very brief summary, see \cite{Men'shchikov_etal2010}.} in a wider context.

\subsection{Existing methods of source extraction}
\label{existing.methods}

Most of the algorithms trying to solve the same (non-trivial) problem of source extraction originated from different ideas.

\cite{StutzkiGuesten1990}'s \textsl{gaussclumps} (developed for position-velocity data cubes in molecular-line studies of molecular
clouds) performs least-squares fits of a Gaussian shape to the brightest peak constrained to keep the position and amplitude of the
fitted shape close to the image maximum. Then it subtracts the fit from the image, producing the residuals image, and fits a new
Gaussian shape to the brightest peak in the residuals. The iterations continue until the total intensity of all subtracted clumps is
equal to the integrated intensity of the original image or there are no significant peaks left.

\cite{WilliamsdeGeusBlitz1994}'s \textsl{clumpfind} (aimed also for molecular clouds data and position-velocity cubes) contours an
image at a number of levels, starting from the brightest peak in the image. It descends down to a minimum contour level, marking as
clumps along the way all connected areas of pixels that are above the contour level. This technique was ``motivated by how the eye
decomposes the maps into clumps'' and it ``mimics what an infinitely patient observer would do'' \citep{WilliamsdeGeusBlitz1994}.
One should be aware that this method ignores the backgrounds in which sources are observed.

\cite{BertinArnouts1996}'s \textsl{sextractor} (designed for use with optical and near-IR images in extragalactic astronomy)
estimates and subtracts background using sigma clipping and spline interpolation, then uses thresholding to find sources in the
background-subtracted image, deblends them if they overlap, and measures their positions and sizes using intensity moments. A very
useful property of this versatile algorithm is that it allows using a detection image that differs from the observed image and it
can match sources with previously obtained catalogs.

CUPID\footnote{CUPID is a source extraction software package developed by the STARLINK team for use with the SCUBA2 surveys; it is a
general wrapper to which additional methods can be added. See documentation:
http://docs.jach.hawaii.edu/star/sun255.htx/sun255.html}'s \textsl{reinhold} (oriented primarily to analyzing clumps in
submillimeter data cubes) identifies ``pixels within the image which mark the edges of emission clumps, producing a set of rings
around the clumps. However, these structures can be badly affected by noise in the data and so need to be cleaned up. This is done
using cellular automata which first dilate the rings or shells, and then erode them. After cleaning, all pixels within each ring or
shell are assumed to belong to a single clump'' (see the reference in the footnote).

CUPID's \textsl{fellwalker} (also oriented towards clumps in submillimeter data cubes) finds image peaks by tracing the line of the
steepest ascent, considering every pixel with a value above a specified threshold as a starting point for a walk to a peak along the
steepest gradient. Having reached a peak, it searches for an even higher pixel intensity in a neighborhood; when found, the
algorithm switches to that pixel and continues uphill. If a peak is found that is higher than all pixels in its neighborhood, a
clump has been detected and the algorithm marks all pixels visited along the way as belonging to the clump.

\cite{Motte_etal2003,Motte_etal2007}'s \textsl{mre-gcl} (created for studies of star formation using ground-based submillimeter
continuum imaging) combines \textsl{gaussclumps} with the image filtering based on a wavelet decomposition. The algorithm decomposes
an image in spatial scales using an isotropic wavelet decomposition with the multi-resolution code \textsl{mr{\_}transform}
\citep{StarckMurtagh2006}, subtracts all scales larger than the largest scale of interest from the original image, and uses
\textsl{gaussclumps} to detect and measure sources in the filtered image. Then the user defines each source's largest extent as
twice its measured size and repeats the decomposition and filtering steps for each source, runs \textsl{gaussclumps} again with the
aim to improve the measurements of sizes and fluxes.

\cite{Molinari_etal2011}'s \textsl{cutex} (developed for studying star formation with \emph{Herschel}) attempts to overcome the
difficulty of thresholding of an entire image: highly-variable backgrounds were expected in star-forming regions and indeed observed
with \emph{Herschel}. It analyzes multi-directional second derivatives of the original image and performs curvature thresholding to
isolate compact sources out of extended emission, then fits variable-size elliptical Gaussians (adding also a planar background) at
their positions. The algorithm can fit up to 8 Gaussians simultaneously in crowded areas, if sources are closer than two
observational beam sizes. This method works only for compact sources with sizes up to approximately 3 times the beam size.

Kirk et al. (2012, in prep.)'s \textsl{csar} (developed for use with the BLAST and \emph{Herschel} images) is another method that
defines clumps in terms of connected pixels. A source is defined as a region of connected pixels bound by a closed isophotal contour
that contains at least one pixel that is at 3\,${\sigma}$ (where ${\sigma}$ is the standard deviation) above the bounding contour
level. The algorithm starts contouring just below the peak on the image and walks down until some predefined background is reached;
sources are considered finished just before they become connected to others. Sources are not allowed to overlap and no attempt is
made to assign flux outside of the closed contours to any source. The technique was designed with a ``purpose of replicating what a
(trained) human would do with an image if extracting sources manually'' (J. Kirk, private comm.).

Crowded regions that are frequently observed with \emph{Herschel}; the deblending of overlapping sources is the origin of major
uncertainties. Whereas \textsl{clumpfind}, \textsl{reinhold}, \textsl{fellwalker}, and \textsl{csar} merely partition the image
between sources not allowing them to overlap, \textsl{gaussclumps}, \textsl{sextractor}, \textsl{mre-gcl}, and \textsl{cutex} can
deblend overlapping sources, which is quite an essential property for obtaining accurate results in crowded regions. We feel the
need to stress here that the observed images are only projections of the complex \emph{three-dimensional} reality onto the plane of
the sky and, as such, it is a fundamental source of major uncertainties in the interpretation of observations and in the derived
properties of objects. As we know from \emph{Herschel} observations \citep[e.g.,][]{Men'shchikov_etal2010, Arzoumanian_etal2011},
the interstellar medium is highly filamentary, thus the filaments' orientations play a very significant role in the appearance of
the regions we observe.

\subsection{Introducing a new approach}
\label{new.approach}

In this paper, we present the source extraction algorithm \textsl{getsources} developed with the aim to overcome the shortcomings of
the existing methods and provide researchers in this \emph{Herschel} era with a better extraction tool. For details on the
astrophysical context of this work, we refer to Appendix~\ref{astrophysical.objects}.

To clarify our terminology, we shall use the term \emph{noise} to refer to the statistical instrumental noise including possible
contributions from any other signals that are not astrophysical in nature, i.e. which are not related to the emission of the areas
in space we are observing. In contrast, the term \emph{background} will refer to the (filamentary) astrophysical backgrounds, such
as cirruses or molecular clouds, containing the \emph{sources} we are observing and \emph{objects} we are studying (e.g., stars,
protostars, cores, etc.)\footnote{The backgrounds we are dealing with are known to be structured at all scales, spatially
fluctuating in an unknown way and thus creating the difficult problem of background removal. If the real backgrounds were just
smooth large-scale structures, one would be able to approximate and subtract or filter them quite well.}. In order to reduce
possible confusion, we are going to clearly distinguish between the morphologically-simple (convex, not very elongated)
\emph{sources} of emission as determined by the source extraction algorithms and the \emph{objects} of specific astrophysical nature
that are selected from the entire extraction catalog on the basis of all available information (besides the images) and some
additional assumptions, criteria, and techniques, depending on the specific interests of a researcher.

The problem of detecting sources in continuum images usually reduces to finding significant intensity peaks, as such images provide
us with just complex intensity distributions of the sky over an observed area. At the present state of the art, source extraction
procedures do not know anything about the astrophysical nature or true physical properties of the objects that produced the emission
of those significant peaks. An extraction algorithm can only detect sources (that are possibly harboring our objects of interest)
and determine their \emph{apparent} two-dimensional intensity distributions \emph{above} the variable background and noise, and
measure their apparent properties at each wavelength. This is the only information contained in the images, which is available to
any extraction algorithm. The purpose of source extraction is to detect as many real sources as possible (distinguishing them from
noise and background fluctuations) and to measure their apparent properties as accurately as possible\footnote{Different extraction
algorithms produce varying numbers of sources and estimates of their properties for the same set of images. The quality of the
extraction methods can be assessed by measuring how well they are able to reconstruct known properties of model sources in simulated
images.}.

A catalog of such sources serves as the fundamental basis for all subsequent in-depth studies of the various objects of different
nature. Only after we have detected significant sources of emission and measured all possible apparent properties (peak intensities,
integrated fluxes, sizes, etc.), can we utilize those results to infer the real astrophysical nature and properties of the
objects\footnote{Real physical properties of objects (e.g., their size) may be quite different from the measured apparent properties
of the extracted sources at different wavelengths, whose accuracy, in turn, critically depends on the quality of the extraction
algorithm.}. At this step, one should combine all the information in different wavebands from the extraction catalog and also from
previous or follow-up observations, as well as any other available information, and classify the objects according to their physical
nature and use them to derive new astrophysical knowledge. In what follows, we only consider the sources and leave the definition of
objects to the future papers exploring various astrophysical applications of this new extraction algorithm.

Unlike other source extraction algorithms (except \textsl{mre-gcl}, Sect.~\ref{existing.methods}), the new method analyzes fine
spatial decompositions of original images across a wide range of scales and across all wavelengths
(Sect.~\ref{decomposing.detection.images}). As part of its multi-wavelength design, \textsl{getsources} removes the noise and
background fluctuations from the decomposed images (Sect.~\ref{removing.noise.background}) separately in each band, and constructs a
set of wavelength-independent detection images (Sect.~\ref{combining.clean.single.scales}) that preserve information in both spatial
and wavelength dimensions as well as possible. Sources are detected in the combined detection images by following the evolution of
their segmentation masks across all spatial scales (Sect.~\ref{detecting.sources}). Measurements of the source properties are
performed in the original images at each wavelength after the background has been subtracted by interpolation under the sources'
``footprints'' and after overlapping sources have been deblended (Sect.~\ref{measuring.cataloging}). To facilitate visual analysis
of the extraction results and various steps of the algorithm, a number of useful images are created for each waveband
(Sect.~\ref{visualizing.extractions}). Based on the results of the initial extraction, detection images are ``flattened'' to produce
much more uniform noise and background fluctuations in preparation for the second, final extraction
(Sect.~\ref{flattening.background.noise}). The performance of \textsl{getsources} for \emph{Herschel} images is illustrated on small
sub-fields of the \object{Aquila} and \object{Rosette} star-forming regions (Sect.~\ref{applications.herschel}).

%||||||||||||||||||||||||||||||||||||||||||||||||||||||||||||||||||||||||||||||||||||||||||||||||||||||||||||||||||||||||||||||||||

\section{The \textsl{getsources} extraction method}
\label{getsources.extraction.method}

The fundamental problem in extracting sources from observed images is that all spatial scales are mixed together and the intensity
of any given pixel contains an unknown contribution from the noise, background, and surrounding blended sources. The central problem
in accurate source extraction is to separate those contributions from the signal of the real sources.

The main idea of \textsl{getsources} is to analyze decompositions of original images (at each wavelength) across a wide range of
spatial scales separated by only a small amount (typically $\sim$ 3--5{\%}). Replacing originals with a set of strongly filtered
images brings several significant advantages. Each of the ``\emph{single scales}'' contains non-negligible signals from only a
relatively narrow range of spatial scales, mostly only from those sources (and the noise and background fluctuations) which have
sizes similar to the scale considered. In effect, this automatically filters out all contributions of the noise, background, or
overlapping sources on irrelevant (much smaller and larger) spatial scales. An immediate benefit is that such a filtering allows one
to manipulate entire single-scale images as a whole and use thresholding to separate sources from the background and noise in the
observed images (see Sect.~\ref{removing.noise.background} for details). Furthermore, considering the same spatial scales across a
wide range of wavelengths allows one to sum up single-scale images at all wavelengths in combined (wavelength-independent)
single-scale detection images and thus preserve the high-resolution information across all wavebands, minimizing the effect of
degrading resolutions. Besides providing a substantial ``super-resolution'' effect, this eliminates the need of matching multiple
catalogs obtained with different beams and reduces the matching and measurement errors.

\begin{figure}
\centering
\centerline{\resizebox{0.7\hsize}{!}{\includegraphics{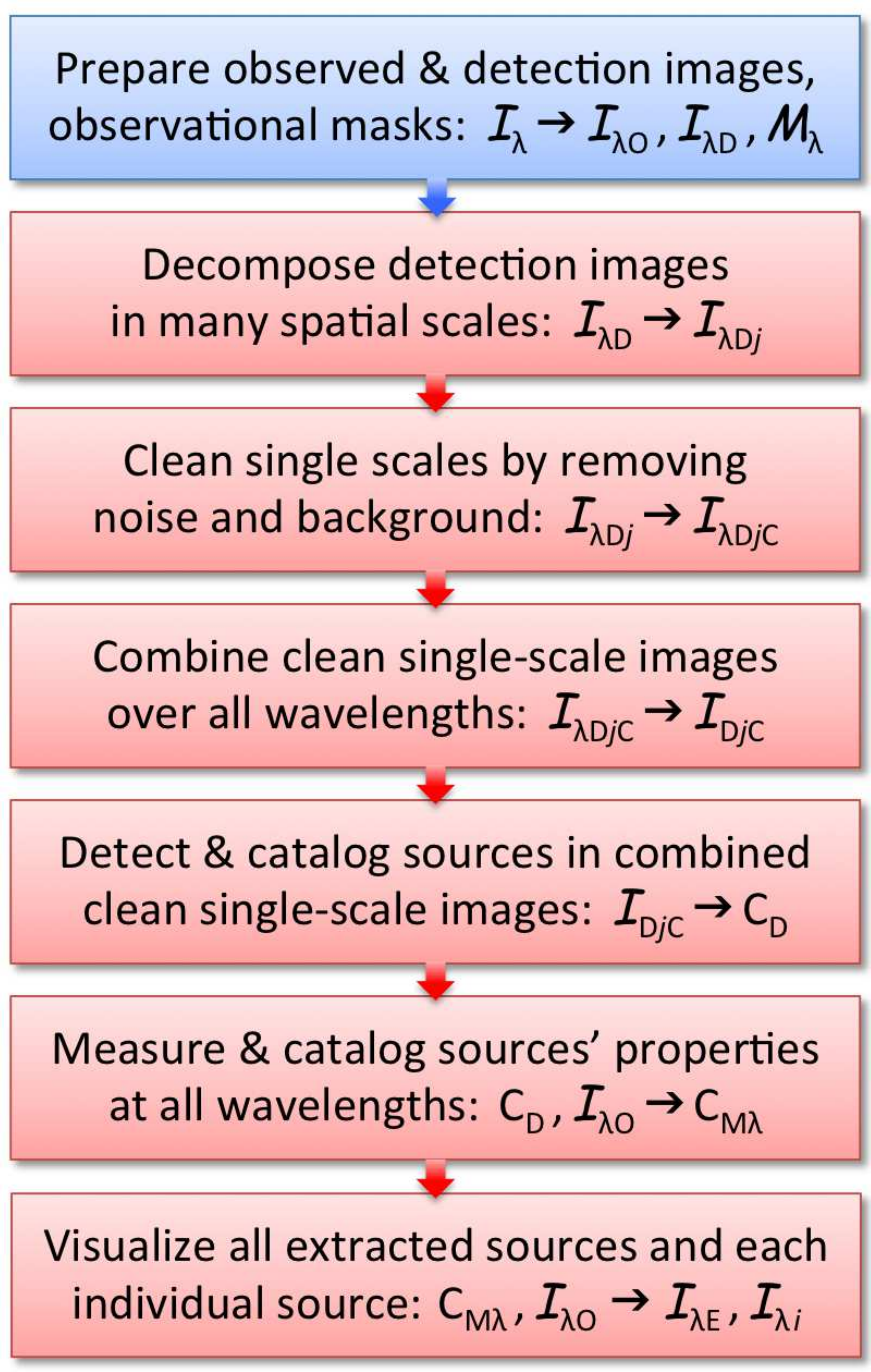}}} 
\caption{
Main processing blocks of the \textsl{getsources} algorithm described in
Sects.~\ref{preparing.images}--\ref{visualizing.extractions}.
}
\label{algorithm}
\end{figure}

The extraction method is represented by 7 processing blocks shown in Fig.~\ref{algorithm}; they will be described below in
Sects.~\ref{preparing.images}--\ref{visualizing.extractions}. In order to make a clear distinction between images and various other
parameters, the images are denoted throughout this paper by the capital calligraphic characters (e.g., $\mathcal{A}, \mathcal{B},
\mathcal{C}$; see Appendix~\ref{list.of.symbols} for a list of all symbols and definitions). The following subsections describe the
algorithm in full detail.

\subsection{Preparing observed and detection images}
\label{preparing.images}

The first step (Fig.~\ref{algorithm}) towards the source extraction is to convert the original images $\mathcal{I}_{\!\lambda}$ at
all wavelengths to the same grid. This means to transform them into the observed images $\mathcal{I}_{{\!\lambda}{\rm O}}$, all
with the same numbers of pixels, the same pixel size, aligned across wavelengths as accurately as possible (covering the same area
on the sky), the same reference pixel and its coordinates. In practice, this is done by resampling all images to the same pixel size
using the astronomical utility \textsl{SWarp} \citep{Bertin_etal2002}.

Note that the alignment of images must be carefully checked before extracting sources with \textsl{getsources}, because of its
multi-wavelength design. Images in all wavebands will be combined together in wavelength-independent detection images
(Sect.~\ref{combining.clean.single.scales}) that will only be good if the originals are aligned within one pixel; significant
misalignments of the images can create spurious sources\footnote{In practice, the most accurate approach to alignment is to use
images containing only small scales (see Sect.~\ref{decomposing.detection.images}), up to $\sim$twice the resolution in each
waveband, as they show misalignments most clearly. One should carefully choose which peaks to align, as the appearance of sources
may be affected by radiative transfer effects or by fluctuating backgrounds or by the close proximity to other sources.}.

\begin{figure*}
\centering
\centerline{\resizebox{0.33\hsize}{!}{\includegraphics{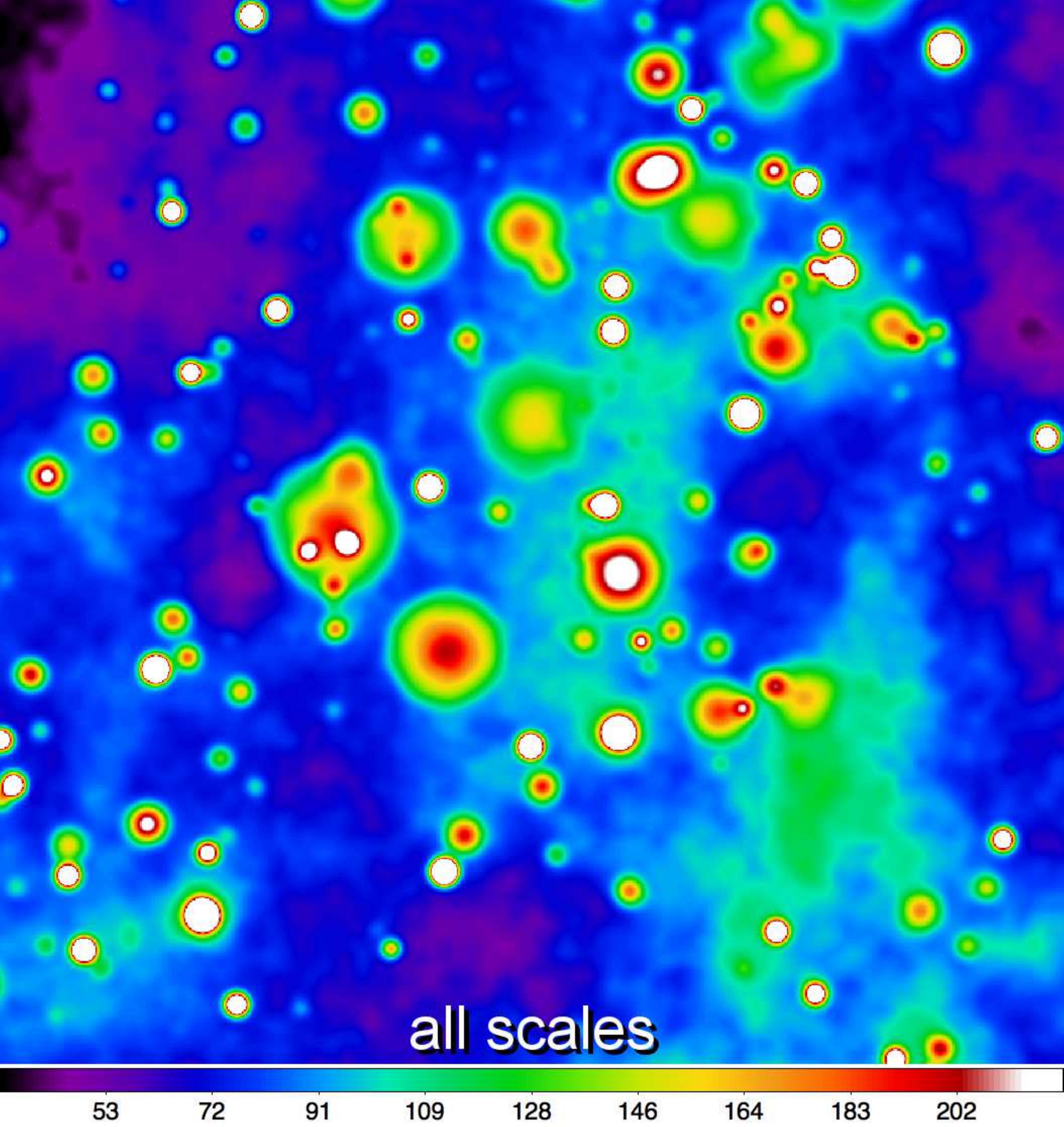}}
            \resizebox{0.33\hsize}{!}{\includegraphics{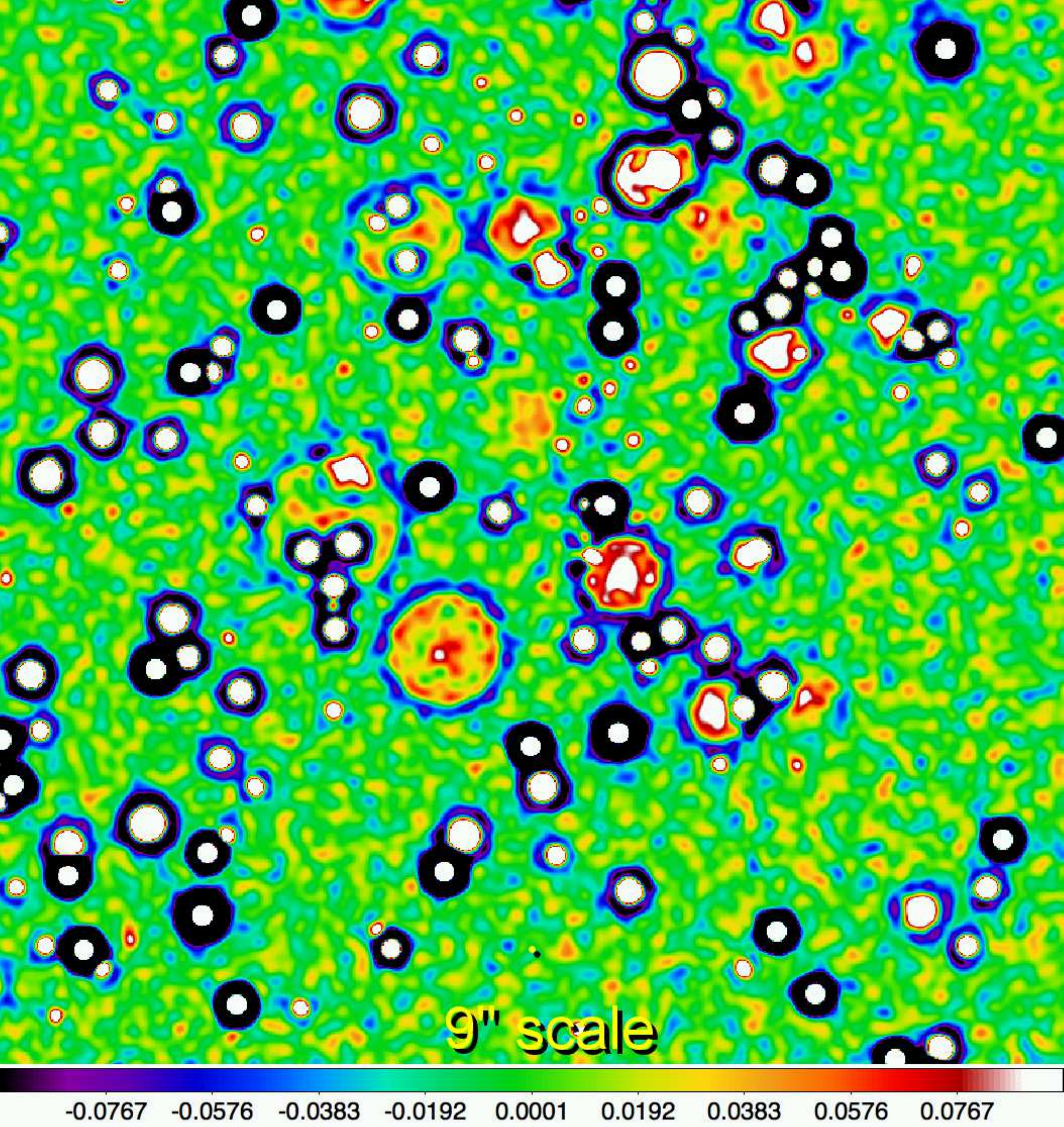}}
            \resizebox{0.33\hsize}{!}{\includegraphics{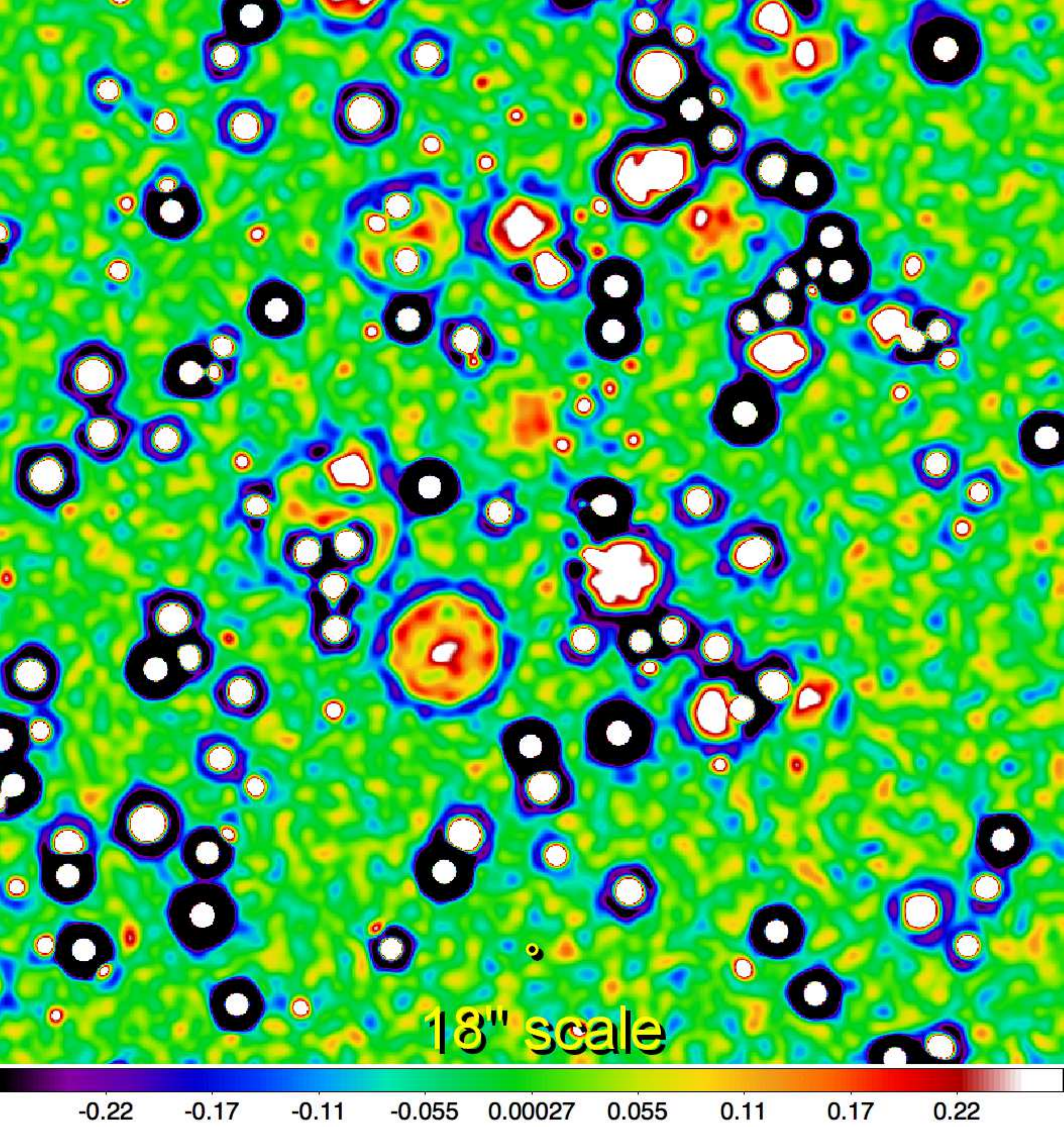}}}
\centerline{\resizebox{0.33\hsize}{!}{\includegraphics{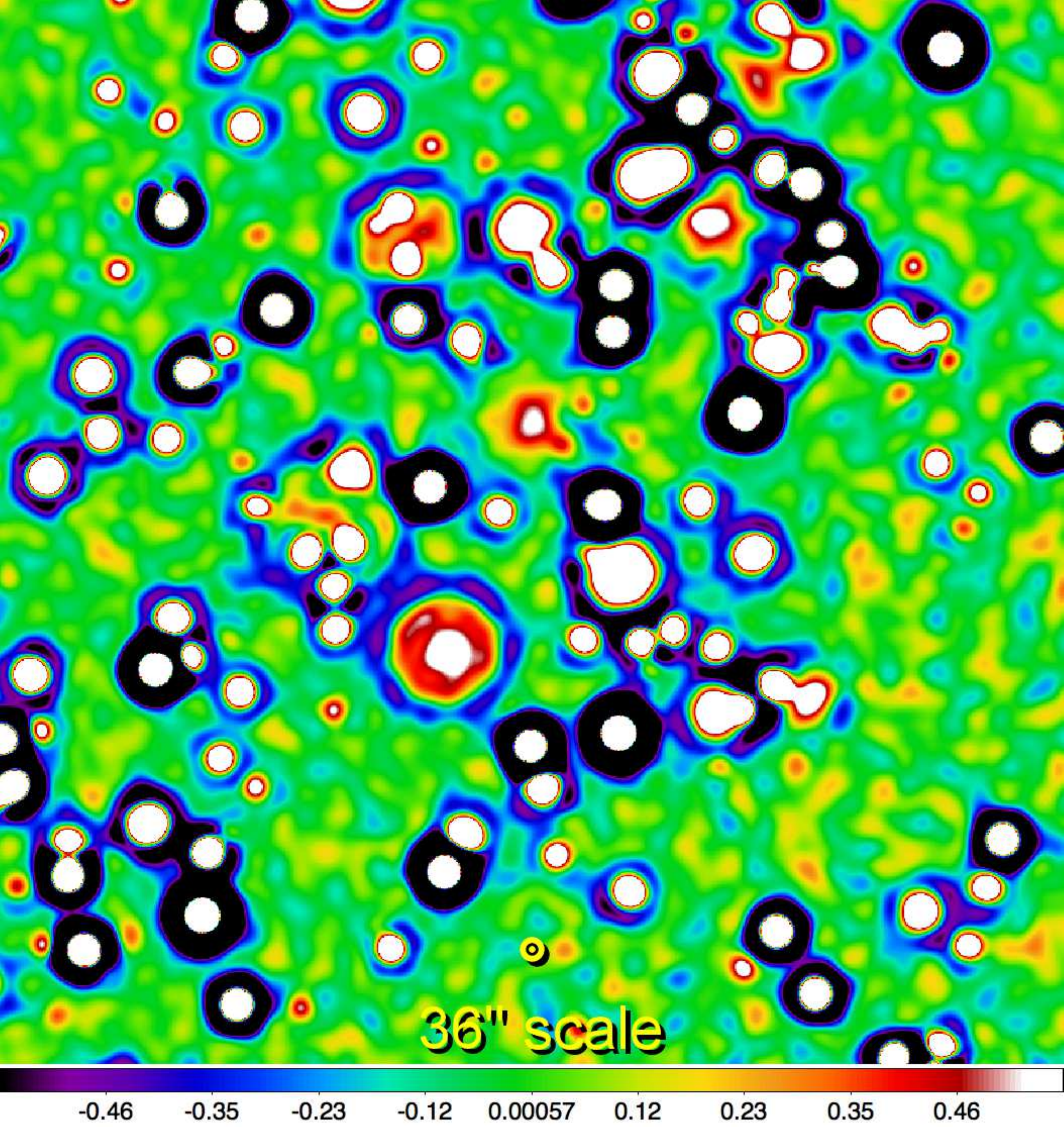}}
            \resizebox{0.33\hsize}{!}{\includegraphics{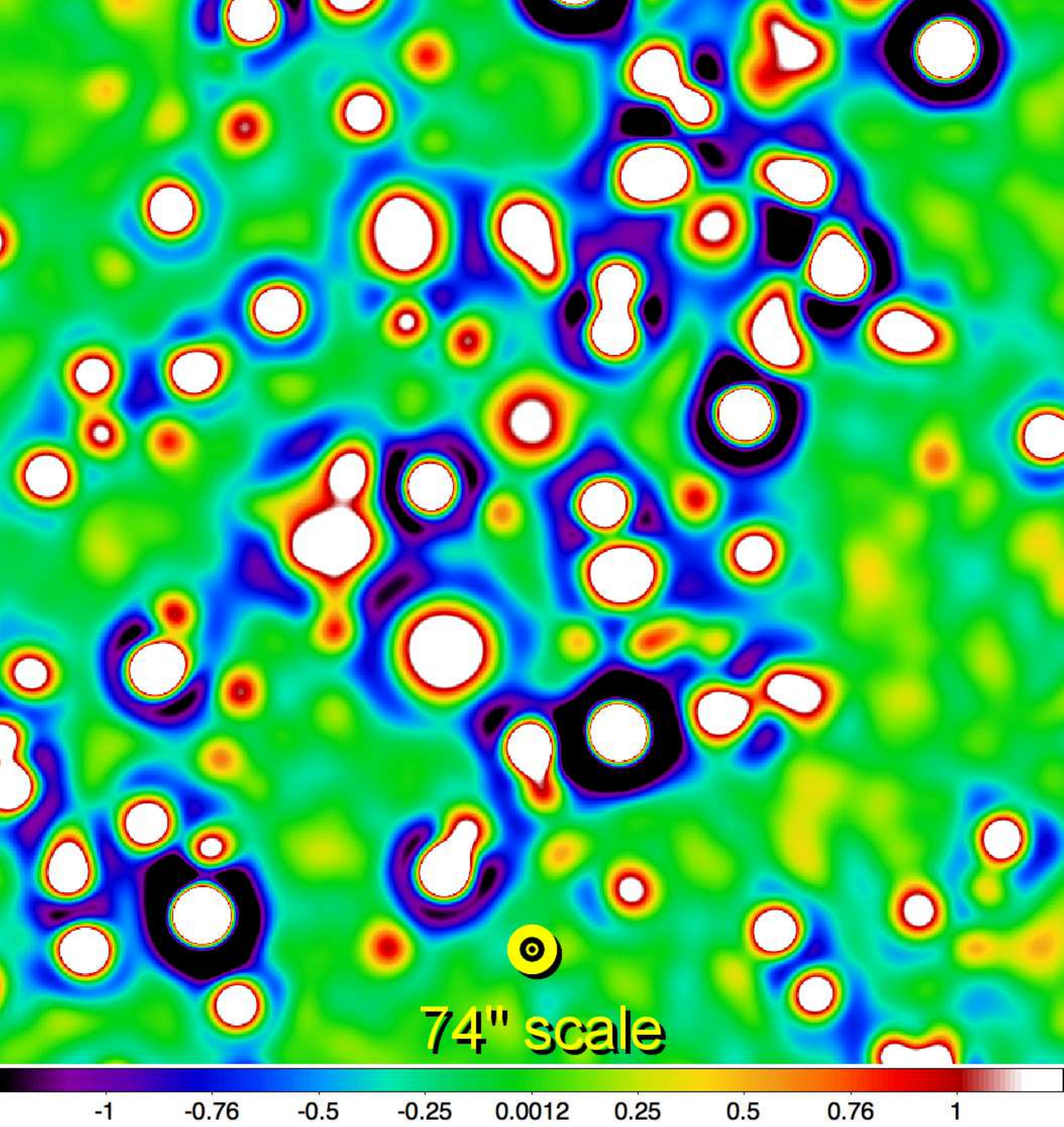}}
            \resizebox{0.33\hsize}{!}{\includegraphics{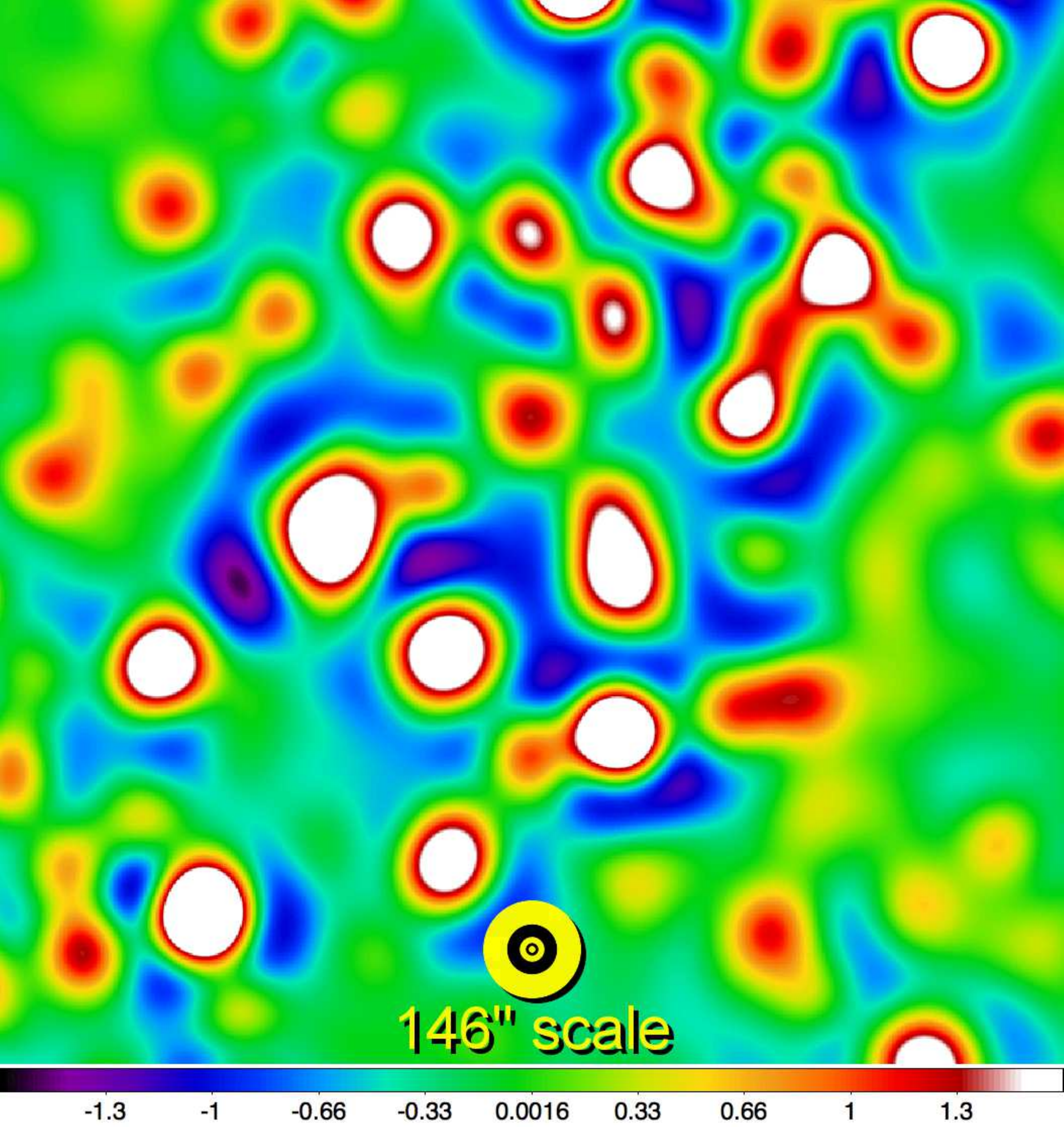}}}
\caption{
Single-scale decomposition (Sect.~\ref{decomposing.detection.images}). The central 0{\fdg}44$\times$0{\fdg}44 sub-field of the
detection image $\mathcal{I}_{{\!\lambda}{\rm D}}$ of a 1{\degr}$\times$1{\degr} simulated star-forming region at 350\,{${\mu}$m}
(\emph{upper left}). Its single-scale images $\mathcal{I}_{{\!\lambda}{\rm D}{j}}$ are shown at the scales indicated (\emph{left to
right}, \emph{top to bottom}) for $j{\,=\,}17, 30, 43, 57, 70$, $N_{\rm S}{\,=\,}$99, $f_{\rm S}{\,=\,}1.053$,
$S_{1}{\,=\,}$4{\arcsec}, $S_{N_{\rm S}}{\,=\,}$660{\arcsec} (see Eq.~\ref{successive.unsharp.masking}). The image dimensions are
1800$\times$1800 pixels and the pixel size $\Delta{\,=\,}$2{\arcsec}. The scales were selected to be separated by a factor of 2 to
illustrate the spatial decomposition. The scale sizes $S_{\!j}$ are visualized by the yellow-black circles and annotated at the
bottom of the panels. For better visibility, the values displayed in the panels are somewhat limited in range; the color coding is a
linear function of intensity in MJy/sr.
} 
\label{single.scales}
\end{figure*}

The observed images $\mathcal{I}_{{\!\lambda}{\rm O}}$ are only used to measure properties of detected sources, at the end of an
extraction (Fig.~\ref{algorithm}, Sect.~\ref{measuring.cataloging}). Most of the processing in the algorithm is done on the
detection images $\mathcal{I}_{{\!\lambda}{\rm D}}$, which in the simplest case may be the same as $\mathcal{I}_{{\!\lambda}{\rm
O}}$, but in general can significantly differ from the latter. Any transformations of the observed images that have a potential to
improve detection quality (such as completeness, reliability) can be used as $\mathcal{I}_{{\!\lambda}{\rm D}}$; this can be
convolution, multiplication by weight images, subtraction of baseline images, etc. For example, one may want to sacrifice a little
bit of the nominal angular resolution in order to reduce the unphysical pixel-to-pixel noise present in the images (on scales
smaller than the observational beam size $O_{\lambda}$) using convolution $\mathcal{I}_{{\!\lambda}{\rm D}} = \mathcal{G}_{\lambda}
* \mathcal{I}_{{\!\lambda}{\rm O}}$, where $\mathcal{G}_{\lambda}$ is the smoothing Gaussian with full width at half maximum (FWHM)
chosen to slightly degrade (by $\sim$5{\%}) the image resolution $O_{\lambda}$. This suppresses unphysical noise and small-scale
artifacts in $\mathcal{I}_{{\!\lambda}{\rm D}}$ that otherwise may become enhanced in the smallest-scale decomposed images. Being
the default way of creating detection images in \textsl{getsources}, such smoothing is not required and may be skipped, if not
deemed beneficial. Another example of a detection image that may be useful in some applications is a column density map produced by
pixel-to-pixel SED fitting of the observed images.

\begin{figure*}
\centering
\centerline{\resizebox{0.33\hsize}{!}{\includegraphics{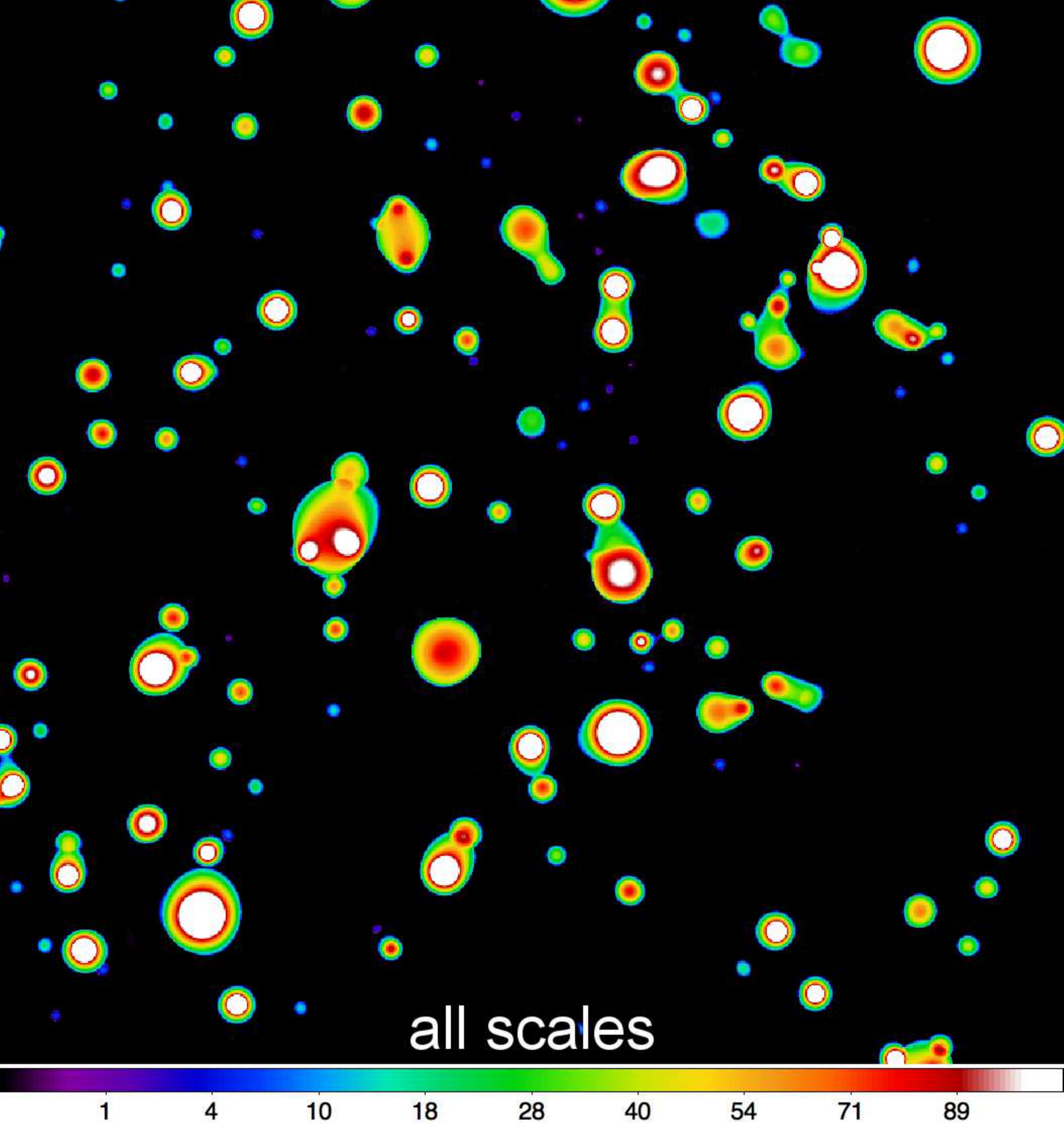}}
            \resizebox{0.33\hsize}{!}{\includegraphics{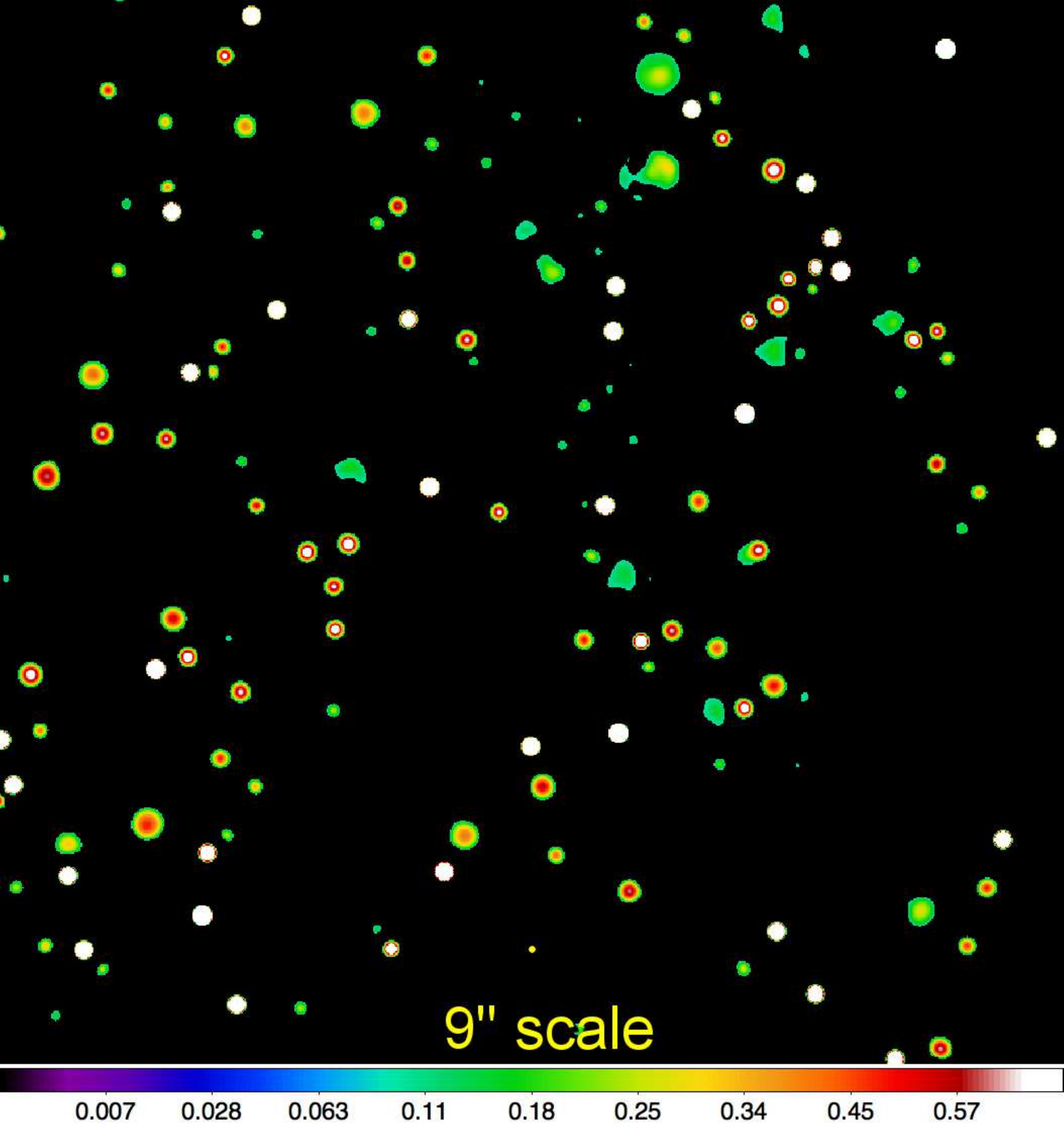}}
            \resizebox{0.33\hsize}{!}{\includegraphics{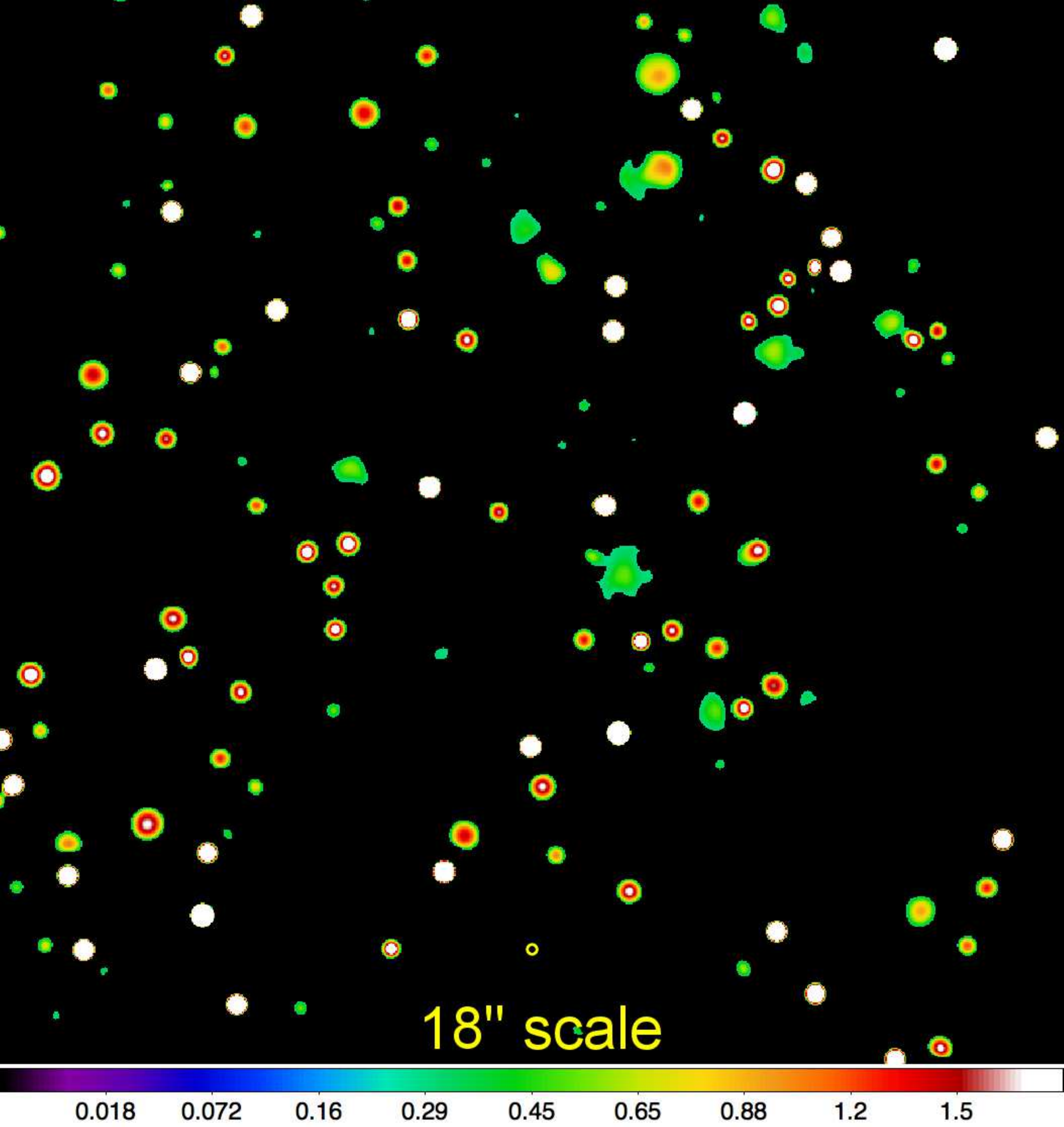}}}
\centerline{\resizebox{0.33\hsize}{!}{\includegraphics{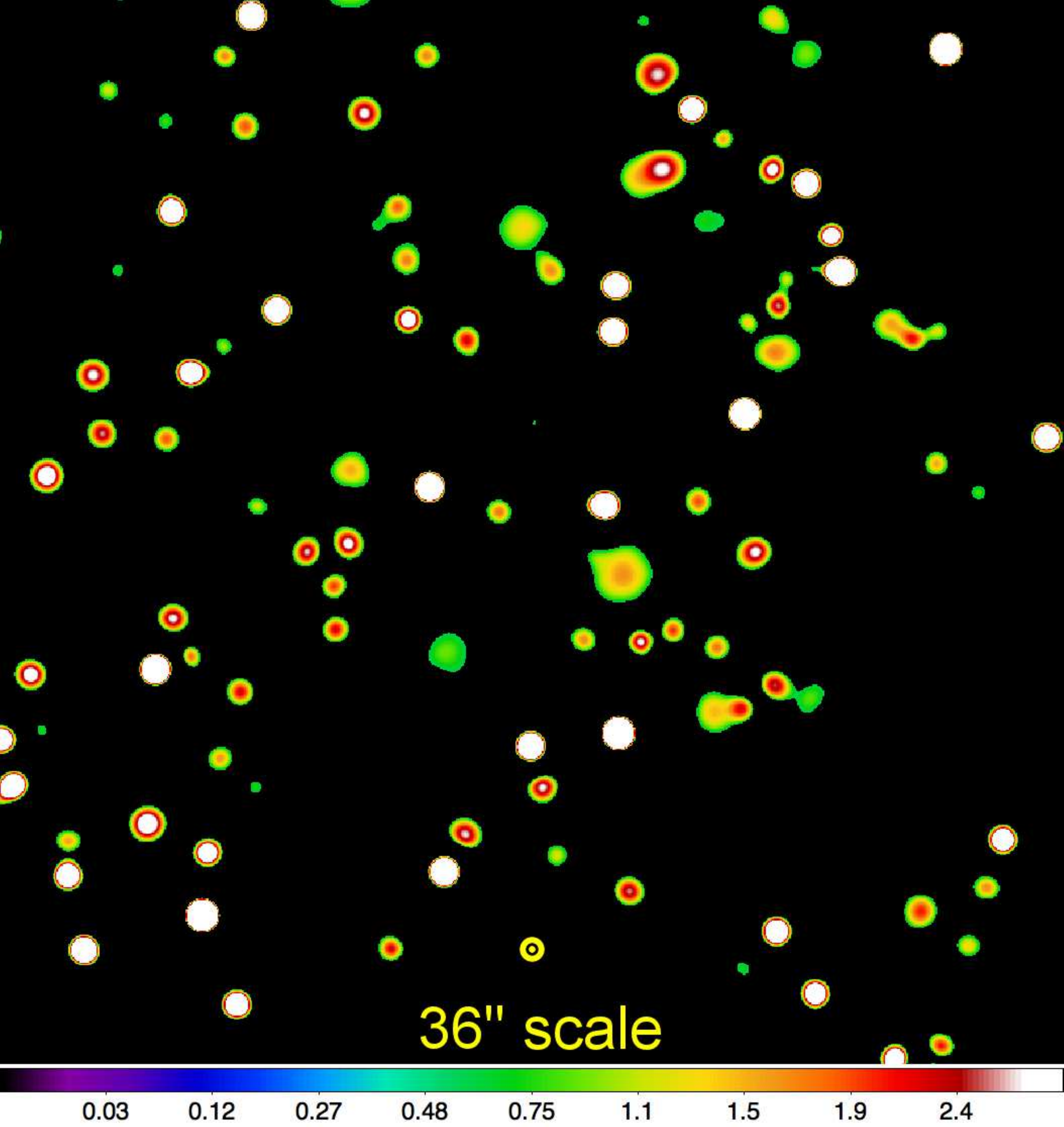}}
            \resizebox{0.33\hsize}{!}{\includegraphics{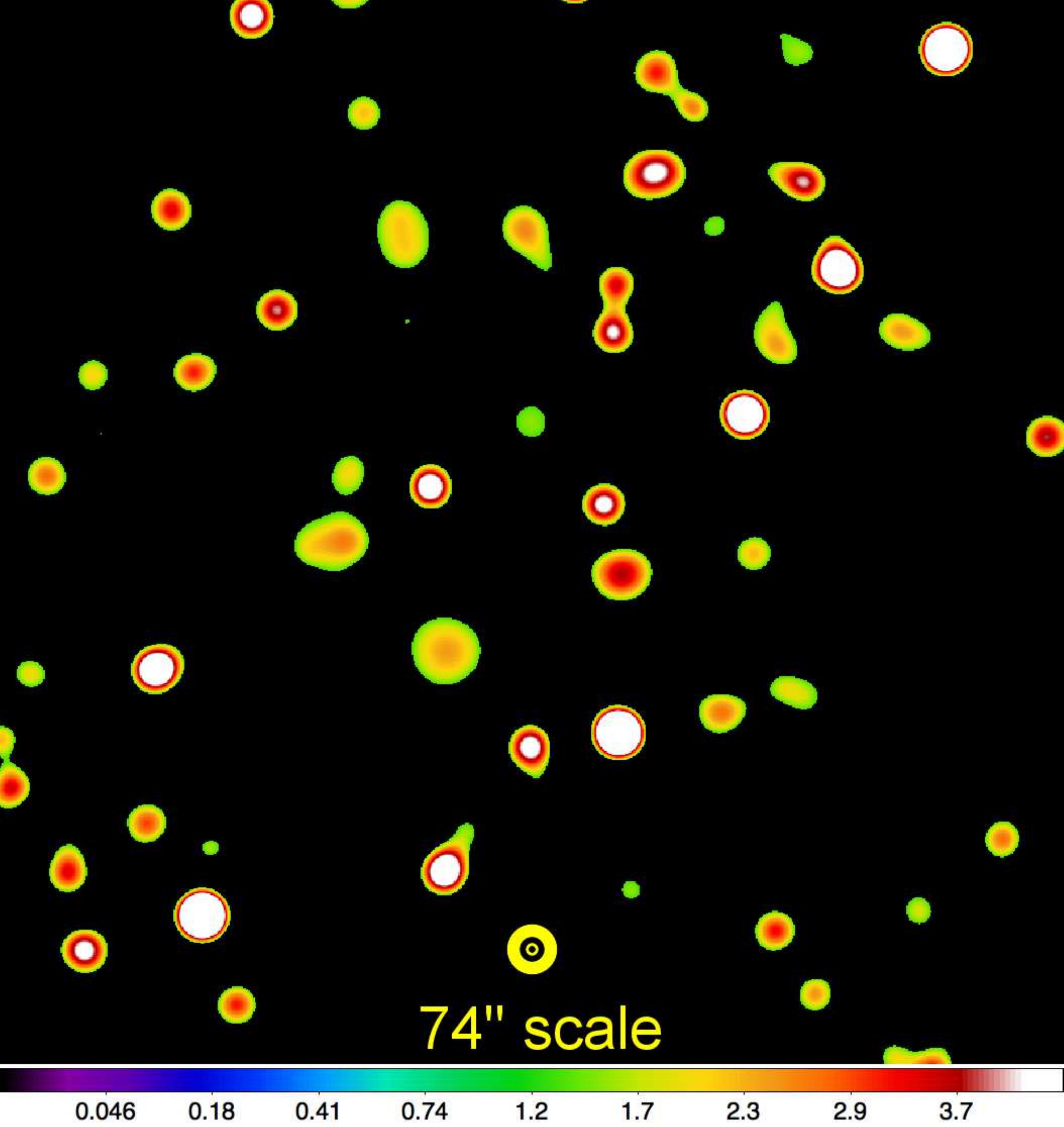}}
            \resizebox{0.33\hsize}{!}{\includegraphics{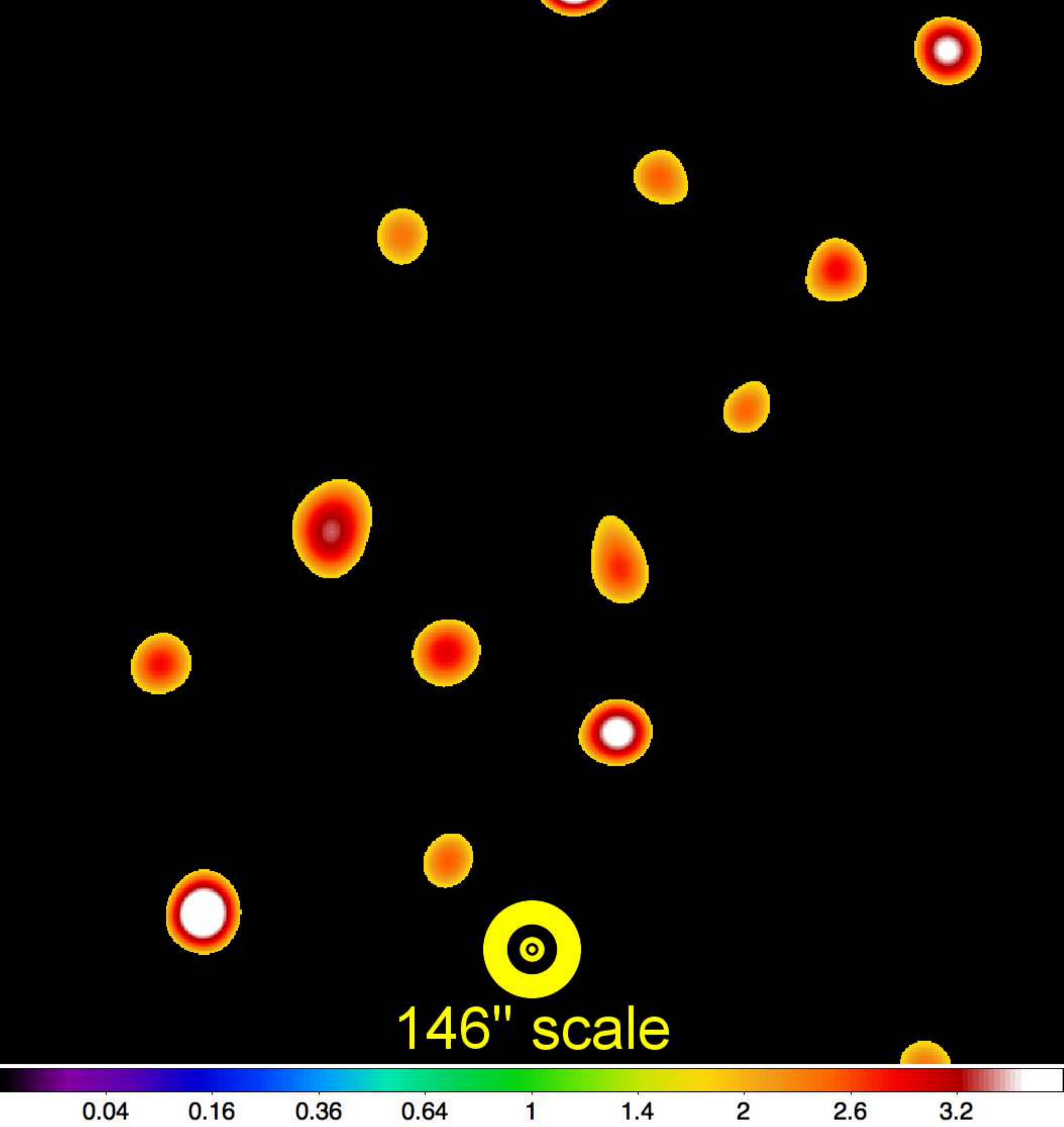}}}
\caption{
Single-scale removal of noise and background (Sect.~\ref{removing.noise.background}). The field of Fig.\,\ref{single.scales} is
shown as the full clean image $\mathcal{I}_{{\!\lambda}{\rm{D\,C}}}$ at 350\,{${\mu}$m} (\emph{upper left}) that accumulates clean
images over all scales. The same set of spatial scales is displayed in the single-scale images $\mathcal{I}_{{\!\lambda}{\rm
D}{j}{\,\rm C}}$ (\emph{left to right}, \emph{top to bottom}), cleaned of noise and background with an iterative procedure described
in Sect.~\ref{removing.noise.background}. All intensity peaks visible in the scales belong to the sources; most of the noise and
background fluctuations (visible in Fig.\,\ref{single.scales}) have been removed. The scale sizes $S_{\!j}$ are visualized by the
yellow-black circles and annotated at the bottom of the panels. For better visibility, the values displayed in the panels are
somewhat limited in range; the color coding is a function of the square root of intensity in MJy/sr.
} 
\label{clean.single.scales}
\end{figure*}

The last part of the preparation is creating the observational masks $\mathcal{M}_{\lambda}$. Those are images with the pixel
values of either 1 or 0, defining the areas in the original images that we are interested in. In practice, some areas of the
observed rectangular images may not have been covered, some other areas may contain high noise or artifacts. The mask images are
used by \textsl{getsources} to exclude from processing any area of $\mathcal{I}_{{\!\lambda}{\rm D}}$ in which the mask has zero
values; in the simplest ideal case, $\mathcal{M}_{\lambda}$ has values of 1 in all pixels. Very noisy areas have the potential to
affect the cleaning and detection algorithms described below and every effort must be made to exclude such areas using carefully
prepared observational masks.

\subsection{Decomposing detection images in spatial scales}
\label{decomposing.detection.images}

The spatial decomposition is done by convolving the original images with circular Gaussians and subtracting them from one another 
(we call this procedure \emph{successive unsharp masking}): 
\begin{equation}
\mathcal{I}_{{\!\lambda}{\rm D}{j}} = \mathcal{G}_{j-1} * \mathcal{I}_{{\!\lambda}{\rm D}} - 
\mathcal{G}_{j} * \mathcal{I}_{{\!\lambda}{\rm D}} \,\,(j{\,=\,}1, 2,\dots, N_{\rm S}),
\label{successive.unsharp.masking}
\end{equation}
where $\mathcal{I}_{{\!\lambda}{\rm D}}$ is the detection image (Sect.~\ref{preparing.images}) at wavelength $\lambda$,
$\mathcal{I}_{{\!\lambda}{\rm D}{j}}$ are its ``single-scale'' decompositions, $\mathcal{G}_{j}$ are the smoothing Gaussian beams
($\mathcal{G}_0$ is a two-dimensional delta-function). The latter have FWHM sizes $S_{\!j}{\,=\,}f_{\rm S}\,S_{\!j-1}$ in the range
$2\,\Delta{\,\la\,}S_{\!j} \la S_{\rm max}$, where $\Delta$ is the pixel size, $f_{\rm S}{\,>\,}1$ is the scale factor, $S_{\rm
max}$ is the maximum spatial scale considered, and the number of scales $N_{\rm S}$ depends on the value of $f_{\rm S}$ (typically
${\approx}1.05$) and $S_{\rm max}$. We adopt $S_{\rm max}{\,=\,}3\,\max\,\{A^{\rm max}_{\lambda}\}$, where $A^{\rm max}_{\lambda}$
is the maximum FWHM sizes of sources to be extracted and an upper limit for $S_{\rm max}$ is the image size\footnote{The
wavelength-dependent maximum sizes of sources are the only user-definable parameters in \textsl{getsources}. The actual maximum
sizes depend on the observed images and the specific interest of a researcher. Before extracting sources, one has to obtain
reasonable guesses of the maximum source sizes from the images and specify them in the configuration file (the parameter 
$A^{\rm max}_{\lambda}$ defaults to $6\,O_{\lambda}$).}.

Equation~\ref{successive.unsharp.masking} implicitly assumes that the convolved images are properly rescaled to conserve their total
flux; therefore, the original image can be recovered by summing up all scales:
\begin{equation} 
\mathcal{I}_{{\!\lambda}{\rm D}} = \sum\limits_{j=1}^{N_{\rm S}}\mathcal{I}_{{\!\lambda}{\rm D}{j}} +
\mathcal{G}_{N_{\rm S}} * \mathcal{I}_{{\!\lambda}{\rm D}}. 
\label{recovered.original.image}
\end{equation}
Before convolution, the input images $\mathcal{I}_{{\!\lambda}{\rm D}}$ are expanded from the edges of the areas covered by the
observational masks $\mathcal{M}_{\lambda}$ towards the image edges and the entire images are further expanded on all sides by a
large enough number of pixels ($3\,S_{\!j}/\Delta$) in order to avoid undesirable border effects. Both expansions are performed
using the pixel values at the edges of the masks and images, respectively; after convolution, the images are reduced back to their
original size.

Small values of $f_{\rm S}$ ensure the best spatial resolution of the single scales, just like fine mesh sizes always better resolve
structures in numerical methods. For practical reasons, the minimum value of $f_{\rm S}$ is 1.03 and the maximum value of $N_{\rm
S}$ is 99. For large $f_{\rm S}$, the single scales actually contain mixture of a wide range of scales, and faint small-scale
structures become completely diluted by the contribution of irrelevant scales. Again, this is very similar to any finite-difference
numerical methods, where structures smaller than a few grid zones disappear within large structures resolved by coarse
grids\footnote{Usually best results are obtained with $N_{\rm S}$ and $f_{\rm S}$ close to their limiting values. For $f_{\rm
S}{\,=\,}2$, the decomposition of Eq.~\ref{successive.unsharp.masking} is identical to that produced by the multi-resolution code
\textsl{mr{\_}transform} \citep{StarckMurtagh2006} with its default linear wavelet transform (``{\`a} trous'' algorithm).}.

\begin{figure*}
\centering
\centerline{\resizebox{0.33\hsize}{!}{\includegraphics{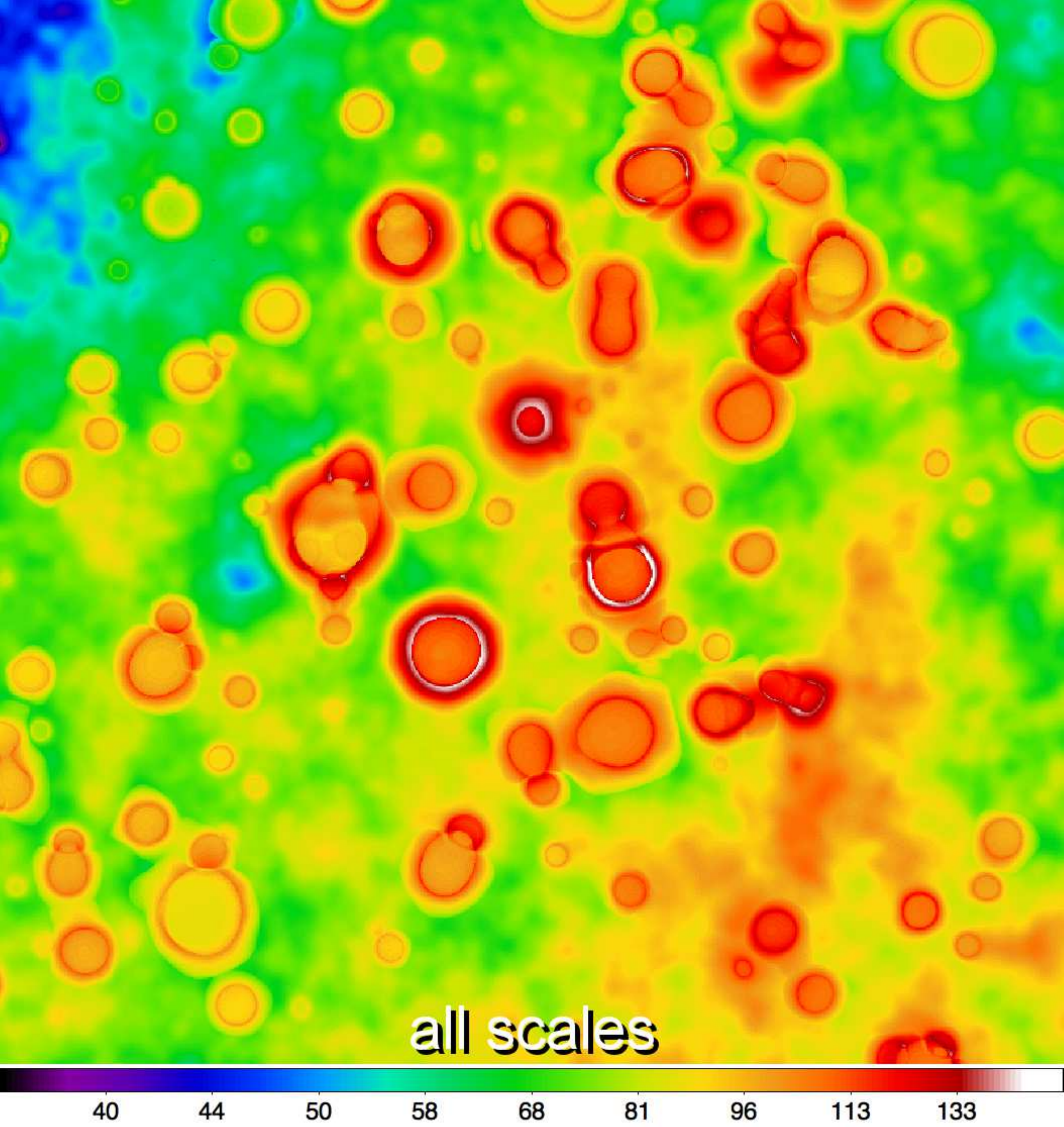}}
            \resizebox{0.33\hsize}{!}{\includegraphics{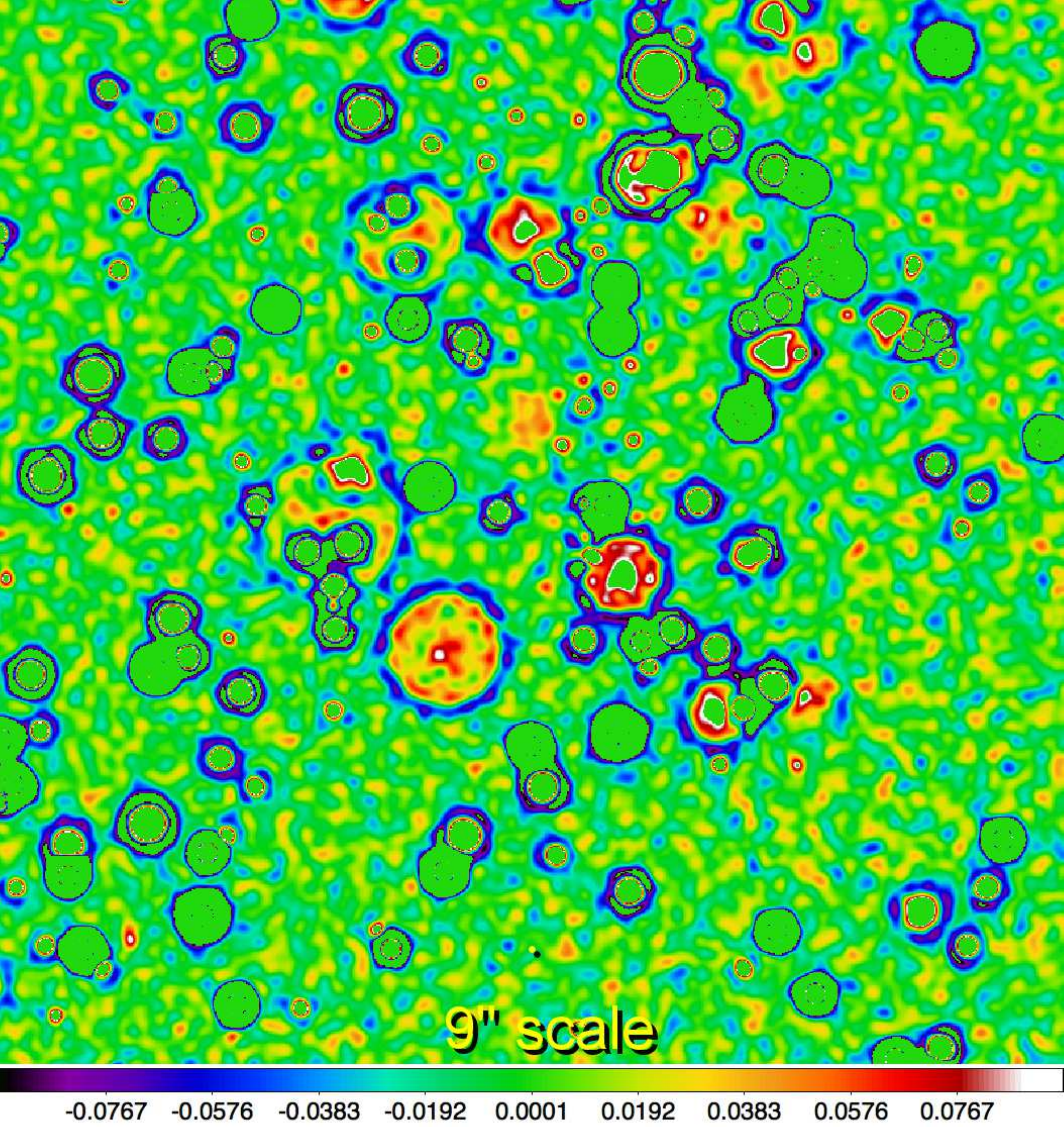}}
            \resizebox{0.33\hsize}{!}{\includegraphics{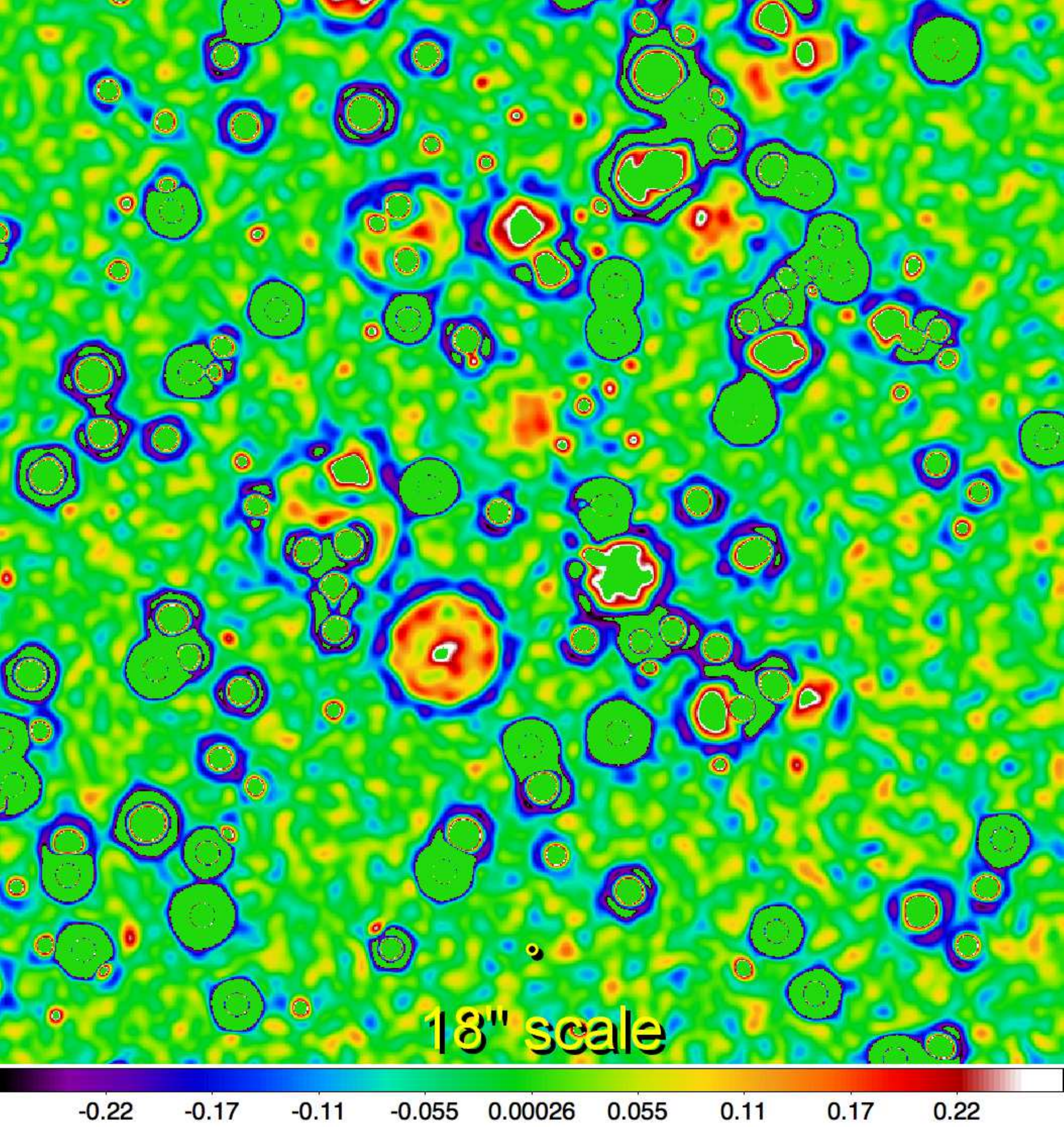}}}
\centerline{\resizebox{0.33\hsize}{!}{\includegraphics{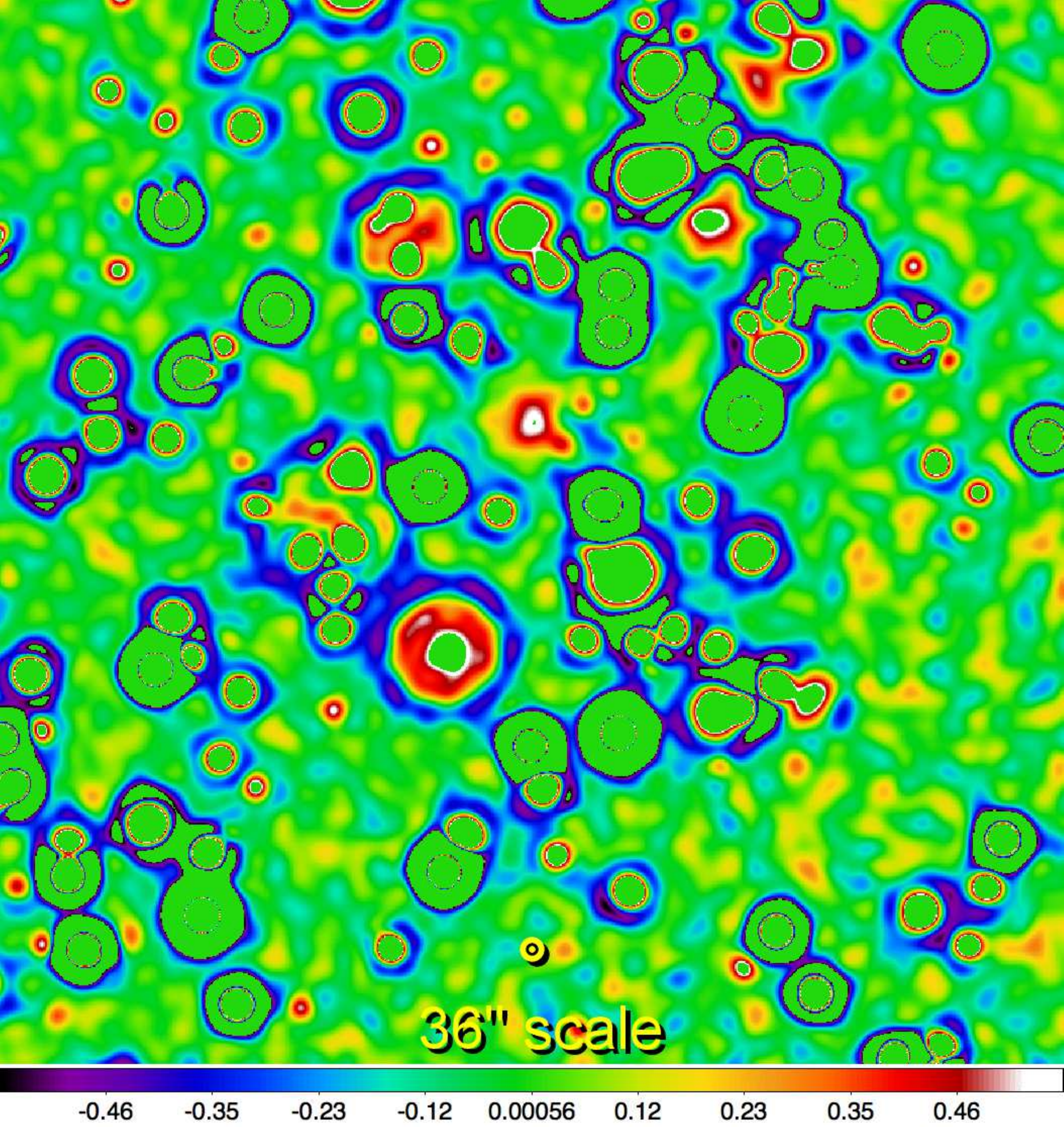}}
            \resizebox{0.33\hsize}{!}{\includegraphics{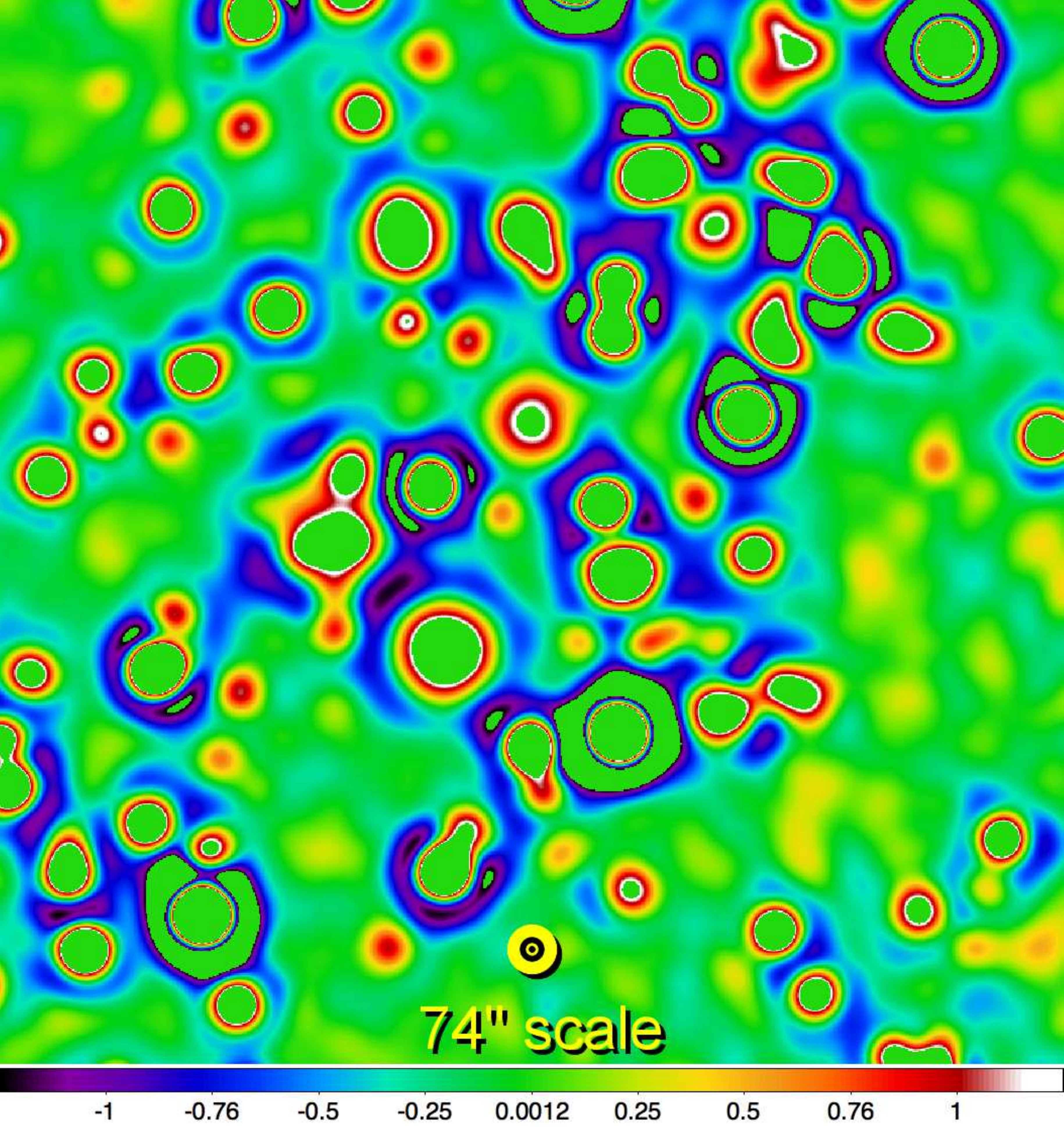}}
            \resizebox{0.33\hsize}{!}{\includegraphics{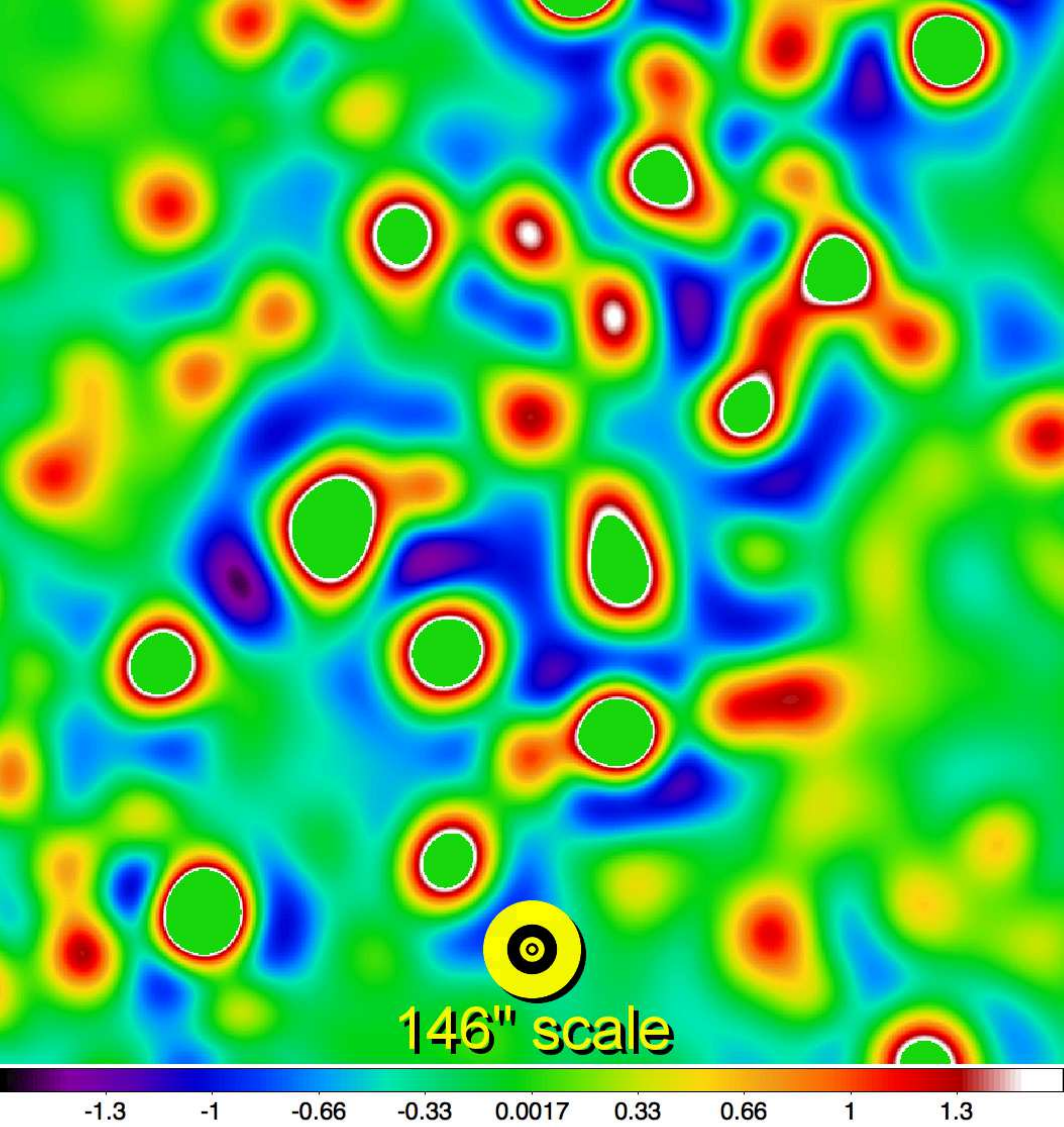}}}
\caption{
Single-scale cleaning residuals (Sect.~\ref{removing.noise.background}). The field of Fig.~\ref{single.scales} is shown as the
reconstructed image $\mathcal{I}_{{\!\lambda}{\rm{D\,R}}}$ of the residuals at 350\,{${\mu}$m} (\emph{upper left}) that accumulates
cleaning residuals from all scales. The same set of spatial scales is displayed in the single-scale images of the residuals
$\mathcal{I}_{{{\!\lambda}{\rm D}}{j}{\,\rm R}}$ (\emph{left to right}, \emph{top to bottom}). The cleaning procedure left no
significant intensity peaks of the simulated objects in the residuals, only the noise- and background-dominated pixels (cf.
Fig.\,\ref{combined.clean.single.scales}). The scale sizes $S_{\!j}$ are visualized by the yellow-black circles and annotated at the
bottom of the panels. The color coding is a linear function of intensity in MJy/sr.
} 
\label{single.scale.residuals}
\end{figure*}

To illustrate the spatial decomposition and all other processing steps of \textsl{getsources}, we shall use images of a simulated
star-forming region that we constructed well before the launch of \emph{Herschel} in order to have a reasonably realistic model for
testing various aspects of our future observational program (the source extraction methods, instruments simulators, etc.); we refer
to Appendix~\ref{simulated.images} for more details. The spatial decomposition of images using convolution has a clear
interpretation in terms of the Fourier transform. Interested readers are referred to Appendix~\ref{fourier.domain}, where we present
the Fourier amplitudes for the individual components of the simulated sky and for a few selected scales of the decomposed images, as
well as examples from the actual \emph{Herschel} observations\footnote{Except the spatial decomposition step, where convolutions
are done using a fast Fourier transform algorithm, our method has been designed to operate in the image space, which is a natural
and intuitive way of source extraction.}.

A sub-field of the simulated region at 350\,{${\mu}$m} (Fig.\,\ref{single.scales}, \emph{upper left}) clearly shows all ingredients:
the cirrus background, protostars (bright compact peaks), and fainter starless cores of various sizes, from completely unresolved to
very extended. Many sources vanish into the background and also many sources are blended with others. However, the single-scale
decompositions filter out emission at all irrelevant scales and display the sources with a much higher contrast than the full image
does. The decomposition of Eq.~\ref{successive.unsharp.masking} thus naturally selects sources of specific sizes, which become best
visible in the images with matching scales. The negative rings around bright sources are the direct consequence of the successive
unsharp masking, the subtraction of an image convolution with a larger smoothing beam from an image convolved with a smaller beam.
All peaks at the first four scales shown in Fig.\,\ref{single.scales} are identifiable with the corresponding peaks in the full
image. However, the situation becomes more problematic as we proceed to larger scales, such as that displayed in the last panel of
Fig.\,\ref{single.scales}. For even larger scales, up to the entire image size, intensities from sources of all sizes become so
heavily mixed and diluted in the large smoothing beams that one cannot disentangle the individual structures anymore.

\begin{figure*}
\centering
\centerline{\resizebox{0.33\hsize}{!}{\includegraphics{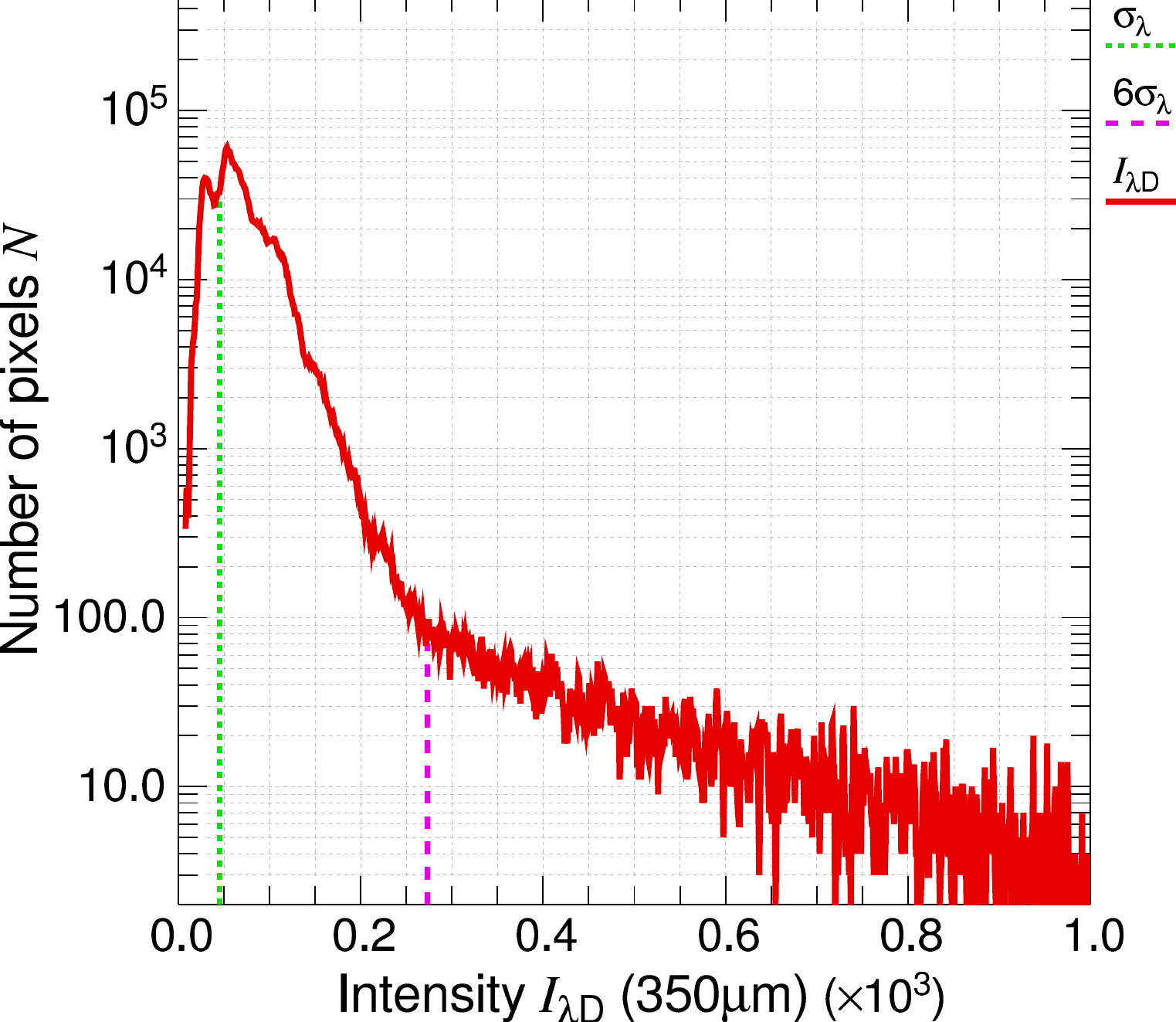}}
            \resizebox{0.33\hsize}{!}{\includegraphics{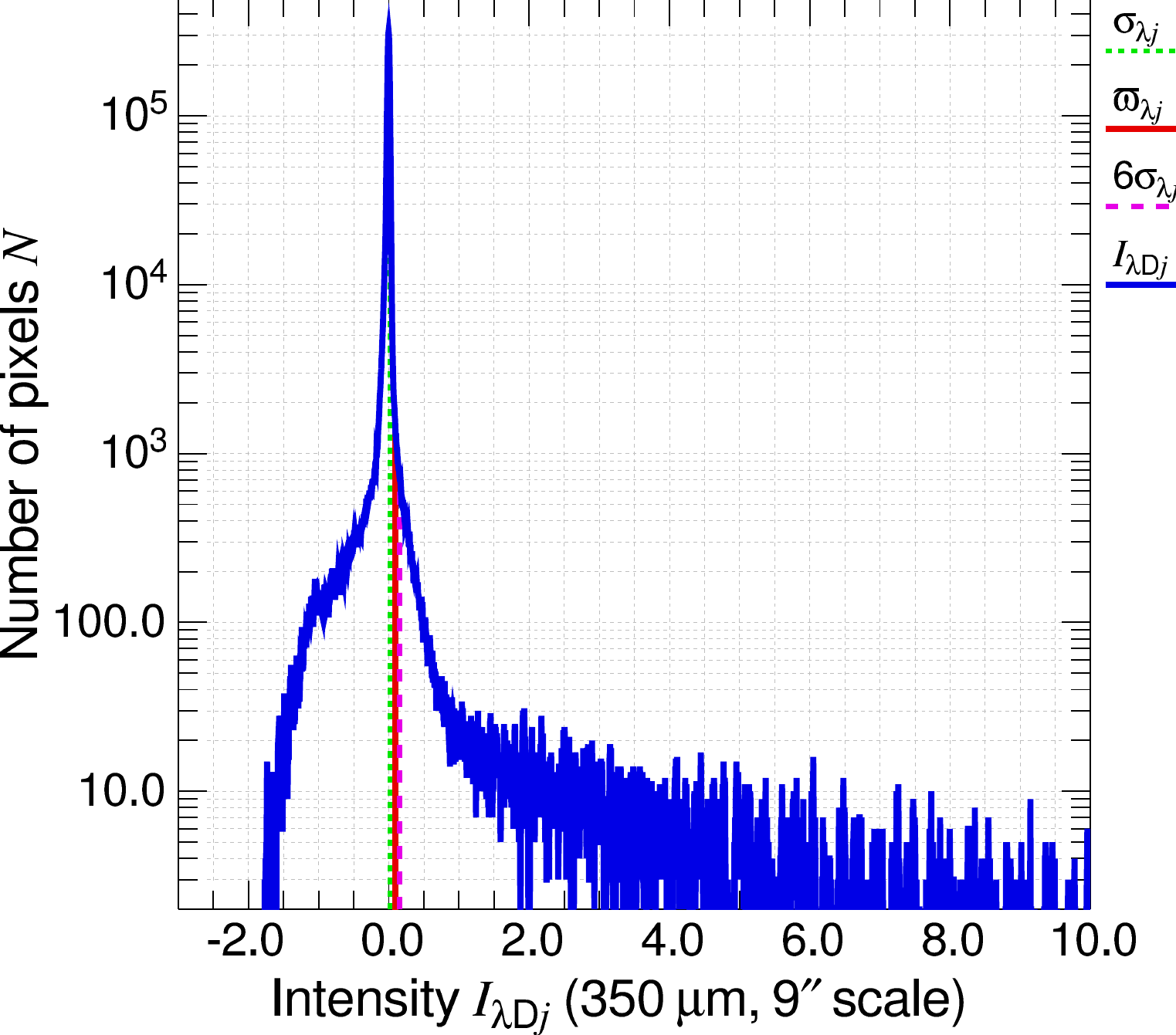}}
            \resizebox{0.33\hsize}{!}{\includegraphics{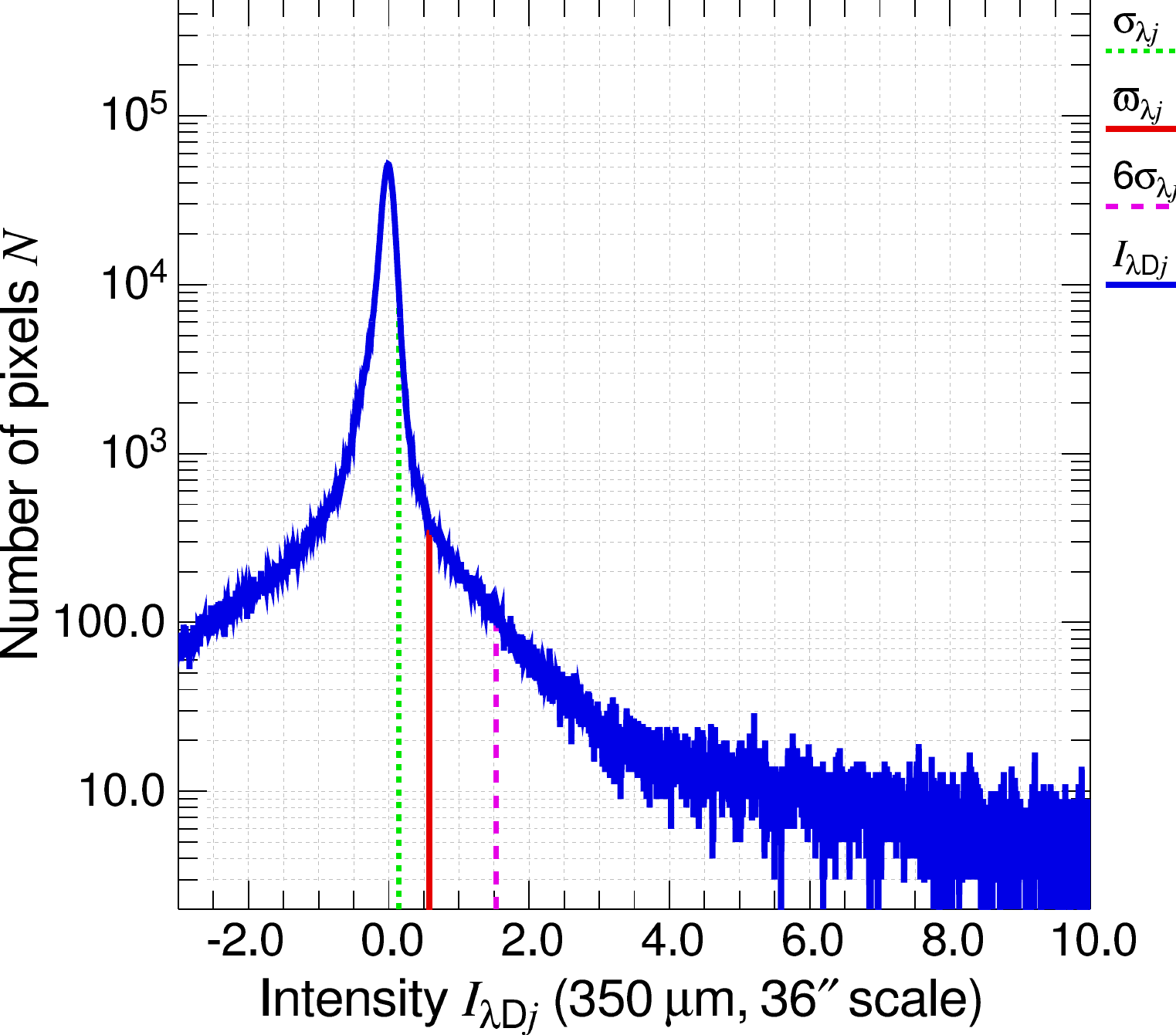}}}
\vspace{1mm}
\centerline{\resizebox{0.33\hsize}{!}{\includegraphics{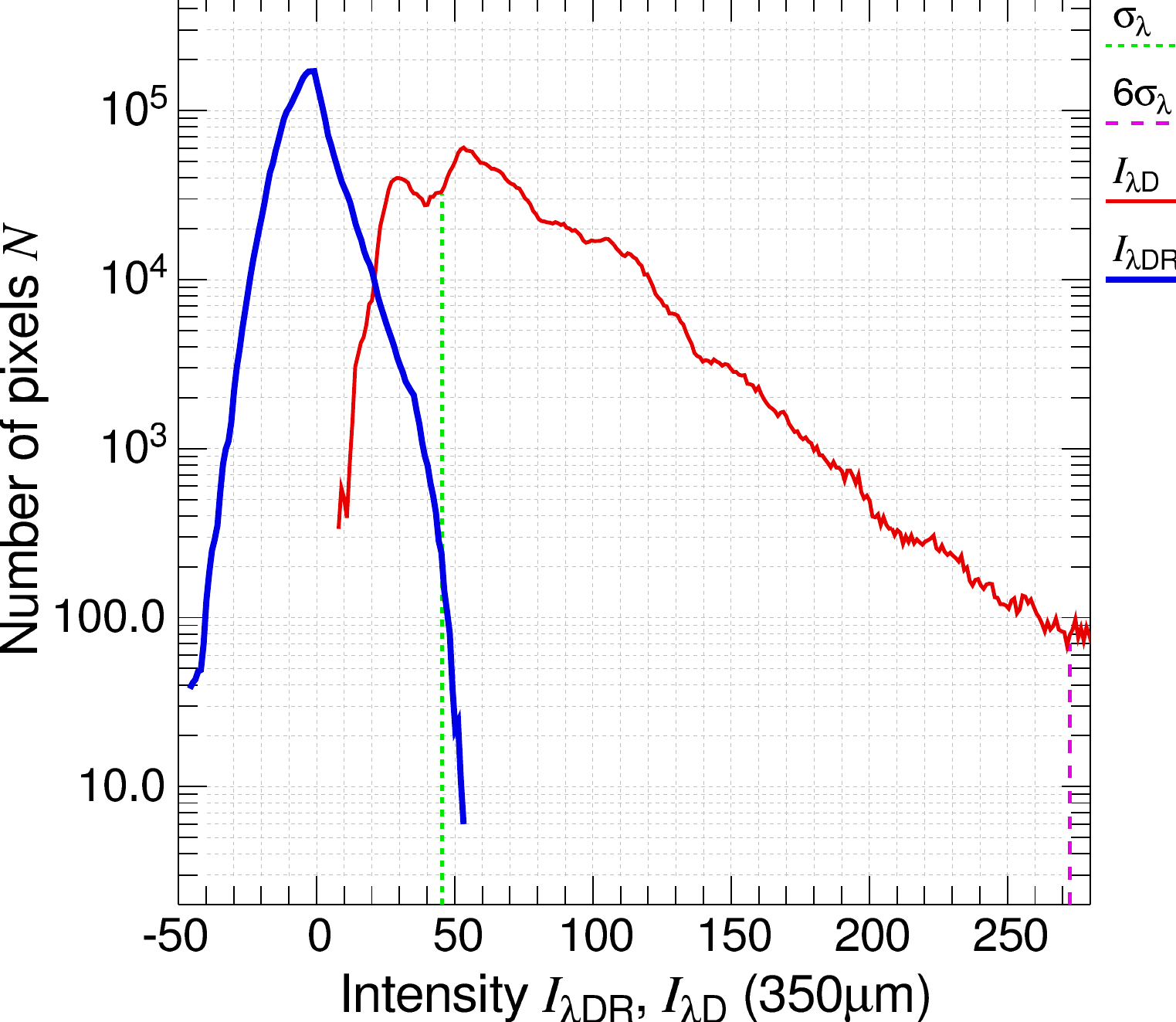}}
            \resizebox{0.33\hsize}{!}{\includegraphics{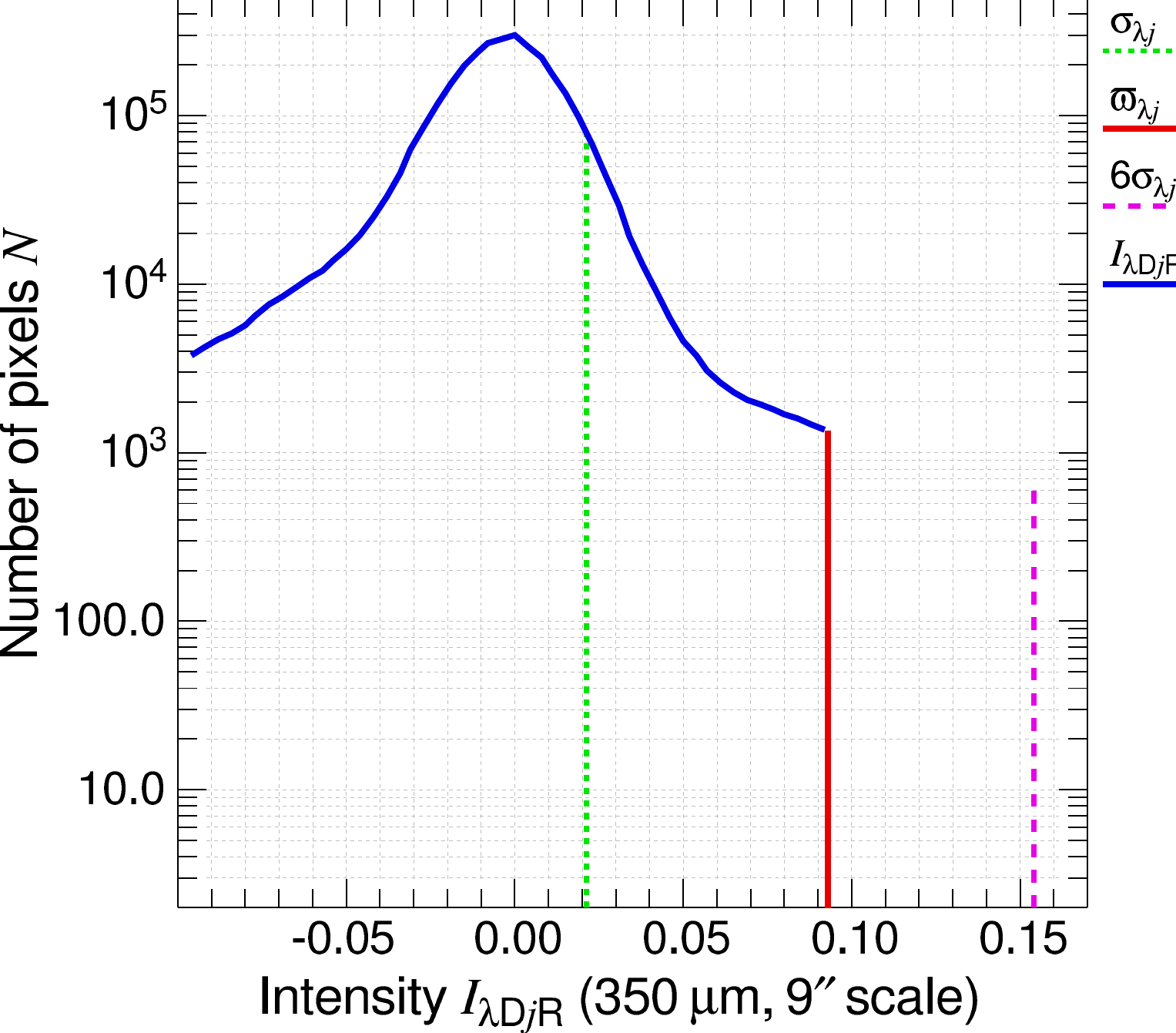}}
            \resizebox{0.33\hsize}{!}{\includegraphics{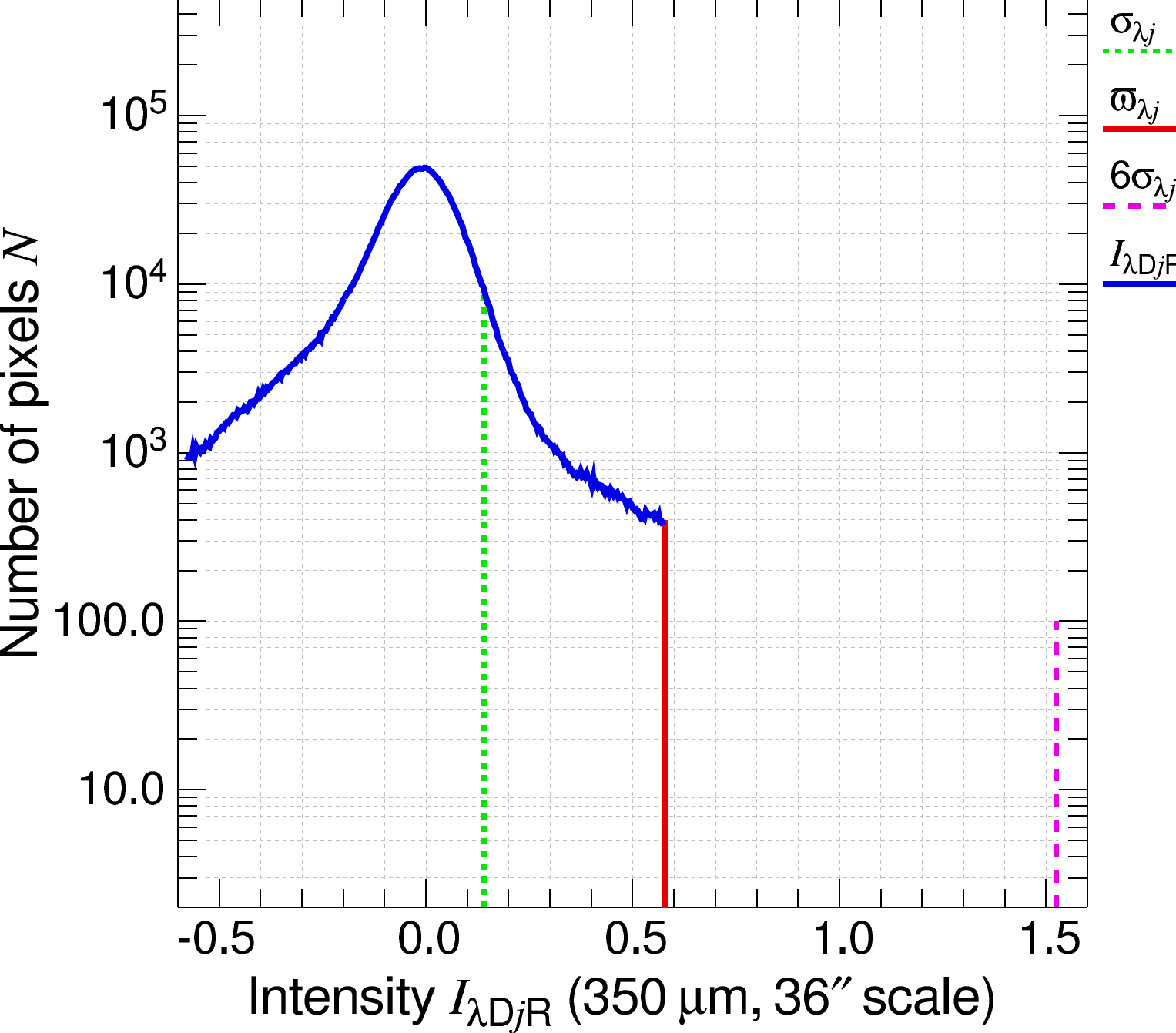}}}
\caption{
Using skewness and kurtosis for iterating accurate thresholds $\varpi_{{\lambda}{j}}$ in the cleaning process
(Sect.~\ref{removing.noise.background}). The \emph{upper} panels show histograms for the full image
$\mathcal{I}_{{\!\lambda}{\rm D}}$ at 350\,{${\mu}$m} (\emph{left}) and for the decomposed images
$\mathcal{I}_{{\!\lambda}{\rm D}{j}}$ at the 9{\arcsec} and 36{\arcsec} scales (\emph{middle, right}) before cleaning. The
\emph{lower} panels display histograms for the reconstructed residuals $\mathcal{I}_{{\!\lambda}{\rm{D\,R}}}$ (\emph{left}) and for
the residuals $\mathcal{I}_{{\!\lambda}{\rm D}{j}{\,\rm R}}$ of the two decomposed images of the upper panels (\emph{middle,
right}). The vertical lines in the left panels indicate the standard deviation $\sigma_{\lambda}$ (\emph{short dash, green}) and
$6\,\sigma_{\lambda}$ (\emph{long dash, magenta}) computed in the full image $\mathcal{I}_{{\!\lambda}{\rm D}}$. In the other
panels they indicate the converged values of the standard deviation $\sigma_{{\lambda}{j}}$ and $6\,\sigma_{{\lambda}{j}}$ in the
single-scale images $\mathcal{I}_{{\!\lambda}{\rm D}{j}}$, as well as the final thresholds $\varpi_{{\lambda}{j}}$ (\emph{solid,
red}). The histogram of the residuals $\mathcal{I}_{{\!\lambda}{\rm{D\,R}}}$ of the cleaning process, reconstructed from all spatial
scales (\emph{lower left}) has much greater symmetry and resembles a Gaussian distribution, whereas the histogram of the full image
$\mathcal{I}_{{\!\lambda}{\rm D}}$ (\emph{red}, copied from the upper-left panel) is highly asymmetric. Both 
$s^{\rm max}_{\lambda}$ and $k^{\,\rm max}_{\lambda}$ have a value of 3.17 and the corresponding variable factors $n_{{\lambda}{j}}$ 
have the values of 4.52 and 4.09 for the two single-scale images. The width of the intensity bins is 1 MJy/sr in the left panels and 
0.004 MJy/sr in all other panels.
} 
\label{single.scale.histograms}
\end{figure*}

This demonstrates the need to set a reasonable upper limit for the sources to be extracted\footnote{Note that \textsl{getsources}
has no fundamental limitation on the spatial scales or source sizes except they must be smaller than the image size.}. Indeed, if
huge structures were to be allowed (with sizes by orders of magnitude larger than $O_{\lambda}$), they would also have to be used in
the background subtraction and deblending, both of which become increasingly inaccurate on very large scales. For accurate removal
of the background, one has to either \emph{approximate} or \emph{interpolate} it; both approaches become highly uncertain when large
distances are involved. Deblending of overlapping structures requires a good \emph{approximation} of their intensity distributions,
which also becomes inaccurate on very large scales. This would also greatly reduce the quality of the detection and measurements
for the majority of ``normal'' sources. Many of the latter would be fully contained within the much larger (sub-structured) sources
and to accurately measure them, one has to consider the large sources as their background\footnote{If one is interested primarily in
extracting very large structures, one could first extract all smaller sources, subtract them from the original image, and then run
the extraction again, targeted specifically at those structures.}.

\subsection{Cleaning single scales of noise and background}
\label{removing.noise.background}

Before one can use the single-scale detection images $\mathcal{I}_{{\!\lambda}{\rm D}{j}}$ for source extraction, they need to be
cleaned of the contributions of noise and background to make sure that most (if not all) non-zero pixels belong to real sources. The
noise and background fluctuations in the far-infrared and submillimeter images of interstellar clouds, such as those coming from the
\emph{Herschel} Gould Belt and HOBYS surveys, do not follow Gaussian statistics. A great advantage of the fine spatial decomposition
employed by \textsl{getsources} (Sect.~\ref{decomposing.detection.images}) is that the emission fluctuations in the
\emph{decomposed} images of interstellar clouds or Galactic cirrus become approximately Gaussian. Thus, significant departures from
Gaussian distribution in the \emph{single-scale} images indicate the presence of sources. For more details and illustrations in
terms of the power spectra of the components of the actual \emph{Herschel} images, we refer to Appendix~\ref{power.spectra}; see
also Fig.~\ref{single.scale.histograms} below.

Cleaning can be done by global intensity thresholding of the single-scale images, as the larger-scale background has been
effectively filtered out by the spatial decomposition. Unlike the original images $\mathcal{I}_{{\!\lambda}{\rm O}}$ or
$\mathcal{I}_{{\!\lambda}{\rm D}}$ that often have a very strong and highly-variable background, the entire single-scale images are
``flat'' in the sense that all signals on considerably larger scales have been removed (see Fig.~\,\ref{single.scales}). Another
advantage of this \emph{single-scale cleaning} is that the noise contribution depends very significantly on the scale. For example,
the small-scale noise gets heavily diluted at large scales, where extended sources become best visible. In effect, in a
reconstructed clean image $\mathcal{I}_{{\!\lambda}{\rm{D\,C}}}{\,=}\sum_{j} \mathcal{I}_{{\!\lambda}{\rm D}{j}{\,\rm C}}$ one can
see large structures better (deeper) than in $\mathcal{I}_{{\!\lambda}{\rm D}}$.

To clean the single-scale images, we designed an iterative algorithm\footnote{A procedure similar to what is usually called ``sigma
clipping''.} that automatically finds at each scale a cut-off level that separates the signal of significant sources from those of
the noise and background. At the first scale ($j{\,=\,}1$) it computes the cut-off (threshold)
$\varpi_{{\lambda}{j}}{\,=\,}n_{{\lambda}{j}}\,\sigma_{{\lambda}{j}}$ for the image
$\mathcal{I}_{{\!\lambda}{\rm D}{j}}\,\mathcal{M}_{\lambda}$, where $\sigma_{{\lambda}{j}}$ is the standard deviation over the
entire image and $n_{{\lambda}{j}}$ is a variable factor having an initial value of $n_{{\lambda}{1}}{\,=\,}$6 (this $j{\,=\,}1$
value was found to be optimal in our tests). Then the procedure masks out all pixels with the values
$|I_{{\lambda}{j}}|{\,\ge\,}\varpi_{{\lambda}{j}}$ and repeats the calculation of $\sigma_{{\lambda}{j}}$ over the remaining pixels,
estimating a new threshold, which is generally lower than the one at the previous iteration. The procedure masks out bright pixels
again and iterates further, always computing $\sigma_{{\lambda}{j}}$ at $|I_{{\lambda}{j}}|{\,<\,}\varpi_{{\lambda}{j}}$,
\emph{outside} the peaks and hollows, until $\varpi_{{\lambda}{j}}$ converges ($\Delta\varpi_{{\lambda}{j}}{\,<\,}$1{\%}). To
produce a clean single-scale image $\mathcal{I}_{{\!\lambda}{\rm D}{j}{\,\rm C}}$, all pixels with
$I_{{\lambda}{j}}{\,<\,}\varpi_{{\lambda}{j}}$ are zeroed, which (ideally) leaves non-zero only those pixels that belong to
significant peaks from sources. Several examples of clean images are displayed in the last five panels of
Fig.~\ref{clean.single.scales}. 

In addition, the low-intensity pixels $|I_{{\lambda}{j}}|{\,<\,}\varpi_{{\lambda}{j}}$ define the single-scale images of the
residuals $\mathcal{I}_{{\!\lambda}{\rm D}{j}{\,\rm R}}$, as well as the reconstructed image of the residuals
$\mathcal{I}_{{\!\lambda}{\rm{D\,R}}}{\,=}\sum_{j} \mathcal{I}_{{\!\lambda}{\rm D}{j}{\,\rm R}}$. The images are shown in
Fig.~\ref{single.scale.residuals}, where one can clearly see that the single-scale residuals are \emph{much} ``flatter'' than those
accumulated over all scales. This illustrates why the single-scale cleaning can be performed on the entire images using
thresholding, in contrast to the full images (for more illustrations, see Figs.~\ref{single.scales}, \ref{single.scale.histograms}).
Furthermore, the flattening step of our method (Sect.~\ref{flattening.background.noise}) ensures that also the \emph{standard
deviations} of the single-scale residuals become as uniform as possible over the entire image.

Starting with the second scale, $j{\,=\,}2$, the factor $n_{{\lambda}{j}}$ is allowed to become lower than its initial value of
$n_{{\lambda}{1}}{\,=\,}$6 at $j{\,=\,}1$, depending on the image and the scale. Being an important parameter for the iterations to
converge to accurate cut-off levels, $n_{{\lambda}{j}}$ must be accurately chosen. Empirical evidence shows that if
$n_{{\lambda}{j}}$ were smaller than an optimal value, the iterations would converge to $\varpi_{{\lambda}{j}}$ that is too low,
resulting in noise peaks contaminating the clean images. On the other hand, if $n_{{\lambda}{j}}$ were larger than its optimal
value, the iterations would converge to a value of $\varpi_{{\lambda}{j}}$ not deep enough, thus some fainter sources present in
$\mathcal{I}_{{\!\lambda}{\rm D}{j}}$ would be missing in the clean images $\mathcal{I}_{{\!\lambda}{\rm D}{j}{\,\rm C}}$.

To determine the appropriate values of $n_{{\lambda}{j}}$, one can use the higher-order statistical quantities, skewness and 
kurtosis 
\begin{equation}
s_{{\lambda}{j}}{\,=\,}\mu_{3{\lambda}{j}}\,\sigma_{{\lambda}{j}}^{-3},\,\,\,
k_{{\lambda}{j}}{\,=\,}\mu_{4{\lambda}{j}}\,\sigma_{{\lambda}{j}}^{-4}-3,
\label{skewness}
\end{equation}
where $\mu_{3{\lambda}{j}}$ and $\mu_{4{\lambda}{j}}$ are the third and fourth moments about the mean (both $s_{{\lambda}{j}}$ and
$k_{{\lambda}{j}}$ are zero for a standard normal distribution). The idea is that when the pixel distribution of the residuals
$\mathcal{I}_{{\!\lambda}{\rm D}{j}{\,\rm R}}$ becomes too asymmetric (large $|s_{{\lambda}{j}}|$) or too peaked (large
$k_{{\lambda}{j}}$), the optimal value of $n_{{\lambda}{j}}$ must actually have been lower. Thus, having iterated the cut-offs
$\varpi_{{\lambda}{1}}$ at the first scale, \textsl{getsources} computes the upper limits to $s_{{\lambda}{j}}$ and
$k_{{\lambda}{j}}$ given by an empirical formula
\begin{equation}
s^{\,\rm max}_{\lambda}{\,=\,}k^{\,\rm max}_{\lambda}{\,=\,}\max\left\{2.14\,\ln
\left(\frac{I^{\,\rm max}_{{\!\lambda}{\rm D}{1}}}{\sigma_{{\lambda}{1}}}+220\right)-11.3,0.25\right\},
\label{nsigma.limits}
\end{equation}
where $I^{\,\rm max}_{{\!\lambda}{\rm D}{1}}$ is the maximum pixel intensity over
$\mathcal{I}_{{\!\lambda}{\rm D}{1}}\,\mathcal{M}_{\lambda}$.

When iterations at scale $j$ converge to a threshold $\varpi_{{\lambda}{j}}$, \textsl{getsources} computes $s_{{\lambda}{j}}$ and
$k_{{\lambda}{j}}$ in the image of the residuals $\mathcal{I}_{{\!\lambda}{\rm D}{j}{\,\rm R}}$. If
$|s_{{\lambda}{j}}|{\,>\,}s^{\,\rm max}_{\lambda}$ or $k_{{\lambda}{j}}{\,>\,}k^{\,\rm max}_{\lambda}$, the algorithm slightly (by a
few percent) reduces $n_{{\lambda}{j}}$, whose initial value is $n_{{\lambda}{j-1}}$ from the previous scale, and re-iterates
$\varpi_{{\lambda}{j}}$. This procedure ensures that $s_{{\lambda}{j}}$ and $k_{{\lambda}{j}}$ always stay within the empirical
bounds in the process of obtaining the thresholds and cleaning the single-scale images. Extensive experimentation has shown that the
limits of Eq.~\ref{nsigma.limits} work very well for all images that we have tested\footnote{We have applied \textsl{getsources} to
the multi-wavelength images of a dozen of simulated fields, the ground-based (sub-)mm images of \object{NGC 2068}, \object{NGC
2071}, \object{NGC 2264}, \object{W 43}, and the \emph{Herschel} images of \object{Aquila}, \object{Cepheus}, \object{Cygnus X},
\object{IC 5146}, \object{Lupus}, \object{M 16}, \object{NGC 4559}, \object{NGC 7538}, \object{Orion B}, \object{Perseus},
\object{Pipe}, \object{Polaris}, \object{RCW 79}, \object{RCW 82}, \object{RCW 120}, \object{Rosette}, \object{Taurus},
\object{Vela}, \object{W 3}, \object{W 48}.}.

\begin{figure*}
\centering
\centerline{\resizebox{0.48\hsize}{!}{\includegraphics{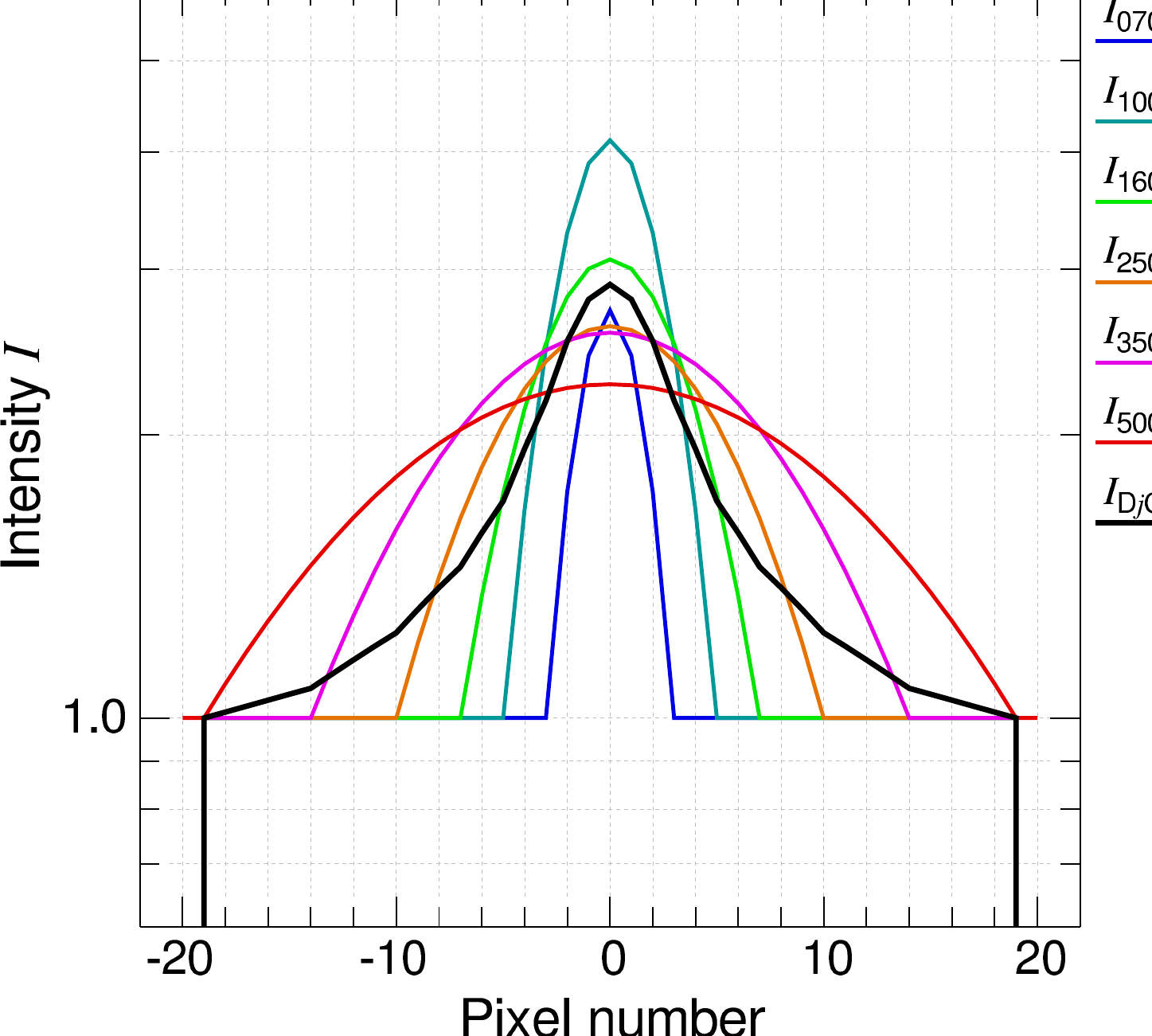}}
            \resizebox{0.48\hsize}{!}{\includegraphics{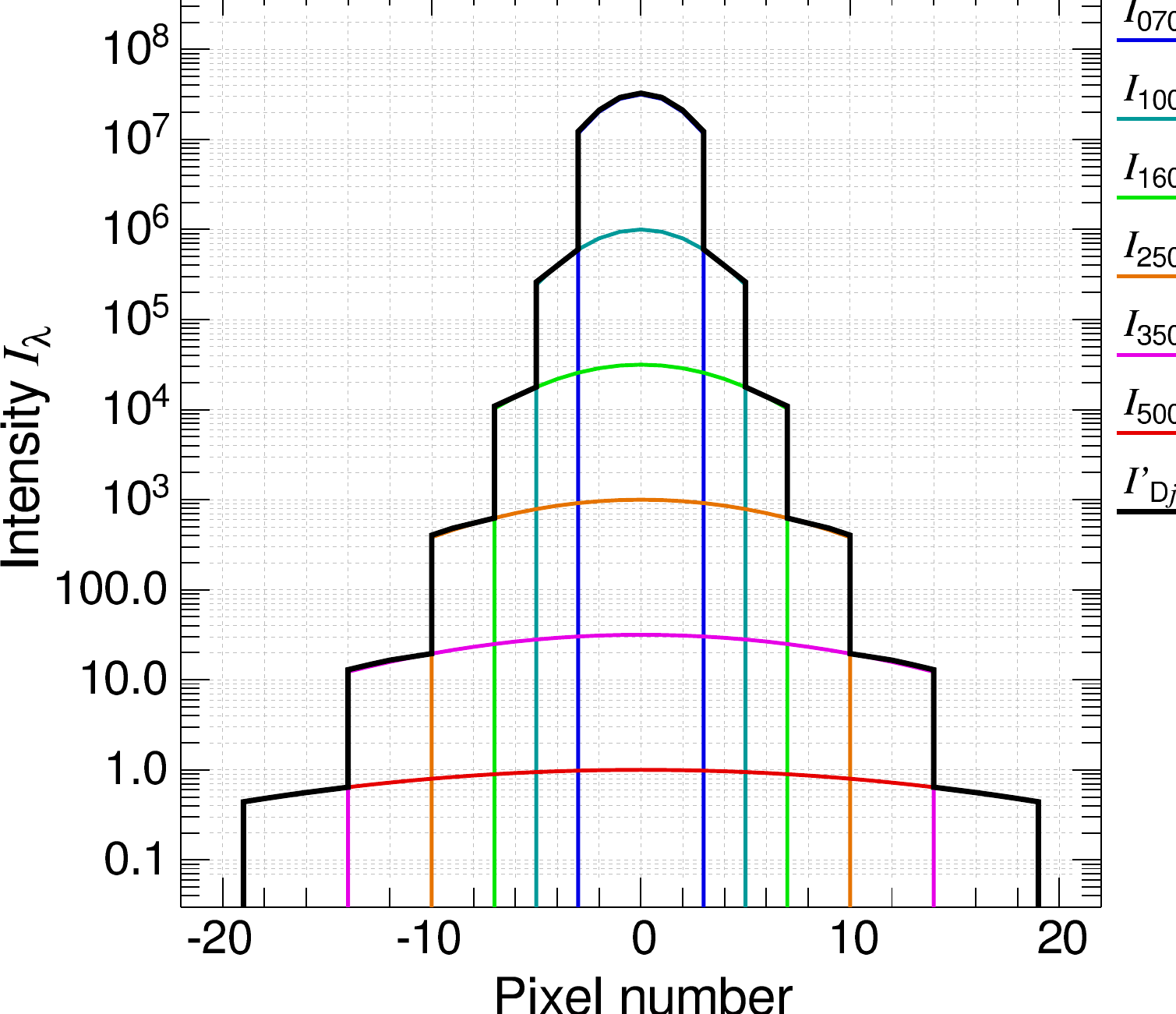}}}
\caption{
Two types of combined single-scale images (Sect.~\ref{combining.clean.single.scales}). Schematically shown are the image
$\mathcal{I}_{{\rm D}{j}{\,\rm C}}$ (\emph{left}) used to detect sources and track the evolution of their shapes and the image
$\mathcal{I}^{\prime}_{{\rm D}{j}{\,\rm C}}$ (\emph{right}) used to determine the characteristic scales and initial footprints.
}
\label{combining}
\end{figure*}

The pixel distributions shown in Fig.~\ref{single.scale.histograms} illustrate the cleaning algorithm. The histogram for the
original image $\mathcal{I}_{{\!\lambda}{\rm D}}$ shown in the upper-left panel contains all spatial scales and is therefore very
wide and asymmetric; it cannot be used to separate sources from noise and background using global thresholding. All scales are
blended together in such images and \emph{any} threshold would enable one to either find only the brightest peaks losing most
fainter sources or create many spurious peaks from the pixels belonging to the variable background or noise. In contrast, the
highly-filtered single-scale images $\mathcal{I}_{{\!\lambda}{\rm D}{j}}$ contain only narrow ranges of spatial scales and thus the
histograms of the residuals $\mathcal{I}_{{\!\lambda}{\rm D}{j}{\,\rm R}}$ (representing the background and noise fluctuations) are
\emph{much} narrower and symmetric, resembling a Gaussian distribution (Fig.~\ref{single.scale.histograms}). Having no signals from
all irrelevant larger scales, the images are much ``flatter'' and therefore well suited for the cleaning algorithm. The single-scale
histograms show that using the upper limits of Eq.~\ref{nsigma.limits} for skewness and kurtosis helps in correcting the variable
factor $n_{{\lambda}{j}}$ and thus in getting deeper thresholds $\varpi_{{\lambda}{j}}$ and much better detection of fainter sources
while avoiding creation of spurious sources.

\subsection{Combining clean single scales over all wavelengths}
\label{combining.clean.single.scales}

The cleaning algorithm outlined in Sect.~\ref{removing.noise.background} is applied to the single-scale detection images
$\mathcal{I}_{{\!\lambda}{\rm D}{j}}$ independently in each waveband. Clearly, \textsl{getsources} also works the same way for just
one image at a single wavelength. It is necessary to fully complete ``monochromatic'' extractions as they provide the footprints of
all sources and also the estimates of the sizes $A^{\rm max}_{\lambda}$ of the largest sources, the information used by our
flattening procedure (Sect.~\ref{flattening.background.noise}). Combining wavelengths means utilizing all information across all
bands and this should be used to improve the detection and measurement qualities over the simple approach of matching separate
catalogs obtained in each waveband independently. Whereas combining independent catalogs on the basis of the association radius is
possible for a few images obtained with similar angular resolutions, this usual approach of \emph{associating} sources between
wavelengths would introduce large and unknown errors when applied to \emph{Herschel} images whose resolutions differ by as much as a
factor of $\sim$7.

In general, it is impossible to combine images at different wavelengths in a meaningful way using full images containing signals on
\emph{all} spatial scales. The fine spatial decomposition employed by \textsl{getsources} (Sect.~\ref{decomposing.detection.images})
that filters out signals from all irrelevant scales, together with the single-scale cleaning (Sect.~\ref{removing.noise.background})
enable us to create wavelength-independent clean images that accumulate only significant intensity peaks from all wavelengths,
representing potential sources. The combined images must be normalized because of highly varying intensities in different bands;
there is no need to preserve the spectral behavior of sources in the images used only for detection (Sect.~\ref{detecting.sources}).
Indeed, the wavelength-dependent properties of all sources will be measured in the observed images
$\mathcal{I}_{{\!\lambda}{\rm O}}$ (Sect.~\ref{measuring.cataloging}) after all sources have been detected.

The cleaning procedure described in Sect.~\ref{removing.noise.background} works well when the small-scale noise and background
fluctuations are relatively uniform across the image and there are no strong artifacts. It is not unusual, however, for the observed
images to contain quite variable noise and various types of artifacts, including those from the map-making process. In order to
reduce possible contamination of the clean images with the pixels belonging to the noise peaks or artifacts, \textsl{getsources}
additionally employs a lower limit on the number of pixels $N_{{\Pi}{\lambda}}$ in a cluster of connected pixels that may remain in
single-scale images:
\begin{equation}
N^{\,\rm min}_{{\Pi}{\lambda}} = \max \left\{ \frac{1}{3}\,\pi \left(\frac{O_{\lambda}}{2 \Delta}\right)^{2}\!,
N^{\,\rm min}_{\Pi} \right\},
\label{minpix}
\end{equation}
where $O_{\lambda}$ is the observational beam size, $\Delta$ is the pixel size, and the default value of the parameter $N^{\,\rm
min}_{\Pi}$ is 4. Small clusters with $N_{{\Pi}{\lambda}} < N^{\,\rm min}_{{\Pi}{\lambda}}$ are removed from the single-scale images
$\mathcal{I}_{{\!\lambda}{\rm D}{j}{\,\rm C}}$ before the latter are combined.

\begin{figure*}
\centering
\centerline{\resizebox{0.33\hsize}{!}{\includegraphics{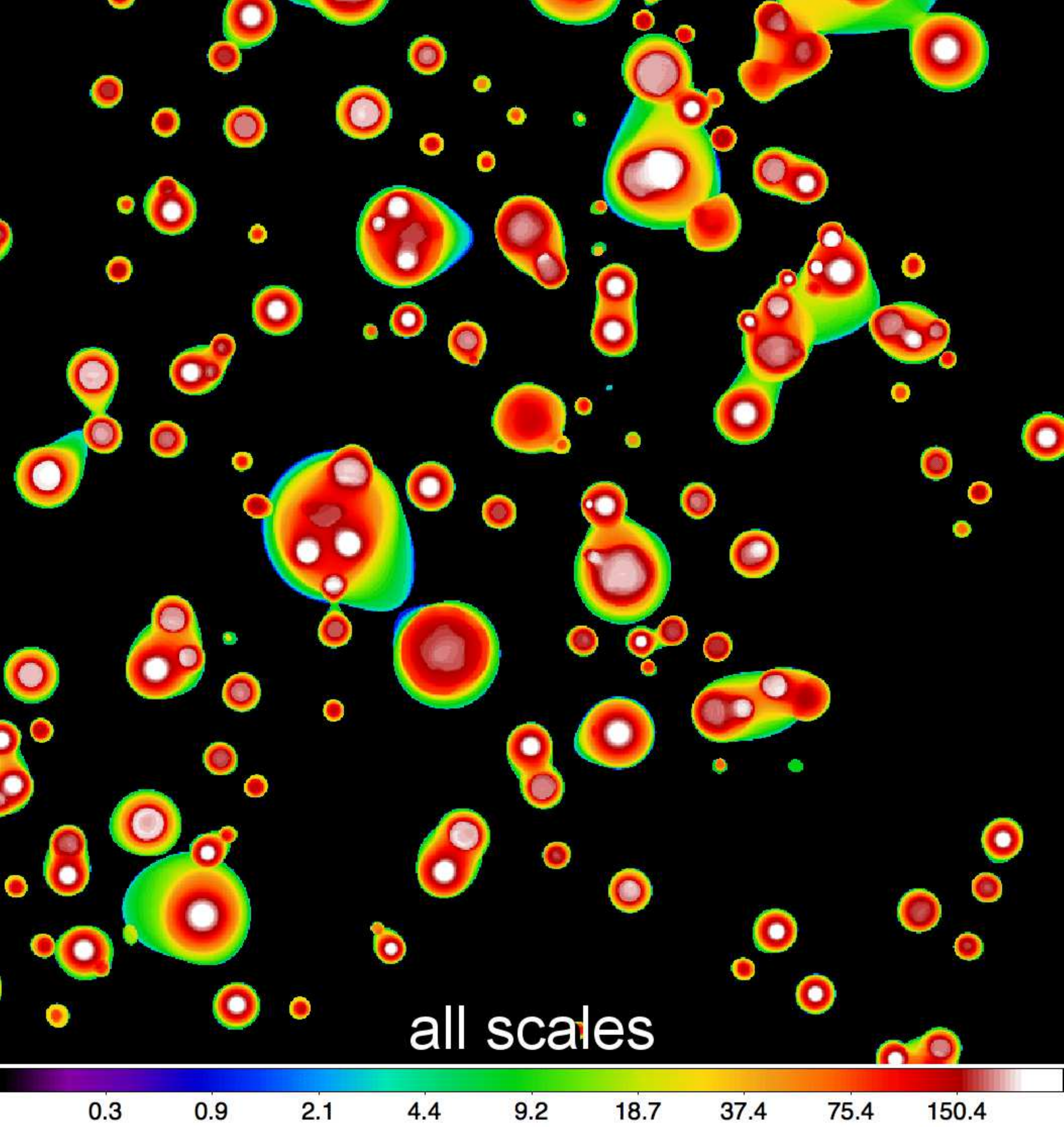}}
            \resizebox{0.33\hsize}{!}{\includegraphics{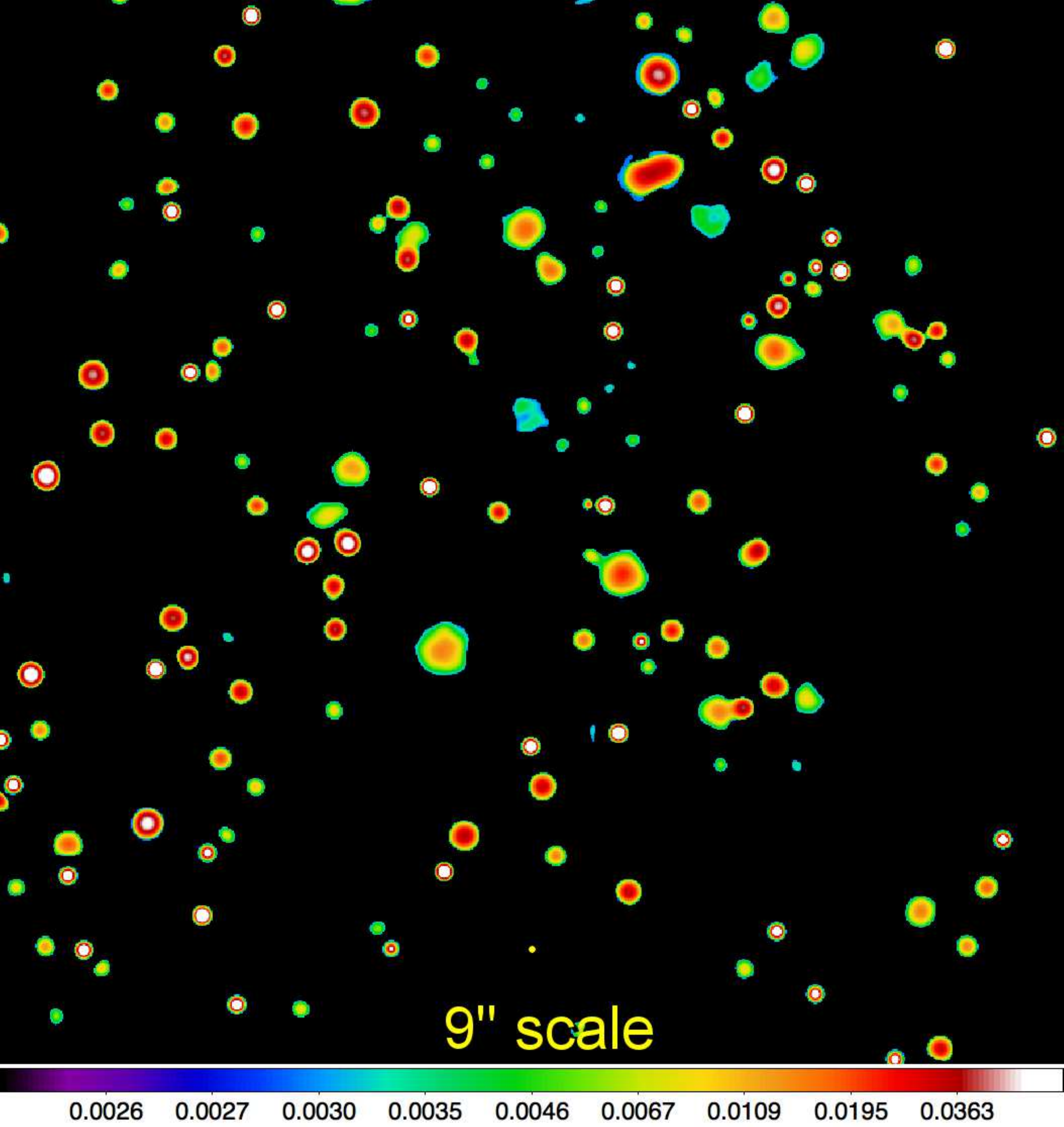}}
            \resizebox{0.33\hsize}{!}{\includegraphics{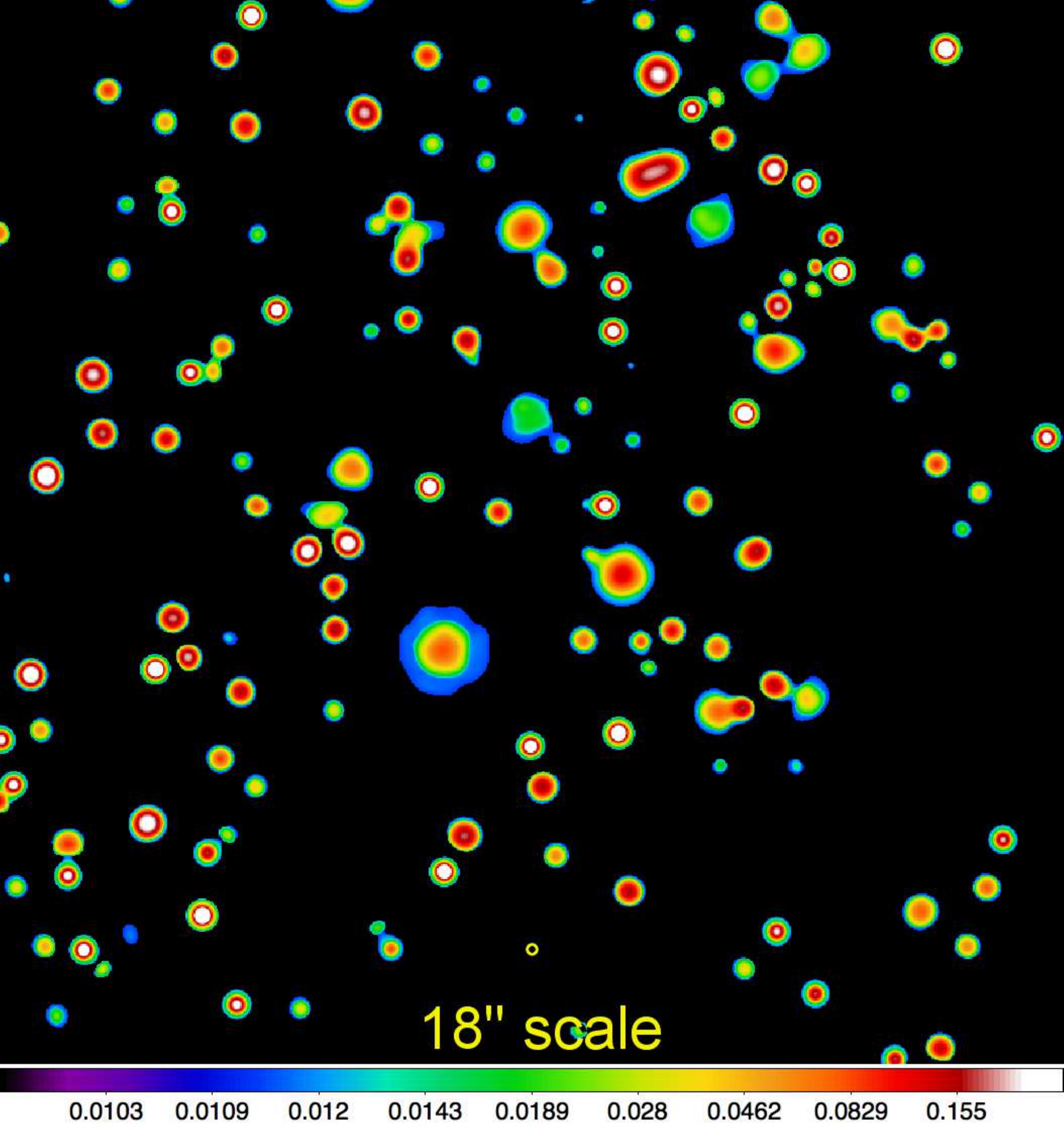}}}
\centerline{\resizebox{0.33\hsize}{!}{\includegraphics{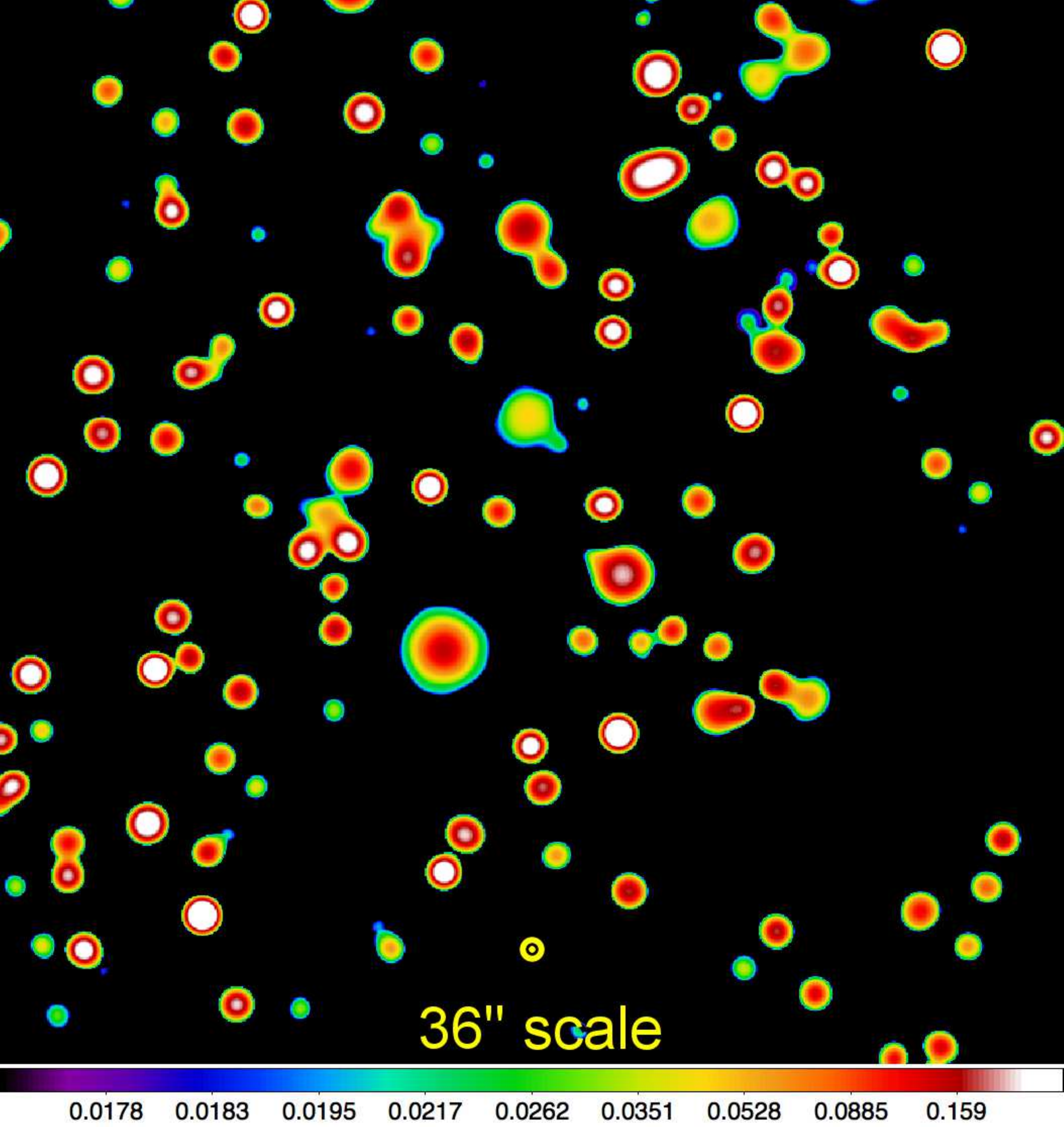}}
            \resizebox{0.33\hsize}{!}{\includegraphics{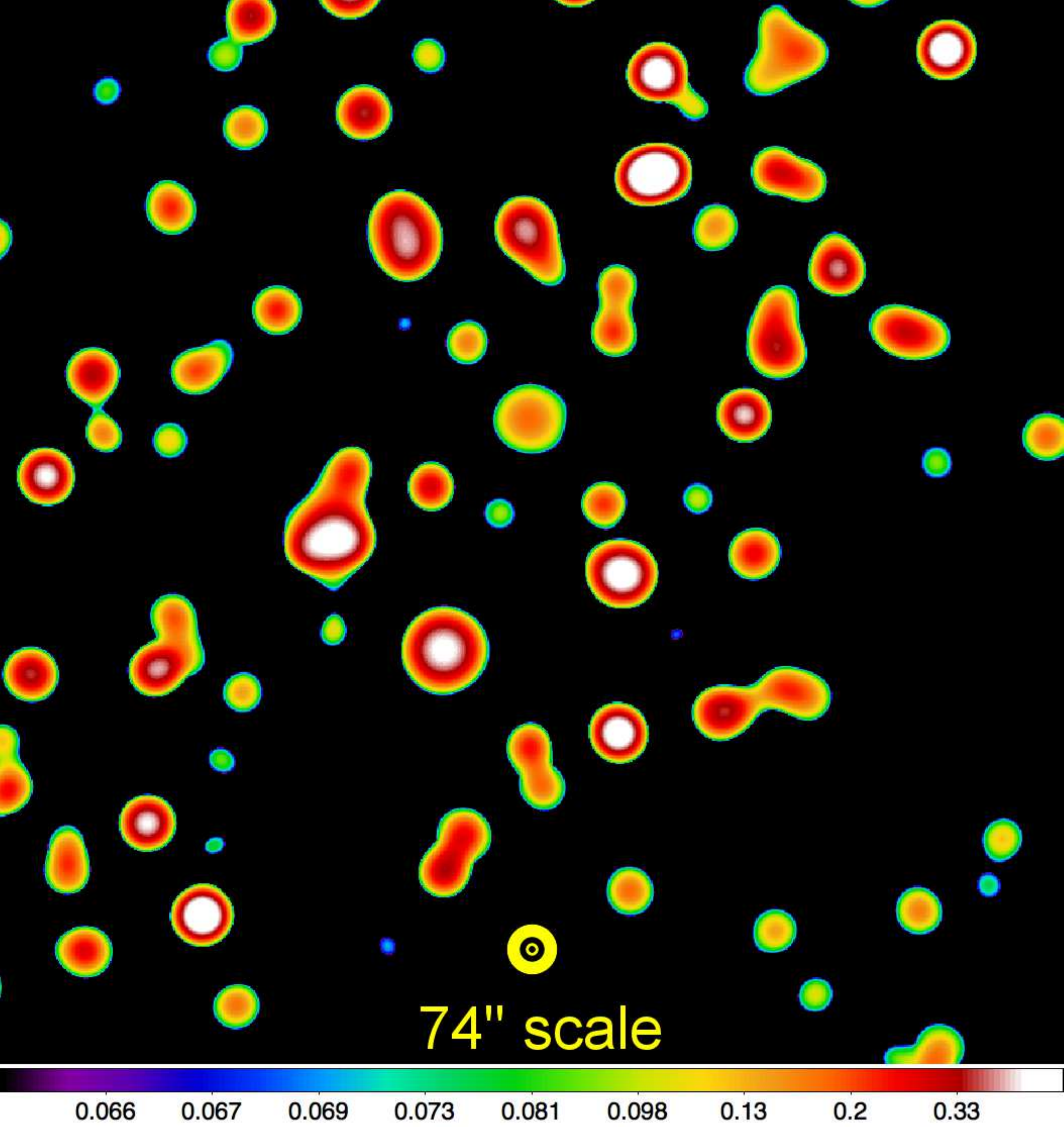}}
            \resizebox{0.33\hsize}{!}{\includegraphics{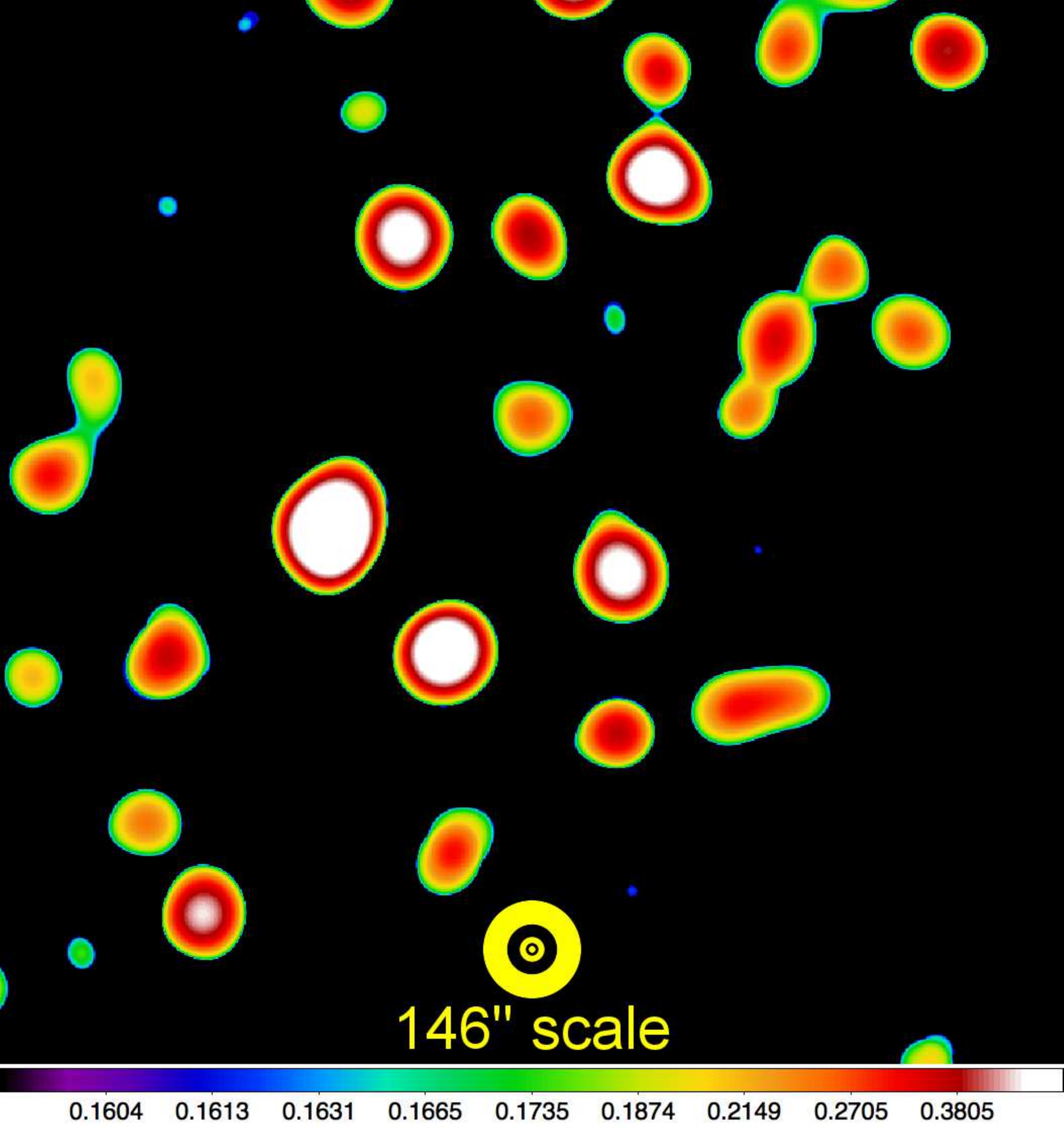}}}
\caption{
Combination of single-scale images (Sect.~\ref{combining.clean.single.scales}). The field of Fig.~\ref{single.scales} is shown as
source masks $\mathcal{M}_{{\rm D}{\rm C}}$ accumulated over all scales and wavebands (\emph{upper left}) that give a summary view
of how the sources, made visible by the cleaning (cf. Fig.~\ref{clean.single.scales}), change their shapes and sizes. The same set
of spatial scales is displayed in the combined clean single-scale images $\mathcal{I}_{{\rm D}{j}{\,\rm C}}$ (\emph{left to right},
\emph{top to bottom}) that accumulate information at those scales from all wavelengths. For better visibility, the values displayed
in the masks image are limited to 300 and in the normalized images $\mathcal{I}_{{\rm D}{j}{\,\rm C}}$ they are limited to $0.07$,
$0.3$, $0.3$, $0.6$, and $0.6$. The scale sizes $S_{\!j}$ are visualized by yellow-black circles and annotated at the bottom of the
panels. The color coding is a logarithmic function of intensity.
} 
\label{combined.clean.single.scales}
\end{figure*}

\begin{figure*}
\centering
\centerline{\resizebox{0.196\hsize}{!}{\includegraphics{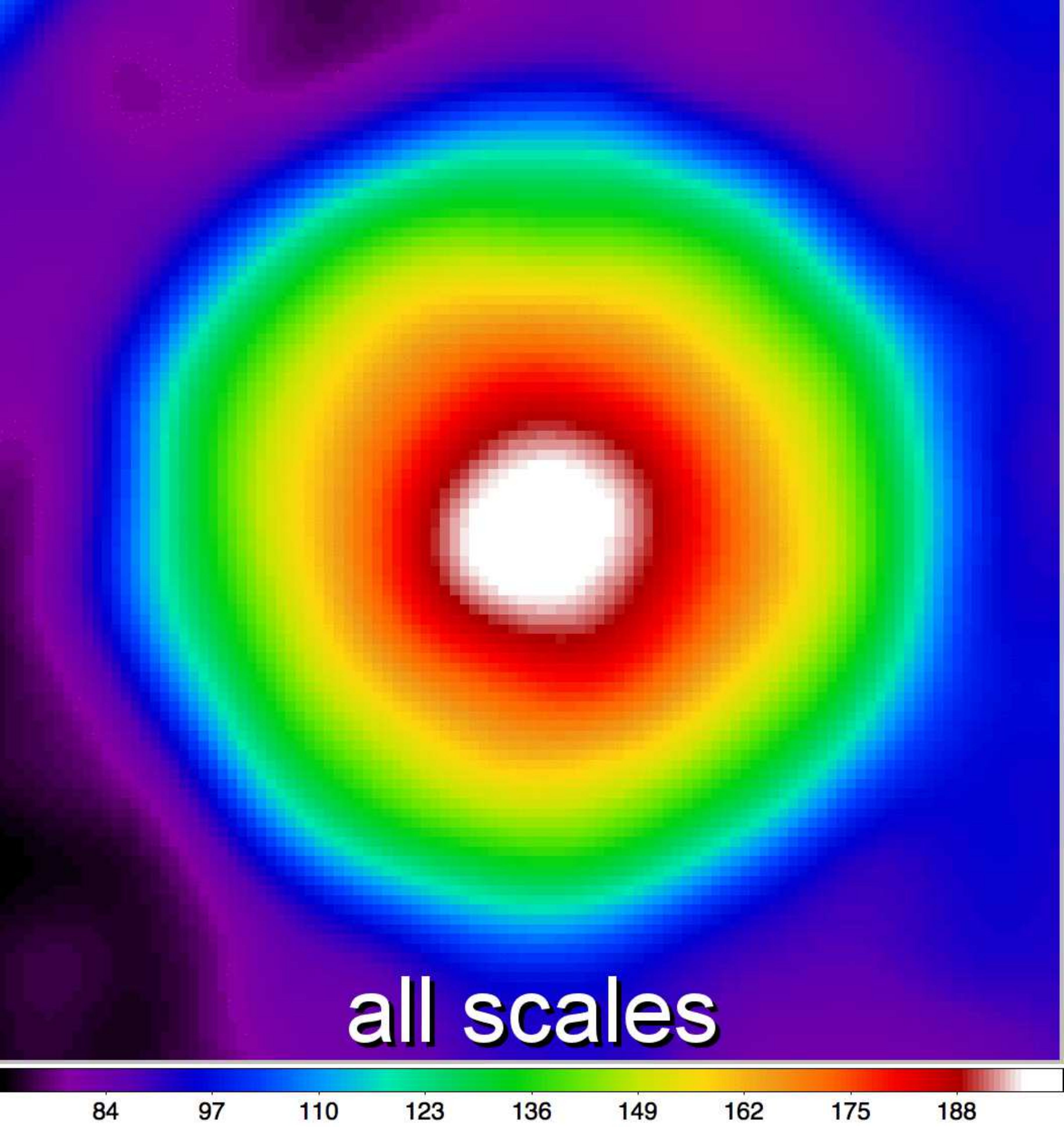}}
            \resizebox{0.196\hsize}{!}{\includegraphics{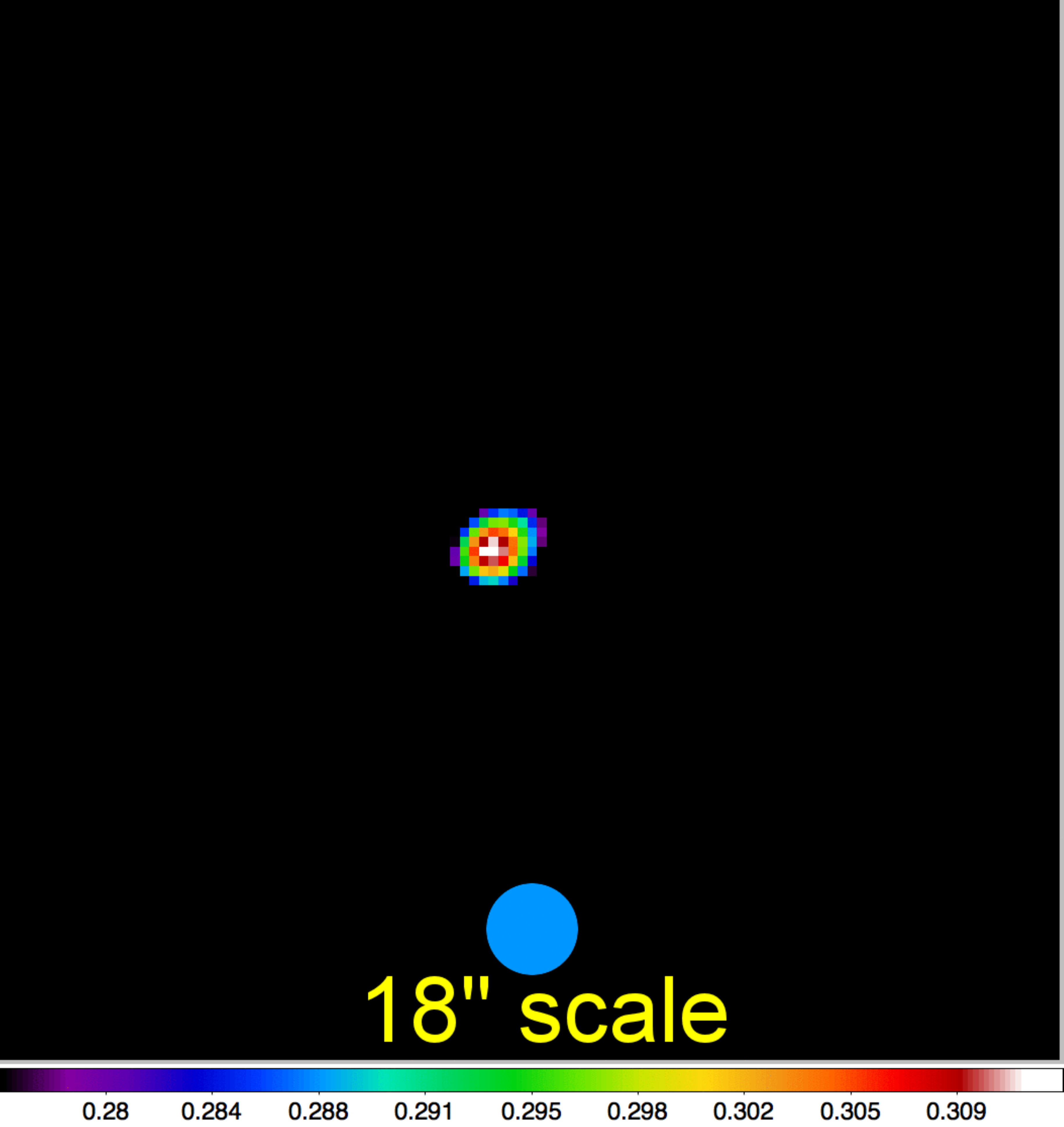}}
            \resizebox{0.196\hsize}{!}{\includegraphics{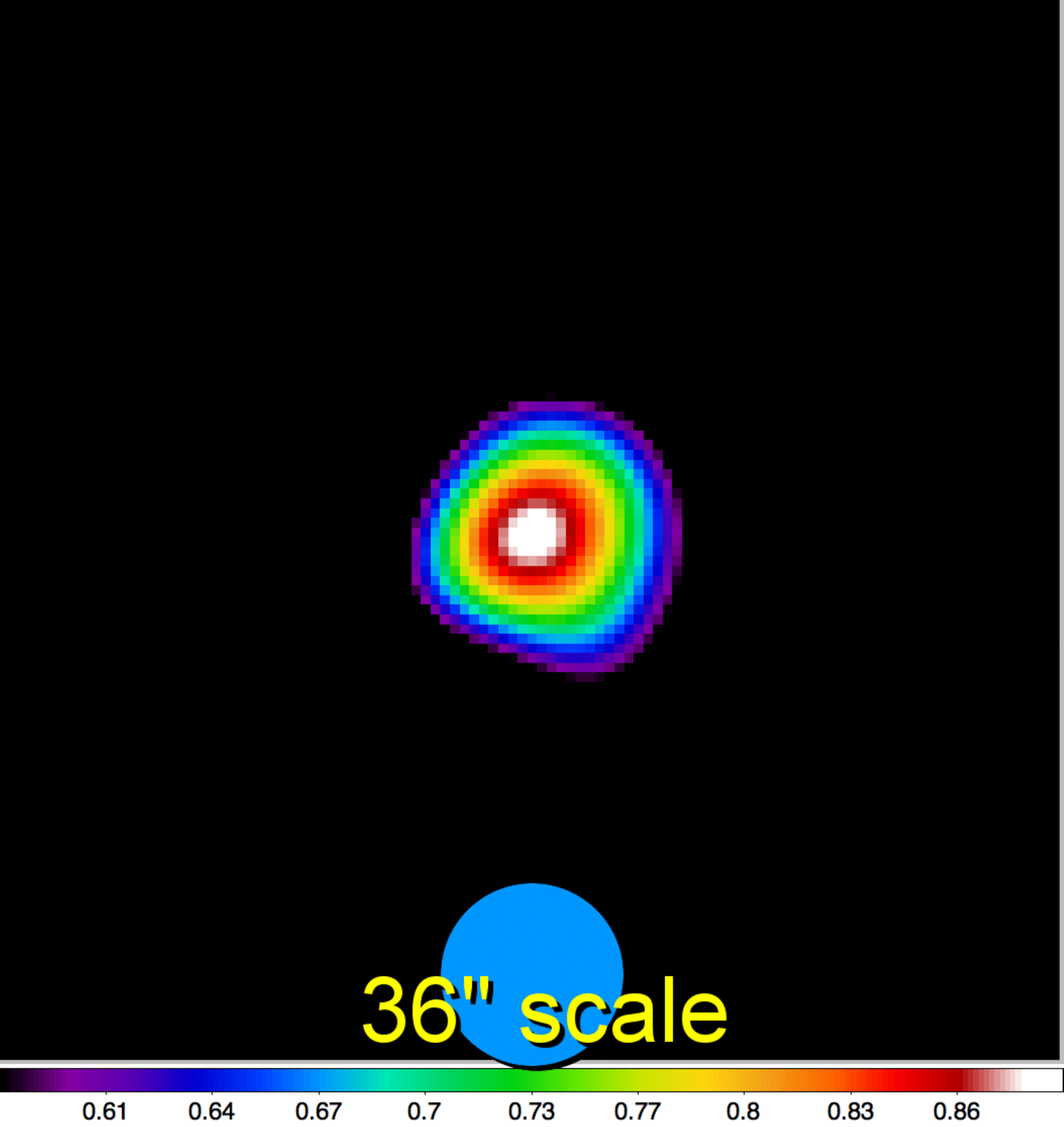}}
            \resizebox{0.196\hsize}{!}{\includegraphics{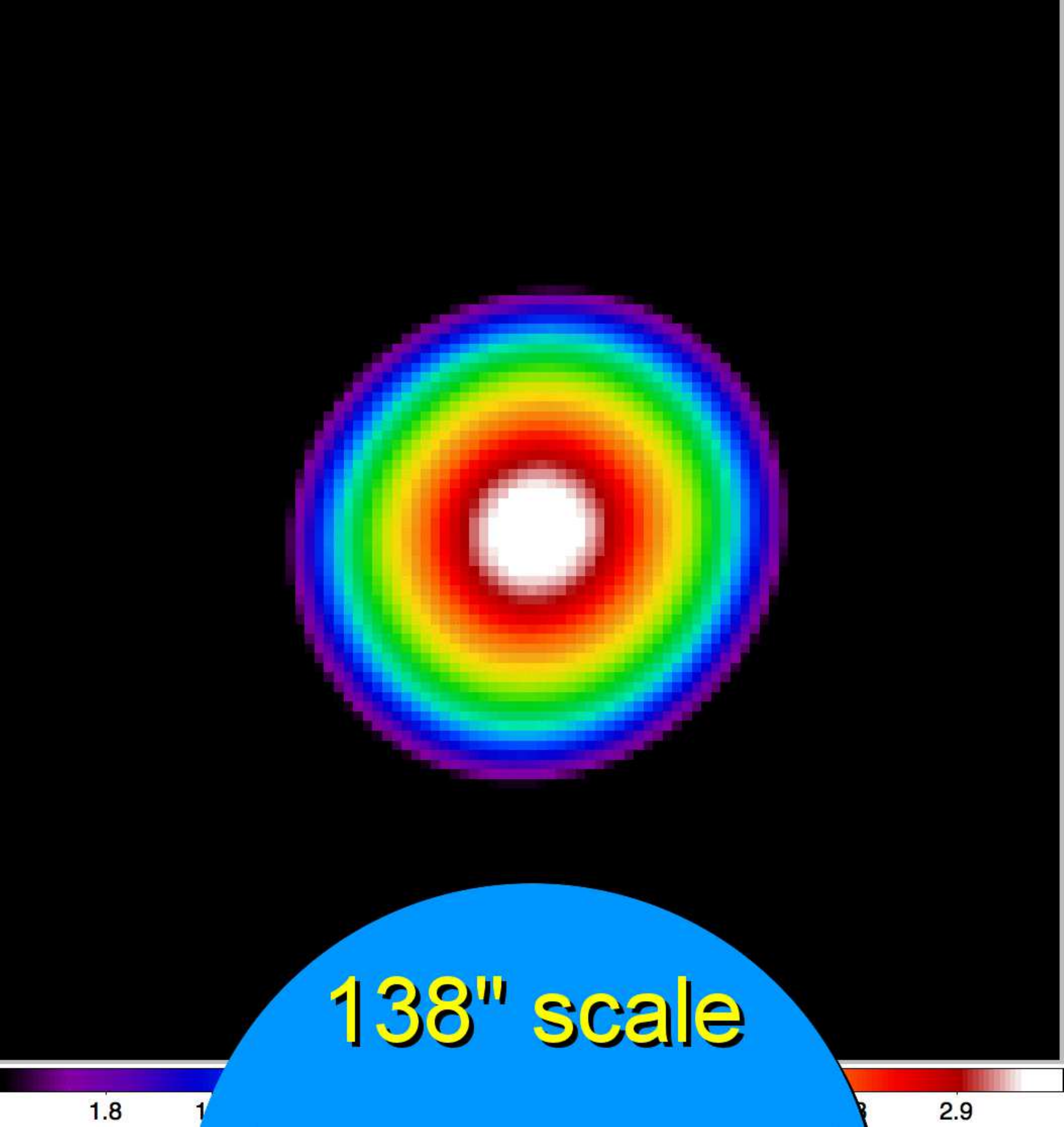}}
            \resizebox{0.196\hsize}{!}{\includegraphics{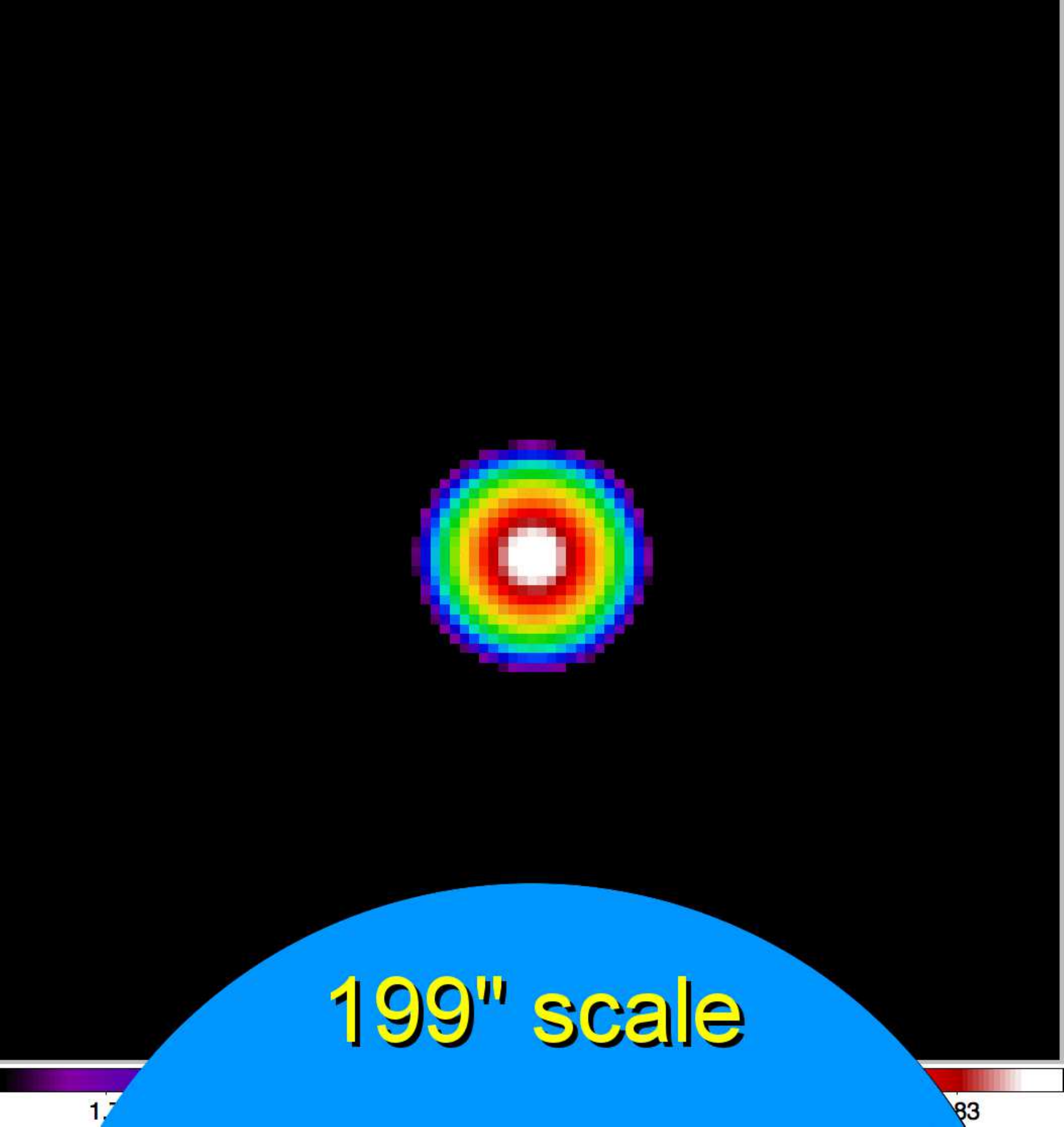}}}
\caption{
Evolution of sources in the clean single-scale images (Sect.~\ref{detecting.sources}). The full image of the source at
350\,{${\mu}$m} with a size of 3{\farcm}7$\times$3{\farcm}7 (\emph{left}) has been cut out of the corresponding panel in
Fig.\,\ref{single.scales} (the source is located south-east of the image center, it is resolved in all wavebands). The other panels
show the source in single-scale images $\mathcal{I}_{{\!\lambda}{\rm D}{j}{\,\rm C}}$ at the scales of 18, 36, 138, and
199{\arcsec}, maximum intensities in the panels being 0.31, 0.89, 3.09, and 1.84 MJy/sr, respectively. The scale sizes $S_{\!j}$ are
visualized by the blue circles and annotated at the bottom of the panels. The color coding is a linear function of intensity.
} 
\label{zoomed.clean.images}
\end{figure*}

When combining single scales from different wavebands, \textsl{getsources} only sums up limited ranges of spatial scales, depending
on the angular resolution and the sizes of sources that can be found in the images. The range of scales is limited from below
because the smallest scales may contain decreasing (and progressively noisier) contribution for the images obtained at longer
wavelengths with poorer resolutions. Single scales within a factor of 3 below the nominal resolution in each band may still contain
considerable signal from the sources. Accordingly, the lower limit of scales $S_{\!j}$ being combined defaults to $\max
\left\{S_{\!1}, 0.33\,O_{\lambda}\right\}$, allowing for a substantial ``super-resolution'' in \textsl{getsources}. On large scales,
the range of scales is also limited by the (wavelength-dependent) maximum sizes $A^{\rm max}_{\lambda}$ of sources (see
Sect.~\ref{decomposing.detection.images}). Initial guesses for $A^{\rm max}_{\lambda}$ are given as input to \textsl{getsources},
however, they are accurately determined during the initial extraction and used in the final extraction
(Sect.~\ref{flattening.background.noise}).

\begin{figure*}
\centering
\centerline{\resizebox{0.33\hsize}{!}{\includegraphics{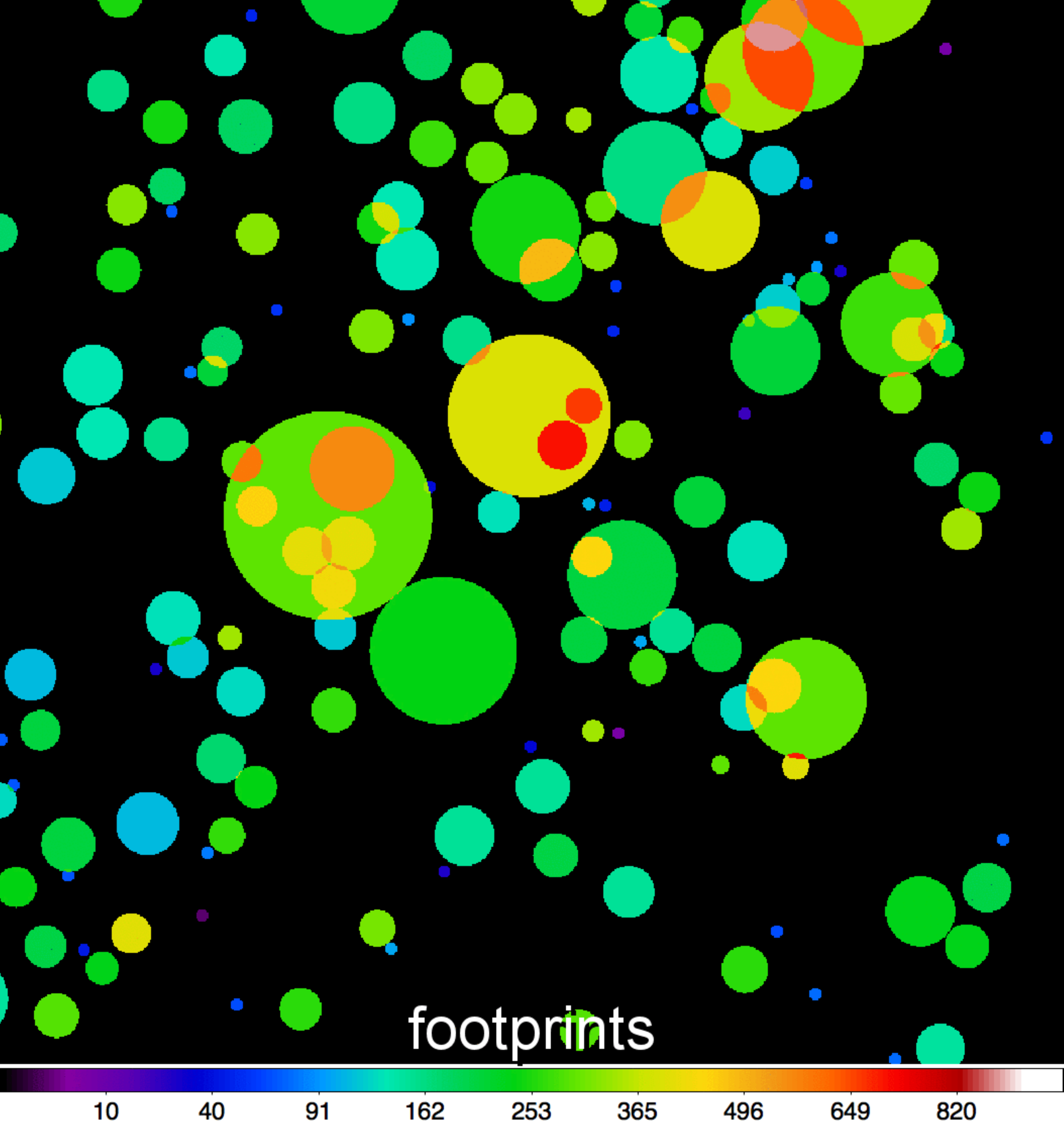}}
            \resizebox{0.33\hsize}{!}{\includegraphics{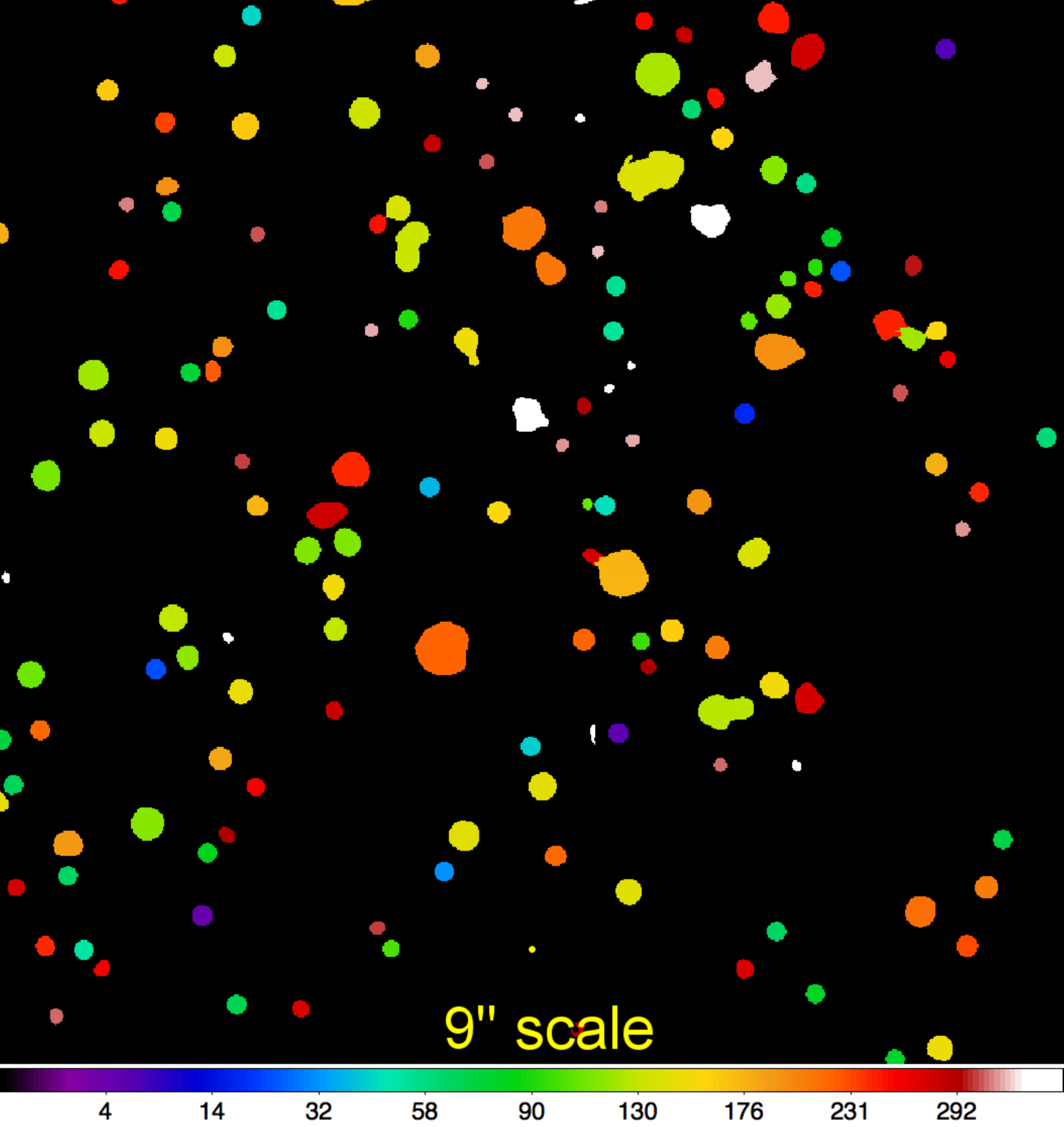}}
            \resizebox{0.33\hsize}{!}{\includegraphics{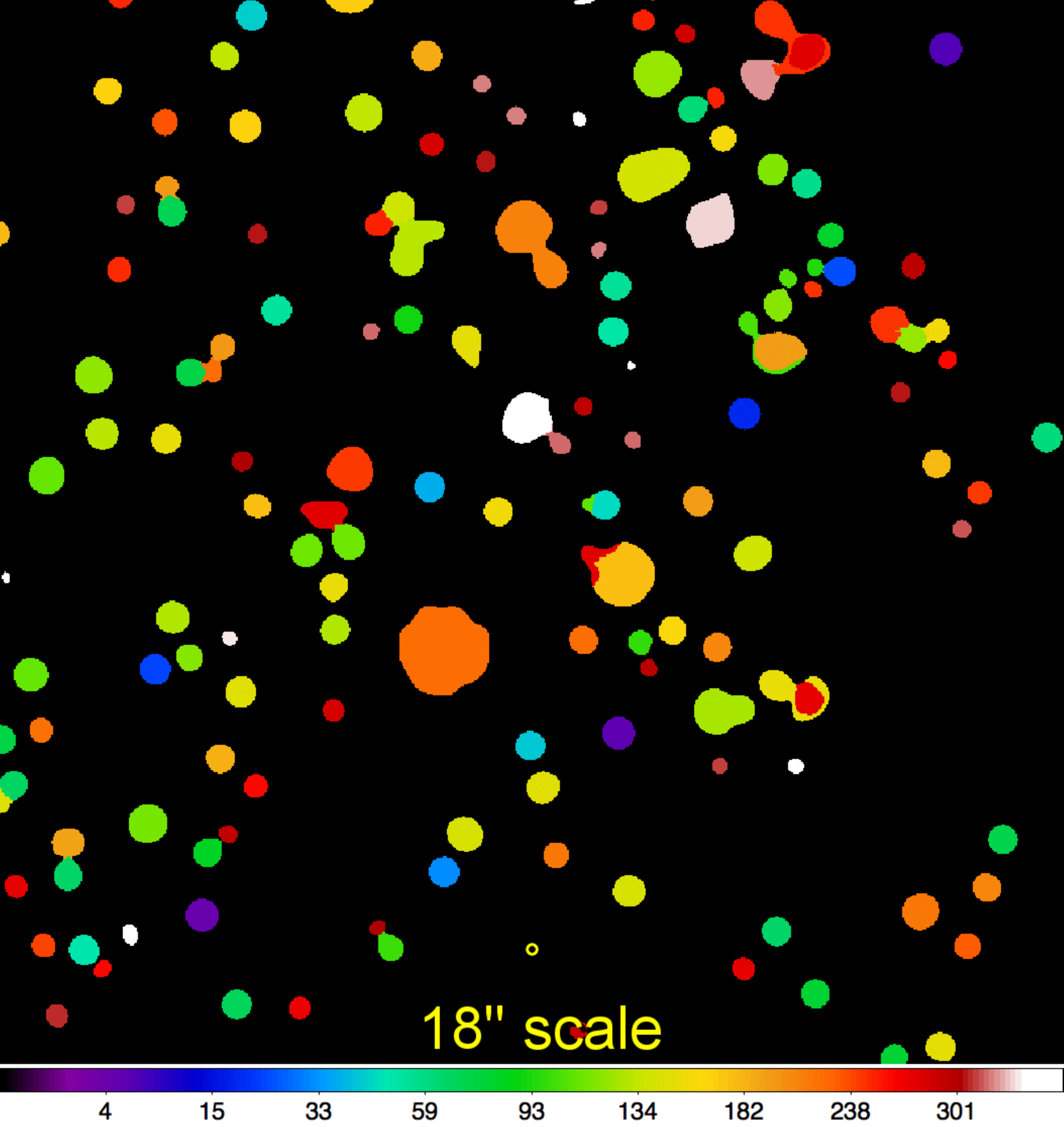}}}
\centerline{\resizebox{0.33\hsize}{!}{\includegraphics{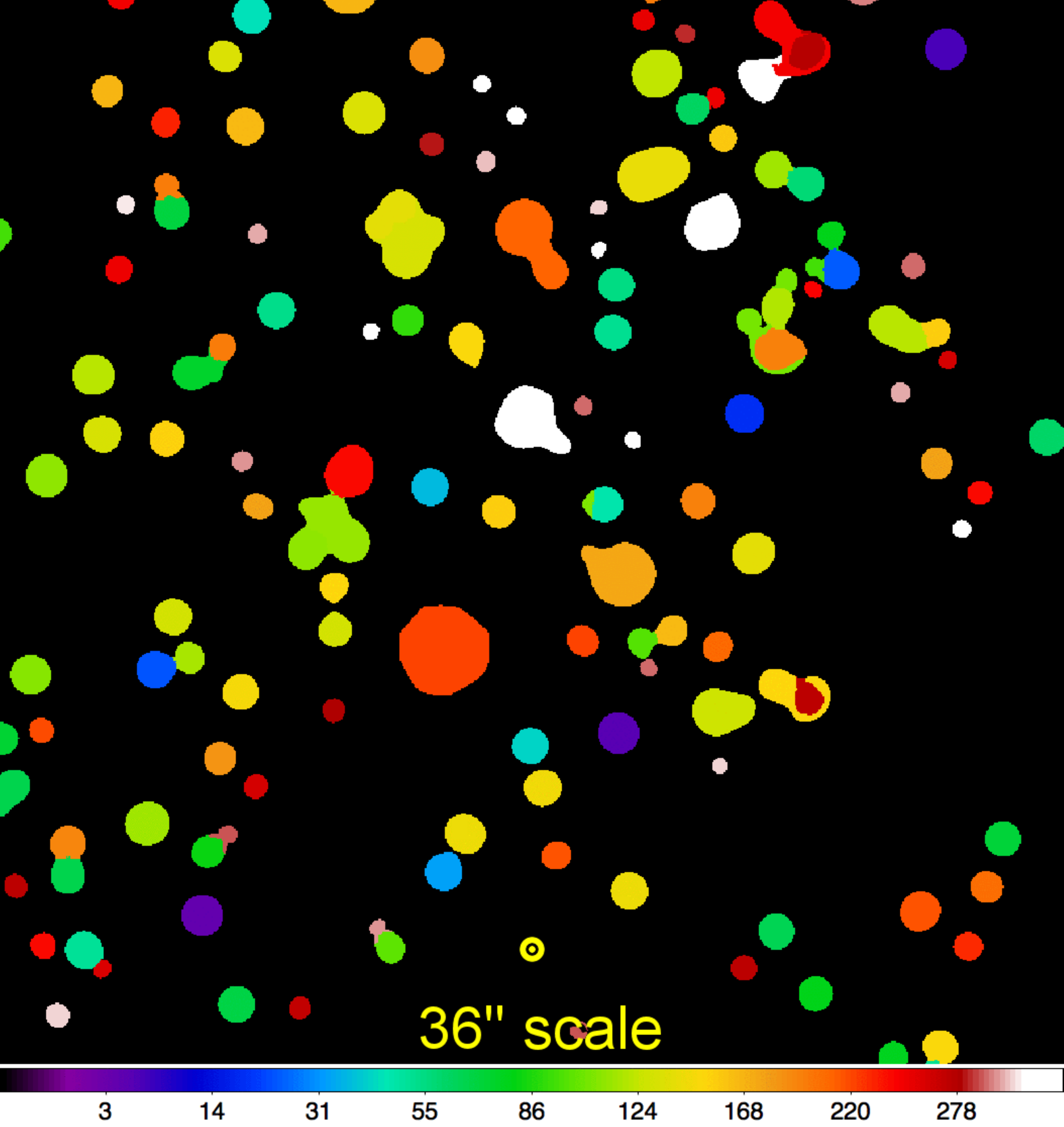}}
            \resizebox{0.33\hsize}{!}{\includegraphics{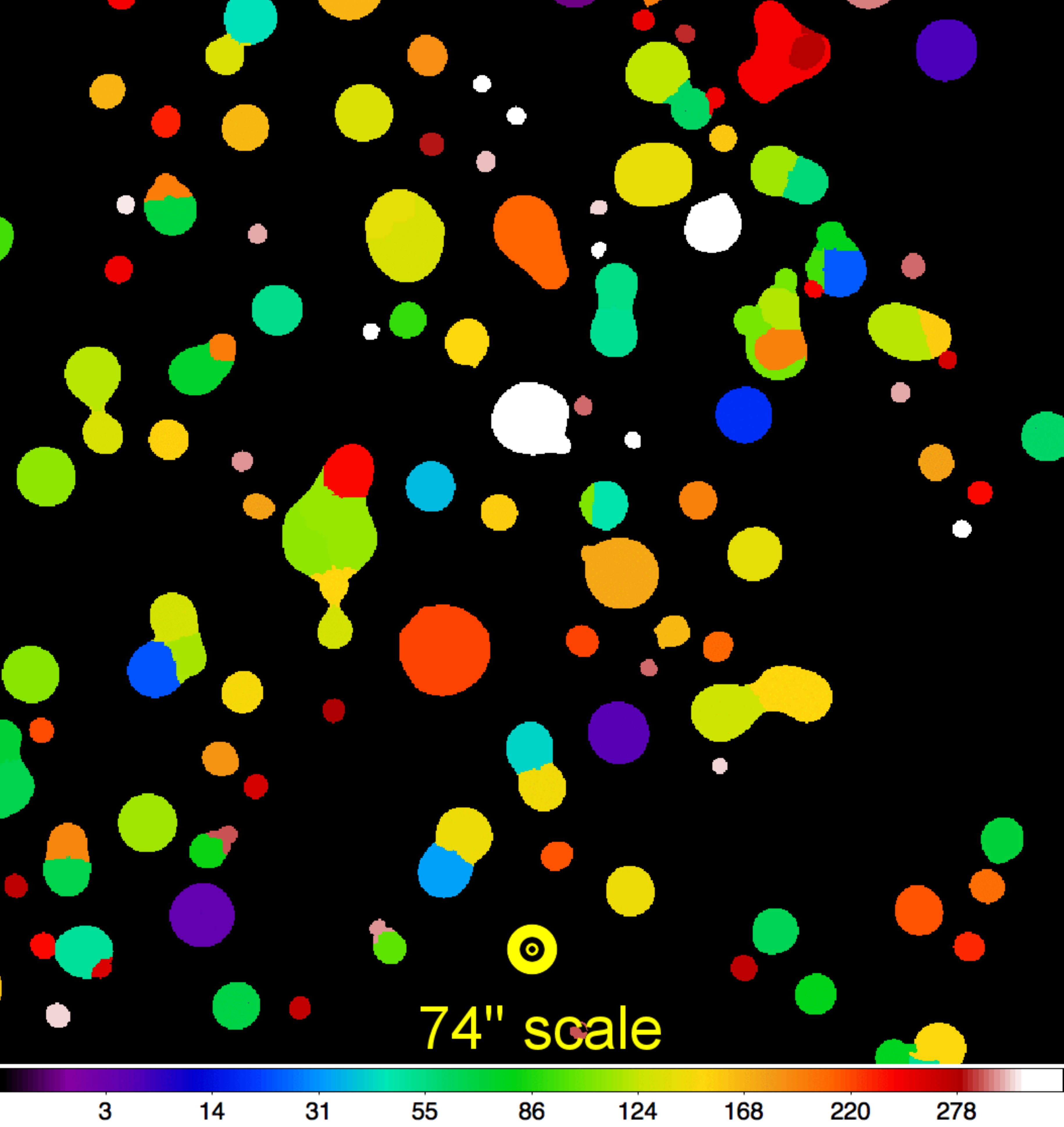}}
            \resizebox{0.33\hsize}{!}{\includegraphics{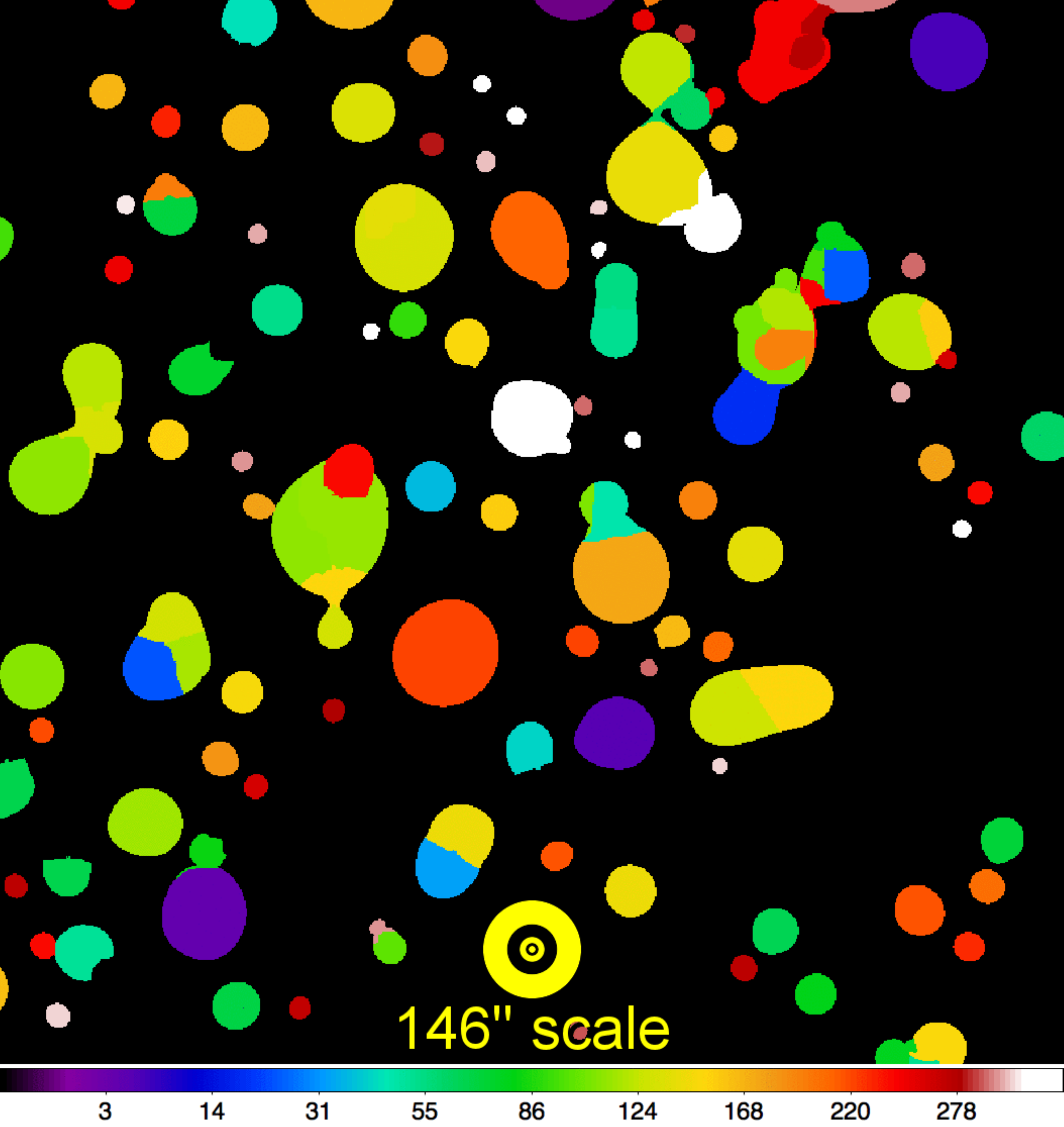}}}
\caption{
Single-scale source detection (Sect.~\ref{detecting.sources}). The field of Fig.~\ref{single.scales} is shown as the initial
footprint image $\mathcal{F}_{\rm D}$ after the source detection (\emph{upper left}). The same set of spatial scales is displayed in
the single-scale segmentation images $\mathcal{I}_{{\rm D}{j}{\,\rm S}}$ (\emph{left to right}, \emph{top to bottom}) showing the
source segmentation masks determined from $\mathcal{I}_{{\rm D}{j}{\,\rm C}}$ (Fig.~\ref{combined.clean.single.scales}). The masks
were obtained and analyzed by the detection procedure described in Sect.~\ref{detecting.sources}. The scale sizes $S_{\!j}$ are
visualized by the yellow-black circles and annotated at the bottom of the panels. The color coding is a function of the square root
of the source number (which makes up the actual pixel values in these segmentation images).
} 
\label{segmentation.images}
\end{figure*}

It is necessary to define two sets of wavelength-combined single-scale images (illustrated in Fig.~\ref{combining}) for different
purposes. The images $\mathcal{I}_{{\rm D}{j}{\,\rm C}}$ are used to \emph{detect} sources and track the evolution of their
\emph{shapes} across all spatial scales (see Sect.~\ref{detecting.sources}) and the images $\mathcal{I}^{\prime}_{{\rm D}{j}{\,\rm
C}}$ are used to follow the dependence of the \emph{peak intensities} of detected sources on scales. The first set of combined
\emph{detection} images is normalized such that all cleaning thresholds become equal to 1 in each band:
\begin{equation}
\mathcal{I}_{{\rm D}{j}{\,\rm C}}{\,=\,}\frac{1}{N}\sum_{\lambda}\frac{f_{{\lambda}{j}}\,}{\varpi_{{\lambda}{j}}}
\max \left\{ \mathcal{I}_{{\!\lambda}{\rm D}{j}{\,\rm C}}, \mathcal{T}_{{\!\lambda}{j}} \right\},
\label{combining.waves.1} 
\end{equation} 
where $\mathcal{T}_{{\!\lambda}{j}}$ is the threshold image (all pixels of which are equal to $\varpi_{{\lambda}{j}}$), $N$ is the
number of wavelengths, and $f_{{\lambda}{j}}$ gradually ``turns on'' the scales smaller than the observational beam $O_{\lambda}$
(for larger scales, $f_{{\lambda}{j}}{\,=\,}$1):
\begin{equation}
f_{{\lambda}{j}}{\,=\,}\left(\frac{S_{\!j}}{O_{\lambda}}\right)^3, \,\,\,
\max \left\{S_{\!1}, 0.33\,O_{\lambda}\right\}{\,\le\,}S_{\!j}{\,\le\,}O_{\lambda}.
\label{normalization}
\end{equation}
The renormalization of images, the threshold $\varpi_{{\lambda}{j}}$ in the curly brackets, and the turn-on factor
$f_{{\lambda}{j}}$ ensure that the images from different wavebands become smoothly combined in $\mathcal{I}_{{\rm D}{j}{\,\rm C}}$
(Eq.~\ref{combining.waves.1}) with no discontinuities and that there are no large changes in the combined images between any two
adjacent scales (Fig.~\ref{combining}, \emph{left}).

The normalization of $\mathcal{I}_{{\!\lambda}{\rm D}{j}{\,\rm C}}$ to a common threshold before summing them up is the most
natural way of combining wavelengths to maximize sensitivity. It removes, however, the natural dependence of the source brightness
on scales, which is used by our detection algorithm to determine the characteristic scale and initial footprint for each source
(Sect.~\ref{detecting.sources}). Our second set of combined images allows that, as they are normalized only at the smallest scale
(there is no reason to use the factor $1/N$ here):
\begin{equation}
\mathcal{I}^{\prime}_{{\rm D}{j}{\,\rm C}}{\,=\,}\sum_{\lambda}
\frac{w_{\lambda}}{I^{\,\rm max}_{{\!\lambda}{\rm D}{1}{\,\rm C}}}\,\mathcal{I}_{{\!\lambda}{\rm D}{j}{\,\rm C}},
\label{combining.waves.2}
\end{equation}
where $I^{\,\rm max}_{{\!\lambda}{\rm D}{j}{\,\rm C}}$ is the maximum intensity over
$\mathcal{I}_{{\!\lambda}{\rm D}{j}{\,\rm C}}$ and the weight $w_{\lambda}$ enhances contributions of the images with higher
angular resolutions:
\begin{equation}
w_{\lambda}{\,=\,}\left(\frac{\bar{O}}{O_{\lambda}}\right)^{\gamma},
\label{weighting}
\end{equation}
where $\bar{O} = N^{-1} \sum_{\lambda} O_{\lambda}$ is the average observational beam size and $\gamma{\,>\,}1$ is a weighting
exponent with a default value of 6. The aim of such weighting is to effectively separate the contributions of different wavebands
over the intensities when computing $\mathcal{I}^{\prime}_{{\rm D}{j}{\,\rm C}}$ (Eq.~\ref{combining.waves.2}). After the
weighting, the summation of $\mathcal{I}_{{\!\lambda}{\rm D}{j}{\,\rm C}}$ practically does not alter their individual intensity
profiles (Fig.~\ref{combining}, \emph{right}). If a source exists in some high-resolution single-scale images at a certain
wavelength, the dependence of its peak intensity across scales is fully preserved in $\mathcal{I}^{\prime}_{{\rm D}{j}{\,\rm C}}$.
In the detection process, the source position, characteristic size, and footprint are largely determined by the high-resolution
images and the spatial scale where it becomes the brightest (Sect.~\ref{detecting.sources}). The initial size and footprint obtained
in the detection step are recomputed during the measurement iterations (Sect.~\ref{measuring.cataloging}).

Figure~\ref{combined.clean.single.scales} illustrates our method of combining wavelengths. The upper-left panel shows an image of
single-scale masks accumulated over scales and all \emph{Herschel} wavebands, obtained from the clean single-scale images:
\begin{equation}
\mathcal{M}_{{\rm D}{\rm C}}{\,=\,}\sum_{\lambda}\sum_{j} 
\frac{1}{\varpi_{{\lambda}{j}}}\min\,\left\{\mathcal{I}_{{\!\lambda}{\rm D}{j}{\,\rm C}},\mathcal{T}_{{\!\lambda}{j}}\right\}
= \sum_{\lambda}\sum_{j} \frac{\mathcal{I}_{{\!\lambda}{\rm D}{j}{\,\rm C}}}{\mathcal{I}_{{\!\lambda}{\rm D}{j}{\,\rm C}}},
\label{accumulated.masks}
\end{equation}
where $\mathcal{T}_{{\!\lambda}{j}}$ is the threshold image and, of course, the division of the image by itself can only be done in
non-zero pixels. The accumulated mask image presents a cumulative view of how the sources made visible by cleaning change their
shape and size across all scales and wavelengths. All sources that can possibly be found in the clean images have left their mark
in $\mathcal{M}_{{\rm D}{\rm C}}$. The other panels of Fig.~\ref{combined.clean.single.scales} display the combined detection
images $\mathcal{I}_{{\rm D}{j}{\,\rm C}}$ for several single scales, which accumulate information from all wavebands. Comparing
them with the corresponding panels of Fig.~\ref{clean.single.scales}, one can verify that each combined scale is indeed populated by
more sources than any of the clean single scales at individual wavelengths.

\subsection{Detecting sources in combined single-scale images}
\label{detecting.sources}

\begin{figure}
\centering
\centerline{\resizebox{0.7\hsize}{!}{\includegraphics{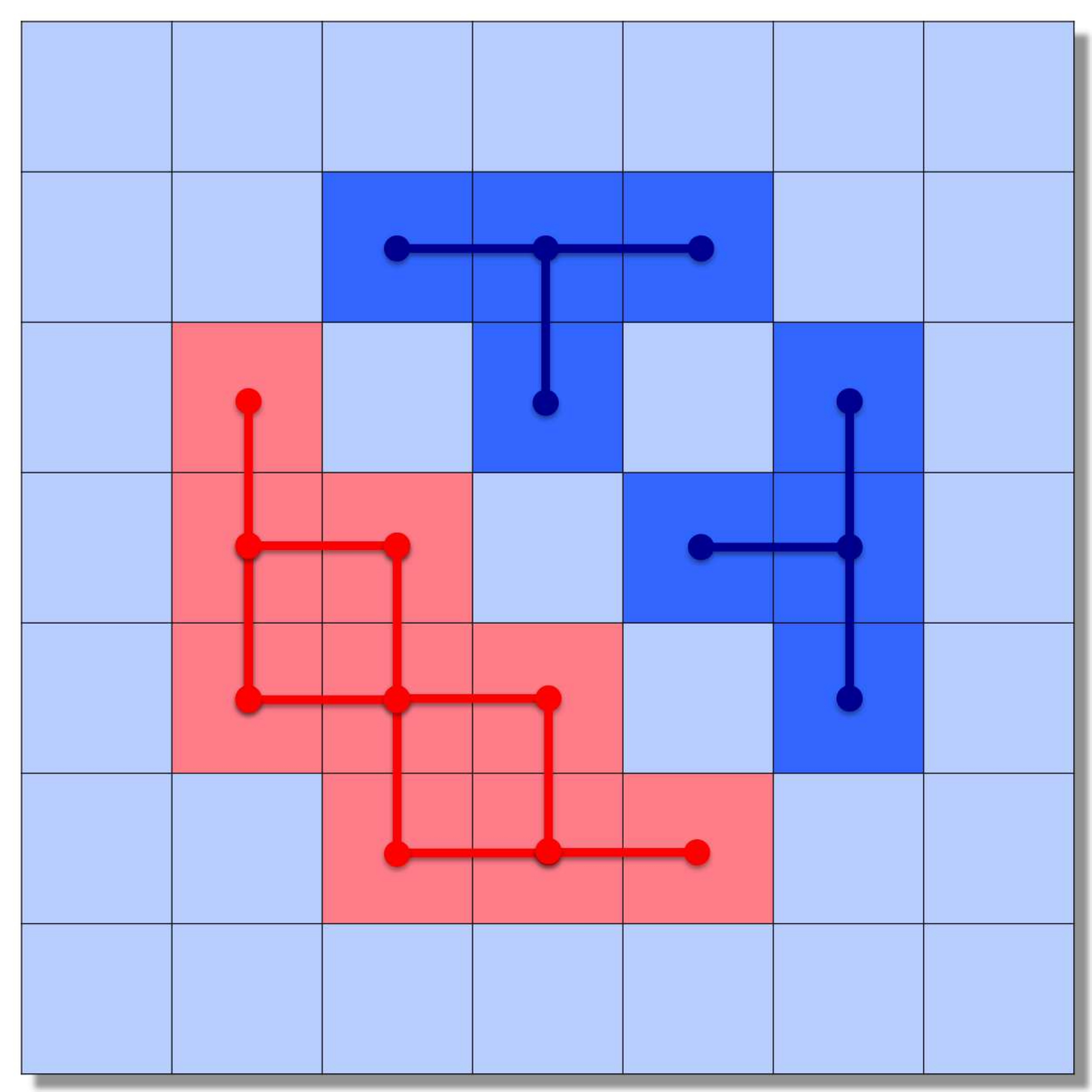}}} 
\caption{
Topology of the pixel connections (Sect.~\ref{detecting.sources}). Pixels of the red shape are 4-connected and thus the shape may
represent a source. Pixels of the dark blue shape are \emph{not} 4-connected: there are no paths connecting any pair of the blue
pixels such that any neighboring pixels in the paths are connected to each other only by their sides. Note, however, that there are
two 4-connected sub-areas in the blue shape, which could represent sources.
}
\label{connected.pixels}
\end{figure}

As can be seen in Figs.~\ref{clean.single.scales} and \ref{combined.clean.single.scales}, sources ``evolve'' from small to large
scales in single-scale images. In the clean images $\mathcal{I}_{{\rm D}{j}{\,\rm C}}$, the sources appear at some relatively small
scales (smaller than their actual size), become the brightest at a scale roughly equal to their size, and eventually vanish at
significantly larger scales (see Fig.~\ref{zoomed.clean.images} for an illustration). The idea of our source detection scheme is to
analyze all $\mathcal{I}_{{\rm D}{j}{\,\rm C}}$ ($j{\,=\,}1,\dots, N_{\rm S}$), identifying the sources and tracking the
``evolution'' of their shapes from small to large scales. For that purpose, \textsl{getsources} employs the \textsl{Tint Fill}
algorithm \citep{Smith_1979}\footnote{Available at http://portal.acm.org/citation.cfm?id=800249.807456}, developed for coloring
arbitrary shapes in digital images. The algorithm assumes that the sets of pixels belonging to any shape are \emph{4-connected},
i.e. that for any pair of pixels $\Pi_l$ and $\Pi_m$ in the shape, there is a path from $\Pi_l$ to $\Pi_m$ through pixels in that
shape, such that neighboring pixels in the path are connected to each other only by their sides (Fig.~\ref{connected.pixels}). Given
a set of 4-connected (side-connected) pixels, each one having the same property (e.g., color), the algorithm fills all (and only)
pixels of that shape with a new value\footnote{Identification of distinct connected regions in similar algorithms is also known as
\emph{connected-component labeling}.}. The algorithm has been slightly generalized to be suitable for the source detection in
single-scale images by replacing color with another pixel property of having a non-zero intensity.

The modified version of the \textsl{Tint Fill} algorithm looks, at each scale $(j{\,=\,}1, 2,\dots, N_{\rm S})$, for 4-connected
areas of non-zero pixels in $\mathcal{I}_{{\rm D}{j}{\,\rm C}}$ and assigns the value of the running source number $i$ to each of
those connected pixels, producing a \emph{segmentation} image $\mathcal{I}_{{\rm D}{j}{\,\rm S}}$. The latter consists of the
segmentation \emph{masks} of the sources found at scale $j$ and previous smaller scales (\textsl{getsources} always analyzes
single-scale images from small to large scales). A source mask is the area of 4-connected pixels (with values $i$) in
$\mathcal{I}_{{\rm D}{j}{\,\rm S}}$, all those pixels having non-zero intensity in $\mathcal{I}_{{\rm D}{j}{\,\rm C}}$ at $j$ or
at smaller scales. The masks uniquely identify the sources and they allow tracking them across all single scales
(Fig.~\ref{segmentation.images}).

In order to better disentangle and follow the evolution of blended sources in $\mathcal{I}_{{\rm D}{j}{\,\rm C}}$, the algorithm
splits the images between their maximum and $\varpi_{j}$ by a number of intensity levels $\omega_{j\,l}$ with a spacing
$\delta\ln\omega_{j\,l}{\,=\,}0.1$ or smaller. At each scale $j$, our source detection procedure works on a sequence of ``partial''
images $\mathcal{I}_{{\rm D}{j}{\,\rm C}{\,l}} = \max\,\{\mathcal{I}_{{\rm D}{j}{\,\rm C}},\omega_{j\,l}\}$, from top to the
bottom, where the last partial image is the entire single-scale image $\mathcal{I}_{{\rm D}{j}{\,\rm C}}$. For better efficiency,
we only consider those levels $\omega_{j\,l}$ that increase the number of pixels in $\mathcal{I}_{{\rm D}{j}{\,\rm C}{\,l}}$ with
respect to $\mathcal{I}_{{\rm D}{j}{\,\rm C}{\,l-1}}$ by at least
\begin{equation}
{\delta}N^{\,\rm min}_{{\Pi}{j}} = \max \left\{ \frac{1}{3}\,\pi \left(\frac{S_{\!j}}{2 \Delta}\right)^{2}\!,
{\delta}N^{\,\rm min}_{\Pi} \right\},
\label{minpix.level}
\end{equation}
where $\Delta$ is the pixel size and ${\delta}N^{\,\rm min}_{\Pi}{\,=\,}$9. The levels $\omega_{j\,l}$ with
${\delta}N_{{\Pi}{j}}{\,<\,}{\delta}N^{\,\rm min}_{{\Pi}{j}}$ are skipped by the detection algorithm. Processing a partial image
$\mathcal{I}_{{\rm D}{j}{\,\rm C}{\,l}}$ within scale $j$, \textsl{getsources} first gives the coordinates of the sources already
found at previous scales and levels to the modified \textsl{Tint Fill} algorithm and the latter fills the evolved shapes of the
``old'' sources with their numbers $i$. Then it checks all remaining pixels of $\mathcal{I}_{{\rm D}{j}{\,\rm C}{\,l}}$ to find new
4-connected areas of non-zero pixels that first appear at the current scale $j$; when found, the segmentation mask of each ``new''
source is filled with its new number $i$.

When an isolated source vanishes at a certain scale, its pixels are freed and may eventually become occupied by its neighbor or
another (significantly larger) source that may appear there at a larger scale. One of two touching sources can also disappear from
larger scales, when it becomes fainter at progressively larger scales and finally merges with the brighter neighbor (i.e., when the
saddle point between the peaks disappears). When two or more isolated sources become connected in a single-scale image, their
segmentation masks are not allowed to overlap. A boundary between two sources is maintained along the normal to the straight line
connecting their centers (the normal being defined at a position along the line where intensity is at minimum); the sources can
still grow at larger scales in the perpendicular direction. This boundary exists, however, only within a limited range of scales,
until one of the touching sources vanishes or merges with another. The algorithm is perfectly able to handle hierarchical
structures, as the segmentation masks \emph{can} overlap at \emph{different} scales for sources of significantly dissimilar
dimensions. Whether a larger source containing smaller sources is detected as such or considered as the background, depends on the
relative difference in the characteristic scales between the large and small sources, on the location of the small sources within
the larger structure, and on the relative brightness of the peaks. If the source sizes are significantly different, the hierarchical
structures are detected; a number of examples can easily be found in Fig.~\ref{segmentation.images} (\emph{upper-left}),
Fig.~\ref{composite.obsbs} (\emph{right}), and Fig.~\ref{extraction.ellipses} (\emph{bottom}).

One can show that a resolved isolated circular source $i$ with the FWHM sizes $A_{i}{\,=\,}B_{i}$ would have its maximum peak
intensity $I_{ij}$ in a single-scale image with a smoothing beam $S_{\!j}{\,\approx\,}A_{i}$. Recall that the spatial decomposition
(Eq.~\ref{successive.unsharp.masking}) is based on convolution; the latter acts as a natural selector of scales in the decomposed
images (cf. Sect.~\ref{decomposing.detection.images}). Indeed, convolving with small beams ($S_{\!j}{\,\ll\,}A_{i}$) would have
almost no effect on the source, whereas using extended beams ($S_{\!j}{\,\gg\,}A_{i}$) would greatly dilute the source. In both
these extremes, sucessive unsharp masking produces decreasing peak intensities (Fig.~\ref{zoomed.clean.images}) while creating the
strongest signal for the sources with sizes $A_{i}{\,\approx\,}S_{\!j}$; completely unresolved sources are the brightest at spatial
scales $S_{\!j}{\,\la\,}O_{\lambda}$.

The scale $j_{\rm F}$, where a source is the brightest, provides the best initial estimate for its actual FWHM size
$A_{i}\,({=\,}B_{i}){\,=\,}S_{\!j_{\,\rm F}}$. This \emph{footprinting} scale defines a source \emph{footprint}, i.e. the entire
area of the image pixels making non-negligible contribution to the total flux of the source. For unresolved sources, the footprints
cover circular areas $\approx {\pi} O_{\lambda}^2$, whereas for well-resolved sources with intrinsically Gaussian intensity
distributions, they cover elliptical areas $\approx {\pi} A_{i} B_{i}$. During the detection process, \textsl{getsources} creates
initial footprints (cf. Fig.~\ref{segmentation.images}, upper-left) with full sizes of $A_{{i}{\,\rm F}}\,({=\,}B_{{i}{\,\rm
F}}){\,=\,}1.15\, (2\,S_{\!j_{\rm F}}){\,=\,}2.3\,S_{\!j_{\rm F}}$, where the empirical factor 1.15 was found to improve results in
our tests. The footprints generally become elliptically-shaped in the measurement iterations (Fig.~\ref{converged.footprints}),
reflecting the source shapes obtained from intensity moments (Sect.~\ref{measuring.cataloging}).

In a multi-wavelength extraction, the combined images are sums of rescaled images over all individual wavebands
(Eq.~\ref{combining.waves.2}) using the weighting (Eq.~\ref{weighting}) that effectively separates
$\mathcal{I}_{{\!\lambda}{\rm D}{j}{\,\rm C}}$ in the combined detection images $\mathcal{I}^{\prime}_{{\rm D}{j}{\,\rm C}}$.
Although the weighting biases $S_{\!j_{\rm F}}$ towards higher-resolution images (makes the initial footprints smaller), this does
not affect the results, as the sizes found in the detection process are replaced with the values obtained from the monochromatic
images $\mathcal{I}_{{\!\lambda}{\rm D}{j}{\,\rm C}}$ during the measurement iterations (Sect.~\ref{measuring.cataloging}).

Coordinates of a source are computed from the moments of intensities (Appendix~\ref{intensity.moments}) in a limited range of
scales, only up to a scale 4 times larger than the one it first appeared at or up to the footprinting scale $j_{\rm F}$, whichever
is smaller. This gives more accurate positions than if recomputed at even larger scales, since the latter tend to mix in more and
more of the signals from the surroundings and thus distort the single-scale intensity distribution of sources. To improve the
positional accuracy even further, \textsl{getsources} uses only the pixels with values greater than $I_{ij} / 1.4$, where $I_{ij}$
is the peak intensity of a source $i$ and the number in the denominator has been found empirically.

In general, observed images contain structures at all scales and there is a real danger of creating spurious sources by mistakenly
detecting noise peaks or background fluctuations that happen to lie on top of larger structures. Although the latter may be
relatively faint, they tend to ``amplify'' the small-scale noise and background and make the small structures appear as significant
sources. This can be especially problematic if the \emph{local} uncertainties of the peak intensities
(Sect.~\ref{measuring.cataloging}) are not possible to estimate due to crowding, as the fluctuations outside the large structures
may be noticeably smaller and thus the signal-to-noise ratio (S/N) may easily be overestimated. 

There is a mechanism in \textsl{getsources} to greatly reduce this possibility. The idea is that small-scale peaks on top of larger
structures tend to survive up to larger scales than they would do otherwise, without the ``support'' of the underlying structures.
The latter tend to make the peaks eventually touch each other or merge at some spatial scale and this is used by the algorithm to
identify and discard insignificant intensity peaks. Noise peaks on top of larger structures become diluted if considered relative to
the intensity level contributed by the larger structure.

\begin{figure}
\centering
\centerline{\resizebox{0.7\hsize}{!}{\includegraphics{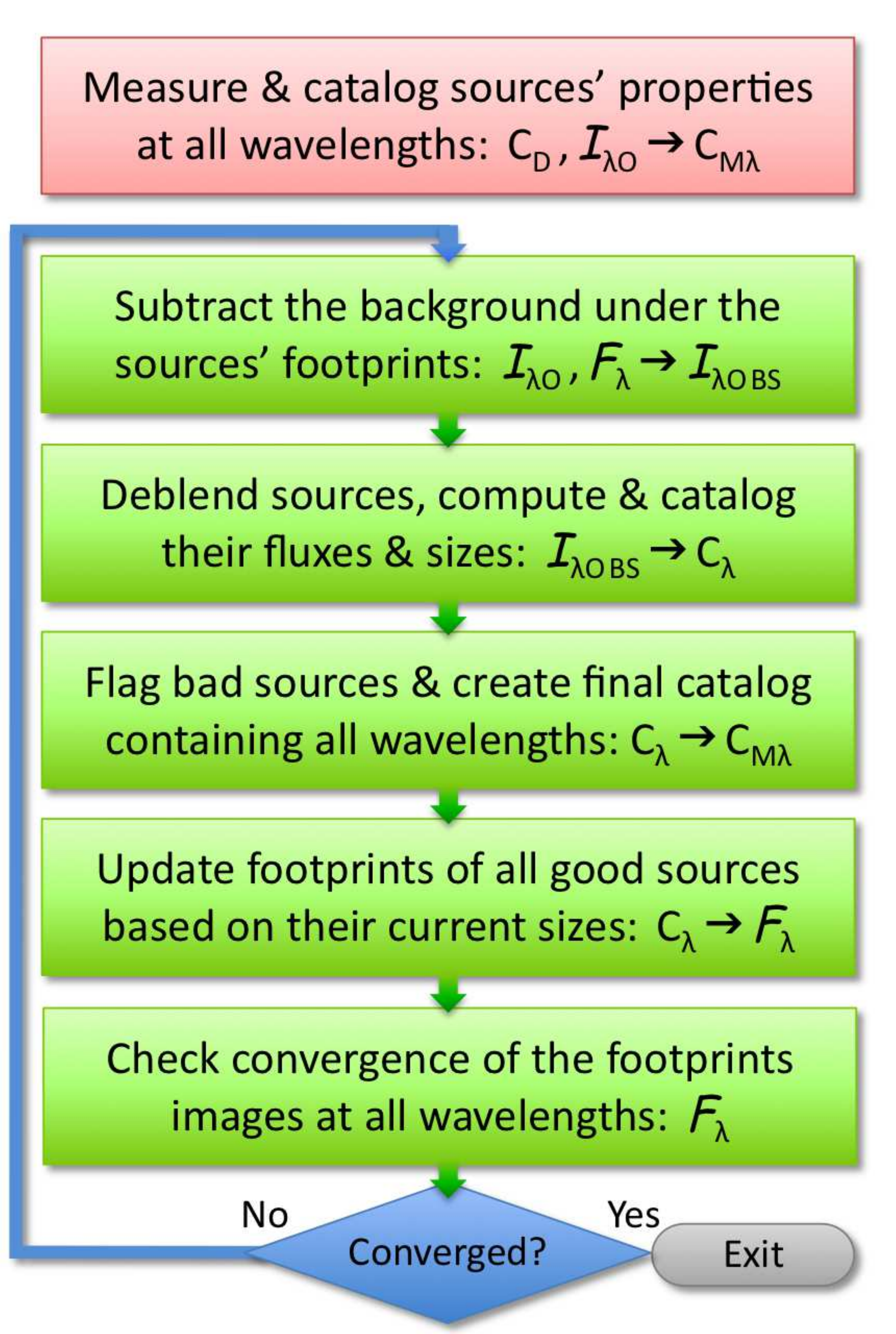}}} 
\caption{
Measurement iterations (Sect.~\ref{measuring.cataloging}). The processing block of measurements from Fig.~\ref{algorithm} is shown
at the top. It is sub-divided here in 5 algorithmic steps that are executed in iterations, until the footprint images
$\mathcal{F}_{\lambda}$ have converged, i.e. stable distributions of source footprints at each wavelength have been obtained. The
deblending block itself includes iterations to disentangle contributions of many overlapping sources to the intensity of a pixel
that belongs to all of them.
}
\label{measurements}
\end{figure}

When \textsl{getsources} analyzes the intensity levels $\omega_{j\,l}$ of the partial images
$\mathcal{I}_{{\rm D}{j}{\,\rm C}{\,l}}$, it finds the level where two or more sources first touch each other and computes the
contrast $C_{i}{\,=\,}I_{ij}/\omega_{j\,l}$, where $I_{ij}$ is the source maximum intensity in the image. The contrast of
touching sources is checked to decide whether the sources should be treated as significant ones. The basis for this decision
is the fact that when an insignificant source touches another source (either real or spurious), its contrast becomes quite low due
to the underlying intensity of a larger structure. In practice, \textsl{getsources} requires that any real source must have its
contrast $C_{i}$ above the minimum value
\begin{equation}
C^{\,\rm min}_{i}{\,=\,}\eta \max \left\{ C_{i\,\rm A}{\,}C_{i\,\rm E}, 1 \right\},
\label{minimum.contrast}
\end{equation}
where $\eta$ is the configuration parameter of \textsl{getsources} (with the default value of 1.35), $C_{i\,\rm A}$ is the
\emph{amplification} factor, and $C_{i\,\rm E}$ is the \emph{elongation} factor:
\begin{equation}
C_{i\,\rm A}{\,=\,}\max \left\{ \frac{1}{2} \left(\frac{F_{i\,\rm lo}}{F_{i\,\rm hi}}\right), 1 \right\}, \,\,
C_{i\,\rm E}{\,=\,}\max \left\{ \frac{1}{2.5} \left(\frac{a_{i}}{b_{i}}\right), 1 \right\},
\label{amplification.elongation}
\end{equation}
where $F_{i\,\rm lo}$ is the flux integrated over the source segmentation mask below the current intensity level $\omega_{j\,l}$ (at
which the source first touches another source) and $F_{i\,\rm hi}$ is the flux integrated above that level; $a_{i}$ and $b_{i}$ are
the major and minor lengths of the source segmentation mask. These factors describe two aspects of the behavior of noise peaks on
top of larger structures: the amplification factor increases with a stronger contribution of the underlying structure, whereas the
elongation factor becomes large when the structure has filamentary shape. For $C_{i\,\rm A}\,C_{i\,\rm E}{\,\le\,}1$, the source
is considered real if its contrast $C_{i} \ge \eta$ (Eq.~\ref{minimum.contrast}); larger values of the product $C_{i\,\rm
A}\,C_{i\,\rm E}$ raise the ``barrier'' higher. For $C_{i} < C^{\,\rm min}_{i}$ the current source is flagged as spurious and its
segmentation pixels are freed to be used by more prominent sources. If some or all of the touching peaks have contrasts above
$C^{\,\rm min}_{i}$, they survive the test and their evolution is followed further to larger scales.

\begin{figure}
\centering
\centerline{\resizebox{0.7\hsize}{!}{\includegraphics{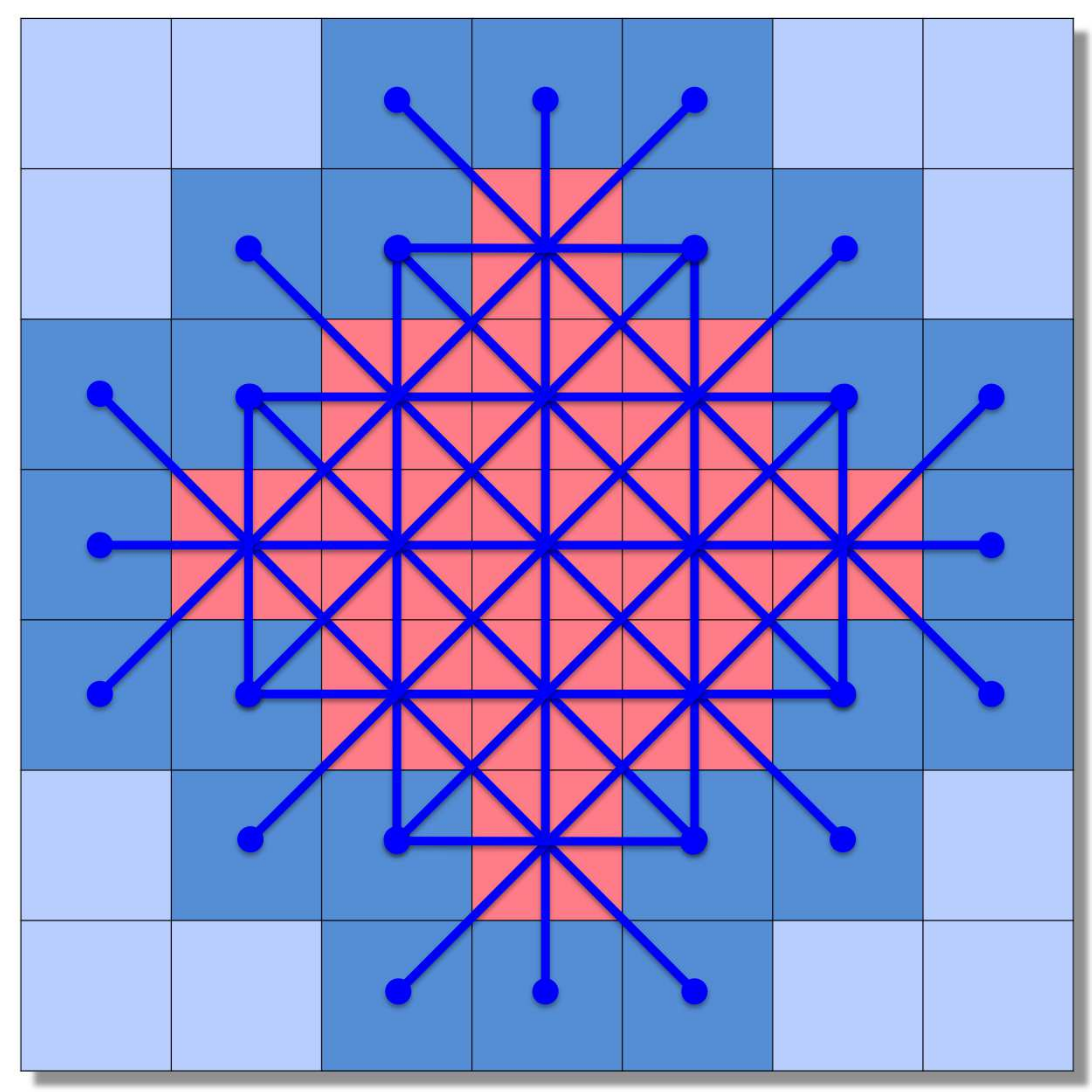}}} 
\caption{
Background interpolation scheme (Sect.~\ref{measuring.cataloging}). The central red pixels belong to the source, defining its
footprint in this example, whereas the surrounding blue pixels are those of the background. At each pixel of the source, the
background is linearly interpolated in the four main directions based on the values of the pixels just outside the footprint
(highlighted in darker blue), the resulting four values per pixel being averaged. Such interpolation probes the actual background
variations around the sources, and thus the interpolated background is more complex and realistic than just a planar surface.
}
\label{interpolation}
\end{figure}

Having detected all sources in the combined images $\mathcal{I}_{{\rm D}{j}{\,\rm C}}$ over all spatial scales of interest and
collected the information in a detection catalog, \textsl{getsources} now returns to the actual observed images
$\mathcal{I}_{{\!\lambda}{\rm O}}$ in each waveband for measurements of the basic properties of the sources, such as their peak
intensity (flux), total (integrated) flux, and size.

\subsection{Measuring and cataloging properties of sources}
\label{measuring.cataloging}

All measurements are performed in the observed images $\mathcal{I}_{{\!\lambda}{\rm O}}$ for known positions $(x_i,y_i)$ of all
sources, obtained in the detection process (Sect.~\ref{detecting.sources}) and not recomputed anymore\footnote{Positions were
derived using filtered detection images and recomputing them from the observed images contaminated by irrelevant spatial scales
would not make them more accurate.}. The measurements must be done together with the background subtraction, deblending, and
improvements of the footprints; these are non-trivial interconnected problems that require iterations
(Fig.~\ref{measurements}).

\begin{figure*}
\centering
\centerline{\resizebox{0.45\hsize}{!}{\includegraphics{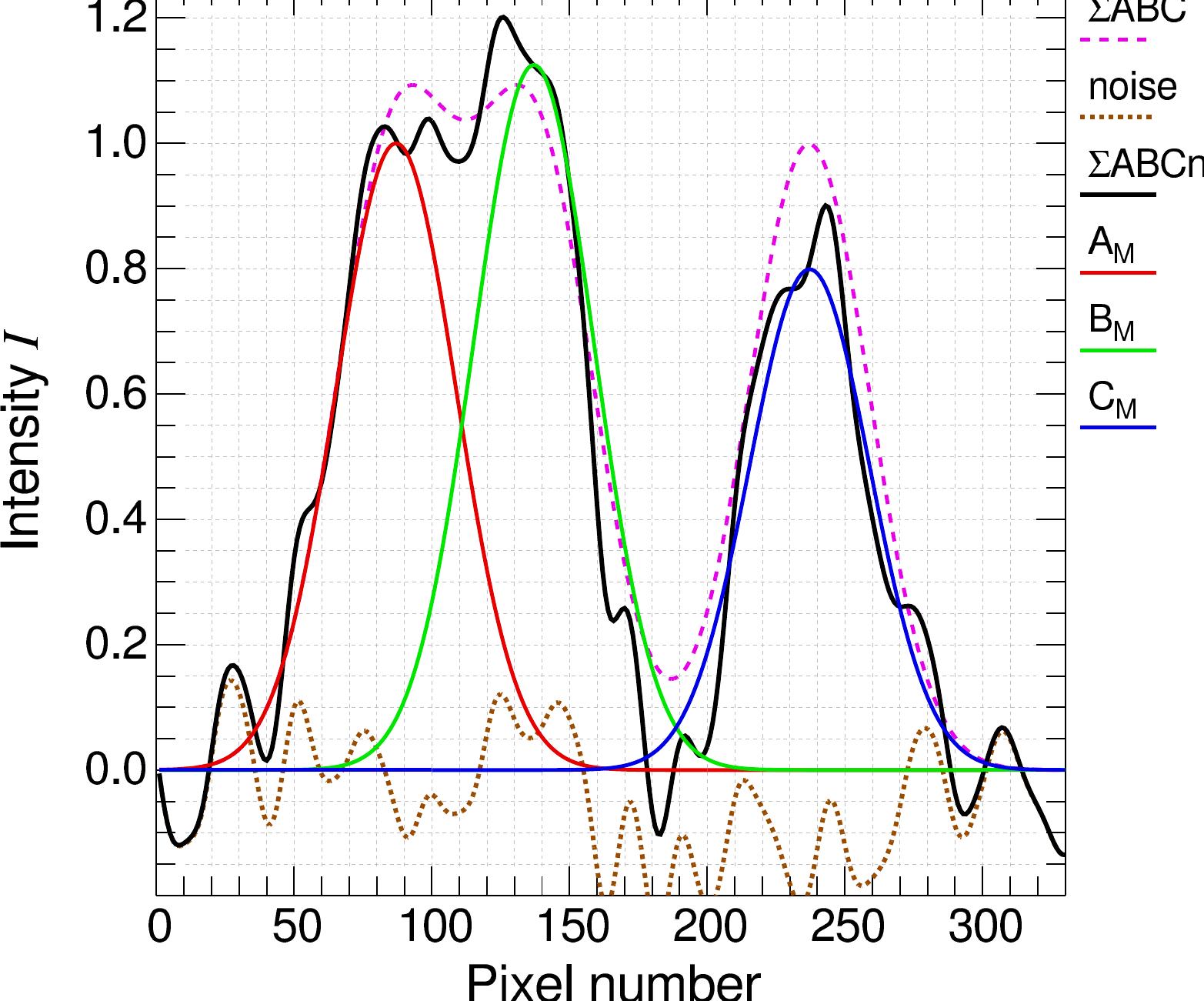}}
            \resizebox{0.45\hsize}{!}{\includegraphics{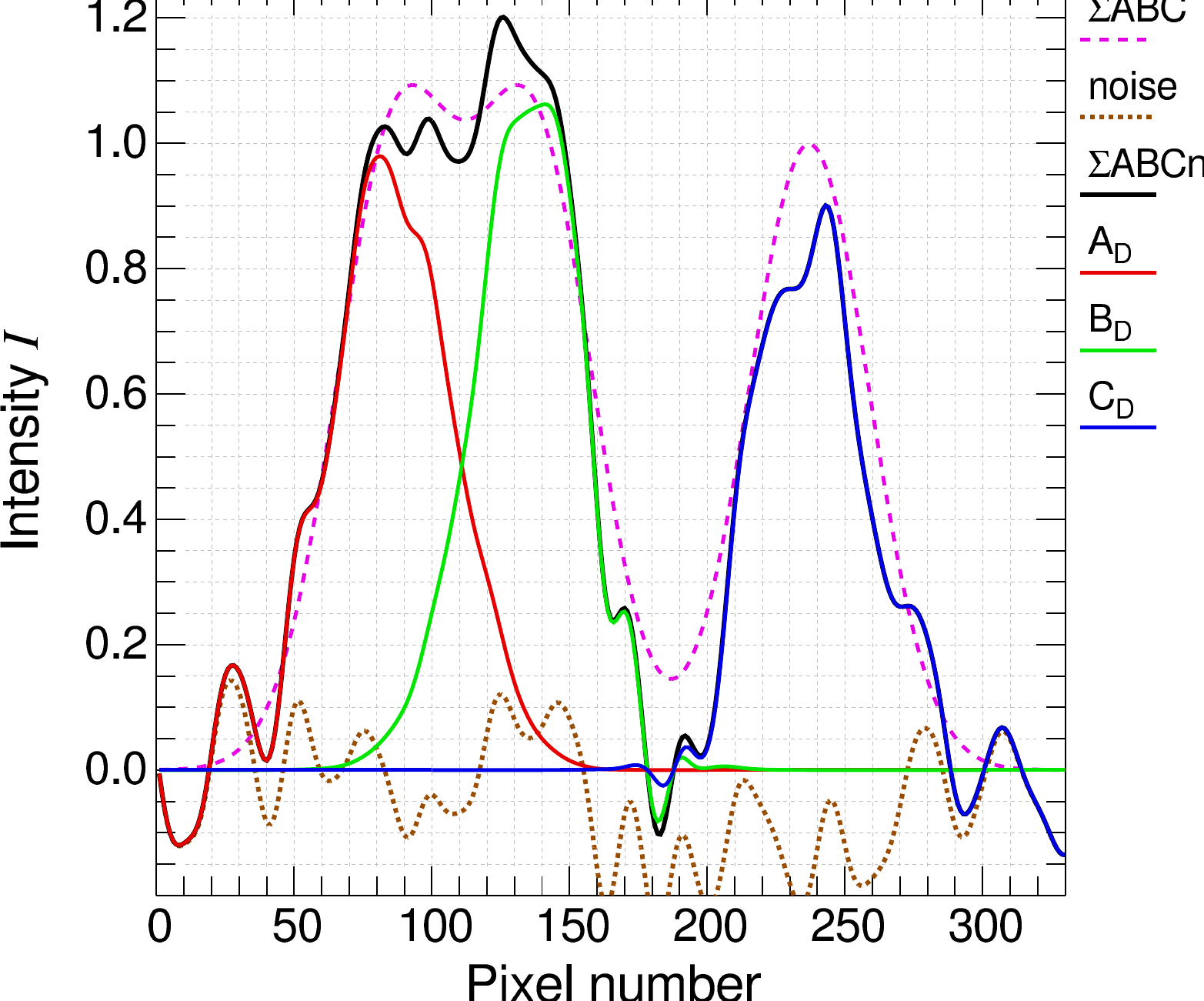}}}
\caption{
Deblending overlapping sources (Sect.~\ref{measuring.cataloging}). Three identical sources (A, B, C) with the same sizes and with
peak intensities normalized to unity are overlapping with their footprints. For clarity of the figure, the sources are not shown
with their individual profiles, but rather with their blended intensity distribution $\Sigma{\rm ABC}$; with noise added, the
profile transforms into $\Sigma{\rm ABCn}$. The deblending profiles ${\rm A_M}$, ${\rm B_M}$, and ${\rm C_M}$
(Eq.~\ref{moffat.function}) are defined at the source positions by the sizes and peak intensities estimated from the
background-subtracted image $\mathcal{I}_{{\!\lambda}{\rm O}{\,\rm BS}}$. At each pixel, the deblending algorithm splits its
intensity $I_{\Sigma{\rm ABCn}}$ between each of the overlapping sources, according to the fraction of the shape's intensity
(Eq.~\ref{deblending.formula}). The resulting deblended profiles of each source, ${\rm A_D}$, ${\rm B_D}$, and ${\rm C_D}$, are
shown in the right panel.
}
\label{deblending}
\end{figure*}

\begin{figure*}
\centering
\centerline{\resizebox{0.305\hsize}{!}{\includegraphics{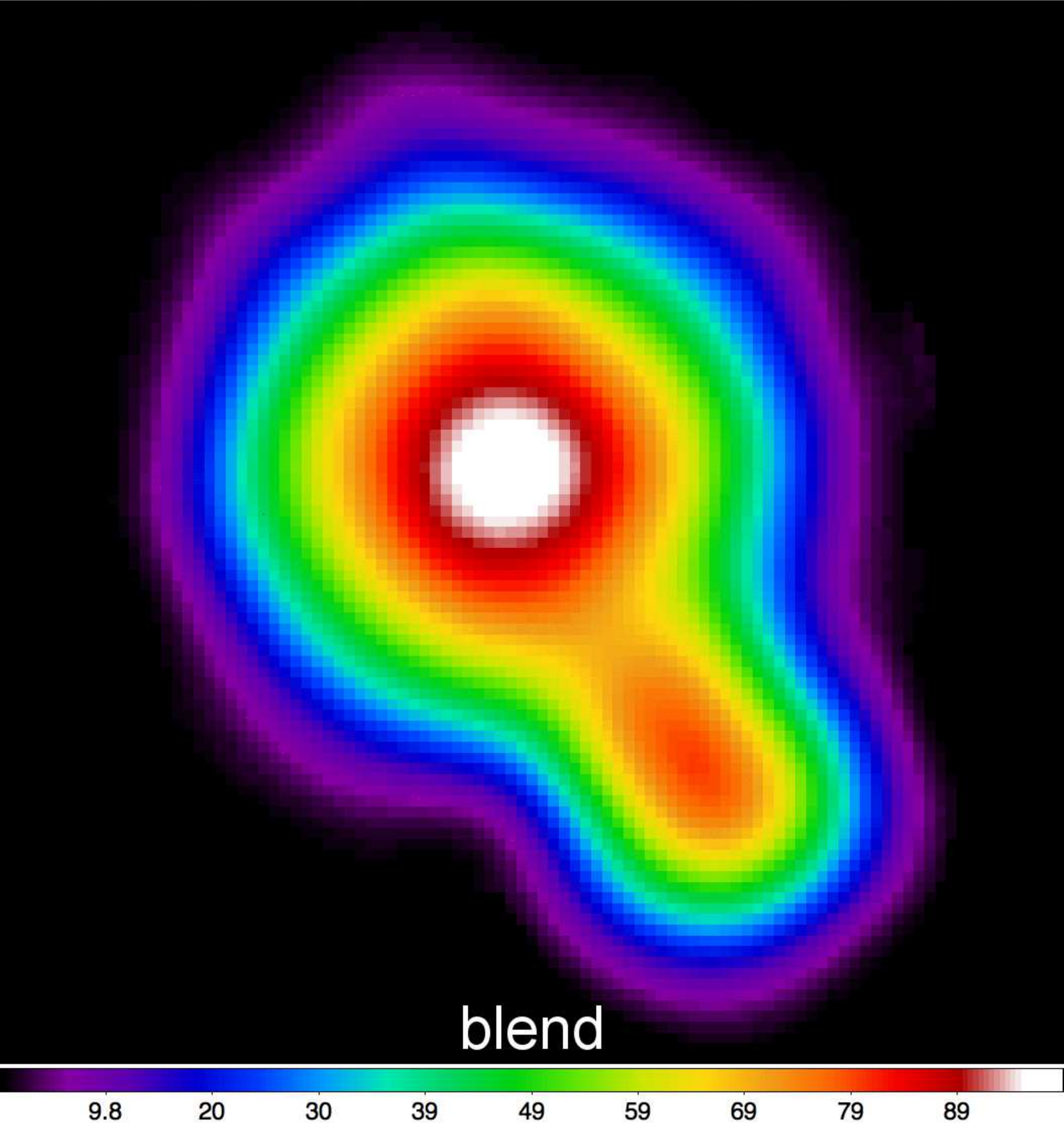}}
            \resizebox{0.305\hsize}{!}{\includegraphics{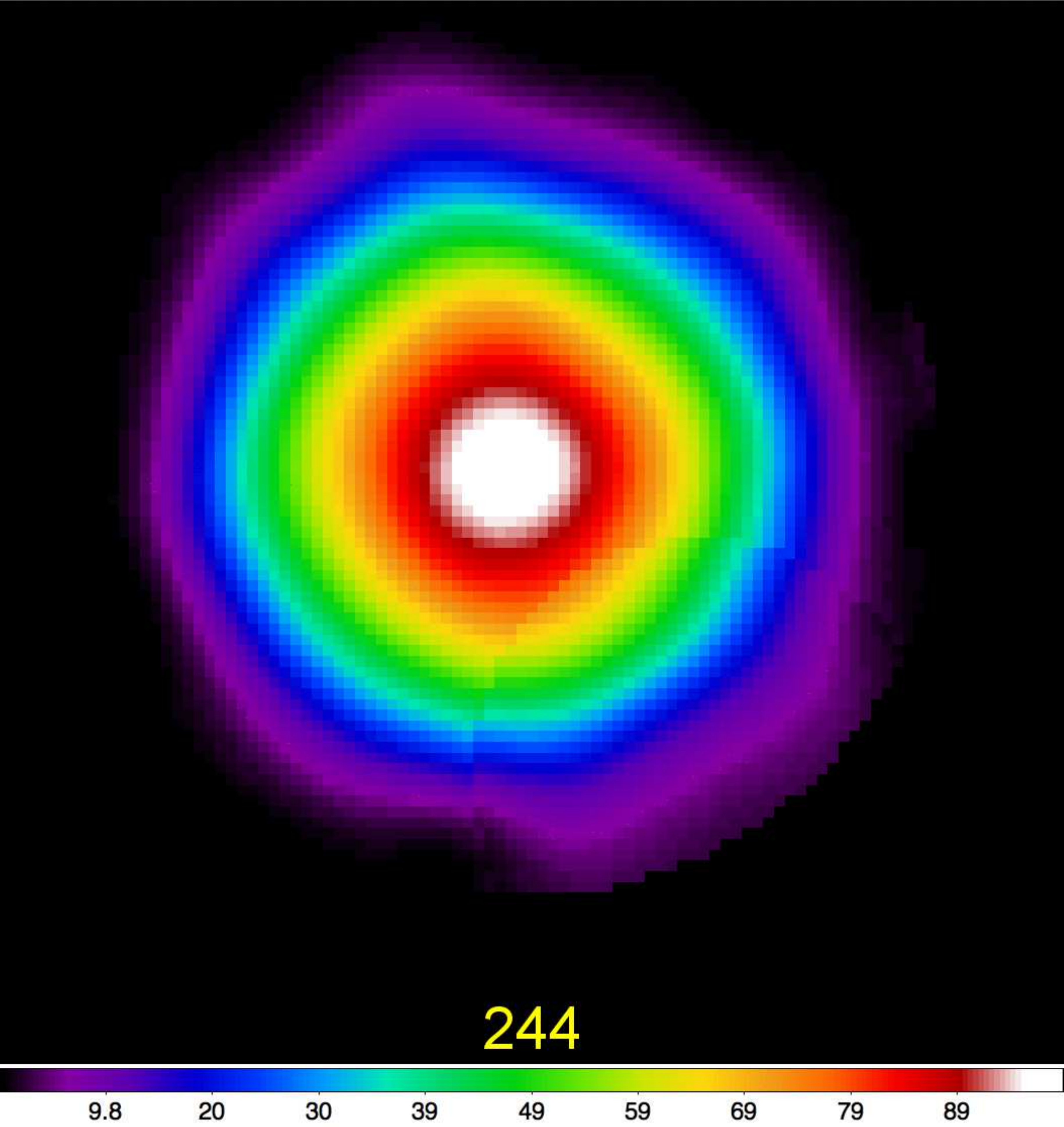}}
            \resizebox{0.305\hsize}{!}{\includegraphics{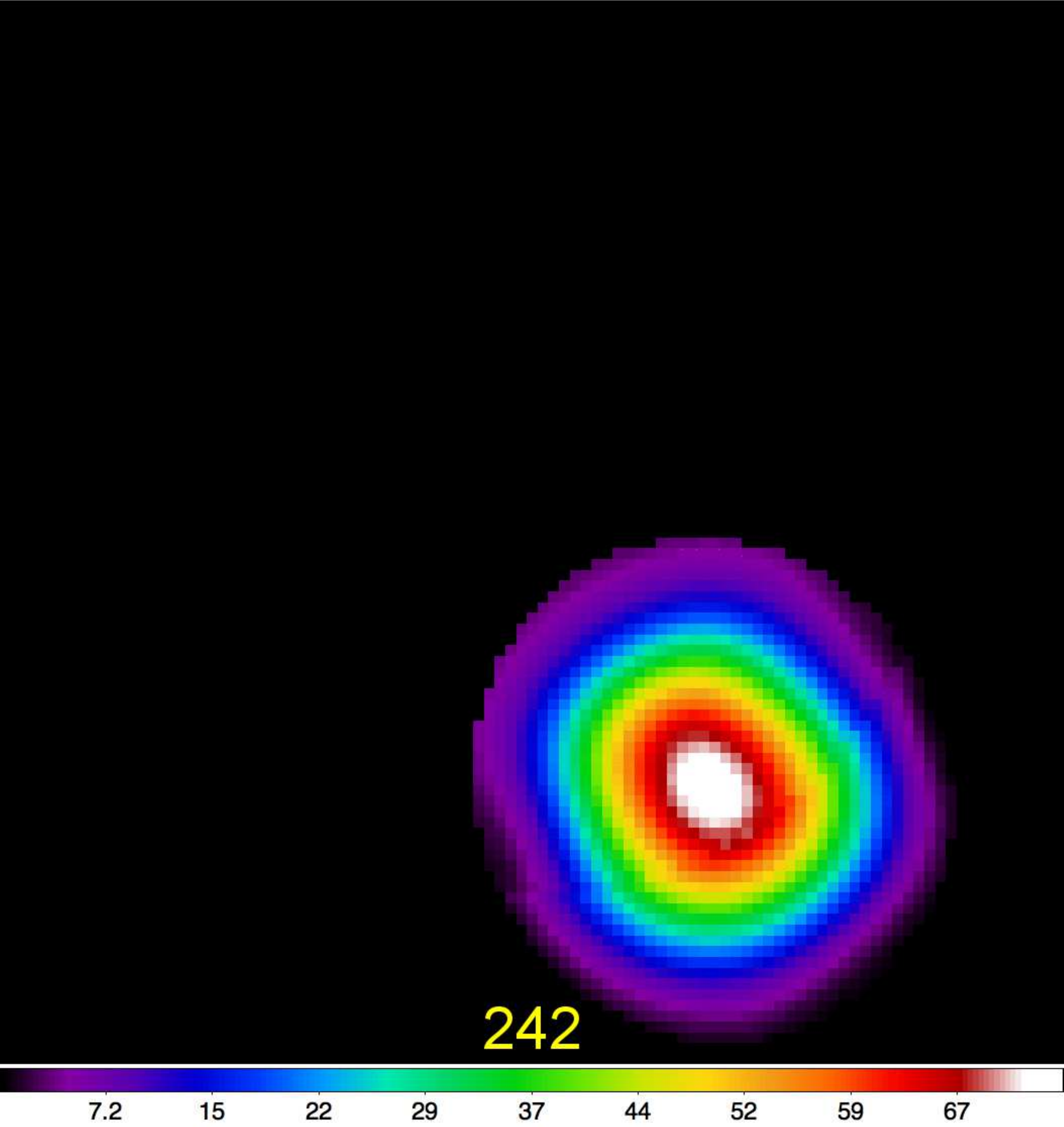}}}
\caption{
Deblending overlapping sources (Sect.~\ref{measuring.cataloging}). Background-subtracted overlapping sources at 350\,{${\mu}$m}
(\emph{left}) are separated into two individual sources, cataloged under the numbers 244 and 242 (\emph{middle}, \emph{right}). 
This blended pair is clearly visible half-way north from the field centers in Figs.~\ref{single.scales}, \ref{clean.single.scales},
\ref{combined.clean.single.scales}, \ref{segmentation.images}, \ref{composite.obsbs}, \ref{extraction.ellipses}; the annulus around
the source 244 is highlighted in Fig.~\ref{bigfoot.images}. Comparison with the known true model parameters shows that the peak
intensities measured for these deblended sources are in error by $-1.1$\% and $-5.1$\% and the total fluxes were calculated with
errors of $-0.5$\% and $-12$\%, respectively. The color coding is a linear function of intensity in MJy/sr.
}
\label{deblended.images}
\end{figure*}

\begin{figure*}
\centering
\centerline{\resizebox{0.33\hsize}{!}{\includegraphics{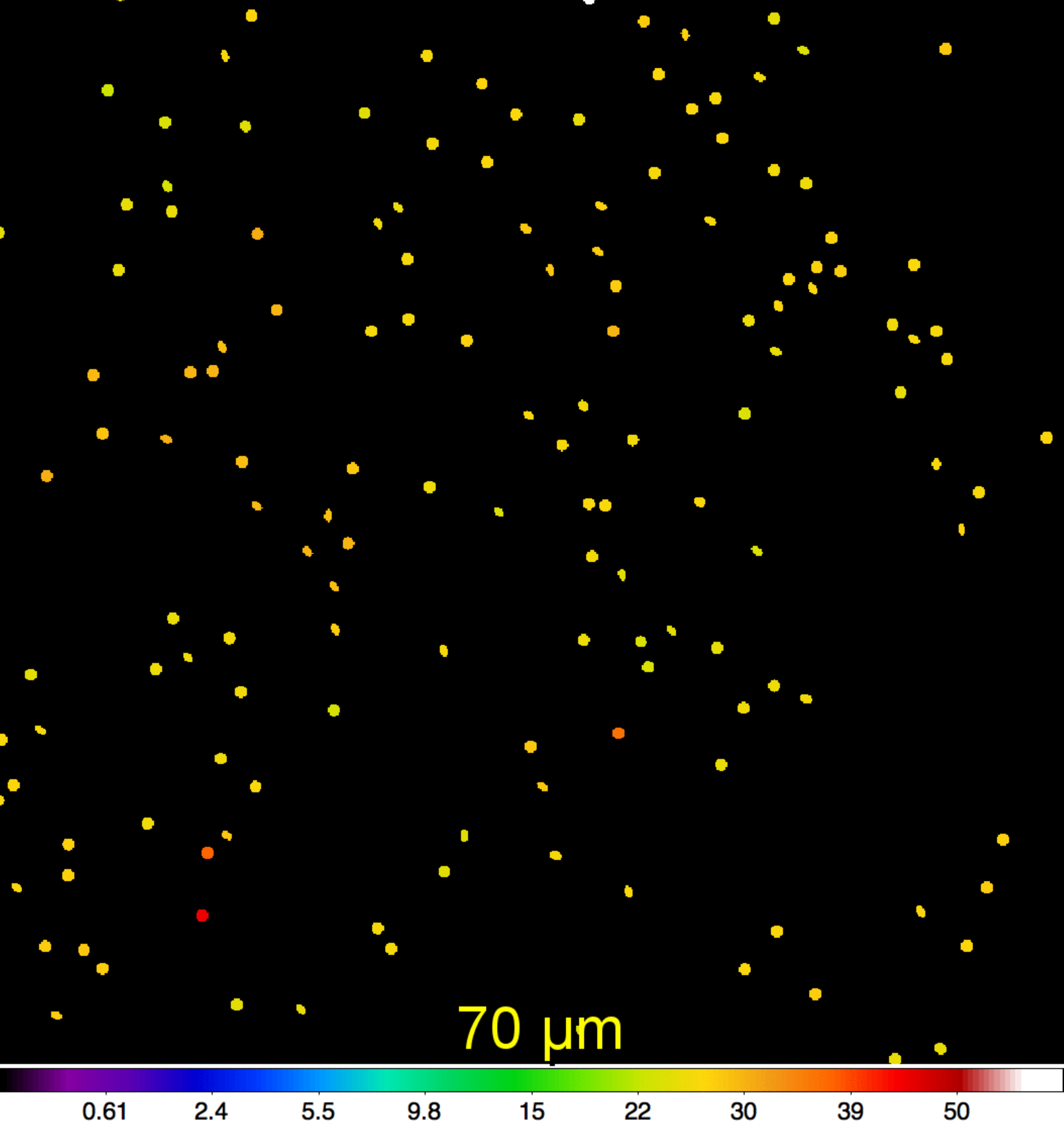}}
            \resizebox{0.33\hsize}{!}{\includegraphics{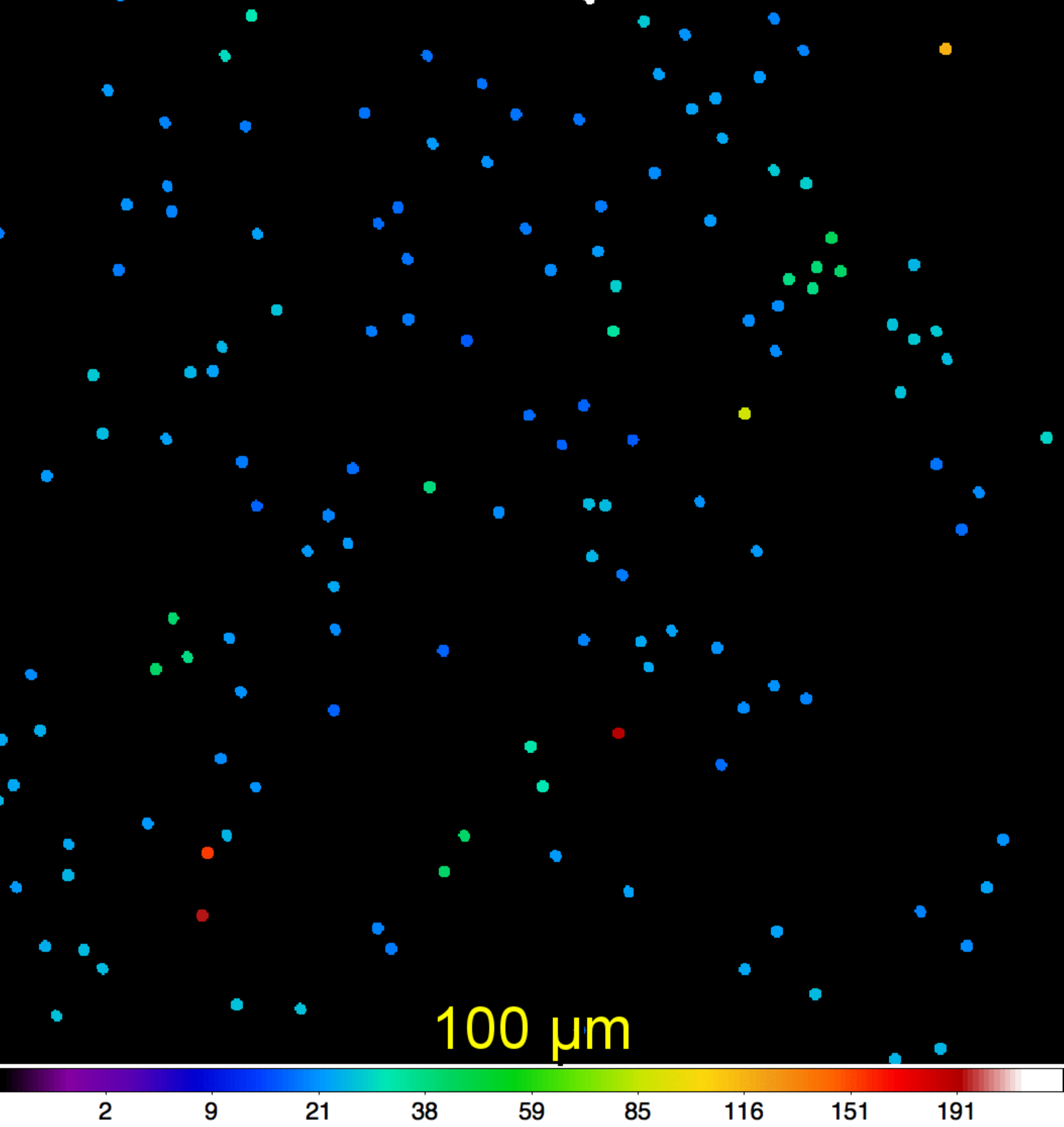}}
            \resizebox{0.33\hsize}{!}{\includegraphics{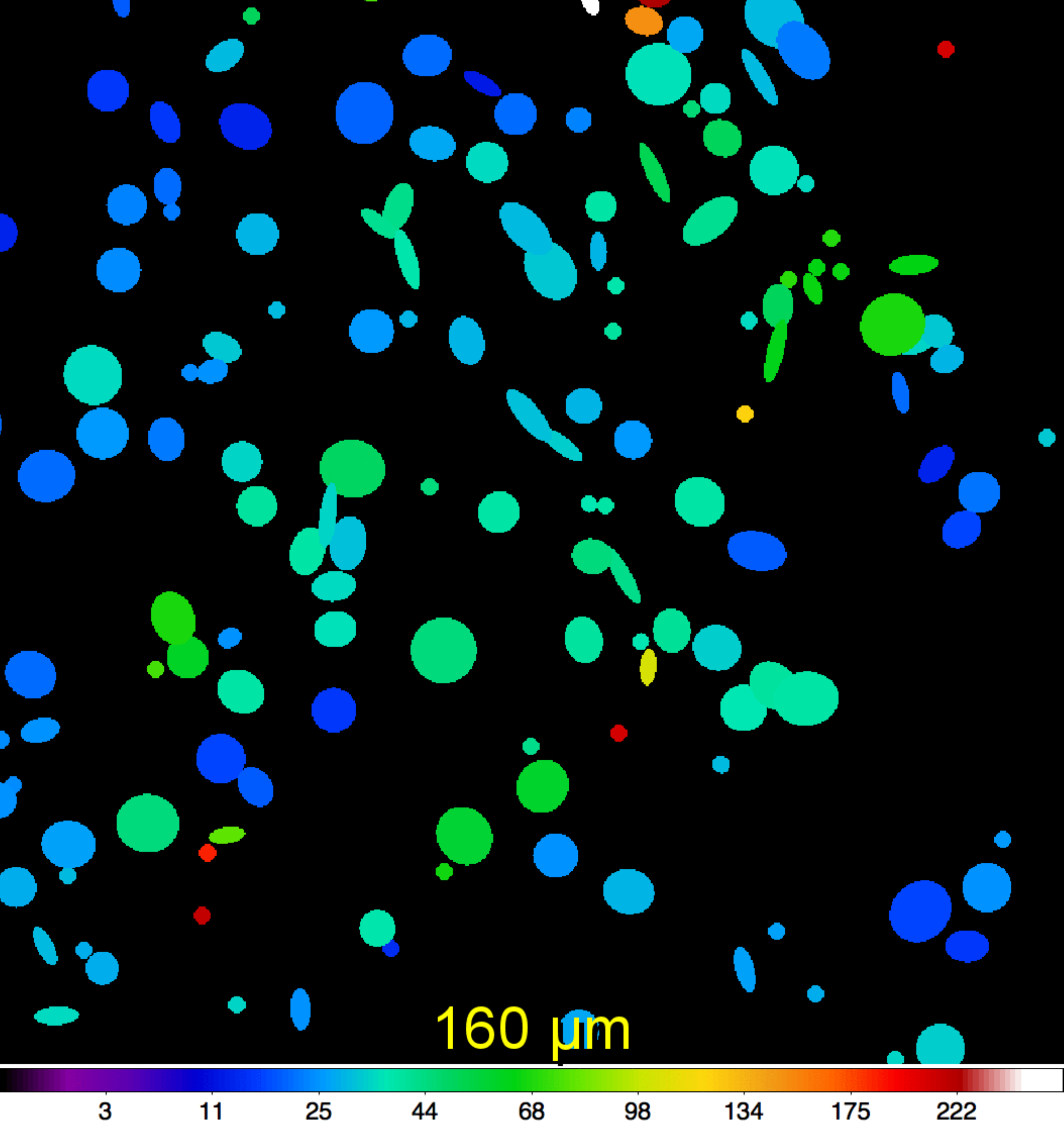}}}
\centerline{\resizebox{0.33\hsize}{!}{\includegraphics{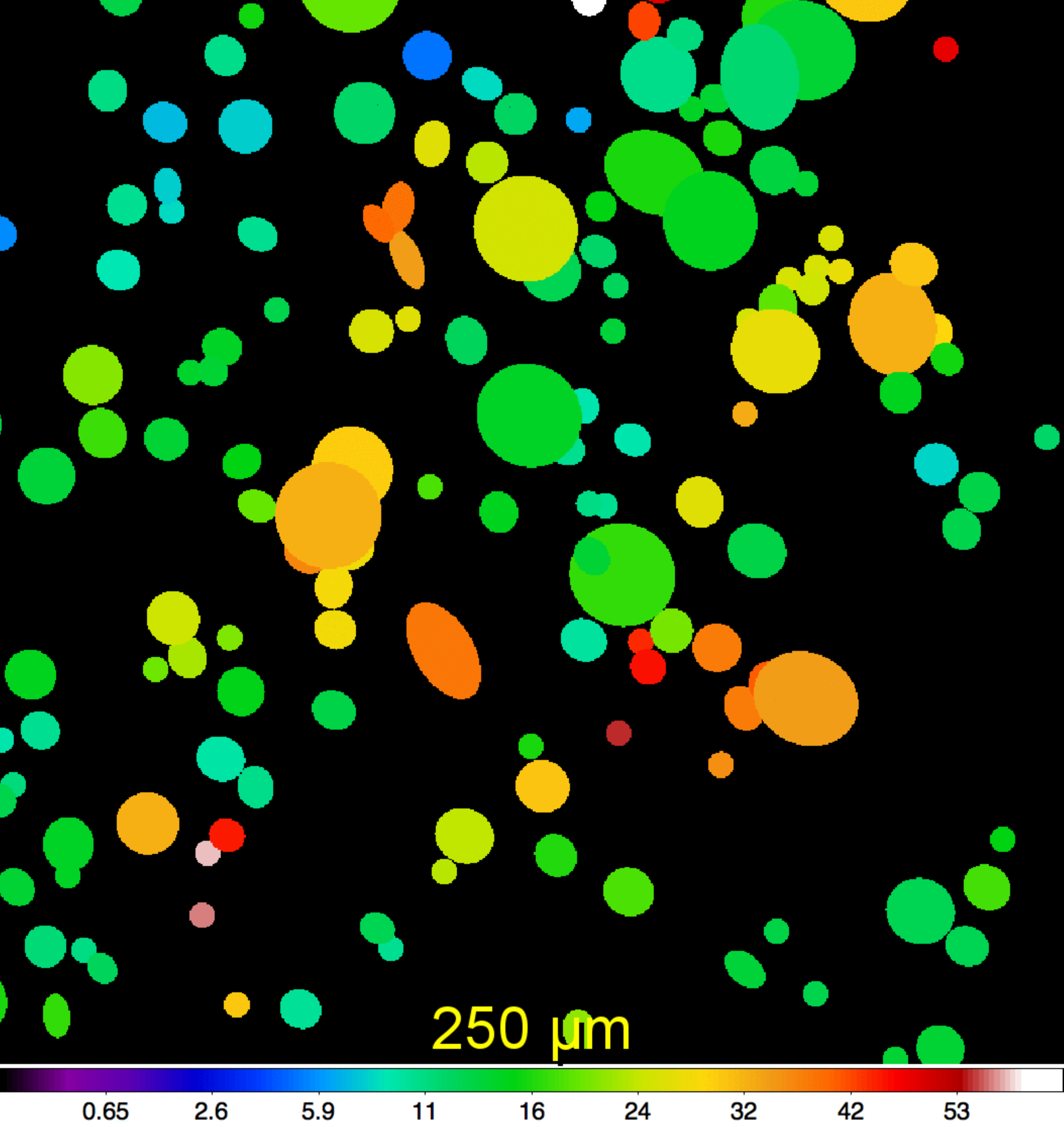}}
            \resizebox{0.33\hsize}{!}{\includegraphics{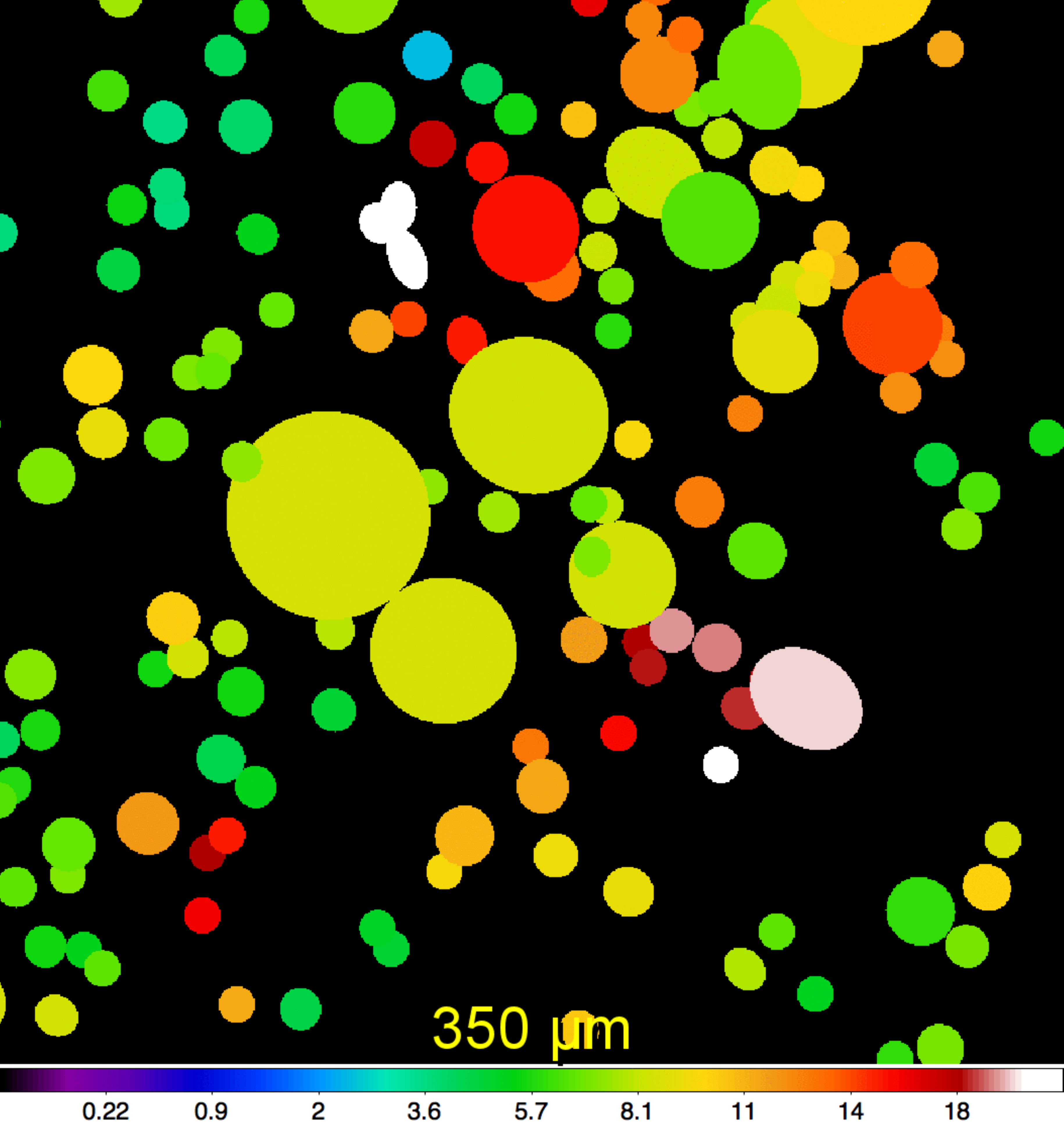}}
            \resizebox{0.33\hsize}{!}{\includegraphics{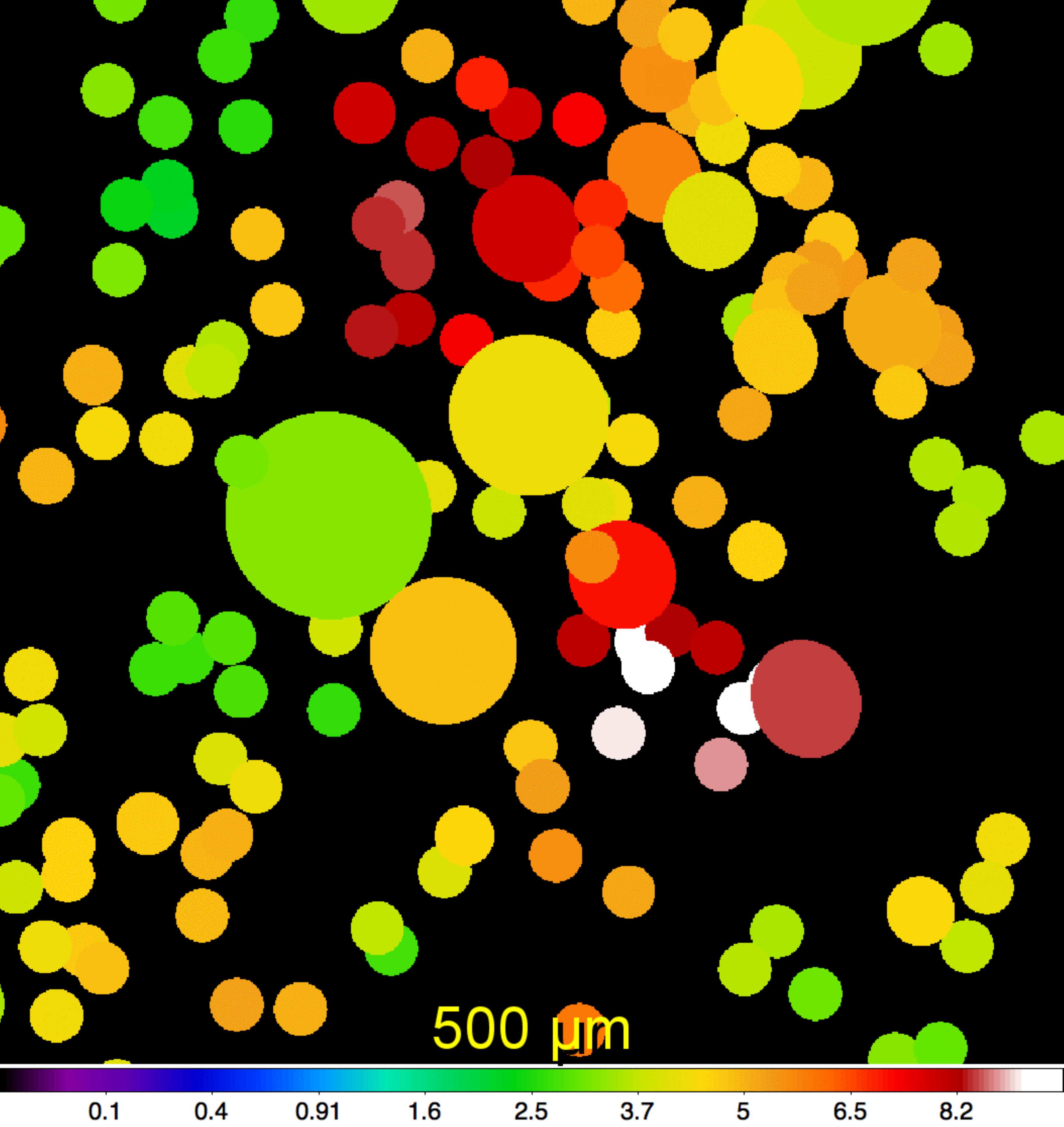}}}                   
\caption{
Converged footprints of the measured sources (Sect.~\ref{measuring.cataloging}). The field of Fig.~\ref{single.scales} is shown at
70, 100, 160, 250, 350, 500\,{${\mu}$m} (\emph{left to right}, \emph{top to bottom}) as the footprints of all detected sources after
the measurement iterations. The pixel values are the standard deviations $\sigma_{{i}{\lambda}{\,\rm P}}$, due to the local noise
and background variations, estimated for each source in an elliptical annulus around its footprint (Fig.~\ref{bigfoot.images}).
Strongly elliptical or too large footprints may appear at those wavelengths where some sources are too faint to be measurable. For
such sources, the information is essentially lost and the intensity moments cannot provide meaningful estimates of their sizes and
orientation. The color coding is a function of the square root of intensity in MJy/sr.
} 
\label{converged.footprints}
\end{figure*}

To properly measure parameters of a source, one has first to determine and subtract the background. As discussed in
Sect.~\ref{introduction}, we define sources as significant intensity peaks detected by the algorithm and whose entire contribution
to the image is bound by their footprints. Based on this definition, \textsl{getsources} determines the background by linearly
interpolating pixel intensities in $\mathcal{I}_{{\!\lambda}{\rm O}}$ under the source footprints
(Figs.~\ref{interpolation},\ref{converged.footprints}). The interpolation is done in the four main directions (two image axes and
two diagonals), based on the pixels just outside the footprints, which do not belong to any source. For each pixel, values from the
4 directions are averaged to produce the background intensity at the pixel. This procedure results in the image of \emph{clean
background} $\mathcal{I}_{{\!\lambda}{\rm O}{\,\rm CB}}$ and the \emph{background-subtracted} image
$\mathcal{I}_{{\!\lambda}{\rm O}{\,\rm BS}}{\,=\,}\mathcal{I}_{{\!\lambda}{\rm O}}-\mathcal{I}_{{\!\lambda}{\rm O}{\,\rm CB}}$
(Figs.~\ref{deblended.images},\ref{composite.obsbs}). In very crowded areas of images with many overlapping sources it is not
possible to probe the background around each of the sources. In this case, the interpolation gives the best possible background
estimate based on the nearest background pixels available around the blended areas\footnote{This simple method works well (as long
as one determines accurate footprints) and it is sufficient, as the background under sources is fundamentally unknown. Estimates of
the background based on more complicated approaches, such as its approximation by some two-dimensional functions, always involve
assumptions and free parameters, and our simulations show that they may well be less accurate than the simple linear
interpolation.}.

\begin{figure*}
\centering
\centerline{\resizebox{0.33\hsize}{!}{\includegraphics{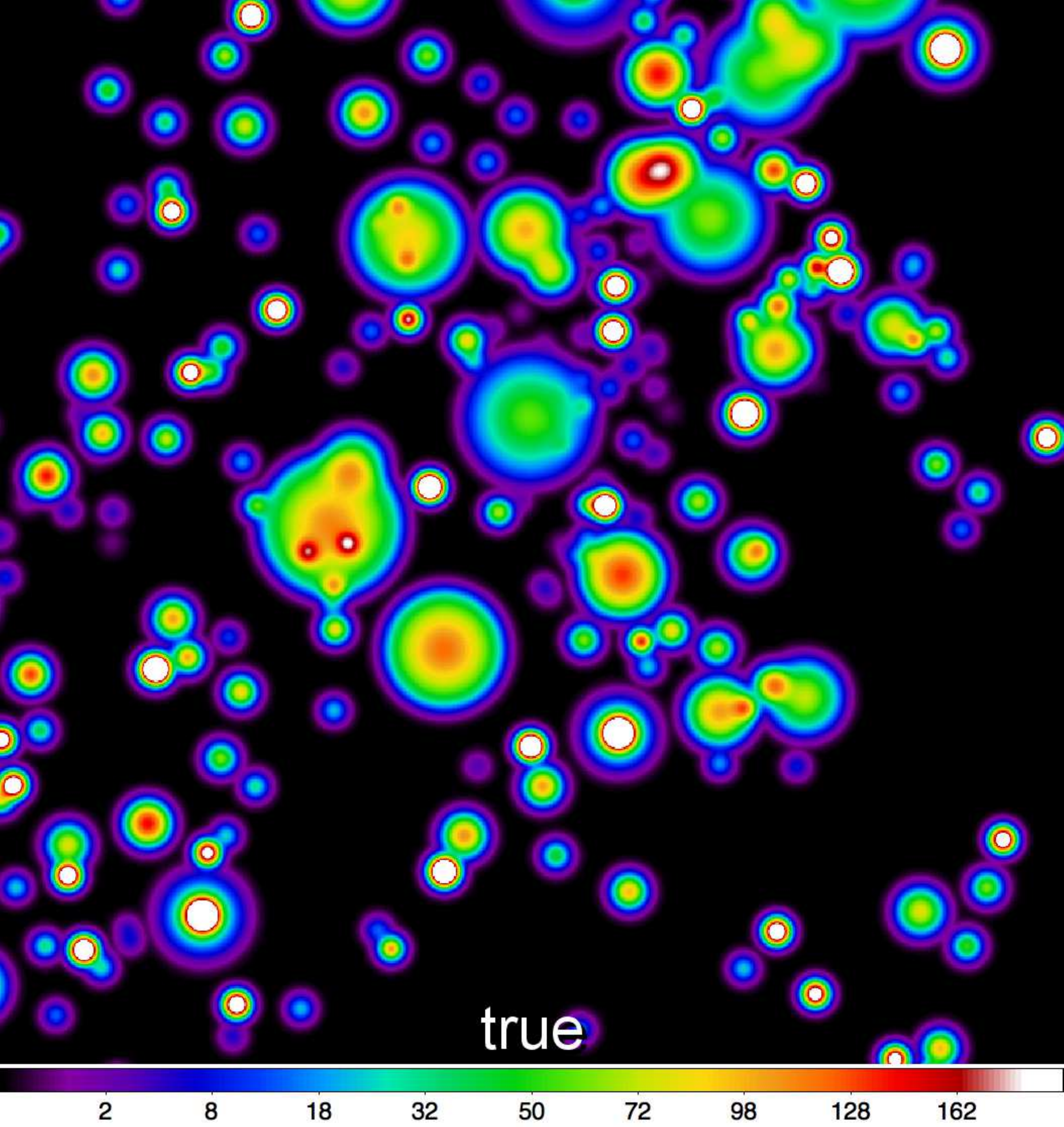}}
            \resizebox{0.33\hsize}{!}{\includegraphics{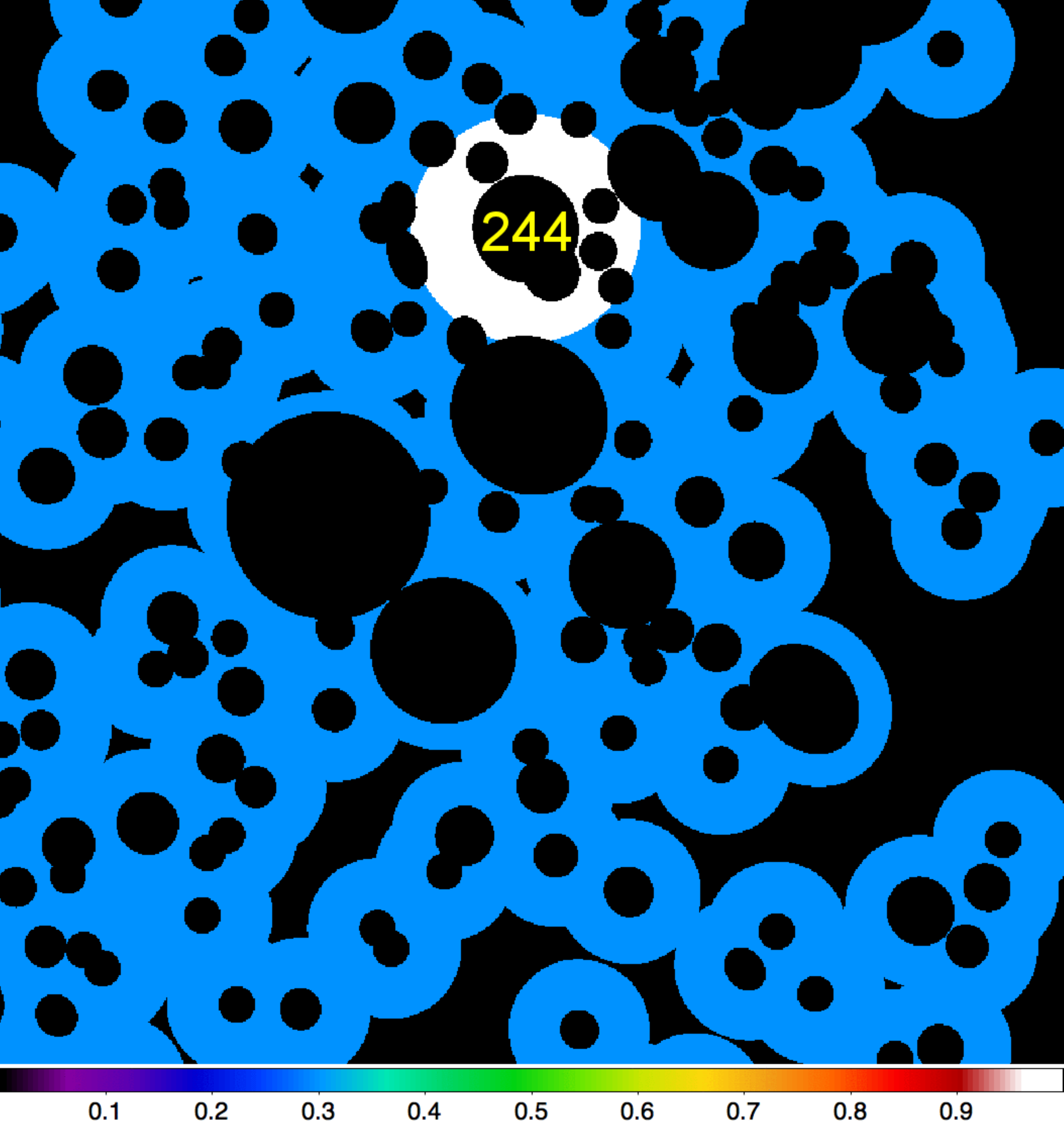}}
            \resizebox{0.33\hsize}{!}{\includegraphics{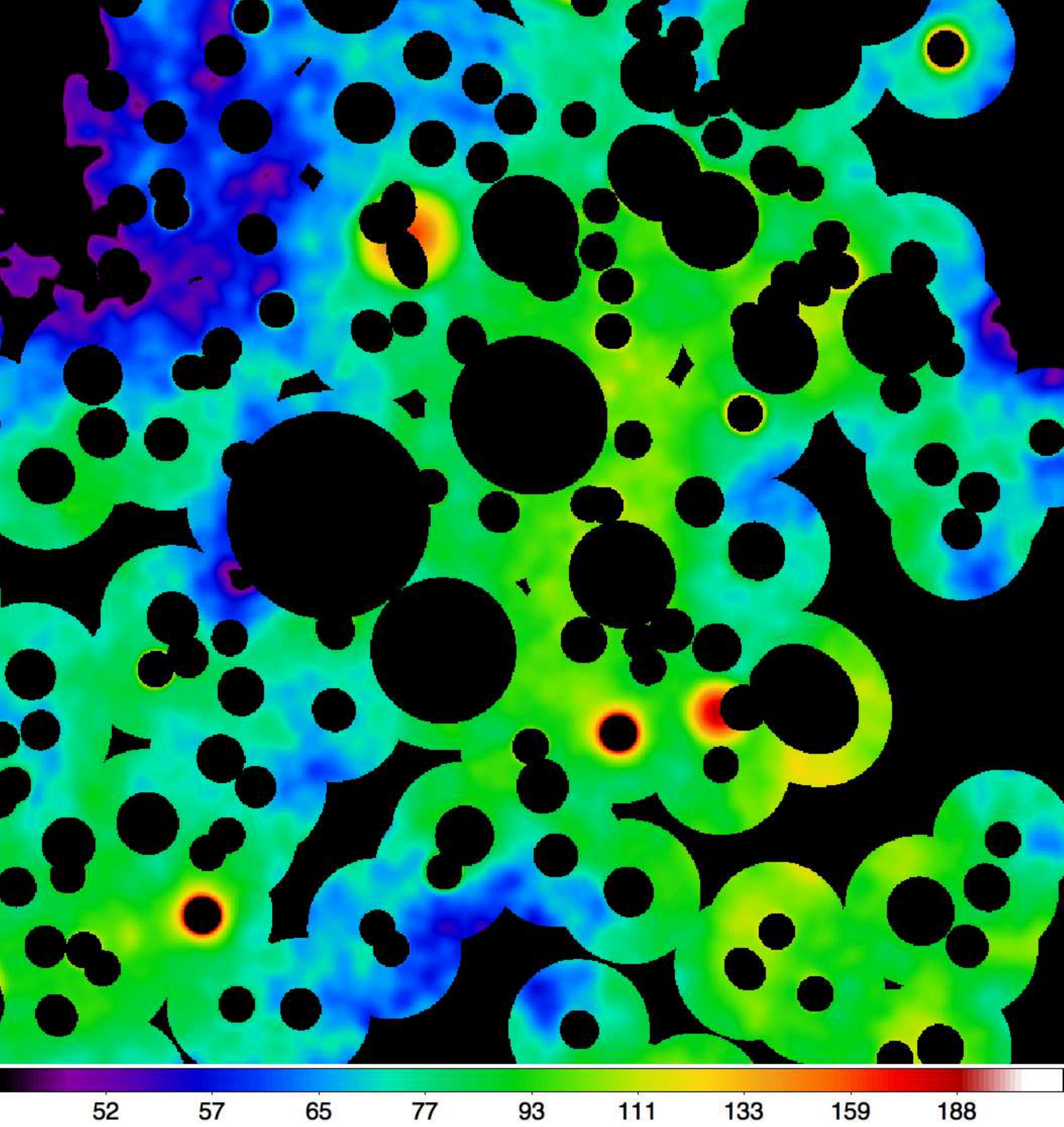}}}
\caption{
Annuli around measured sources (Sect.~\ref{measuring.cataloging}). The field of Fig.~\ref{single.scales} is shown as the true
intensities of the model sources at 350\,{${\mu}$m} convolved to a 17{\arcsec} resolution (\emph{left}), the image of annuli
$\mathcal{A}_{\lambda}$ of all detected sources (\emph{middle}) slightly modified to highlight the annulus area around the source
244 from Fig.~\ref{deblended.images}, and the product $\mathcal{A}_{\lambda}\,\mathcal{I}_{{\!\lambda}{\rm O}}$ (\emph{right}) to
visuaize the actual observed intensities used to compute the flux uncertainties $\sigma_{{i}{\lambda}{\,\rm P}}$ shown in
Fig.~\ref{converged.footprints}. The corresponding footprint images $\mathcal{F}_{\!\lambda}$ are presented in
Fig.~\ref{converged.footprints} and the observed image $\mathcal{I}_{{\!\lambda}{\rm O}}$ is displayed in
Fig.~\ref{extraction.ellipses}. The color coding in the left panel is a function of the square root of intensity, in the other
panels it is a linear function of intensity in MJy/sr.
}
\label{bigfoot.images}
\end{figure*}

Having created the background-subtracted images $\mathcal{I}_{{\!\lambda}{\rm O}{\,\rm BS}}$, the algorithm deblends values of
pixels in overlapping sources and computes peak intensities $F_{{i}{\lambda}{\,\rm P}}$, and total fluxes $F_{{i}{\lambda}{\,\rm
T}}$ for each source $i$. At the first measurement iteration, it uses the initial size estimate $S_{\!j_{\rm F}}$ obtained during
the detection (Sect.~\ref{detecting.sources}), whereas in the subsequent iterations, the size and orientation from a previous
iteration are used. Our iterative algorithm employs \emph{deblending shapes}, the two-dimensional analogs of the function
\citep{Moffat_1969}
\begin{equation}
I_{{i}{\lambda}{\,\rm M}} = F_{{i}{\lambda}{\,\rm P}}\,\left(1+(r/R_0)^2\right)^{-\zeta},
\label{moffat.function}
\end{equation}
where $r$ is the radial distance from the peak and $R_0$ is a function of the FWHM shape ($A_{{i}{\lambda}}$, $B_{{i}{\lambda}}$,
$\Theta_{{i}{\lambda}}$) of a source $i$. The power-law exponent is fixed at $\zeta{\,=\,}10$ to have stronger, more realistic wings
compared to the exponential wings of a Gaussian ($\zeta\!\rightarrow\!\infty$)\footnote{As an example, the intensity of a circular
$I_{{i}{\lambda}{\,\rm M}}$ at the footprint edge is by a factor of 1.565 higher than that of the corresponding Gaussian.}. The
deblending shapes (Eq.~\ref{moffat.function}) are used to split the intensity $I_{{\!\lambda}{\rm O}{\,\rm BS}}$ of a pixel
between the source $i$ and all overlapping sources according to a fraction of the profiles' intensities at that pixel:
\begin{equation}
I_{i{\,\lambda}}{\,=\,}\frac{|I_{{i}{\lambda}{\,\rm M}}|}{\sum\limits_{k} 
|I_{{k}{\lambda}{\,\rm M}}|}\,I_{{\!\lambda}{\rm O}{\,\rm BS}},
\label{deblending.formula}
\end{equation}
where the summation is done over all sources whose footprints contain the pixel (Figs.~\ref{deblending}, \ref{deblended.images}).
The deblending iterations start with the original (blended) intensities $I_{{\!\lambda}{\rm O}{\,\rm BS}}$ at the positions of each
source and perform the splitting of the pixel values until the intensity $I_{{i}{\lambda}}(x_i,y_i)$ at the center of each source
converges to its deblended peak intensity $F_{{i}{\lambda}{\,\rm P}}$. The deblended intensities $I_{{i}{\lambda}}$ within the
footprint ellipses (Fig.~\ref{converged.footprints}) are then used to integrate the total fluxes $F_{{i}{\lambda}{\,\rm T}}$, as
well as their FWHM shapes ($A_{{i}{\lambda}}$, $B_{{i}{\lambda}}$, $\Theta_{{i}{\lambda}}$) from the intensity moments (Appendix
\ref{intensity.moments}).

Local uncertainties of the peak intensities $F_{{i}{\lambda}{\,\rm P}}$ are given by the standard deviations
$\sigma_{{i}{\lambda}{\,\rm P}}$ estimated in the observed images $\mathcal{I}_{{\!\lambda}{\rm O}}$ in an elliptical annulus
defined around each source $i$ just outside its footprint\footnote{This is equivalent to the standard approach of measuring flux
errors for an isolated source. Heavily crowded fields present, however, a serious problem, as no source-free annulus exist around
many of the sources situated within the regions. No relevant \emph{local} values of the uncertainties can be found in that case, as
more distant source-free areas of images are likely to have different properties in case of highly-variable background or noise.}.
To ensure that the uncertainties are statistically meaningful, the images of annuli $\mathcal{A}_{\lambda}$
(Fig.~\ref{bigfoot.images}) are constructed by requiring that the area of any annulus must contain 50 areas of the observational
beam $O_{\lambda}$. In the crowded fields, such as the one used for the illustrations in this paper, the footprints and annuli may
overlap quite heavily (Figs.~\ref{converged.footprints}, \ref{bigfoot.images}) and therefore not all pixels can be used, as many of
them belong to other sources. In such cases, \textsl{getsources} obtains an estimate of $\sigma_{{i}{\lambda}{\,\rm P}}$ as local as
possible by expanding the outer edge of the annulus outwards, until its usable area of non-zero pixels becomes $50\,O_{\lambda}$
(Fig.~\ref{bigfoot.images}, \emph{middle}).

\begin{figure*}
\centering
\centerline{\resizebox{0.33\hsize}{!}{\includegraphics{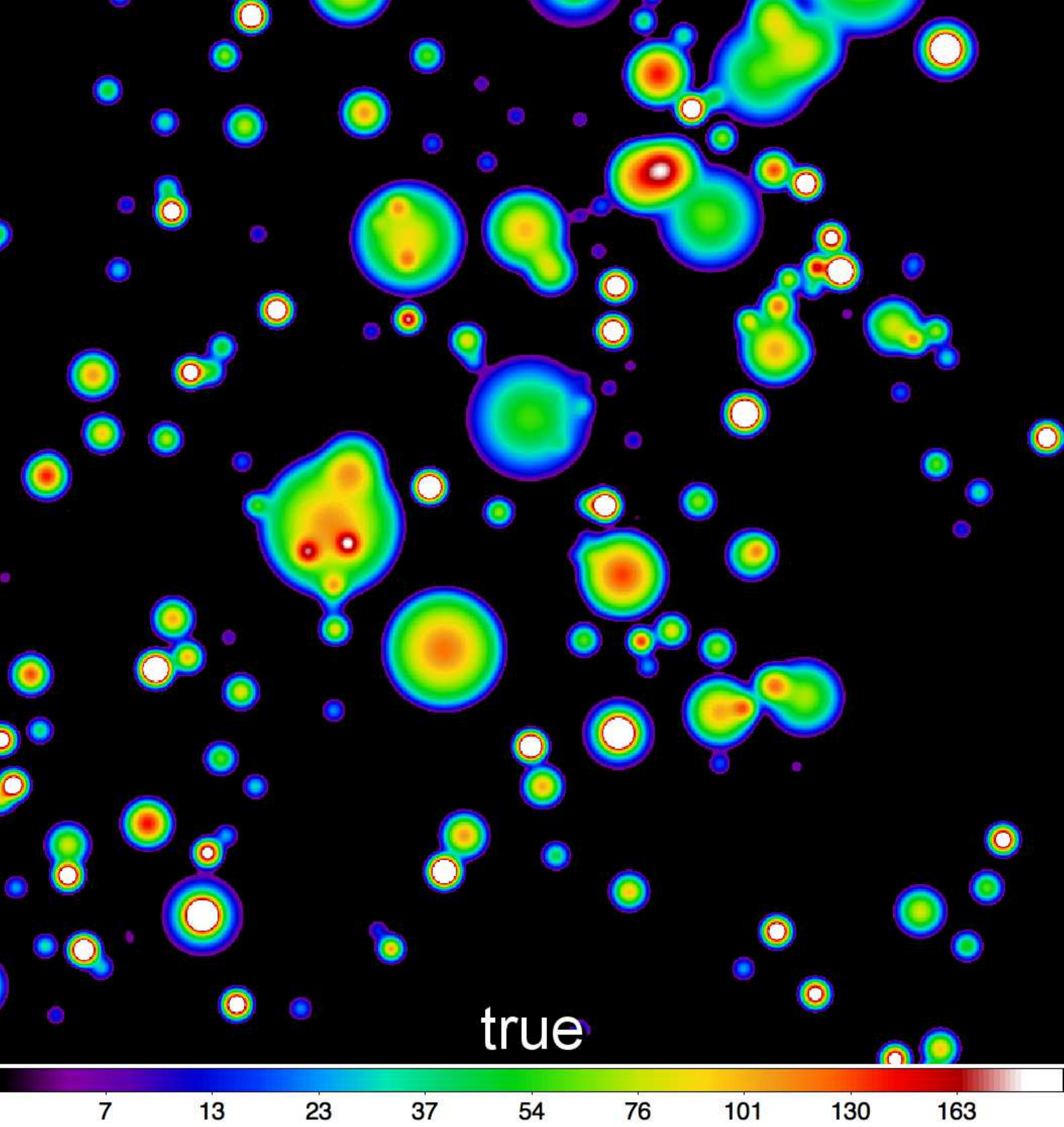}}
            \resizebox{0.33\hsize}{!}{\includegraphics{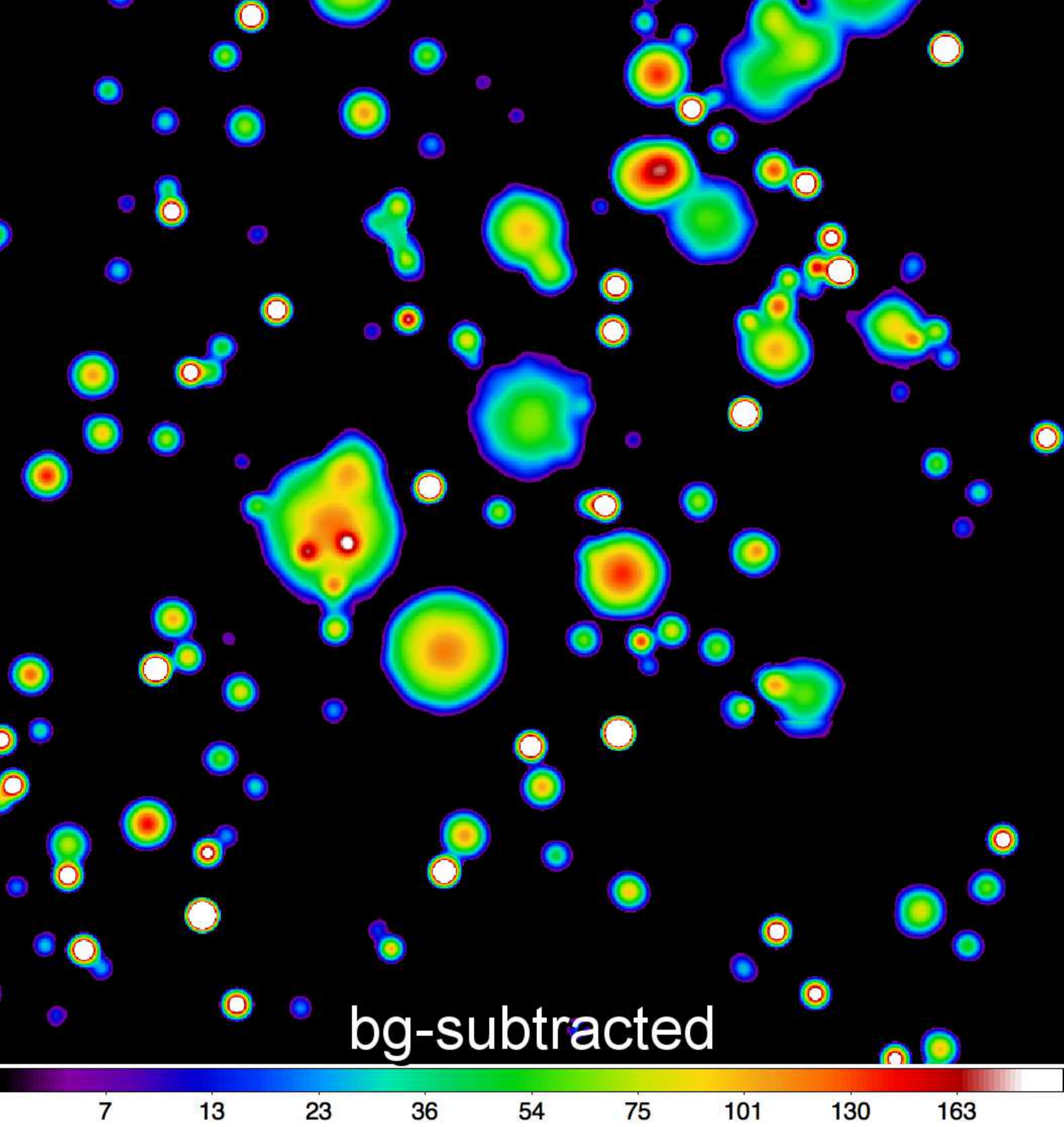}}
            \resizebox{0.33\hsize}{!}{\includegraphics{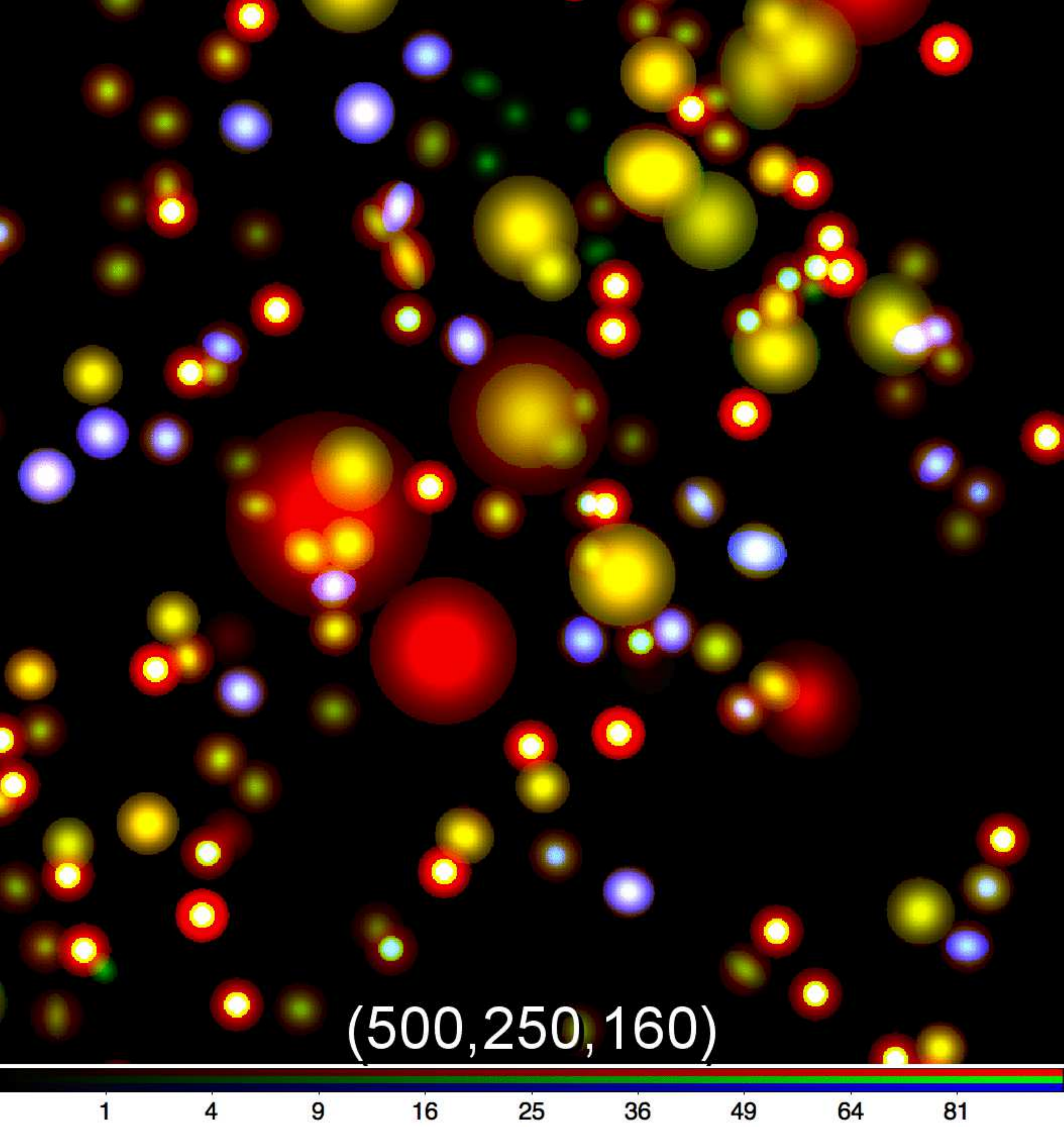}}}
\caption{
Background-subtracted sources (Sect.~\ref{measuring.cataloging}). The field of Fig.~\ref{single.scales} is shown as the true
intensities of the model sources at 350\,{${\mu}$m} convolved to a 17{\arcsec} resolution (\emph{left}), the background-subtracted
image $\mathcal{I}_{{\!\lambda}{\rm O}{\,\rm BS}}$ at 350\,{${\mu}$m} (\emph{middle}), and the composite 3-color RGB image (500,
250, 160\,{${\mu}$m}) created using the images of the deblending shapes of each extracted source (\emph{right}). For the true model
intensity distribution (no background) to be more comparable to the background-subtracted image, it is shown above 5 MJy/sr. The
color coding is a function of the square root of intensity in MJy/sr.
}
\label{composite.obsbs}
\end{figure*}

Uncertainties $\sigma_{{i}{\lambda}{\,\rm T}}$ of the total fluxes $F_{{i}{\lambda}{\,\rm T}}$ are estimated under the following
assumptions. If a source footprint contains $N_{\rm B}$ observational beams $O_{\lambda}$, then the error of the sum of intensities
over the footprint area will be the square root of the quadratic sum of the individual errors, since the beam measurements are
statistically independent. Assuming the individual errors to be identical and equal to $\sigma_{{i}{\lambda}{\,\rm P}}$, the total
flux uncertainty is
\begin{equation}
\sigma_{{i}{\lambda}{\,\rm T}}{\,=\,}\sigma_{{i}{\lambda}{\,\rm P}}\,
\frac{(A_{{i}{\,\rm F}{\lambda}}\,B_{{i}{\,\rm F}{\lambda}})^{1/2}}{1.15\,(2\,O_{\lambda})},
\label{total.flux.error}
\end{equation}
where $A_{{i}{\,\rm F}{\lambda}}$ and $B_{{i}{\,\rm F}{\lambda}}$ are the major and minor axes of the footprint ellipse (cf.
Eq.~\ref{footprints}), $O_{\lambda}$ is assumed to be a circular Gaussian, and the empirical factor 1.15 has been introduced in
Sect.~\ref{detecting.sources}.

With the peak intensities and their uncertainties estimated for each source, one can define the standard S/N ratio
$\Omega_{{i}{\lambda}} = F_{{i}{\lambda}{\,\rm P}} / \sigma_{{i}{\lambda}{\,\rm P}}$, to quantify how prominent the sources are
against the noise and background fluctuations in their immediate environments. Conventional practice is to use the quantity for
defining reliability criteria to avoid contamination of the extraction catalogs with spurious detections; at the same time,
$\sigma_{{i}{\lambda}{\,\rm P}}$ determines the errors of the measured fluxes. However, in contrast to the traditional source
extraction methods, \textsl{getsources} performs detection on highly-filtered images $\mathcal{I}_{{\rm D}{j}{\,\rm C}}$
that are quite different from the measurement images $\mathcal{I}_{{\!\lambda}{\rm O}}$ and it is important to make a clear
distinction between the detection and measurement qualities.

In the wavelength-dependent detection images $\mathcal{I}_{{\!\lambda}{\rm D}{j}{\,\rm C}}$, we define the contrasts
$C_{{i}{\lambda}{j}}{\,=\,}I_{{i}{\lambda}{j}}/\varpi_{{\lambda}{j}}$, similar to those introduced in
Sect.~\ref{detecting.sources}. At the footprinting scale $j_{\rm F}$ the contrast is, within a factor of $n_{{\lambda}{j}}$
(Sect.~\ref{removing.noise.background}), the monochromatic \emph{detection significance}
\begin{equation}
\Xi_{{i}{\lambda}}{\,=\,}\frac{I_{{i}{\lambda}{j_{\rm F}}}}{\sigma_{{\lambda}{j_{\rm F}}}}{\,=\,}n_{{\lambda}{j}}\,C_{{i}{\lambda}{j}}.
\label{siginificance}
\end{equation}
Although $\Xi_{{i}{\lambda}}$ is the single-scale analog of $\Omega_{{i}{\lambda}}$, the quantity $\sigma_{{\lambda}{j_{\rm F}}}$
evaluates the level of uncertainties at the footprinting scale in the single-scale \emph{detection} images, whereas
$\sigma_{{i}{\lambda}{\,\rm P}}$ quantifies the level of fluctuations during \emph{measurements} in the full observed images.

\begin{figure*}
\centering
\centerline{\resizebox{0.33\hsize}{!}{\includegraphics{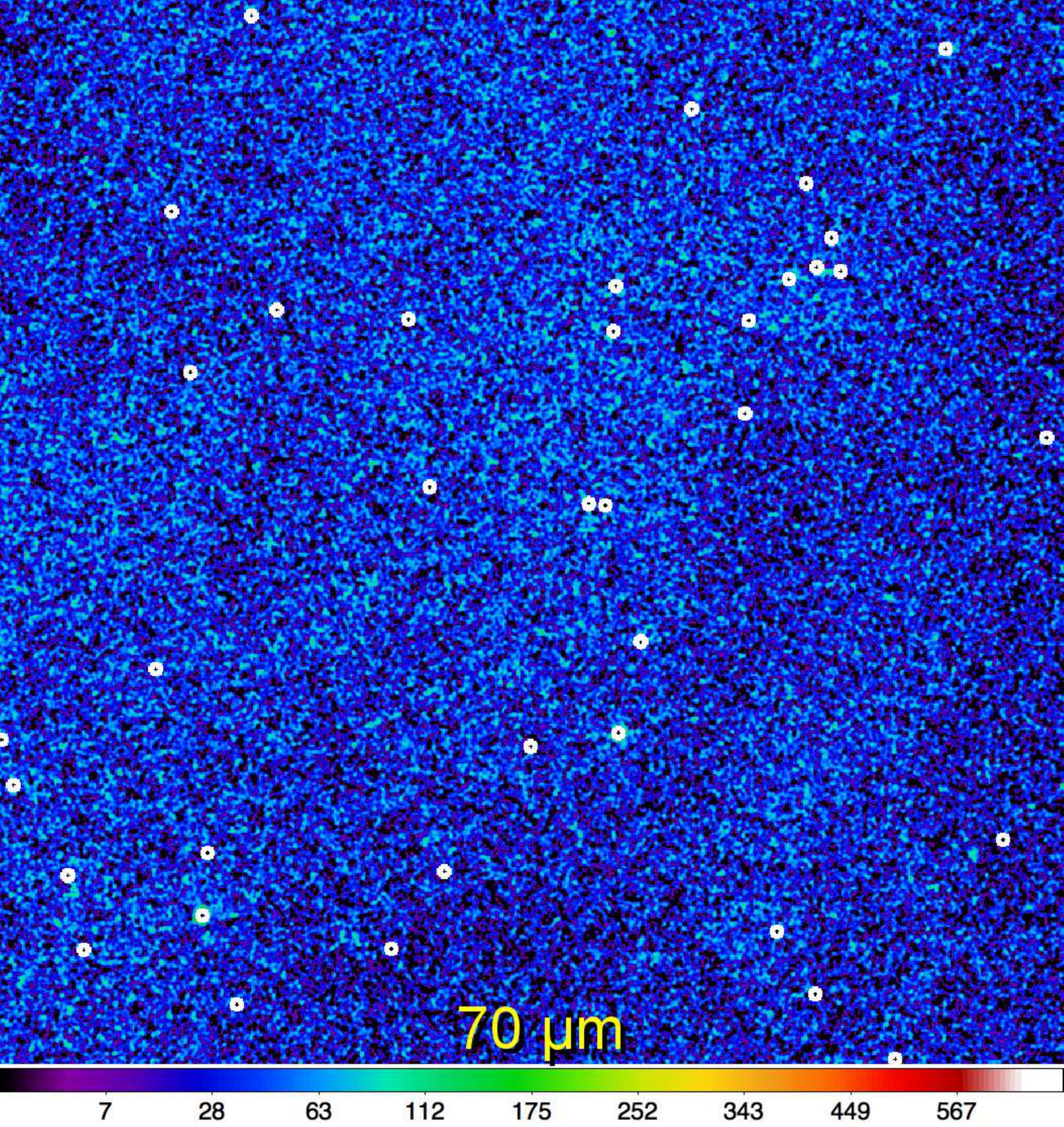}}
            \resizebox{0.33\hsize}{!}{\includegraphics{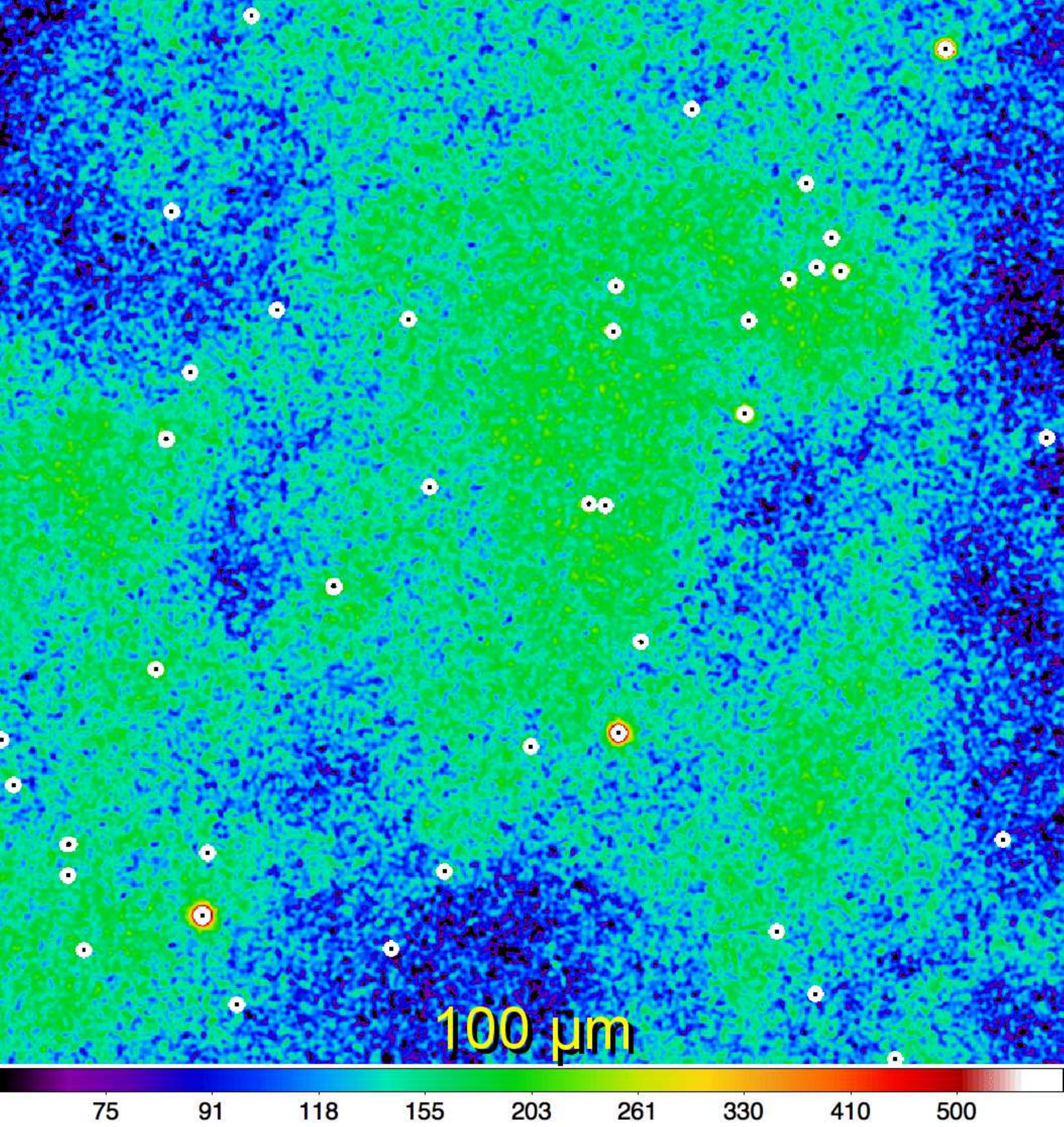}}
            \resizebox{0.33\hsize}{!}{\includegraphics{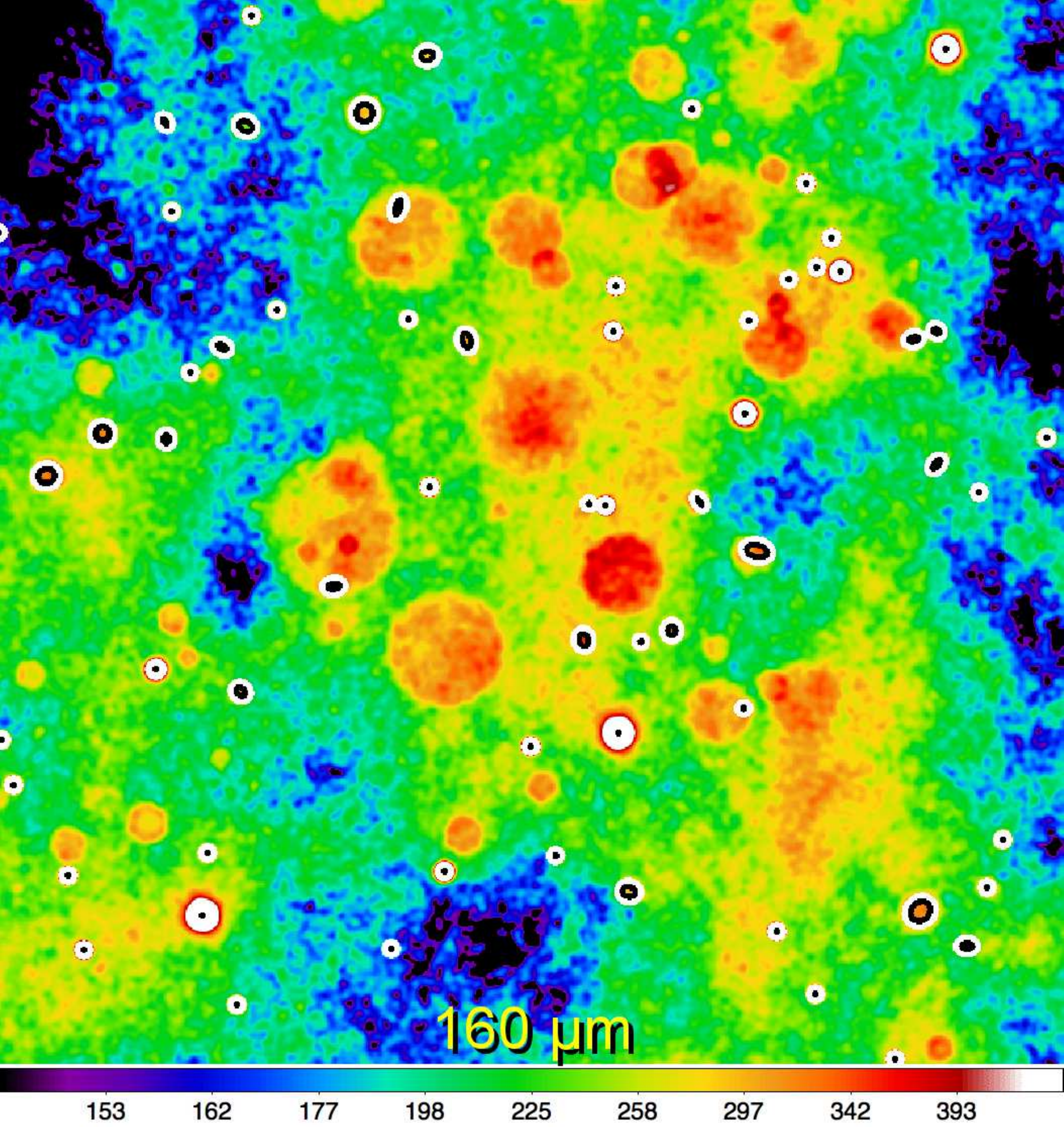}}}
\centerline{\resizebox{0.33\hsize}{!}{\includegraphics{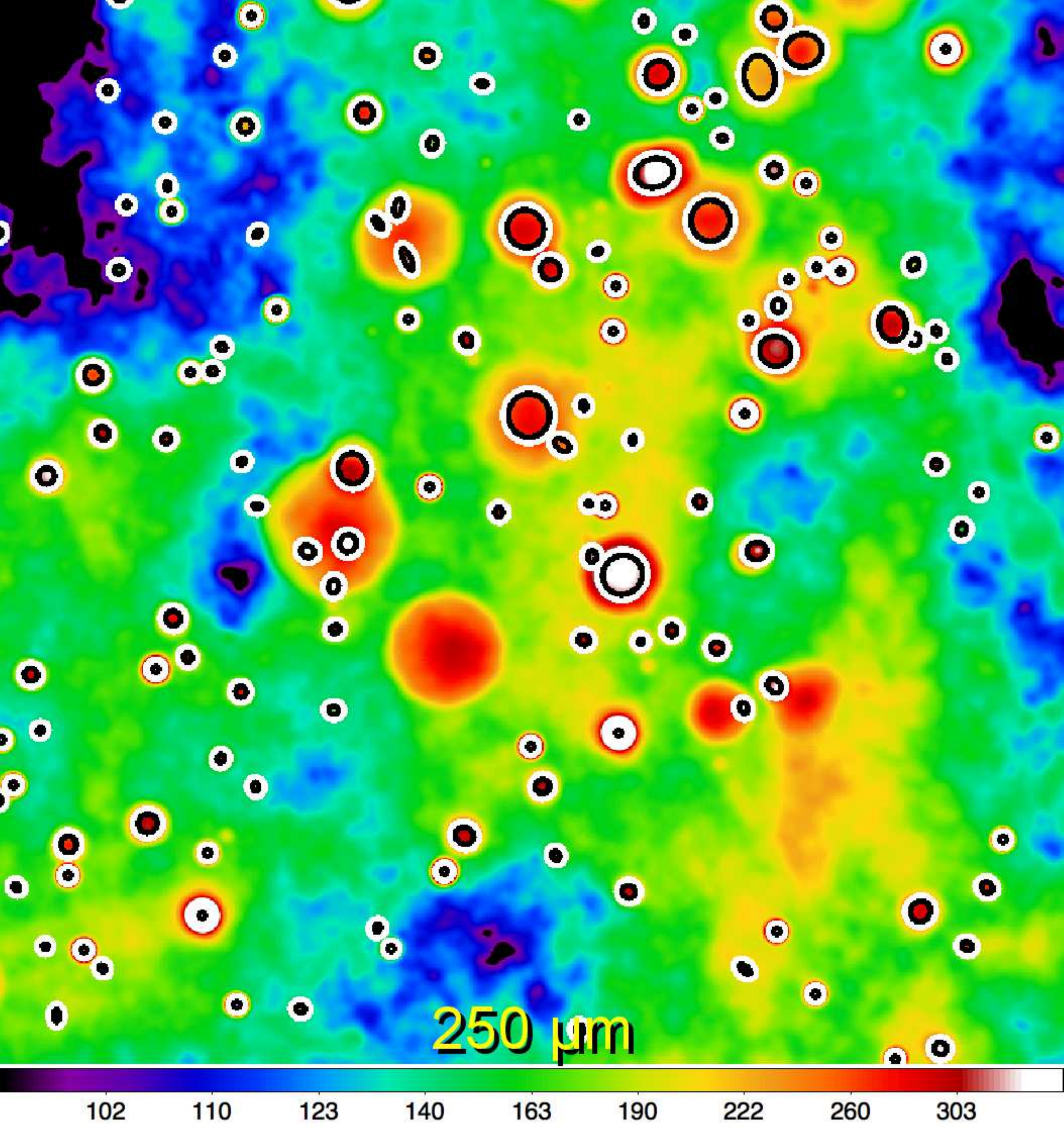}}
            \resizebox{0.33\hsize}{!}{\includegraphics{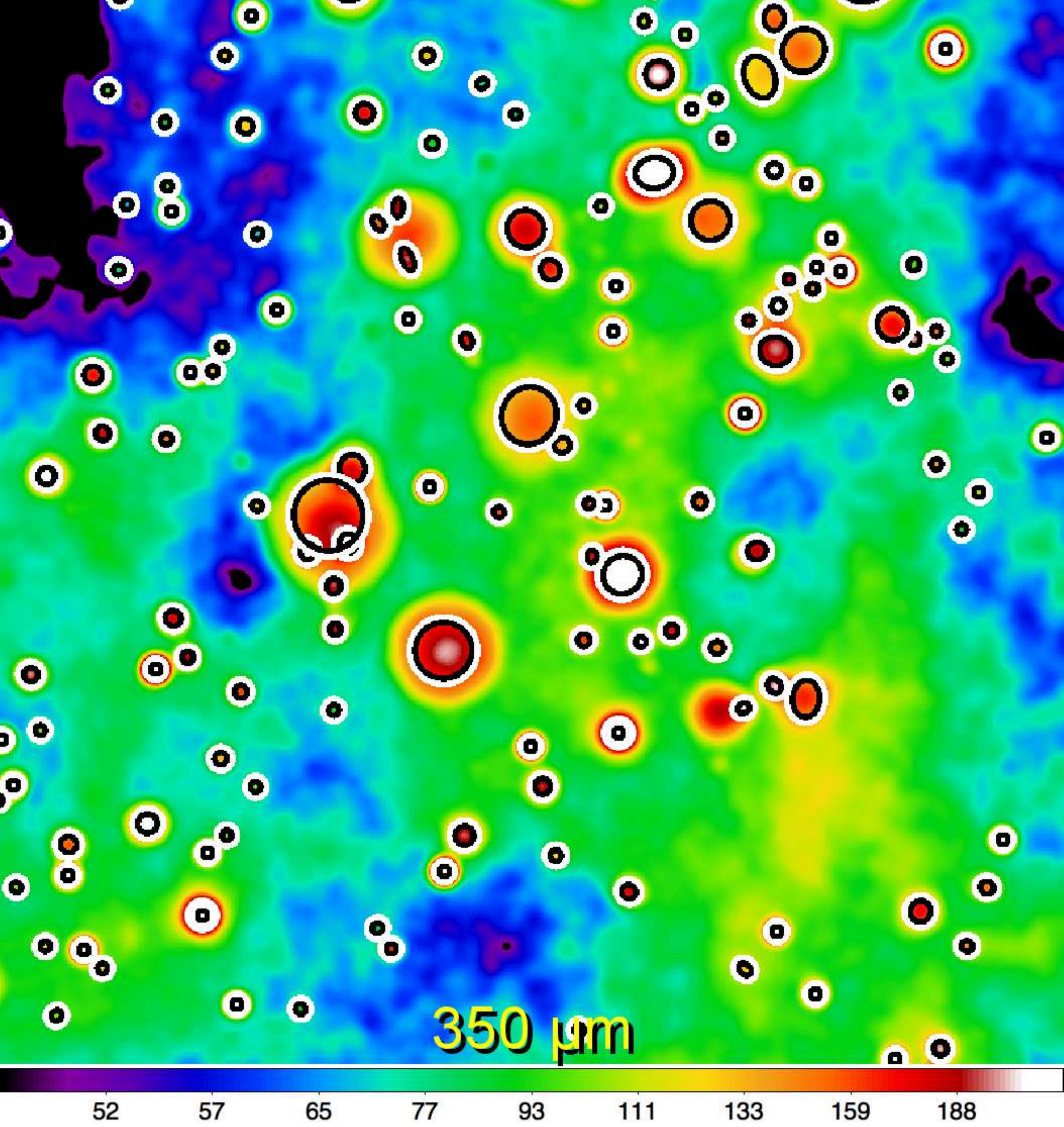}}
            \resizebox{0.33\hsize}{!}{\includegraphics{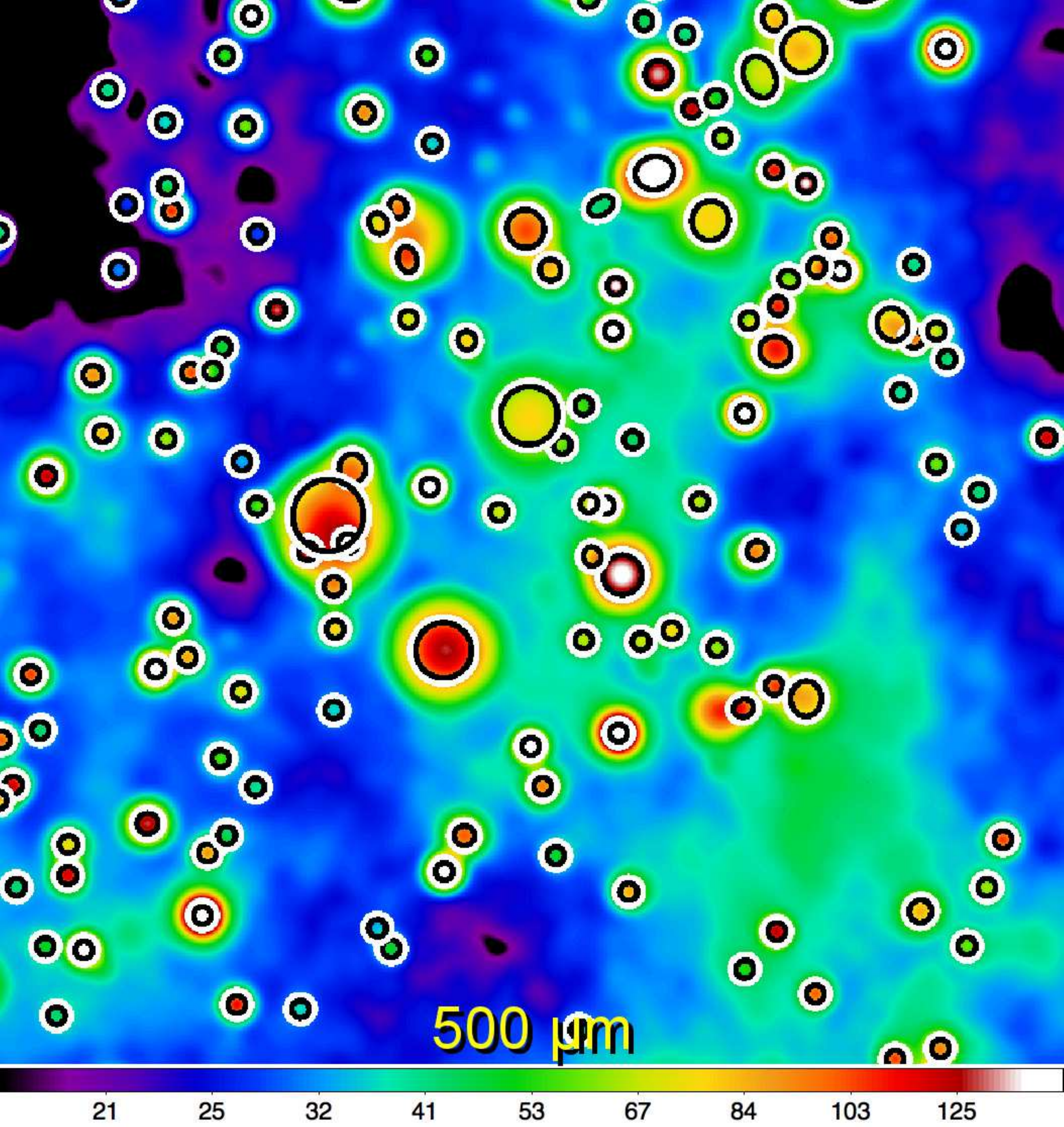}}}                   
\caption{
Visualization of the measured and cataloged sources (Sect.~\ref{visualizing.extractions}). The field of Fig.~\ref{single.scales}
is shown at 70, 100, 160, 250, 350, 500\,{${\mu}$m} (\emph{left to right}, \emph{top to bottom}) with the extraction ellipses (FWHM)
of the measurable sources ($F_{{i}{\lambda}{\,\rm P}} > \sigma_{{i}{\lambda}{\,\rm P}}$ and $F_{{i}{\lambda}{\,\rm T}} >
\sigma_{{i}{\lambda}{\,\rm T}}$) overlaid on top of the observed images $\mathcal{I}_{{\!\lambda}{\rm O}}$. The default condition,
that a tentative source must be detected in at least two bands, was used. Only the protostellar cores are visible at 70 and
100\,{${\mu}$m}, whereas at 160\,{${\mu}$m} cold starless cores starts to appear, becoming clearly visible in the SPIRE bands. For
better visibility, the values displayed in the panels are somewhat limited in range; the color coding is a function of
the square root of intensity in MJy/sr.
} 
\label{extraction.ellipses}
\end{figure*}

Since \textsl{getsources} is a multi-wavelength extraction method and the detection is generally performed in the \emph{combined}
Extraction of an isolated source situated in approximately uniform background and noise results in
$\Xi_{{i}{\lambda}}{\,\approx\,}\Omega_{{i}{\lambda}}$. For sources in more complicated environments, however, $\Xi_{{i}{\lambda}}$
gives a cleaner and more accurate estimate of the detection quality and a better criterion for selecting real sources. The measured
value of $\Omega_{{i}{\lambda}}$ inversely depends on the fluctuation level of highly-variable backgrounds, such as those observed
by \emph{Herschel} in Galactic regions. The single-scale cleaning of $\mathcal{I}_{{\!\lambda}{\rm D}{j}}$
(Sect.~\ref{removing.noise.background}) filters out all irrelevant spatial scales and thus substantially reduces the fluctuations
outside sources. Therefore, the significance of detections by \textsl{getsources} may well be considerably higher than the
conventional $\Omega_{{i}{\lambda}}$ would indicate\footnote{Simulations show that $\Xi_{{i}{\lambda}}$ gives a notably more
accurate representation of the true model S/N and with a much lower dispersion than the conventional estimates do using the full
images containing all spatial scales.}.

Since \textsl{getsources} is a multi-wavelength extraction method and the detection is generally performed in the \emph{combined}
images $\mathcal{I}_{{\rm D}{j}{\,\rm C}}$, it would be useful to define another quantity to measure the significance of source
detection more globally, across all wavebands. To obtain a meaningful quantity, one cannot use the combined images, because they
contain renormalized and summed up monochromatic images (Eq.~\ref{combining.waves.1}). Assuming that we have a set of independent
images $\mathcal{I}_{{\!\lambda}{\rm D}}$, it makes sense to define the \emph{global significance} as
\begin{equation}
\Xi_{i}{\,=\,}\left(\sum_{\lambda}\,\Xi_{{i}{\lambda}}^2\right)^{1/2}.
\label{global.significance}
\end{equation}
We consider two levels of the robustness of source detection: the \emph{reliable} level $\Xi_{\rm rel}$ and \emph{tentative} level
$\Xi_{\rm ten}$ (default values of 7 and 5, respectively). Reliable sources, i.e. those with $\Xi_{{i}{\lambda}}{\,\ge\,}\Xi_{\rm
rel}$ in at least one waveband used for detection, are cataloged without checking whether they are detected at any other wavelength.
Tentative sources, i.e. those with $\Xi_{{i}{\lambda}}{\,<\,}\Xi_{\rm rel}$ in all wavebands used for detection, are kept in the
final catalog only if $\Xi_{{i}{\lambda}}{\,\ge\,}\Xi_{\rm ten}\,n_{\rm det}^{-1/2}$ in at least $n_{\rm det}$ wavebands ($n_{\rm
det}$ defaults to 1 and 2 for the monochromatic and multi-wavelength extractions, respectively). For any cataloged source,
$\Xi_{i}{\,\ge\,}\Xi_{\rm ten}$, and for reliable sources, $\Xi_{i}{\,\ge\,}\Xi_{\rm rel}$.

Having measured the properties of all detected sources at each wavelength, \textsl{getsources} completely removes those that are
likely to be spurious. The algorithm treats removal of such sources with great care, dividing the measurement iterations into three
phases. During the first phase, only non-detections are removed from the catalog, i.e. those sources that do not fulfill the above
requirement of the simultaneous detection in at least $n_{\rm det}$ wavebands, all other detected sources are given a chance to
converge. During the second phase, the algorithm removes sources with extremely low \emph{goodness} ($G_{i}{<\,}0.01$) which we 
define as
\begin{equation}
G_{i}{\,=\,}\min\left\{\frac{1}{2}\left(\frac{R_{i}}{\Xi_{\rm rel}}-1\right),1\right\},
\label{goodness}
\end{equation}
where $R_{i}{\,=\,}\Xi_{i}\,\Omega_{i}\,C^{-1}_{{i\,\rm E}{j_{\rm F}}}$ is the source \emph{reliability}, $\Omega_{i}$ is the
\emph{global signal-to-noise ratio} defined analogously to the global significance $\Xi_{i}$ (cf. Eq.~\ref{global.significance}),
and $C_{{i\,\rm E}{j_{\rm F}}}$ is the elongation factor (Eq.~\ref{amplification.elongation}) at the footprinting scale $j_{\rm F}$.
Finally, during the third phase, sources are also removed when their position is too close to another source (within
$\max\,\{1.5,(O_{\rm min}/\Delta)/3\}$ pixels, where $O_{\rm min} = \min\,\{O_{\lambda}\}$) of almost the same size (within a factor
of 2) and they have a lower value of $\Xi_{i}\,\Omega_{i}$ than the other source.

Since \textsl{getsources} is a multi-wavelength extraction method, globally good sources may well be detectable
($\Xi_{{i}{\lambda}}{\,\ge\,}\Xi_{\rm ten}\,n_{\rm det}^{-1/2}$) or measurable ($\Omega_{{i}{\lambda}}{\,\ge\,}1$) in only a limited
number of wavebands. Indeed, it is quite usual that sources are either insignificant or non-measurable at some wavelengths and this
generally leads to their footprints being very unreliable. In order to produce most accurate measurements, the footprints of such
sources are removed from the corresponding footprint images $\mathcal{F}_{\!\lambda}$ in that waveband. This becomes very important
in crowded regions with many overlapping sources. Since the quality of both the background subtraction and deblending depends
directly on the accuracy of the accumulated footprints, $\mathcal{F}_{\!\lambda}$ must always remain as clean as possible, free of
the footprints of insignificant or non-measurable sources. This wavelength-dependent removal is done starting from the second phase;
however, the measurements of such sources are still kept in the catalog from the first phase of iterations.

After the removal of bad sources, \textsl{getsources} analyzes the spatial distribution of all sources at each wavelength and flags
them to provide some useful information on global properties of a source. The single-digit flag $f_{i}$ is assigned to sources
according to the following definitions. A source is called \emph{isolated} ($f_{i}{\,=\,}0$) if it is \emph{not} blended with any
other source at any wavelength. Two sources are called \emph{blended} ($f_{i}{\,=\,}1$) if the intersection area of their footprints
is greater than 20{\%} of either of the footprints in at least one waveband. A source is called \emph{sub-structured}
($f_{i}{\,=\,}2$) if a footprint of at least one smaller source is fully contained within the inner 56{\%} of the footprint area (or
within $0.75\,A_{{i}{\,\rm F}{\lambda}}$ and $0.75\,B_{{i}{\,\rm F}{\lambda}}$). A source is called \emph{sub-structuring}
($f_{i}{\,=\,}3$) if it causes another source to be flagged as sub-structured. In addition to the global flag $f_{i}$, we also
define the monochromatic flag $f_{{\,i}{\lambda}}$ that carries information on the wavelength-dependent properties for each source.
Among other details, the flag identifies sources that are insignificant or non-measurable in a waveband.

Further, \textsl{getsources} updates the extraction catalog, where each line contains the source number, coordinates, world
coordinates \citep[using the \textsl{xy2sky} utility,][]{Mink2002}, global significance, flag, and goodness, followed by the 
measured properties at each wavelength: 
\begin{displaymath}
i\: x_{i}\: y_{i}\: \alpha_{i}\: \delta_{i}\: \Xi_{\,i}\: f_{i}\: G_{i}\: \left(\, \Xi_{{\,i}{\lambda}}\: f_{{\,i}{\lambda}}\:
F_{{i}{\lambda}{\,\rm P}}\: \sigma_{{i}{\lambda}{\,\rm P}}\: F_{{i}{\lambda}{\,\rm T}}\: \sigma_{{i}{\lambda}{\,\rm T}}\:
A_{{i}{\lambda}}\: B_{{i}{\lambda}}\: \Theta_{{i}{\lambda}}\, \right)_{N}.
\label{catalog.line}
\end{displaymath}
The standard signal-to-noise ratio $\Omega_{{i}{\lambda}}$ is not cataloged, because it can easily be derived from the catalog
entries. The last processing block of each measurement iteration updates the accumulated footprints $\mathcal{F}_{\!\lambda}$ of all
sources, based on the latest values of the source sizes and position angles. Full sizes and orientation of the footprint ellipses
are computed from
\begin{equation}
A_{{i}{\,\rm F}{\lambda}}{\,=\,}2.3\max\,\left\{O_{\lambda}, S_{\!j_{\rm F}{\lambda}}\right\}, \,\,\,
B_{{i}{\,\rm F}{\lambda}}{\,=\,}\frac{B_{{i}{\lambda}}}{A_{{i}{\lambda}}}\,A_{{i}{\,\rm F}{\lambda}}, \,\,\,
\Theta_{{i}{\,\rm F}{\lambda}}{\,=\,}\Theta_{{i}{\lambda}}.
\label{footprints}
\end{equation}

Having obtained the improved footprints, \textsl{getsources} checks whether the total number of non-zero pixels in
$\mathcal{F}_{\!\lambda}$ has noticeably changed with respect to the previous iteration. If so, the measurement iterations continue
until the total area of all footprints changes by less than 0.1{\%}. Convergence properties of the measurement iterations vary for
different fields; somewhere between 5 and 30 iterations may be required to obtain a stable pattern of all footprints for each
waveband (Fig.~\ref{converged.footprints}) and to produce the final catalog. In the last iteration, \textsl{getsources} creates an
additional catalog of colors from all possible combinations of total fluxes $F_{{i}{\lambda}{\,\rm T}}$ over all $\lambda$, a
catalog of peak and background intensities, as well as three versions of the azimuthally-averaged intensity profiles
($\mathcal{I}_{{\!\lambda}{\rm O}}$, $\mathcal{I}_{{\!\lambda}{\rm O}{\,\rm BS}}$, $\mathcal{I}_{{\!\lambda}{\rm O}{\,\rm BSD}}$)
computed from the original, background-subtracted, and deblended images, respectively.

\subsection{Visualizing extractions}
\label{visualizing.extractions}

\begin{figure*}
\centering
\centerline{\resizebox{0.33\hsize}{!}{\includegraphics{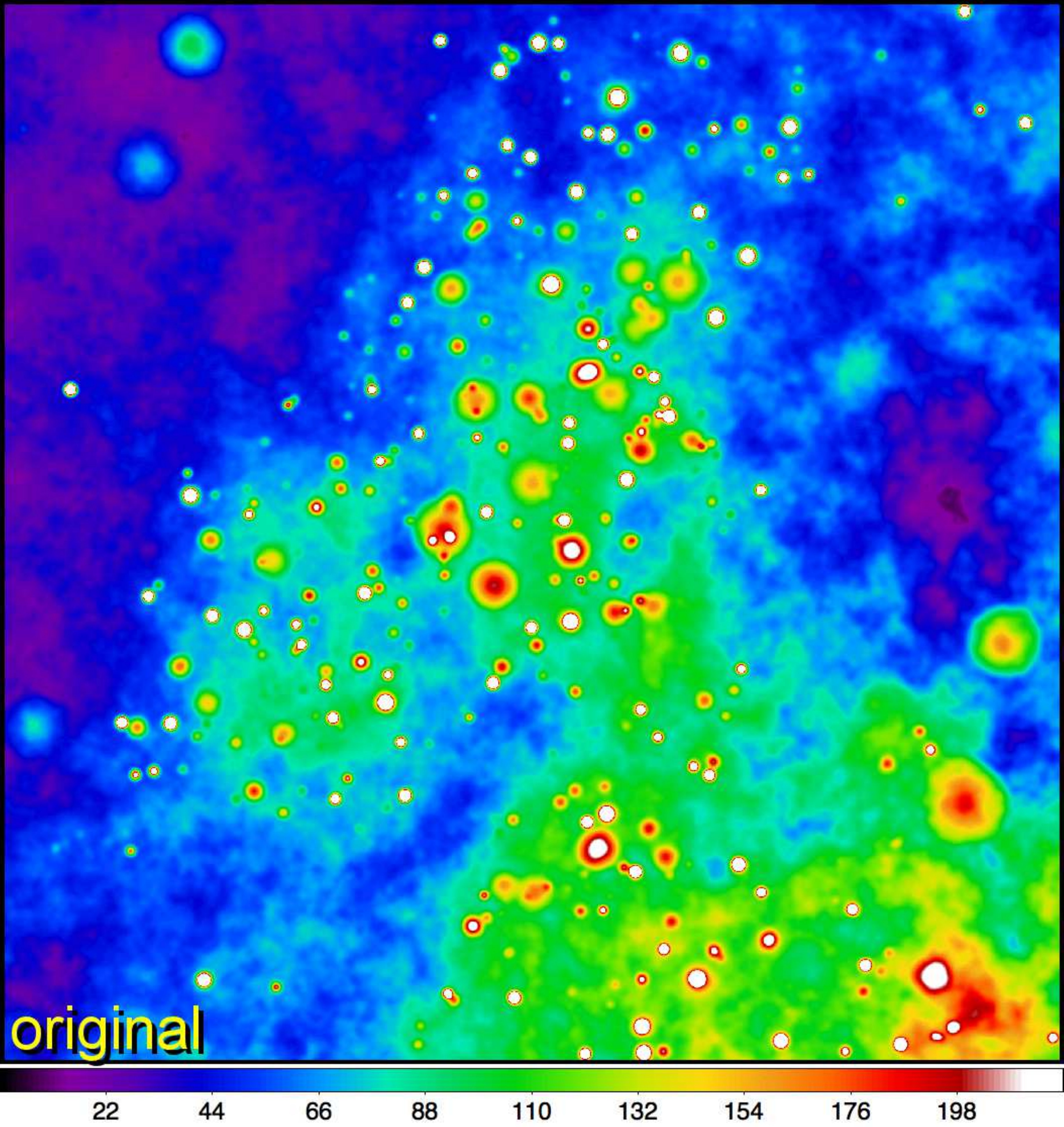}}
            \resizebox{0.33\hsize}{!}{\includegraphics{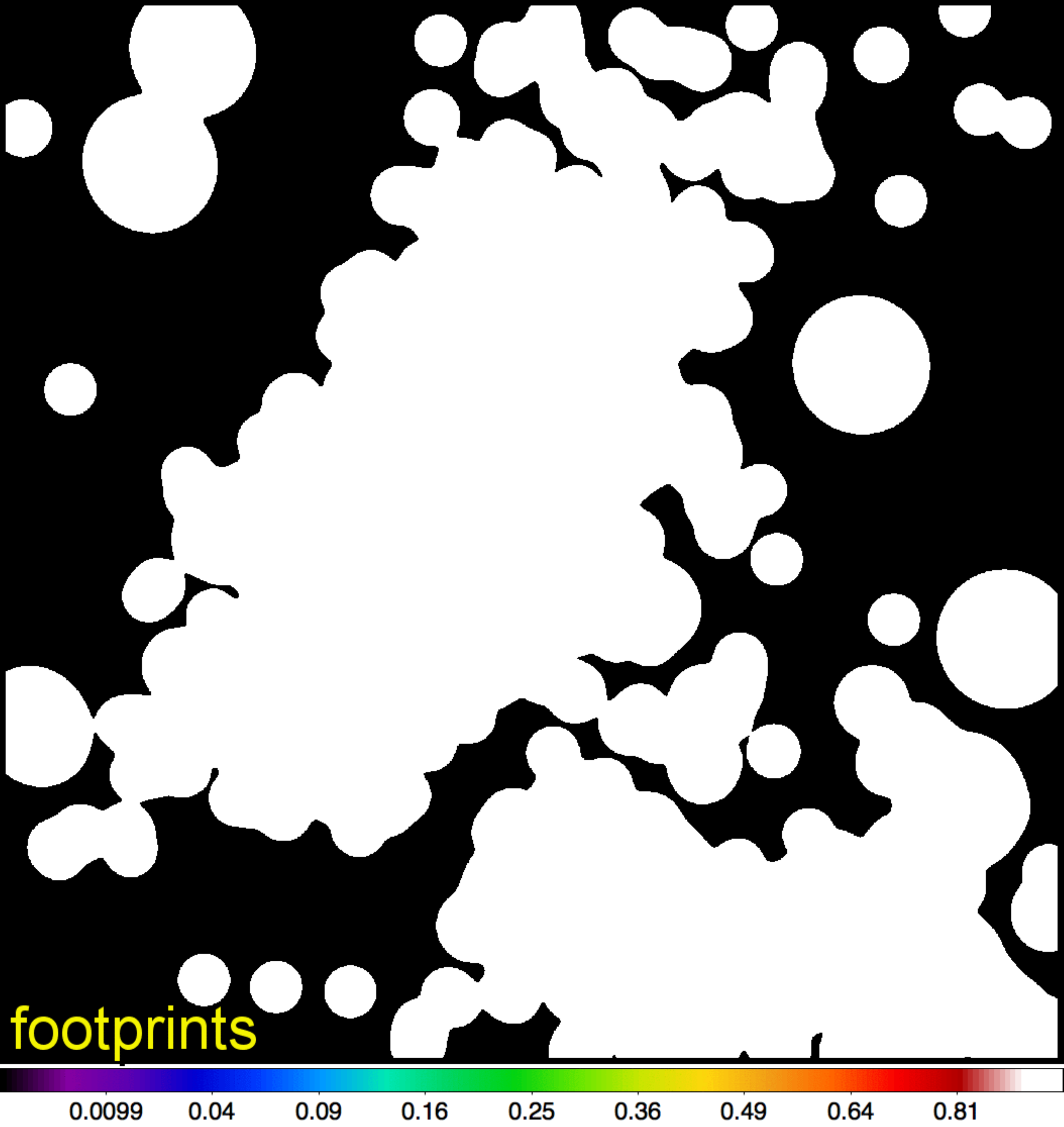}}
            \resizebox{0.33\hsize}{!}{\includegraphics{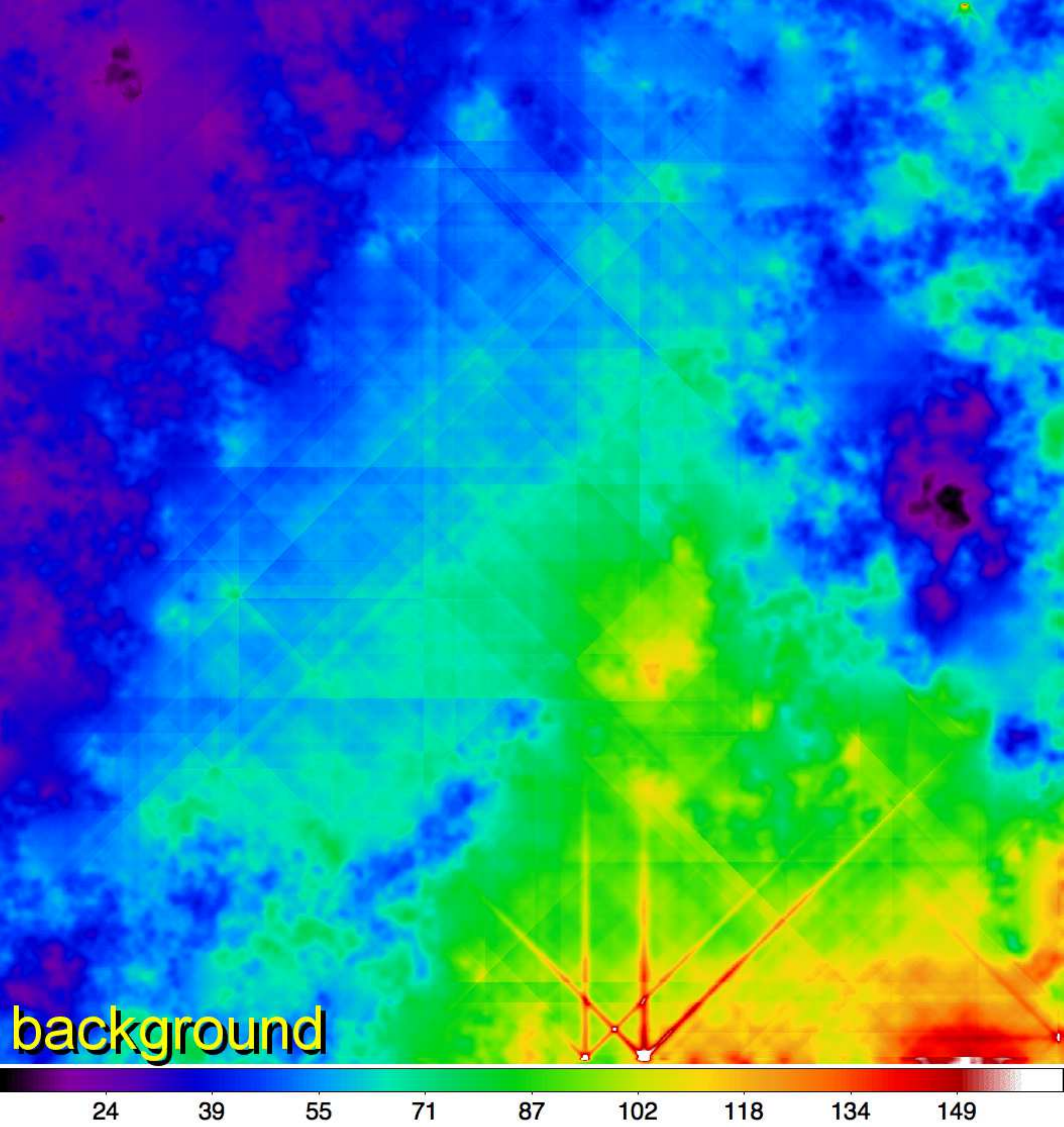}}}
\centerline{\resizebox{0.33\hsize}{!}{\includegraphics{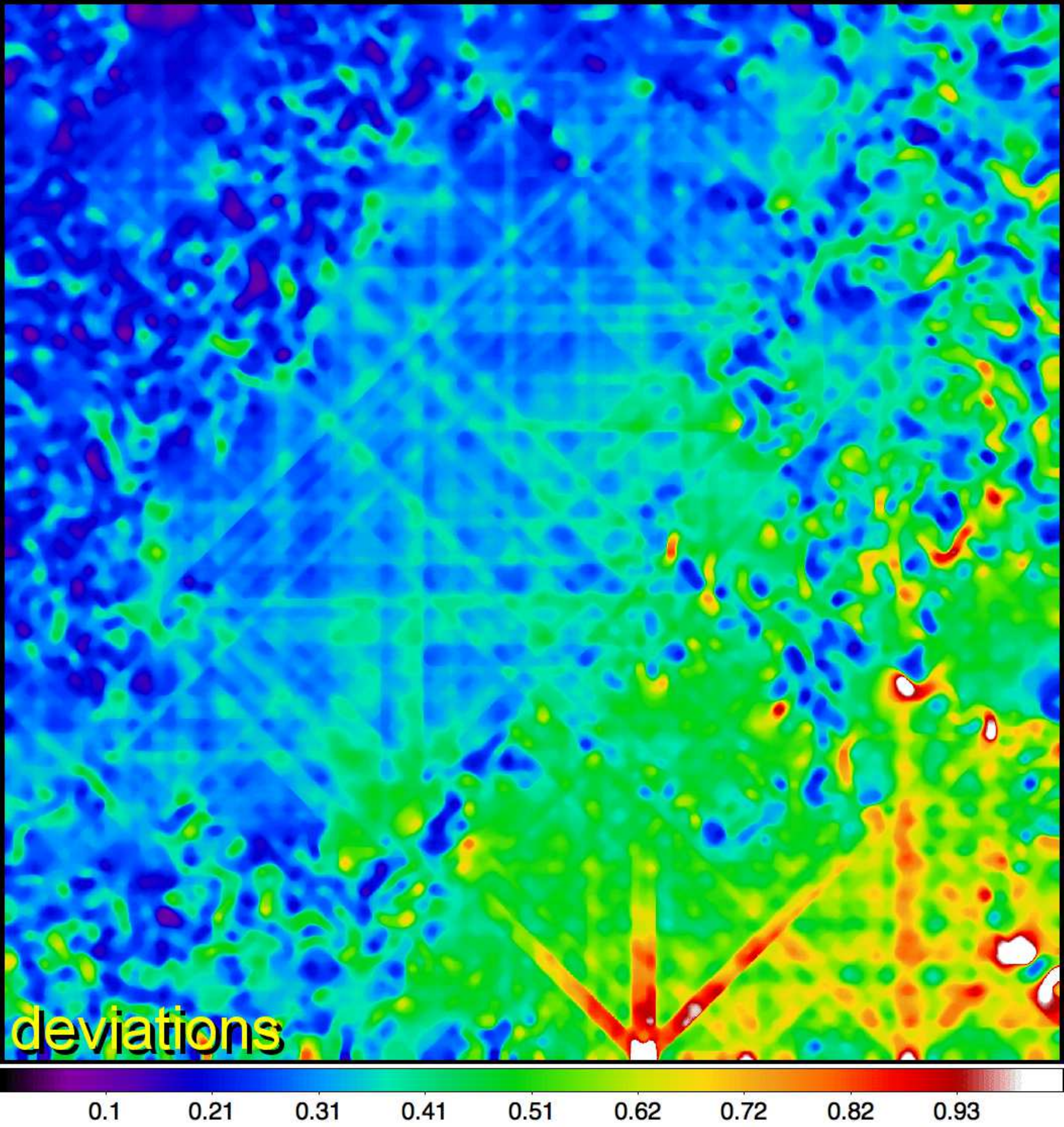}}
            \resizebox{0.33\hsize}{!}{\includegraphics{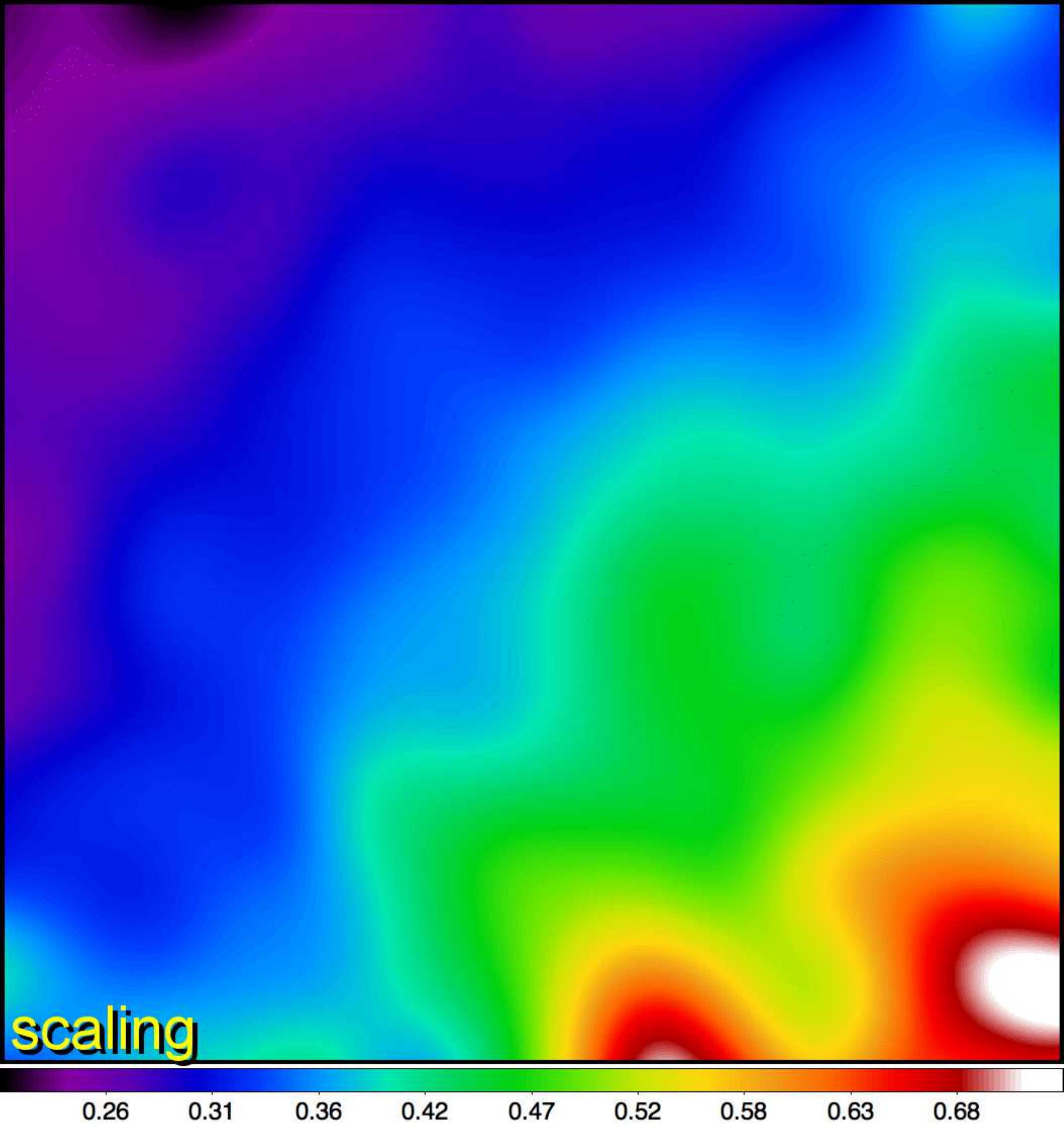}}
            \resizebox{0.33\hsize}{!}{\includegraphics{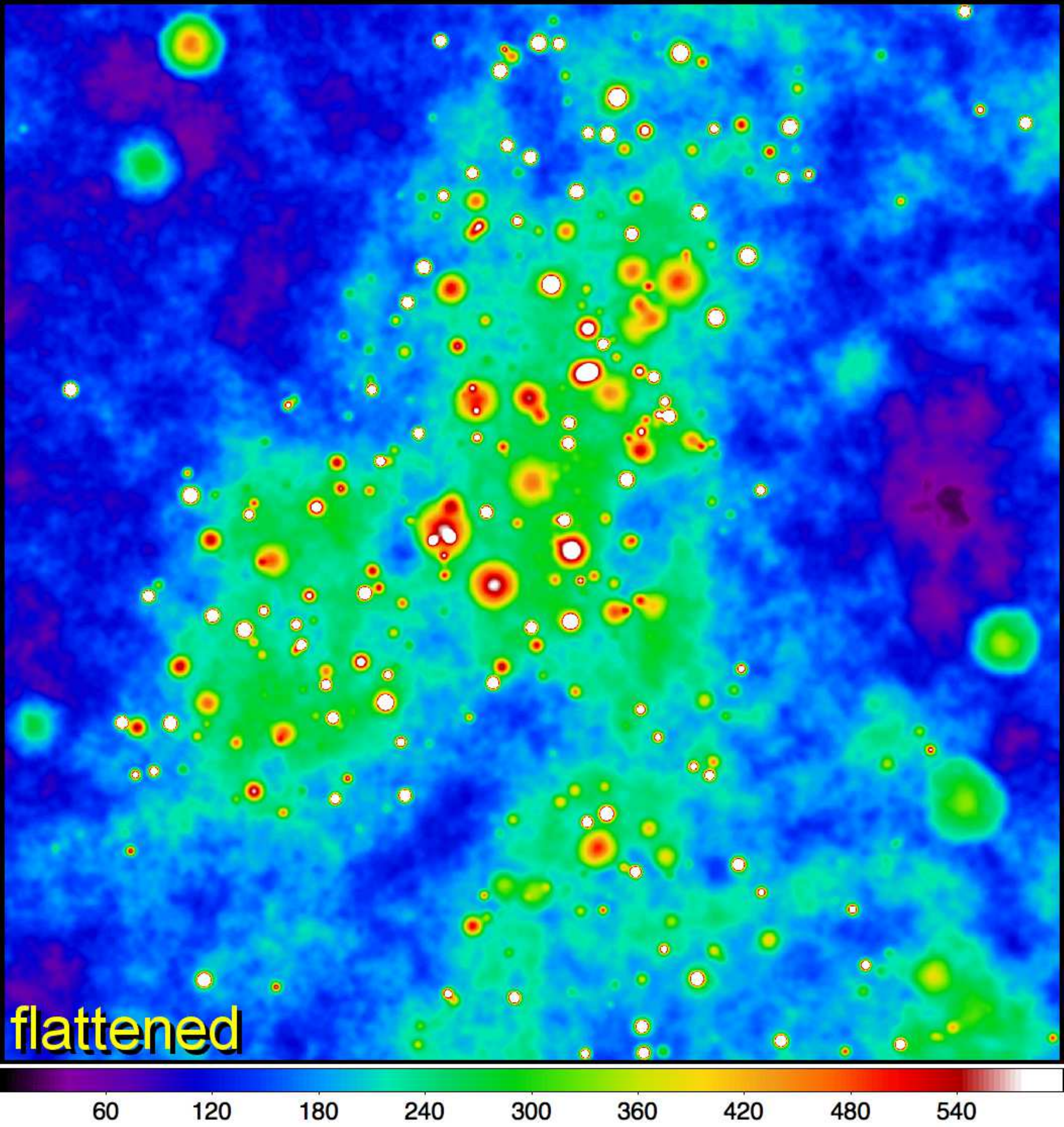}}}
\caption{
Flattening of the detection images (Sect.~\ref{flattening.background.noise}). The entire 1{\degr}$\times$1{\degr} simulated
star-forming region at 350\,{${\mu}$m} is shown (\emph{upper left}), the central area of which was used to illustrate
\textsl{getsources} in this paper. The image $\mathcal{F}^{*}_{\!\lambda}$ of the converged footprints $\mathcal{F}_{\!\lambda}$
somewhat expanded by convolution (\emph{upper middle}) is used to interpolate all detected sources off the image to obtain the clean
background image (\emph{upper right}). The image $\mathcal{D}_{\!\lambda}$ of the standard deviations (\emph{lower left}) is
computed in a very small (3$\times$3 pixels) sliding window and further median-filtered using a 21$\times$21 pixels sliding window
to reduce effects of possible artifacts. The smoothed scaling image $\mathcal{I}_{{\!\lambda}{\rm F}}$ (\emph{lower middle}) is
obtained by convolving the image in the previous panel with a circular Gaussian beam $\mathcal{G}_{{\!\lambda}{\it A}}$ larger than
the maximum size of sources extracted at 350\,{${\mu}$m} in the initial extraction. The resulting flattened image
$\mathcal{I}_{{\!\lambda}{\rm D}{\rm F}}$ (\emph{lower right}) is obtained by dividing the original image
$\mathcal{I}_{{\!\lambda}{\rm D}}$ by the scaling image $\mathcal{I}_{{\!\lambda}{\rm F}}$. For better visibility, the values
displayed in the panels are somewhat limited in range; the color coding is a linear function of intensity in MJy/sr.
}
\label{flattening.images}
\end{figure*}

In order to facilitate visual analysis of the extraction results, \textsl{getsources} produces a number of useful images for each
waveband. These include (but are not limited to) the background-subtracted images $\mathcal{I}_{{\!\lambda}{\rm D}{\,\rm BS}}$,
$\mathcal{I}_{{\!\lambda}{\rm O}{\,\rm BS}}$ (Fig.~\ref{composite.obsbs}), observed images $\mathcal{I}_{{\!\lambda}{\rm O}}$ with
the extraction ellipses overlaid on top (Fig.\,\ref{extraction.ellipses}), and detection images $\mathcal{I}_{{\!\lambda}{\rm D}}$
(Fig.~\ref{flattening.images}). Other images show just the central positions of the detected sources on top of the images
$\mathcal{I}_{{\!\lambda}{\rm D}{\,\rm BS}}$ and $\mathcal{I}_{{\!\lambda}{\rm O}{\,\rm BS}}$; these are useful for visualizing
crowded regions, where very little can actually be seen under the ellipses. Furthermore, the source images
$\mathcal{I}_{{\!i}{{\lambda}{\rm O}}{\,\rm BSD}}$ display the background-subtracted, deblended intensity distribution for each
individual source at each wavelength (Fig.~\ref{deblended.images}).

For an easy visualization of the various steps of the algorithm, \textsl{getsources} produces also a number of useful additional
images. These include (but not limited to) the interpolated backgrounds $\mathcal{I}_{{\!\lambda}{\rm D}{\,\rm CB}}$,
$\mathcal{I}_{{\!\lambda}{\rm O}{\,\rm CB}}$, the images of $\sigma_{{i}{\lambda}{\,\rm P}}$ (Fig.~\ref{converged.footprints}),
the annuli $\mathcal{A}_{\lambda}$ (Fig.~\ref{bigfoot.images}), the deblending shapes $\mathcal{I}_{{\!\lambda}{\rm M}}$
(Fig.~\ref{composite.obsbs}). The clean single-scale images $\mathcal{I}_{{\!\lambda}{\rm D}{j}{\,\rm C}}$
(Fig.~\ref{clean.single.scales}), $\mathcal{I}_{{\rm D}{j}{\,\rm C}}$ (Fig.~\ref{combined.clean.single.scales}), the accumulated
footprints $\mathcal{F}_{\rm D}$ (Fig.~\ref{segmentation.images}), and segmentation images $\mathcal{I}_{{\rm D}{j}{\,\rm S}}$
(Fig.~\ref{segmentation.images}) also remain available for an in-depth analysis and understanding of the extraction process and
results.

\subsection{Flattening background and noise}
\label{flattening.background.noise}

Despite the fine spatial filtering employed by \textsl{getsources}, there is one common problem that still needs special treatment.
It is known that the \emph{Herschel} images of Galactic regions often show highly-variable backgrounds. The standard deviations of
the combined background and noise fluctuations outside sources may differ by orders of magnitude between various areas of a large
image $\mathcal{I}_{{\!\lambda}{\rm D}}$. Any simple thresholding method would have a difficulty in handling such images, as the
global thresholds would not be good for all areas. Although the single-scale decomposition used in our method makes the images
$\mathcal{I}_{{\!\lambda}{\rm D}{j}}$ much ``flatter'' and more suitable for use with global thresholding, it cannot overcome the
problem completely, as the backgrounds fluctuate on all spatial scales, from the smallest to the largest ones. If the background or
noise fluctuations strongly vary between some areas of the image $\mathcal{I}_{{\!\lambda}{\rm D}}$, they will still influence the
intensity distributions of the single-scale images $\mathcal{I}_{{\!\lambda}{\rm D}{j}}$, although to a much lesser degree.

\begin{figure}
\centering
\centerline{\resizebox{0.7\hsize}{!}{\includegraphics{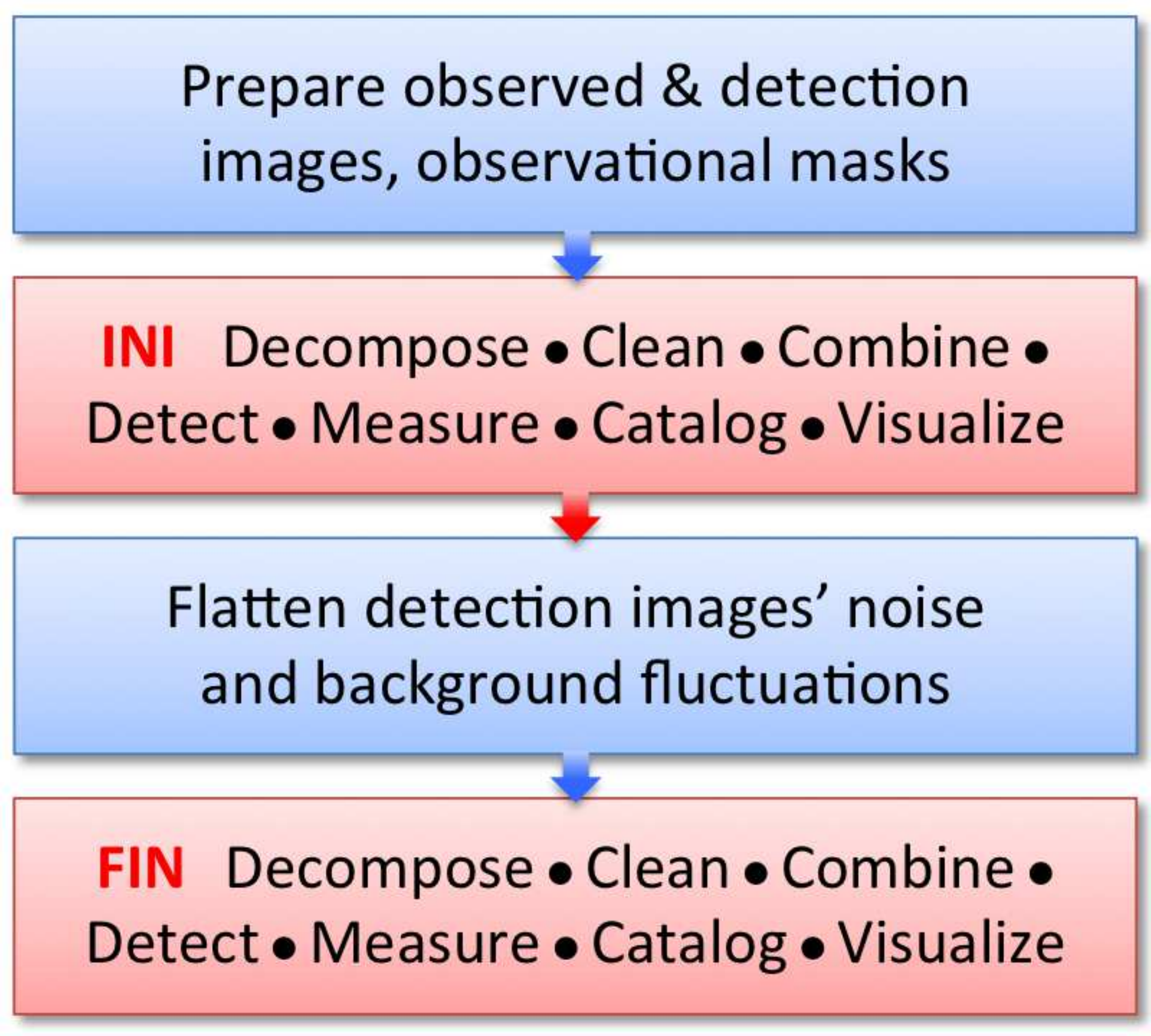}}} 
\caption{
Flattening of the detection images (Sect.~\ref{flattening.background.noise}). The \textsl{getsources} algorithm requires two
complete extractions, the \emph{initial} and the \emph{final} extractions (red blocks, expanded in Fig.~\ref{algorithm}; the
preparation steps are shown in blue).
}
\label{two.step.approach}
\end{figure}

Our method adopts a special two-step approach (Fig.~\ref{two.step.approach}) to overcome the problem; essentially, two complete
extractions are performed instead of a single one. The \emph{initial} deeper extraction\footnote{The depth is automatically adjusted
by lowering 4 configuration parameters to the following values: $\Xi_{\rm rel} = 4$, $\Xi_{\rm ten} = 3$, $n_{\rm det} = 1$, $\eta =
1.1$.} aims at finding all possible candidate sources, in order to remove them and create the cleanest background images, free of
any sources. The background images are then used to create the standard-deviation images $\mathcal{D}_{\!\lambda}$ and convolve them
with a Gaussian beam $\mathcal{G_{{\lambda}{\it A}}}$, producing the scaling (flattening) images $\mathcal{I}_{{\!\lambda}{\rm F}}$
(see below for details). Dividing the detection images by the scaling images, we obtain the detection images
$\mathcal{I}_{{\!\lambda}{\rm D}{\rm F}}$ that are flat, in the sense that the standard deviations in their background areas
(outside of the sources) are approximately the same. The flattening procedure can be expressed as
\begin{equation}
\mathcal{I}_{{\!\lambda}{\rm D}{\rm F}} = \mathcal{I}_{{\!\lambda}{\rm D}}\,\mathcal{I}^{-1}_{{\!\lambda}{\rm F}} 
= \mathcal{I}_{{\!\lambda}{\rm D}} \left(\mathcal{G_{{\lambda}{\it A}}} * \mathcal{D}_{\!\lambda}\right)^{-1}.
\label{flattening}
\end{equation}
The second and \emph{final} extraction is performed the same way and with the same parameters as the initial extraction\footnote{The
three parameters automatically lowered in the initial extraction are now set to their default values: $\Xi_{\rm rel} = 7$, $\Xi_{\rm
ten} = 5$, $n_{\rm det} = 2$, $\eta = 1.35$.}. The only difference is that the \emph{detection} images $\mathcal{I}_{{\!\lambda}{\rm
D}}$ are replaced with their flattened versions $\mathcal{I}_{{\!\lambda}{\rm D}{\rm F}}$ and that, instead of the initial
\emph{guesses} for the maximum sizes of the sources, the actual maximum sizes $A^{\rm max}_{\lambda}$ derived in the initial
extraction are used. In both extractions, the measurements are performed on the same observed images $\mathcal{I}_{{\!\lambda}{\rm
O}}$.

The flattening procedure is illustrated in Fig.~\ref{flattening.images}. For reference, the upper-left panel shows the original
detection image $\mathcal{I}_{{\!\lambda}{\rm D}}$ at 350\,{${\mu}$m} of the simulated star-forming region used in this paper for
illustrations of the method; the entire 1{\degr}$\times$1{\degr} is shown here in order to clearly see the flattening effect. The
simulated background in this field has a temperature gradient along one diagonal of the image, clearly visible in the images. The
dust temperature $T_{\rm d}$ linearly varies from 20\,K in the upper-left corner down to 15\,K in the lower-right corner, where the
background appears much brighter at 350\,{${\mu}$m}. The footprint image $\mathcal{F}^{*}_{\!\lambda}$ shown in the upper-middle
panel is the image $\mathcal{F}_{\!\lambda}$ (Fig.~\ref{converged.footprints}) somewhat expanded by convolution, to make sure that
the resulting clean background has no residual artifacts that would influence the quality of the final flattening image. The image
$\mathcal{F}^{*}_{\!\lambda}$ is used to remove all sources from the detection image with our background interpolation scheme
(Sect.~\ref{measuring.cataloging}). 

The interpolated background in Fig.~\ref{flattening.images} is smooth and clean, except for a couple of artifacts at its lower edge.
The images that are bright and variable \emph{at their edges} may lead to such edge artifacts, because \textsl{getsources} uses
convolution and interpolation. Although the images are sufficiently expanded (extrapolated) by the algorithm before convolution,
they remain essentially unknown beyond their edges. This type of artifacts never happens with the entire images from real-world
observations that produce very low intensities at their edges, due to the baseline subtraction. Such problems may only exist in
simulated images or in sub-fields that have been cut out of full larger images\footnote{A natural remedy is to define the sub-fields 
that are larger than the area one is interested in studying and make sure that the intensities at the edges of the extraction area 
are relatively low.}.

From the background image \textsl{getsources} creates the image of standard deviations $\mathcal{D}_{\!\lambda}$, computed in a very
small (3$\times$3 pixels) sliding window. The aim here is not to produce statistically-meaningful values, but to ensure that the
features of $\mathcal{D}_{\!\lambda}$ remain as local as possible; much larger windows would smooth out the values, which may not be
suitable for the original highest-resolution images. The image $\mathcal{D}_{\!\lambda}$ is further median-filtered using a
21$\times$21 pixels sliding window to reduce the effects of possible residuals or artifacts. The same image
$\mathcal{F}^{*}_{\!\lambda}$ is used again to interpolate the median-filtered pixel values under the footprints, resulting in the
image shown in the lower-left panel of Fig.~\ref{flattening.images}. The scaling (flattening) image
$\mathcal{I}_{{\!\lambda}{\rm F}}$ is obtained by convolving the filtered $\mathcal{D}_{\!\lambda}$ image with a circular Gaussian
beam $\mathcal{G_{{\lambda}{\it A}}}$ of a size 3 times the maximum size of sources $A^{\rm max}_{\lambda}$ measured in the initial
extraction; the smoothing ensures that the flattening does not distort the intensity distribution of even the largest sources. The
scaling image $\mathcal{I}_{{\!\lambda}{\rm F}}$ resembles fairly well the original temperature gradient that was introduced in the
simulated images along their diagonal. The resulting flattened image $\mathcal{I}_{{\!\lambda}{\rm D}{\rm F}}$ is the original
image $\mathcal{I}_{{\!\lambda}{\rm D}}$ divided by $\mathcal{I}_{{\!\lambda}{\rm F}}$. The large-scale background variations
clearly visible in the original detection image $\mathcal{I}_{{\!\lambda}{\rm D}}$ have been mostly removed from the flattened
image $\mathcal{I}_{{\!\lambda}{\rm D}{\rm F}}$, making the latter suitable for the global single-scale thresholding applied by
\textsl{getsources} in the final extraction.

%||||||||||||||||||||||||||||||||||||||||||||||||||||||||||||||||||||||||||||||||||||||||||||||||||||||||||||||||||||||||||||||||||

\section{Applications to \emph{Herschel} images}
\label{applications.herschel}

In Sect.~\ref{getsources.extraction.method}, our multi-wavelength source extraction method was described and illustrated using the
images of a simulated star-forming region. The simulation is one of our suite of benchmarks designed to aid in the development of
\textsl{getsources} and in accurate tests of its performance in various conditions against the model images with fully known
properties of the sources, background, and noise (the benchmarks will be described elsewhere; Men'shchikov et al., in prep.). In
addition to the purely synthetic skies, \textsl{getsources} has been successfully tested on the ground-based millimeter continuum
survey of the \object{Aquila Rift} complex \citep{Maury_etal2011}, where all extracted sources have been carefully verified by eye
inspection and their parameters evaluated manually.

We present here two real-life examples of \textsl{getsources} extractions for our Gould Belt and HOBYS surveys of the nearby and
more distant star-forming regions with \emph{Herschel}. For this purpose, we defined sub-fields of the observed images of
\object{Aquila} and \object{Rosette}\footnote{The respective sub-fields are similar to the areas displayed in Fig.~5 of
\cite{Ko"nyves_etal2010} and in Fig.~1 of \cite{Hennemann_etal2010}.}, small enough to enable readers to verify the extraction
results with their own eyes. We emphasize that the only goal of this presentation is to help the readers clarify various aspects of
this new source extraction method; any astrophysical analysis of the fields is beyond the scope of this paper.

\subsection{A cluster of resolved prestellar cores in \object{Aquila}}
\label{prestellar.cores.aquila}

\begin{figure*}
\centering
\centerline{\resizebox{0.33\hsize}{!}{\includegraphics{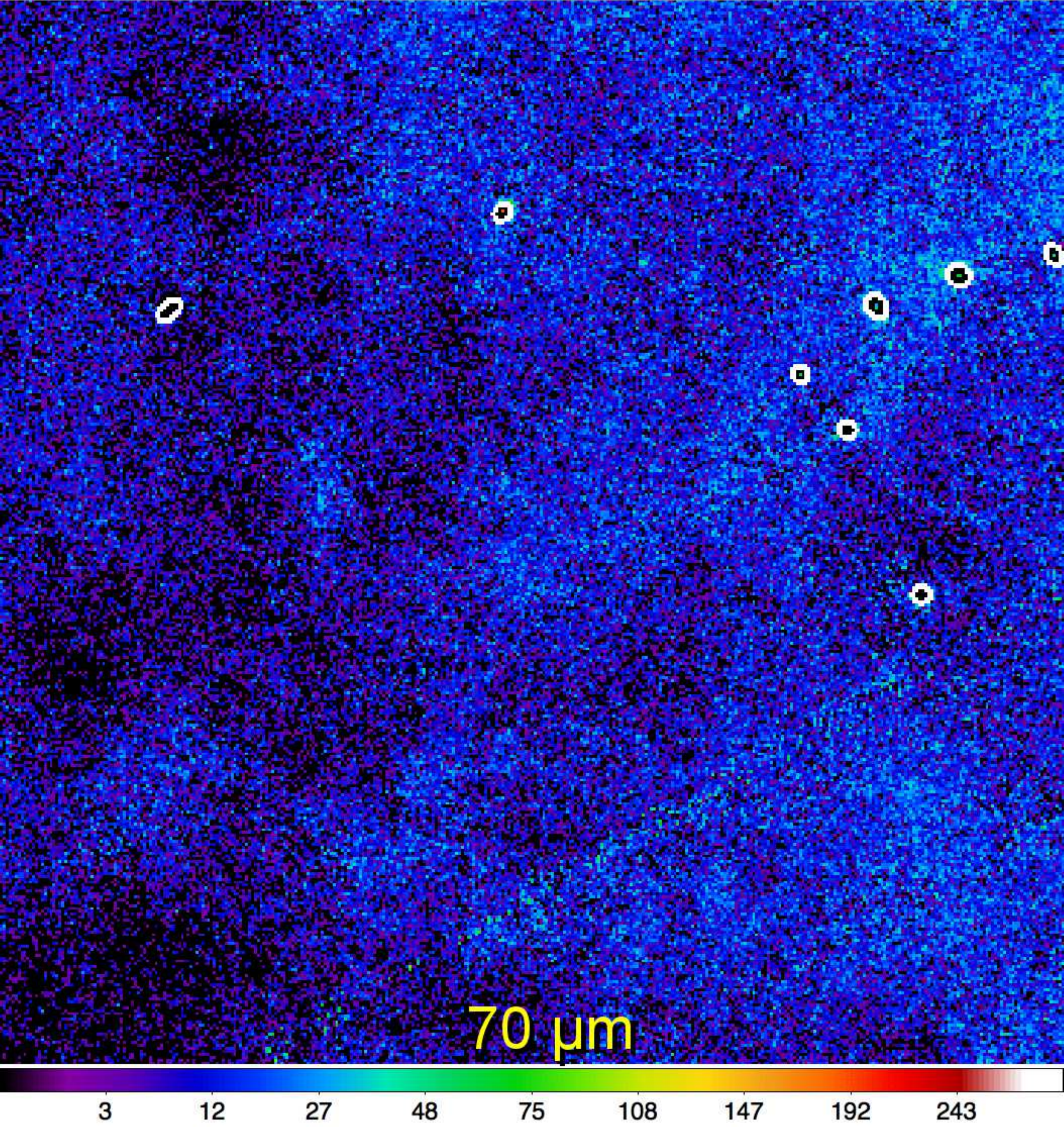}}
            \resizebox{0.33\hsize}{!}{\includegraphics{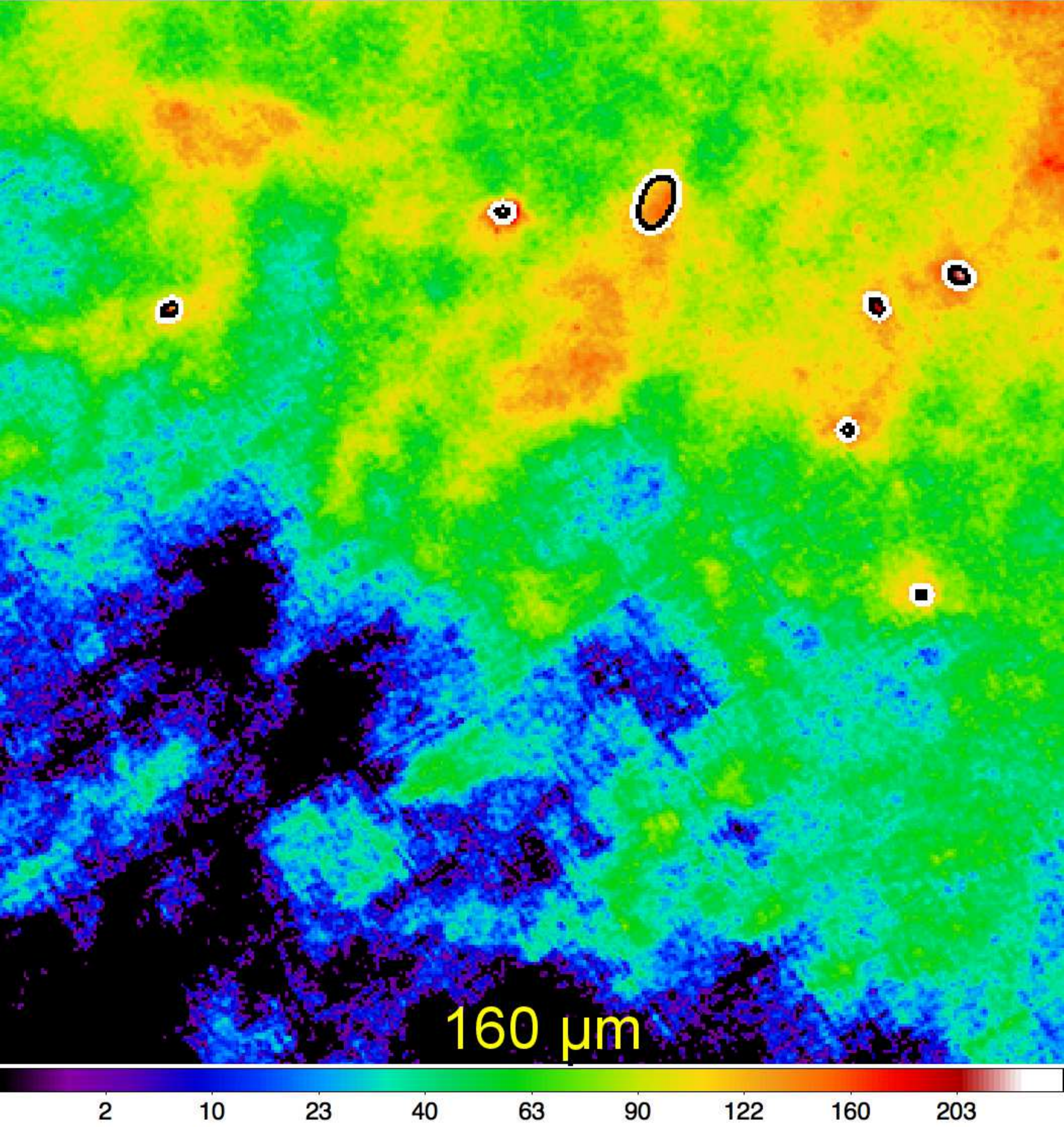}}
            \resizebox{0.33\hsize}{!}{\includegraphics{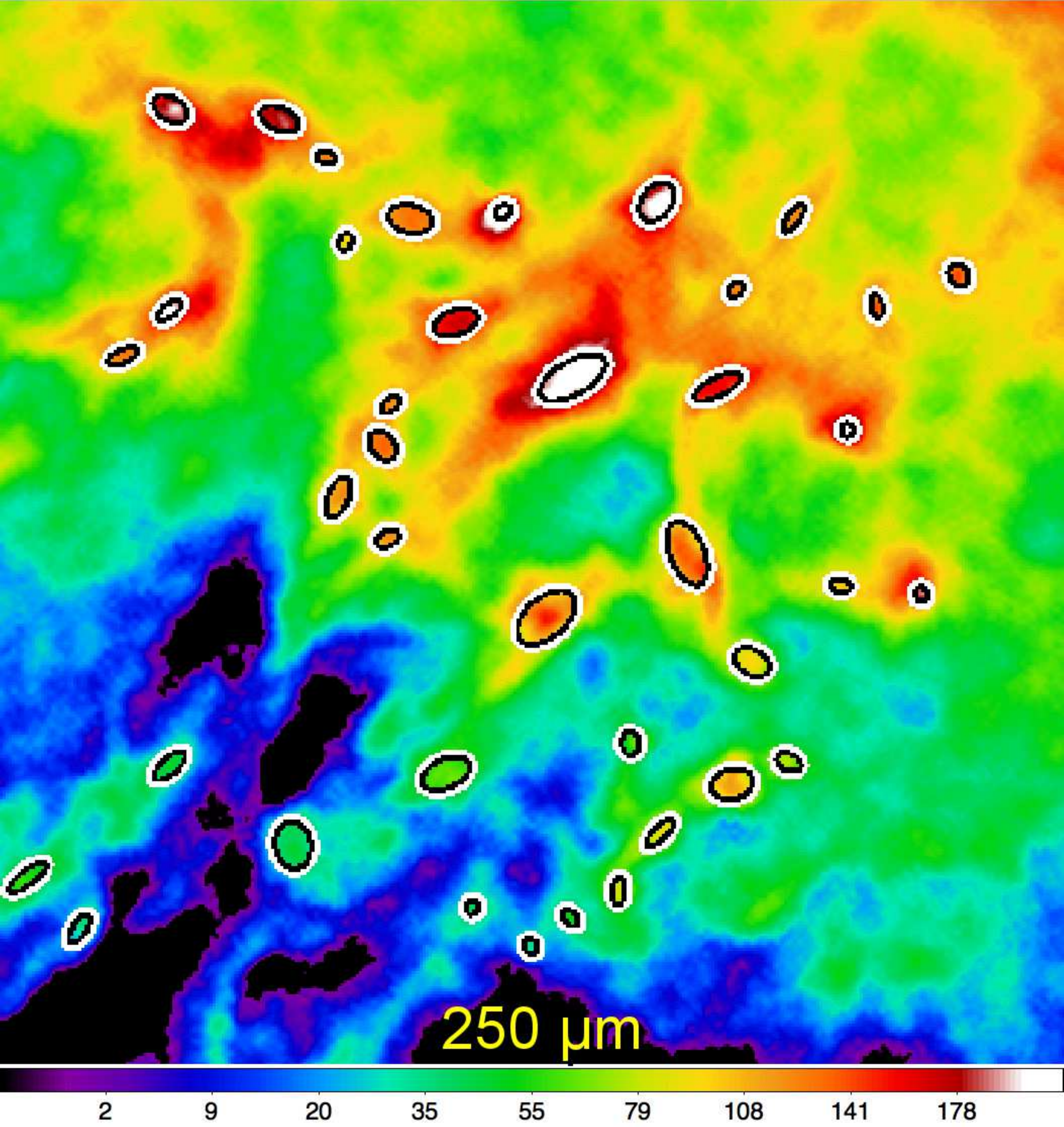}}}
\centerline{\resizebox{0.33\hsize}{!}{\includegraphics{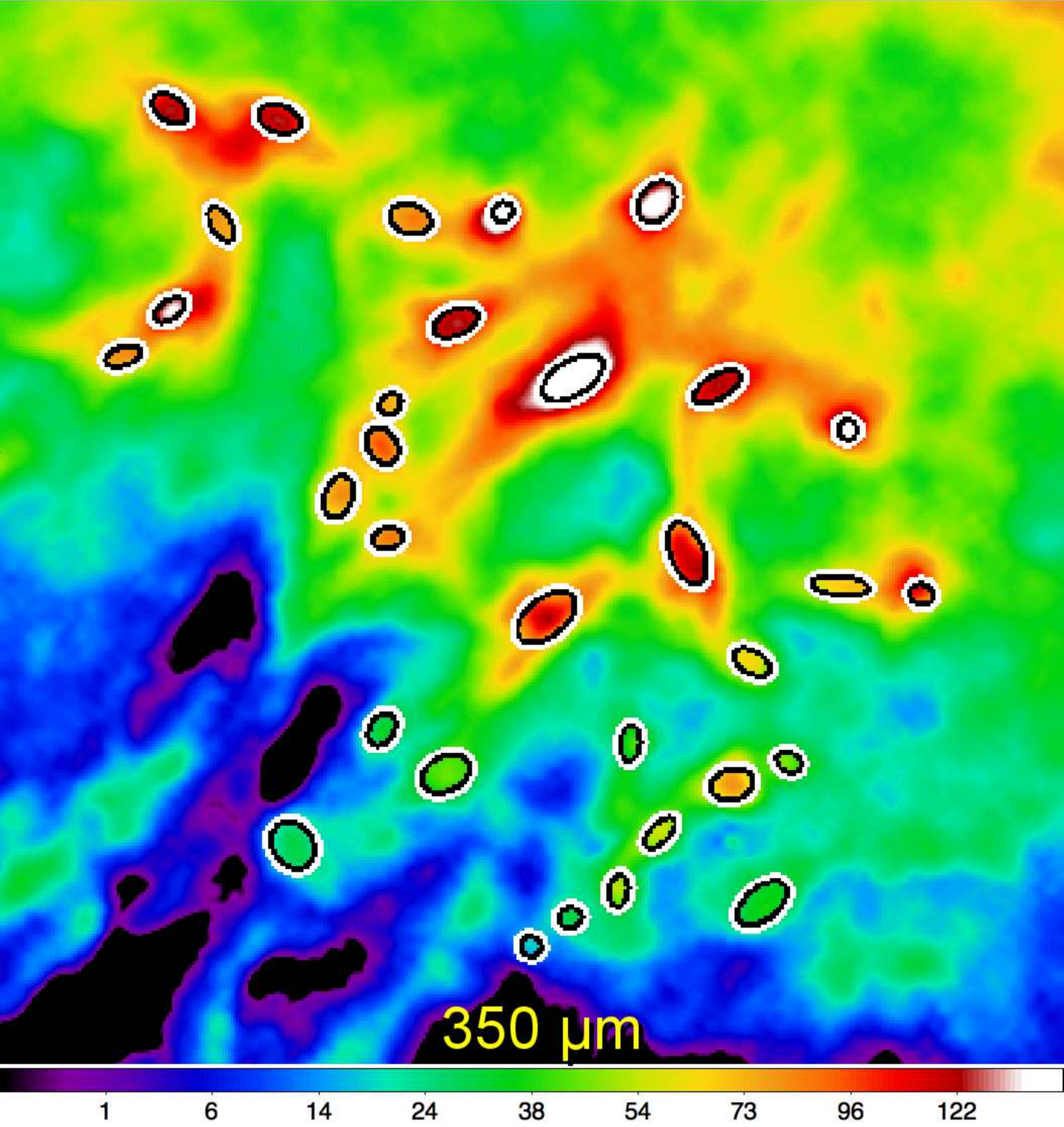}}
            \resizebox{0.33\hsize}{!}{\includegraphics{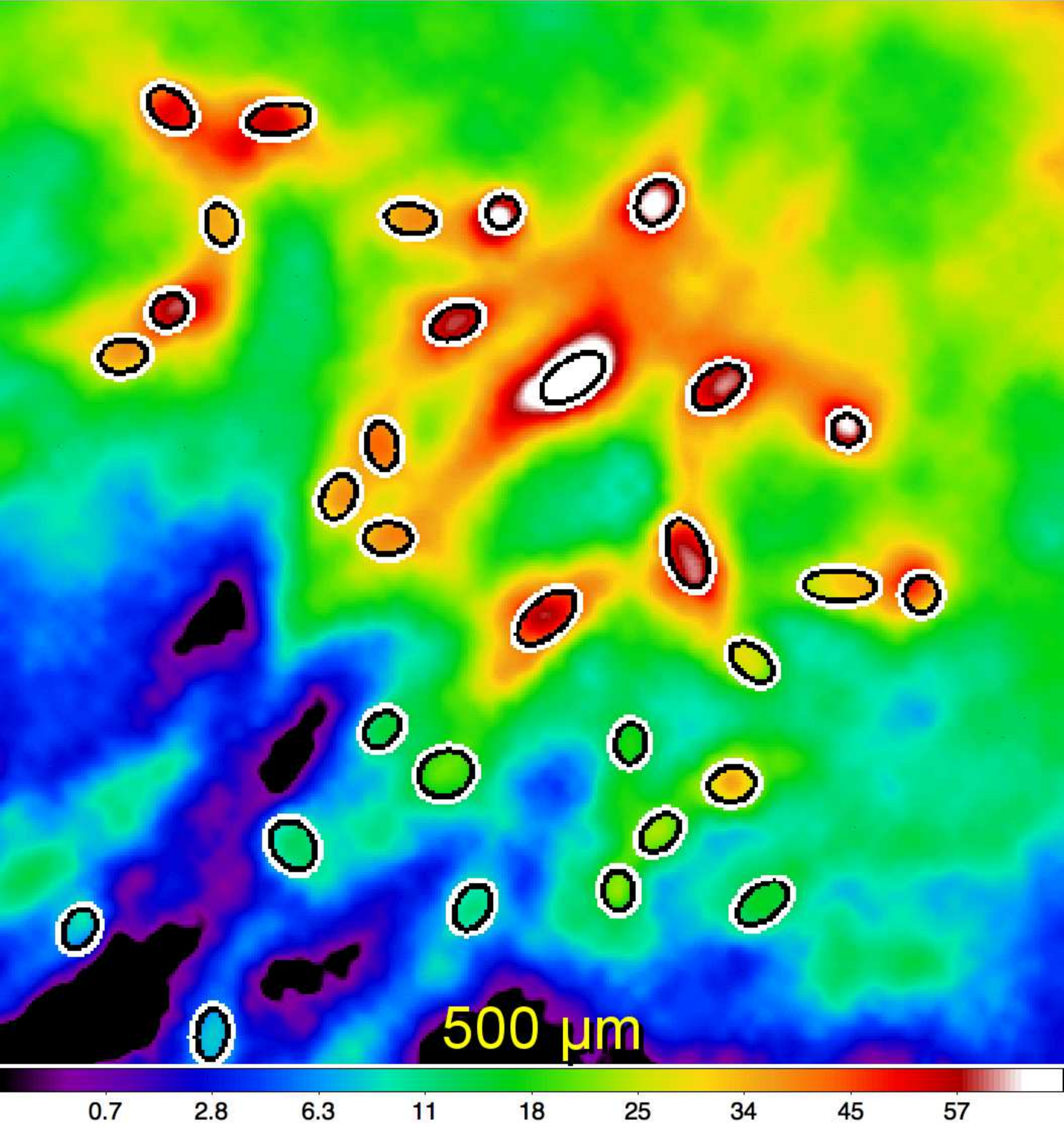}}
            \resizebox{0.33\hsize}{!}{\includegraphics{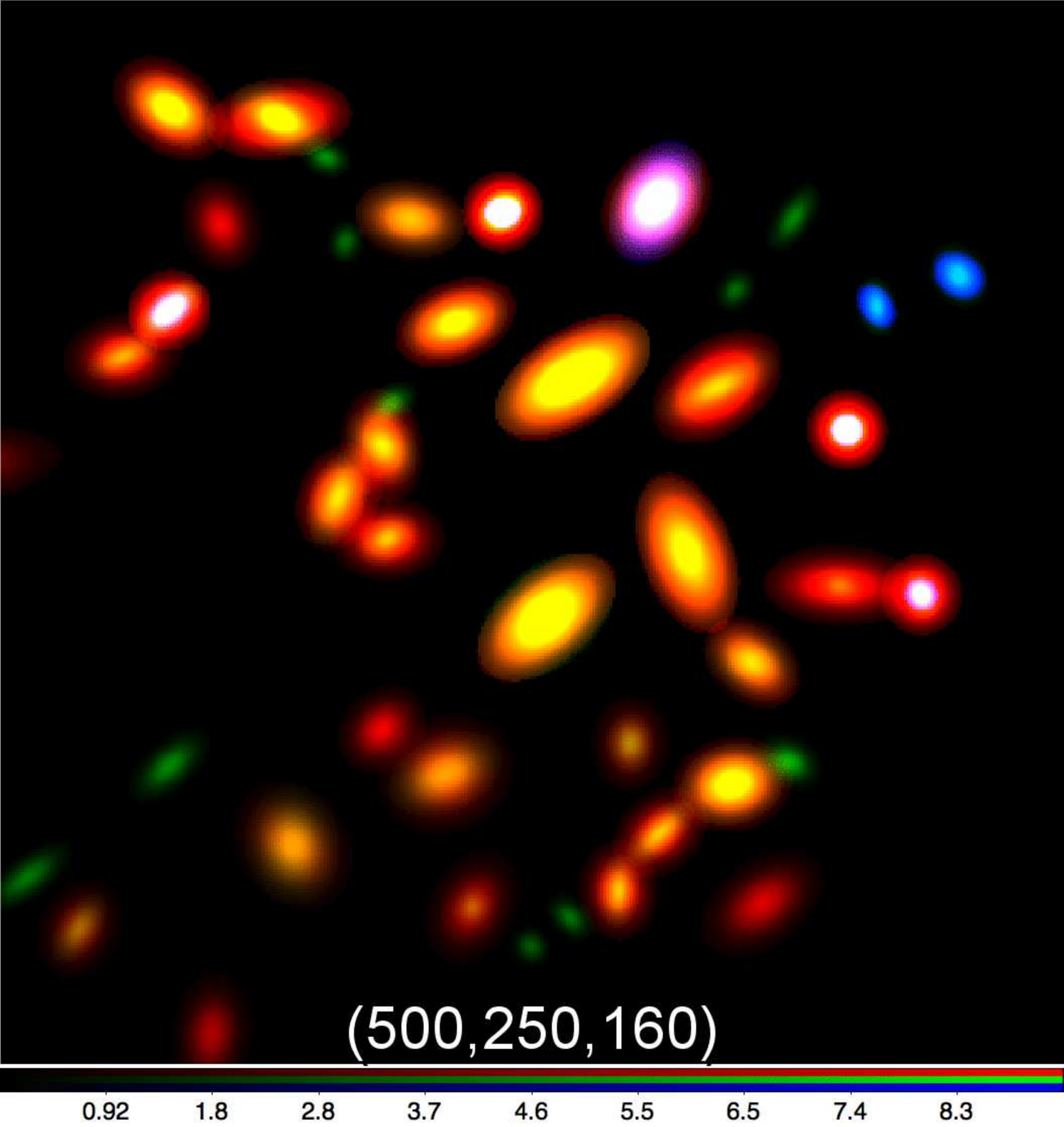}}}                   
\caption{                                              
The \object{Aquila} sub-field (18{\arcmin}$\times$18{\arcmin}) is shown as the observed images $\mathcal{I}_{{\!\lambda}{\rm O}}$ at
70, 160, 250, 350, 500\,{${\mu}$m} (\emph{left to right, top to bottom}) with the extraction ellipses (FWHM) of only measurable
sources ($F_{{i}{\lambda}{\,\rm P}} > \sigma_{{i}{\lambda}{\,\rm P}}$ and $F_{{i}{\lambda}{\,\rm T}} > \sigma_{{i}{\lambda}{\,\rm
T}}$) overlaid, as well as the composite 3-color RGB image (500,\,250,\,160\,{${\mu}$m}) created using the images
$\mathcal{I}_{{\!\lambda}{\rm M}}$ of the deblending shapes of each extracted source (\emph{lower-right}). The default condition,
that a tentative source must be detected in at least two bands, was used. Only the protostars are visible at 70\,{${\mu}$m}, whereas
at 160\,{${\mu}$m} one starless core appears, the other cold sources becoming clearly visible in the SPIRE bands. For better
visibility, the values displayed in the panels are somewhat limited in range; the color coding in the lower-right panel is linear, 
in all other panels it is a function of the square root of intensity in MJy/sr.
}                                                      
\label{aquila.ellipses}                                
\end{figure*}

\begin{figure*}                                        
\centering                                             
\centerline{\resizebox{0.33\hsize}{!}{\includegraphics{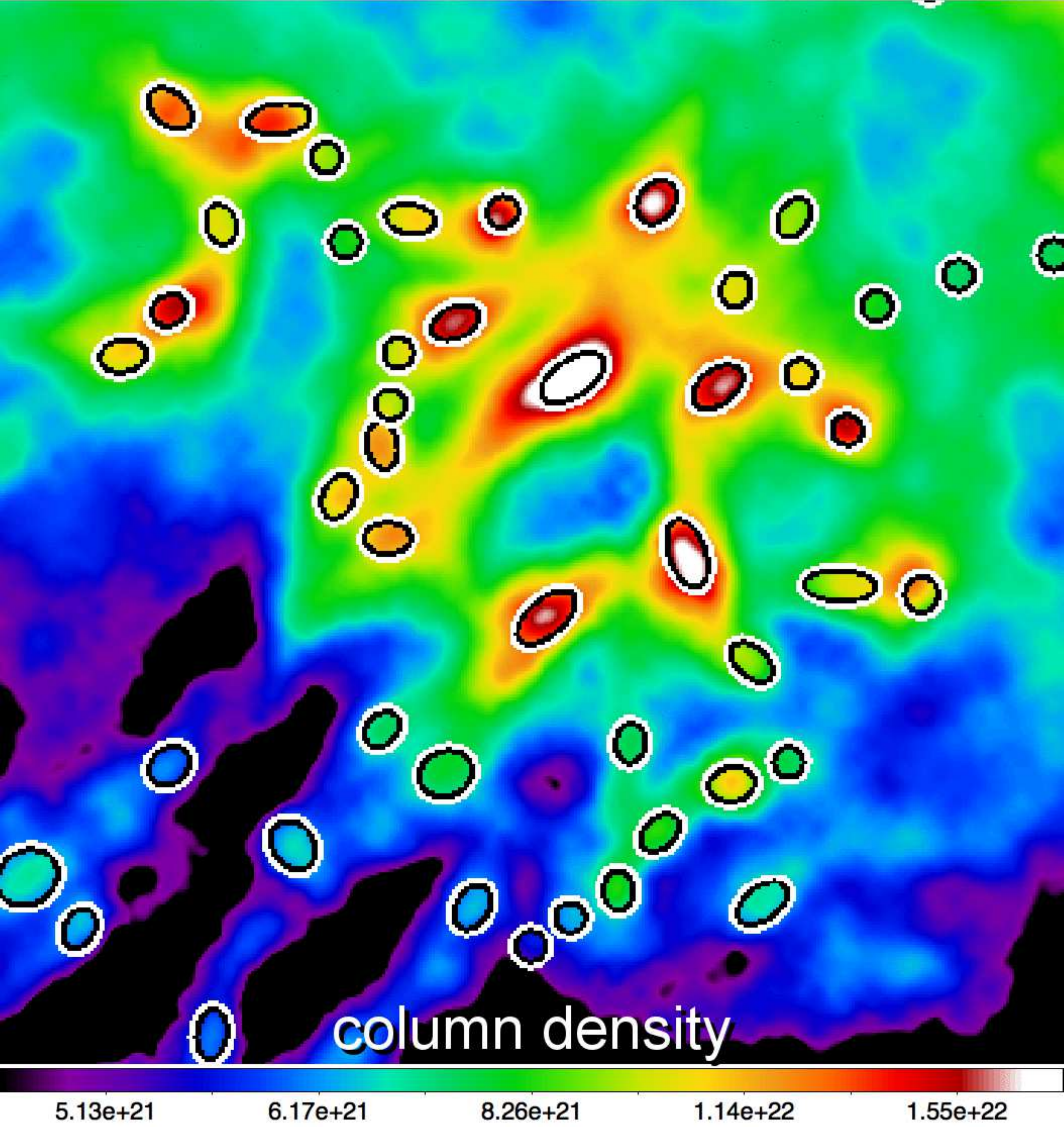}}
            \resizebox{0.33\hsize}{!}{\includegraphics{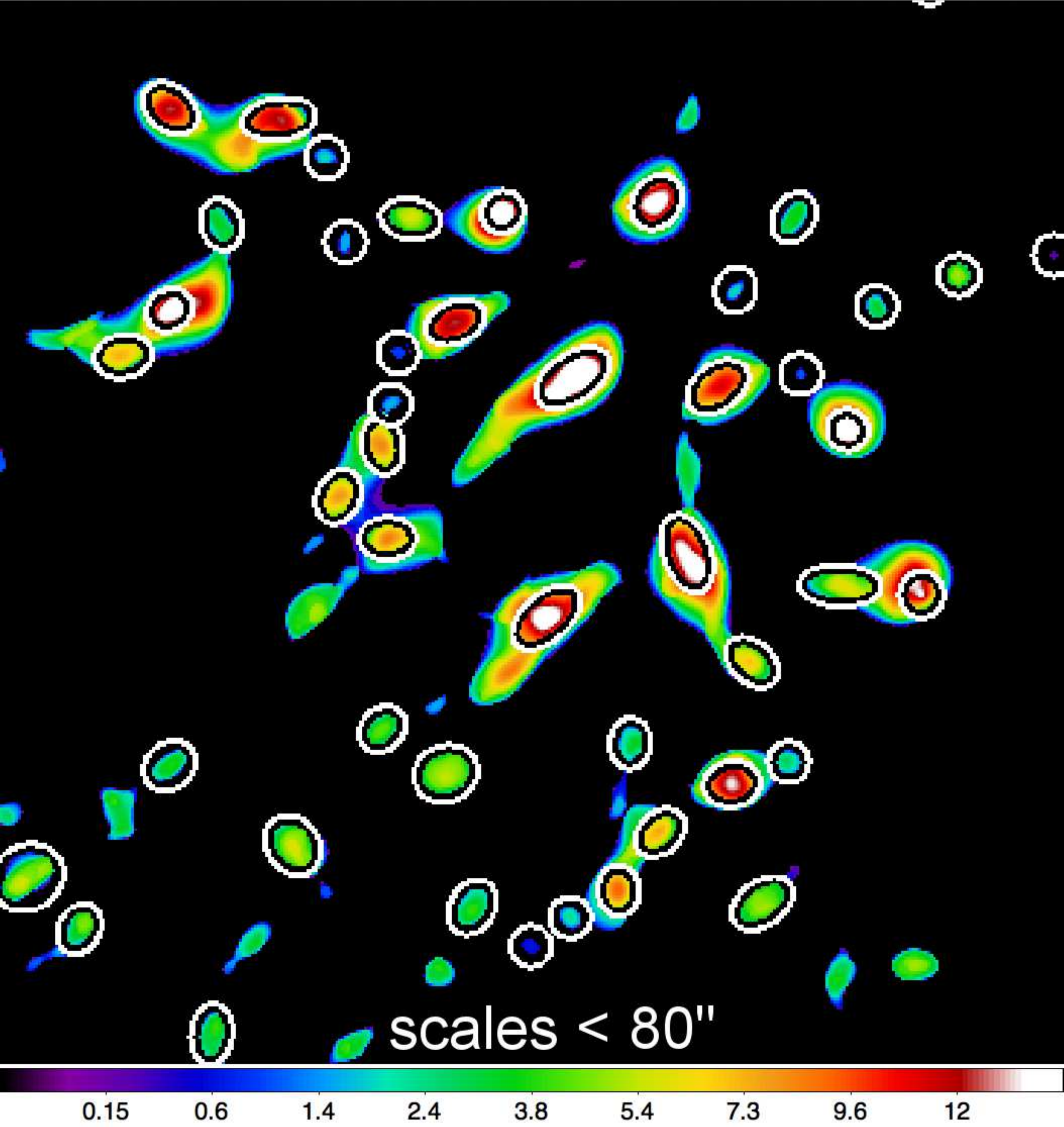}}
            \resizebox{0.33\hsize}{!}{\includegraphics{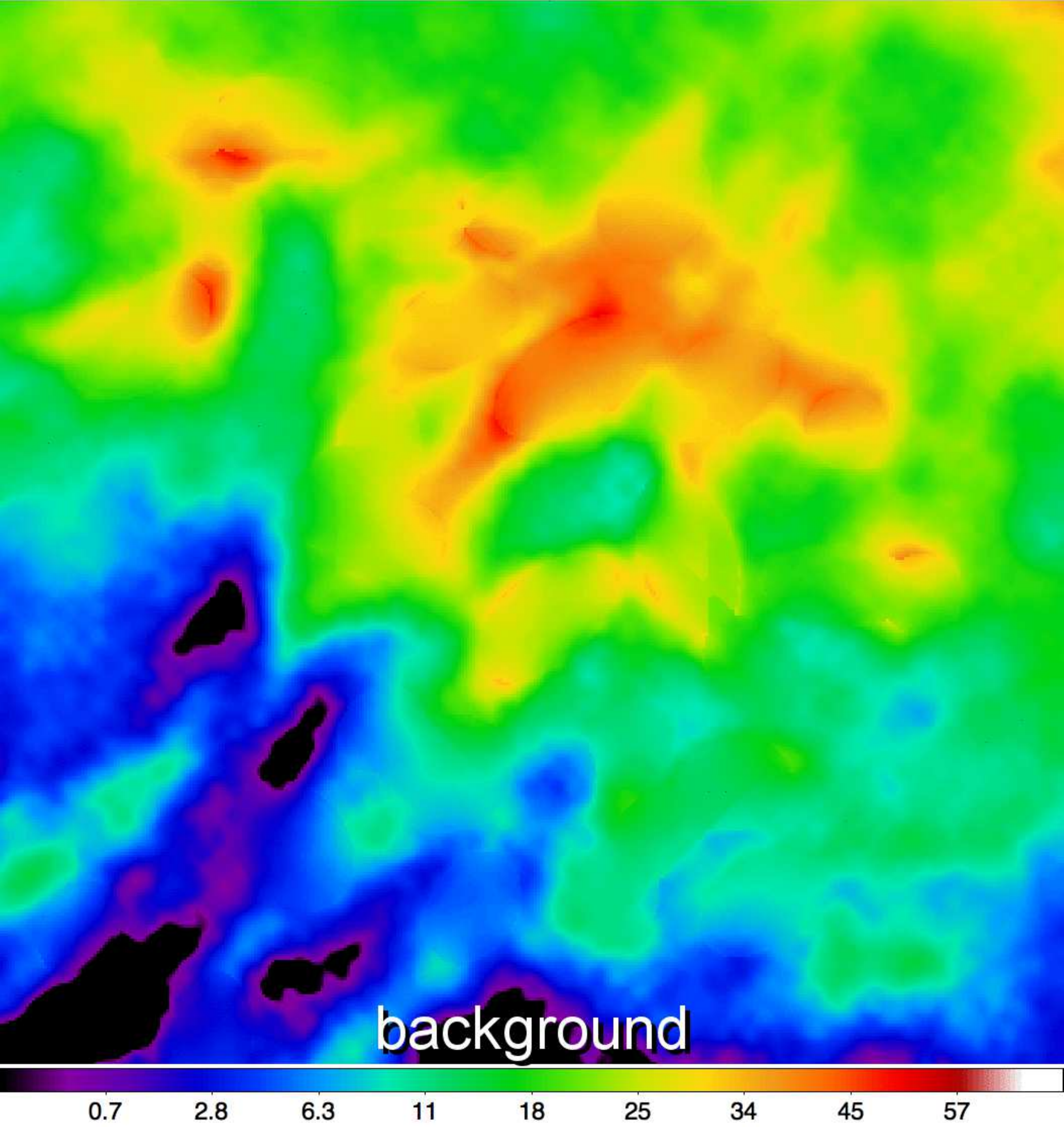}}}
\caption{
The \object{Aquila} sub-field is shown as a column density image with a 37{\arcsec} resolution \citep[\emph{left},][]{Ko"nyves_etal2010},
an accumulated clean combined detection image containing spatial scales of up to 80{\arcsec} (\emph{middle}), obtained by summing up
the single scales $\mathcal{I}_{{\rm D}{j}{\,\rm C}}$ in that range, with the 500\,{${\mu}$m} ellipses of \emph{all} detected
sources overplotted. Also shown is the clean background $\mathcal{I}_{{\!\lambda}{\rm O}{\,\rm CB}}$ (\emph{right}) at
500\,{${\mu}$m}. For better visibility, the values displayed in the panels are somewhat limited in range; the color coding in all
panels is a function of the square root of column density and of intensity in MJy/sr.
}
\label{aquila.visuals}
\end{figure*}

The observations, data reduction, and first results for the \object{Aquila} star-forming region (part of the \emph{Herschel} Gould
Belt survey, adopted distance $D{\,=\,}$260\,pc) have been described by
\cite{Andre_etal2010,Men'shchikov_etal2010,Ko"nyves_etal2010,Bontemps_etal2010}. The \object{Aquila} sub-field (365$\times$365
3{\farcs}0 pixels), shows a cluster of cold prestellar cores, clearly visible in all SPIRE wavebands south-east of the bright W40
region in the \object{Aquila} field. The ``cold cluster'' was chosen to illustrate the performance of our new method for studying
the earliest phases of low-mass star formation in the nearby regions (Figs.~\ref{aquila.ellipses}, \ref{aquila.visuals}).

The lower-right panel of Fig.~\ref{aquila.ellipses} shows a 3-color composite image of the extracted sources in the \object{Aquila}
sub-field, represented by their elliptical deblending shapes $\mathcal{I}_{{\!\lambda}{\rm M}}$ (Eq.~\ref{moffat.function}) with the
measured peak intensities, sizes, and orientations. The image gives an overview of the source properties combining the information
from the 500, 250, and 160\,{${\mu}$m} bands in a single image. With just a few exceptions, all sources are red, yellow-red, and
green, indicating that the radiation is emitted by cold starless objects, without any significant central energy source. Only 8
protostars (blue, white-red, and white-pink sources) are visible in the composite and PACS images, while as many as 39 cold
prestellar cores become detectable and measurable in the SPIRE wavebands, as can be seen in the lower panels of
Fig.~\ref{aquila.ellipses}.

The dense cold prestellar cores are clearly situated within a complex web of filamentary structures at significant intensity peaks;
due to their nature, the cores must also coincide with significant column density peaks. The left panel of Fig.~\ref{aquila.visuals}
displays a column density image of the cold cluster, demonstrating that most extracted sources are indeed centered on the column
density peaks. A couple of compact sources appear to be offset from the column density peaks and intensity peaks at SPIRE
wavelengths. These are the protostars prominent in the PACS bands that happened to either coincide (in projection) with those
locations or form off-center in the dense clumps. We show in Fig.~\ref{aquila.visuals} the ellipses of all 46 detected sources;
there are 7 additional sources that are not measurable at some wavelengths and therefore not visualized by ellipses in
Fig.~\ref{aquila.ellipses}. The middle panel of Fig.~\ref{aquila.visuals} shows the same set of extraction ellipses on top of the
combined detection image (at scales $S_{\!j}{\,\le\,}80${\arcsec}) that clarifies why \textsl{getsources} found the sources at all
those locations. In addition, the right panel shows the clean background image $\mathcal{I}_{{\!\lambda}{\rm O}{\,\rm CB}}$ at
500\,{${\mu}$m} that demonstrates that no significant sources in the \object{Aquila} sub-field were left undetected by
\textsl{getsources}.

These results (visualized in Figs.~\ref{aquila.ellipses}, \ref{aquila.visuals}) demonstrate that \textsl{getsources} handles very
well the multi-wavelength \emph{Herschel} observations of resolved starless cores, the main ingredient of the \object{Aquila}
sub-field. Although all unresolved protostars were also extracted, it is the next example that focuses on protostellar population.

\subsection{Clustered unresolved protostars in \object{Rosette}}
\label{compact.protostars.rosette}

\begin{figure*}                                                               
\centering
\centerline{\resizebox{0.33\hsize}{!}{\includegraphics{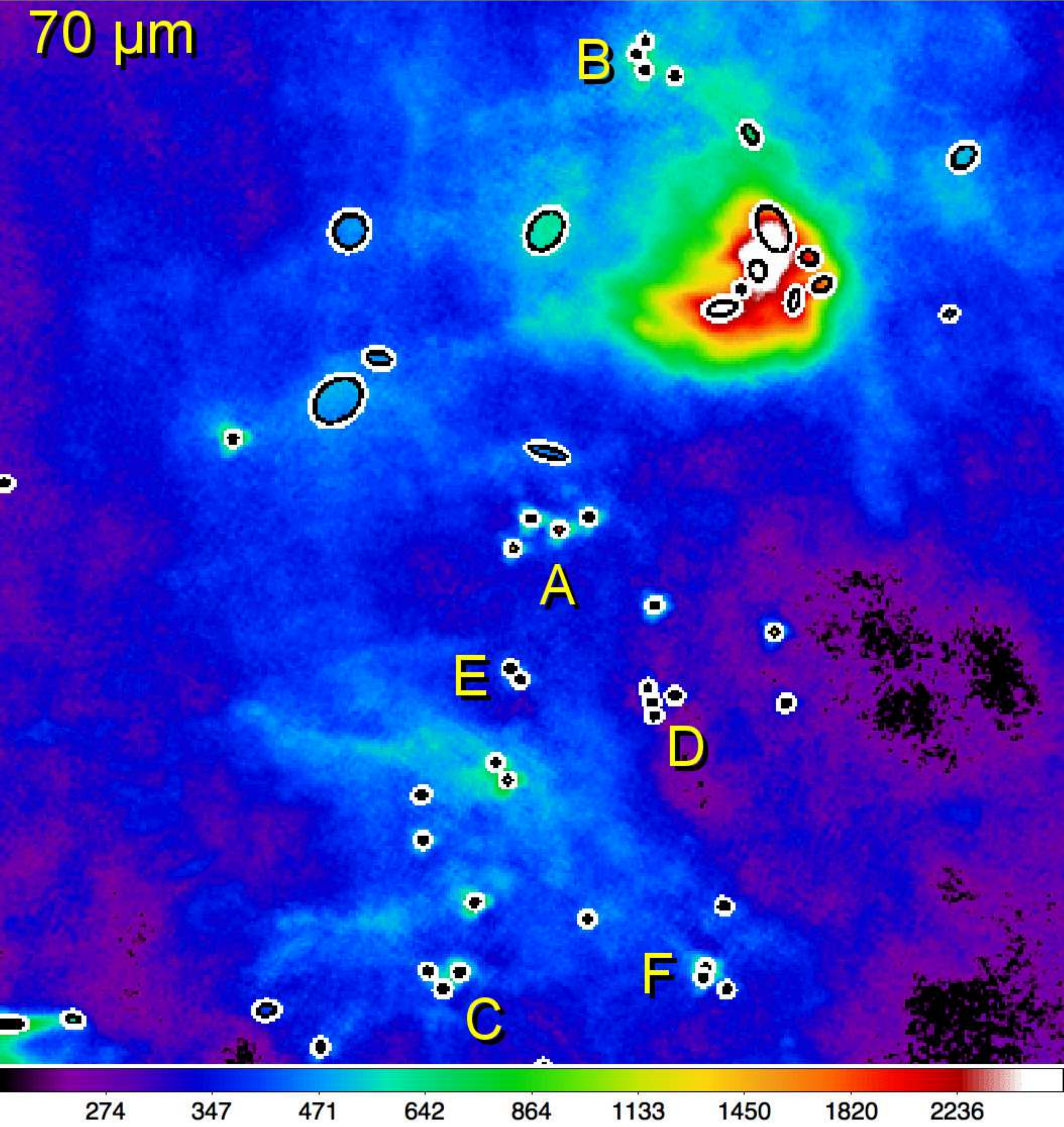}}
            \resizebox{0.33\hsize}{!}{\includegraphics{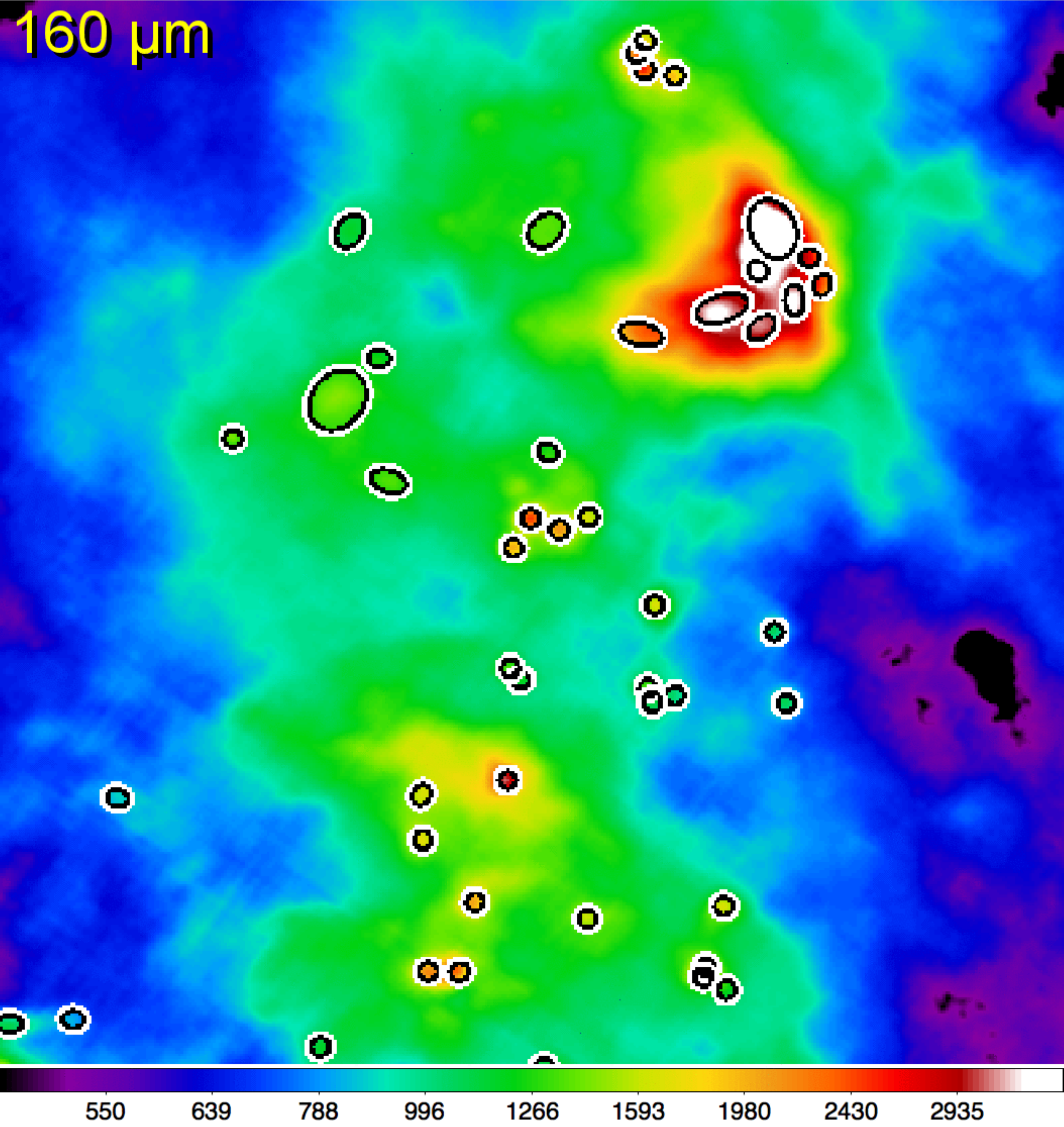}}
            \resizebox{0.33\hsize}{!}{\includegraphics{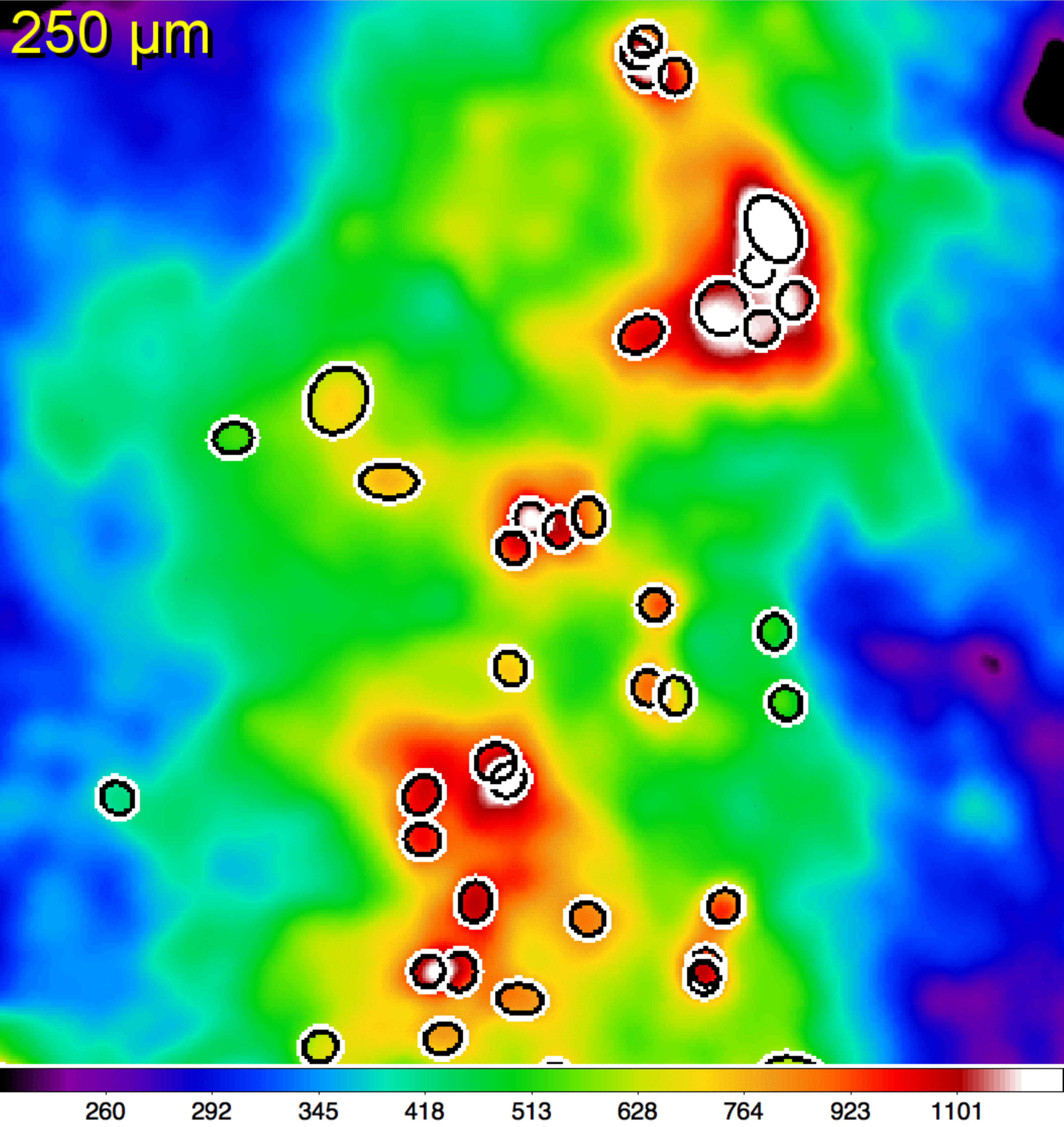}}}
\centerline{\resizebox{0.33\hsize}{!}{\includegraphics{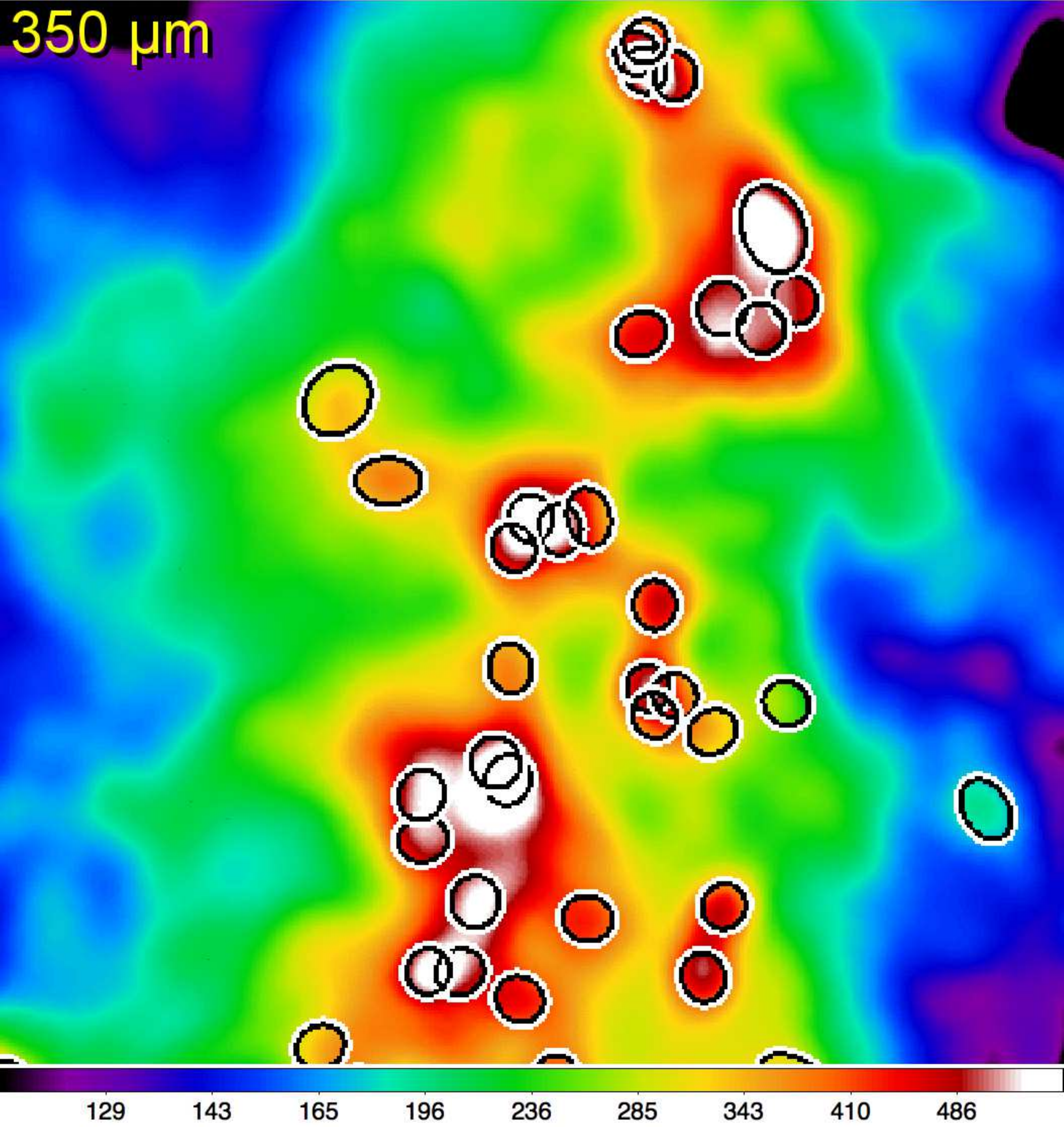}}
            \resizebox{0.33\hsize}{!}{\includegraphics{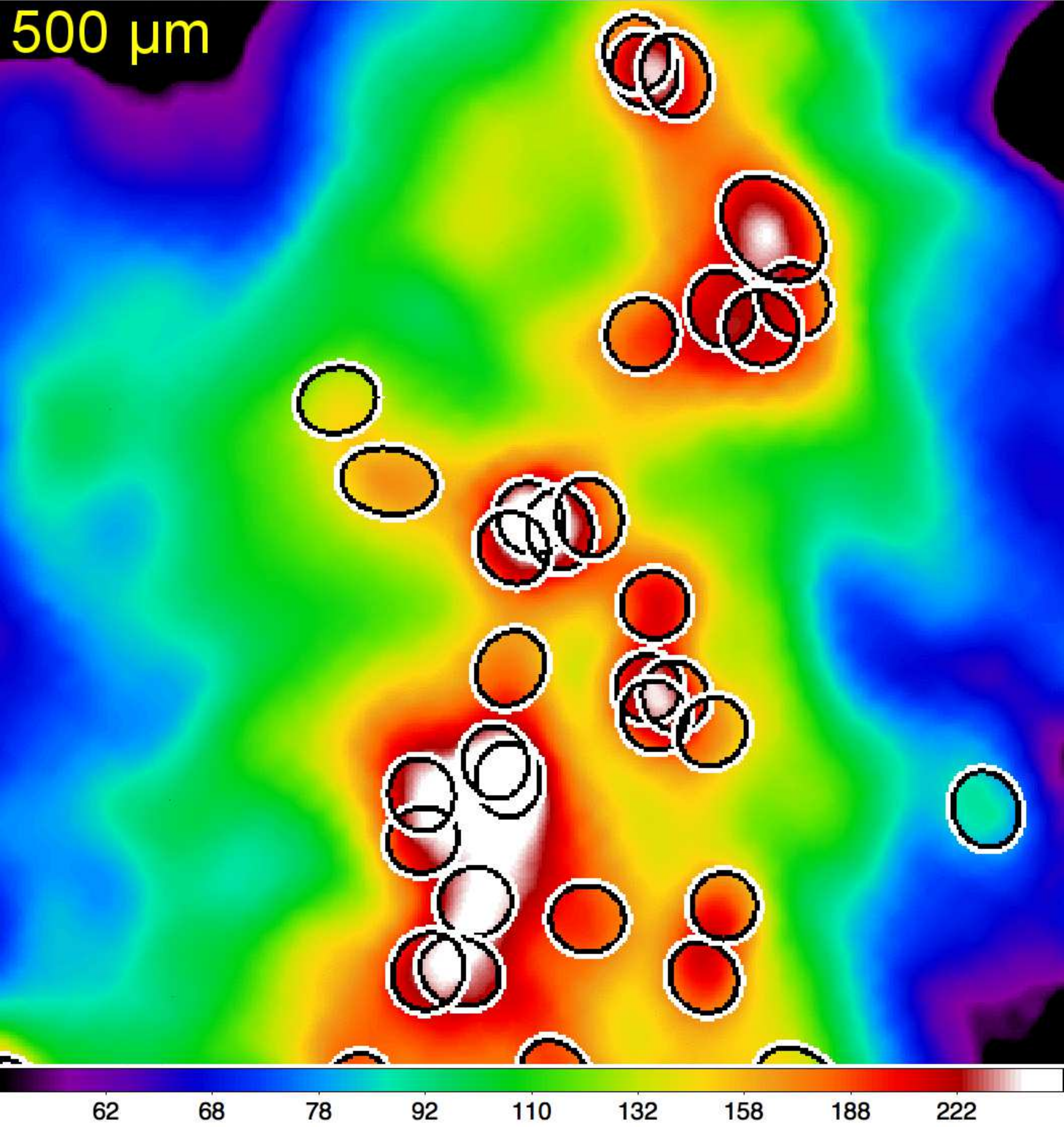}}
            \resizebox{0.33\hsize}{!}{\includegraphics{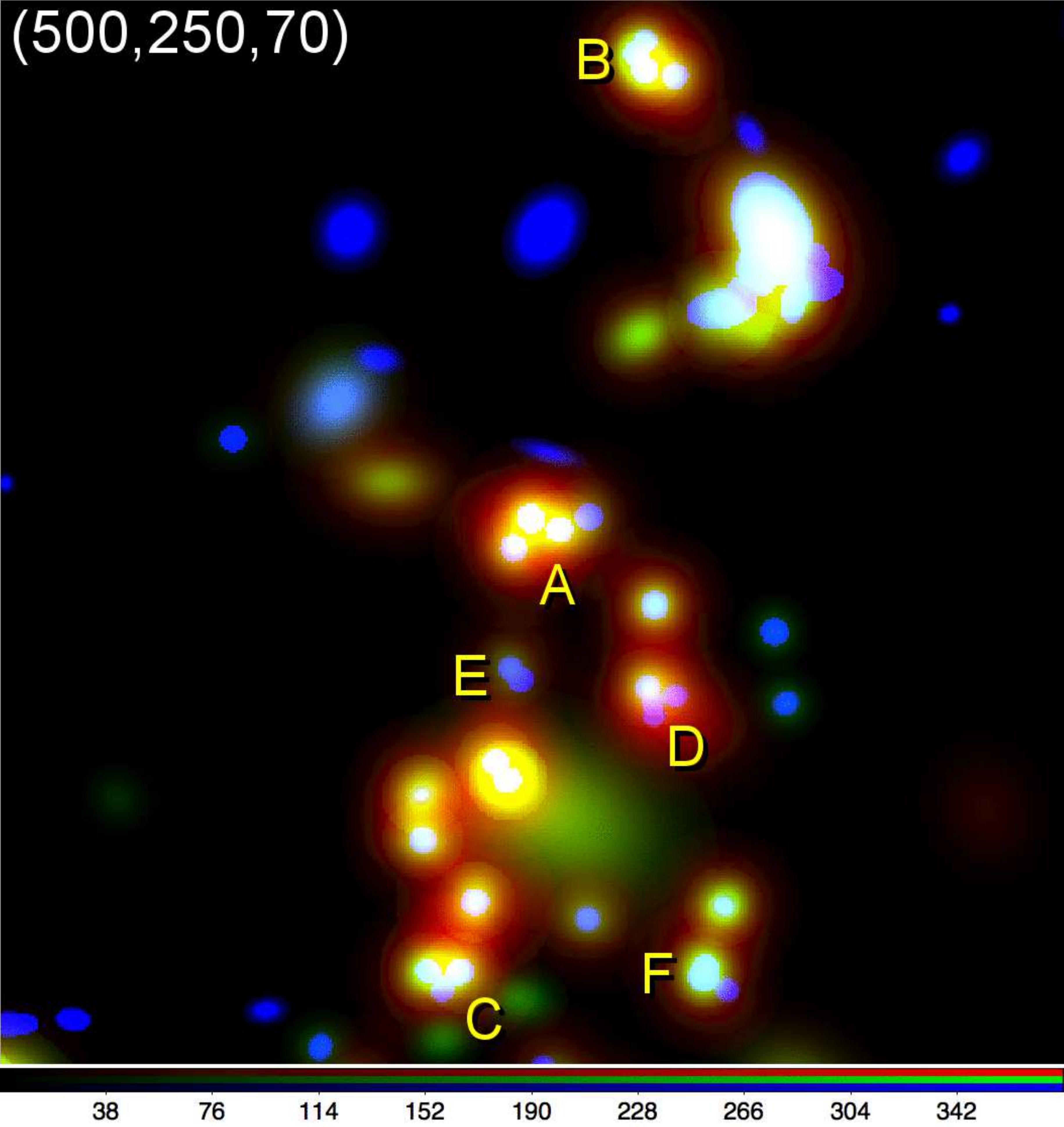}}}
\caption{
The \object{Rosette} sub-field (9{\arcmin}$\times$9{\arcmin}) is shown as as the observed images $\mathcal{I}_{{\!\lambda}{\rm O}}$
at 70, 160, 250, 350, 500\,{${\mu}$m} (\emph{left to right}, \emph{top to bottom}) with the extraction ellipses (FWHM) of only
measurable sources ($F_{{i}{\lambda}{\,\rm P}} > \sigma_{{i}{\lambda}{\,\rm P}}$ and $F_{{i}{\lambda}{\,\rm T}} >
\sigma_{{i}{\lambda}{\,\rm T}}$) overlaid, as well as the composite 3-color RGB image (500, 250, 70\,{${\mu}$m}) created using the
images $\mathcal{I}_{{\!\lambda}{\rm M}}$ of the deblending shapes of each extracted source (\emph{lower-right}). The default
condition, that a tentatve source must be detected in at least two bands, was used. Most of the compact sources visible at
70\,{${\mu}$m} are unresolved protostars; several groups of them, discussed in Sect.~\ref{compact.protostars.rosette}, are labeled
A--F. For better visibility, the values displayed in the panels are somewhat limited in range; the color coding in the lower-right
panel is linear, in the other panels it is a function of the square root of intensity in MJy/sr.
} 
\label{rosette.ellipses}
\end{figure*}

\begin{figure*}
\centering
\centerline{\resizebox{0.33\hsize}{!}{\includegraphics{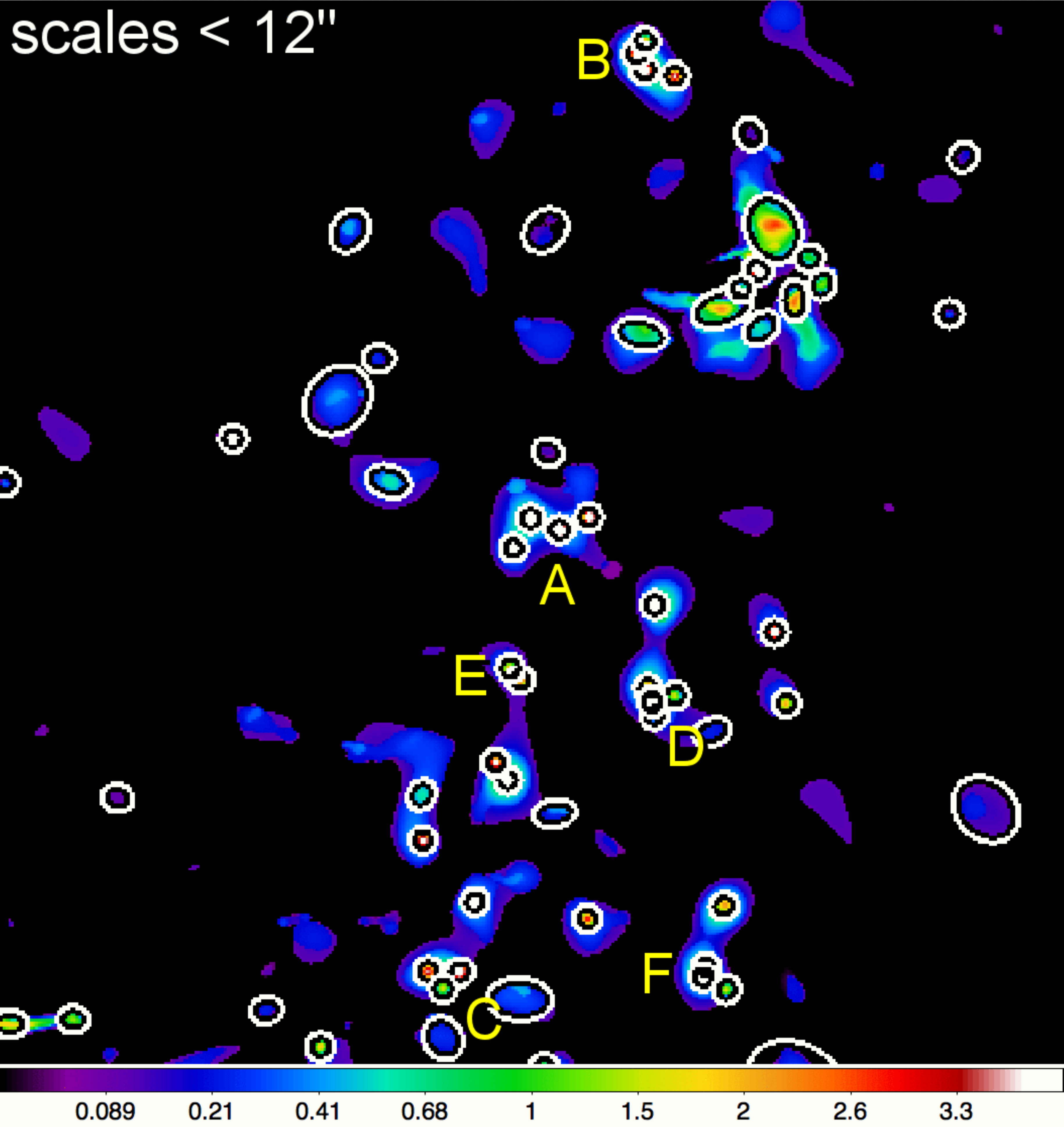}}
            \resizebox{0.33\hsize}{!}{\includegraphics{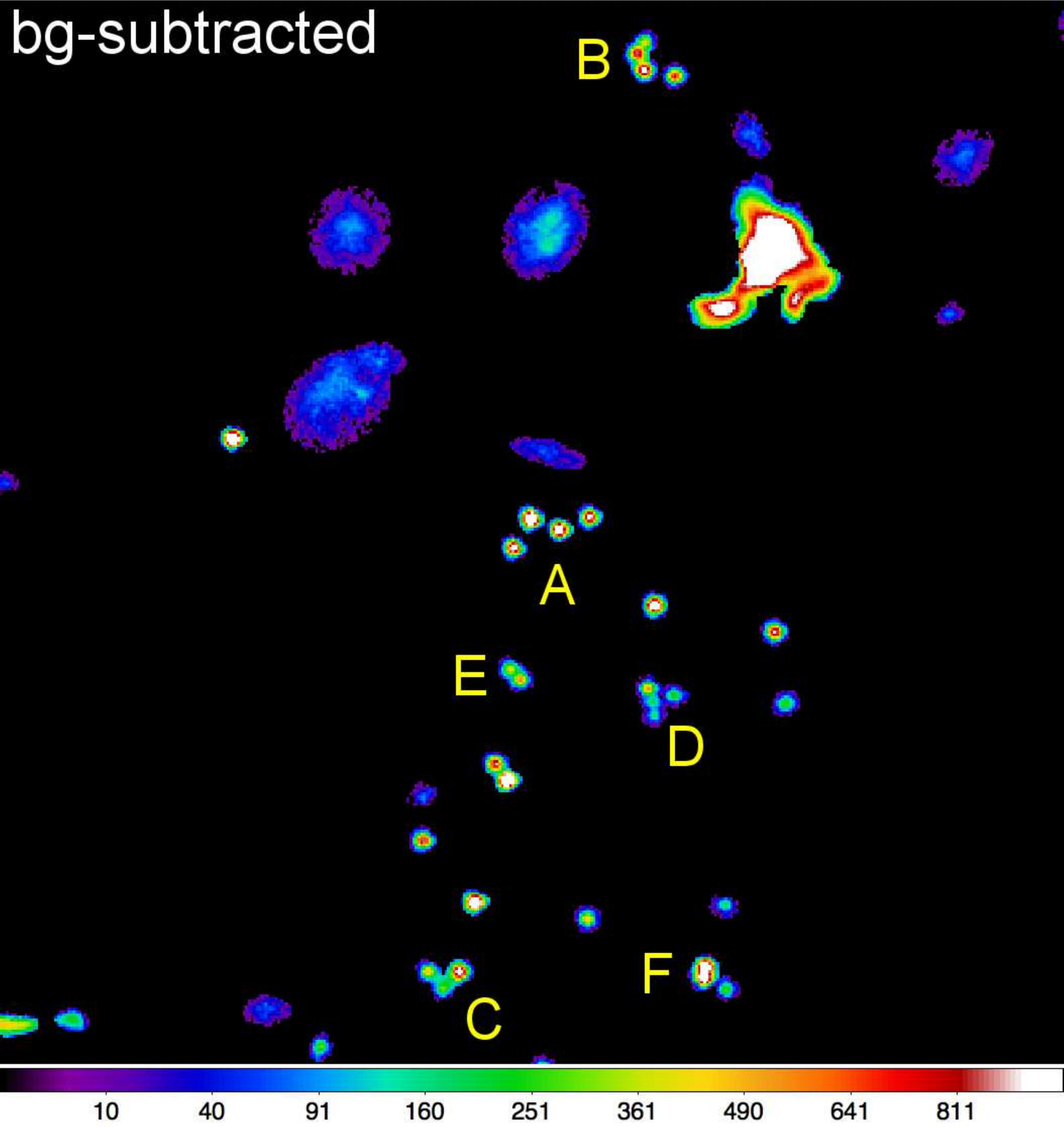}}
            \resizebox{0.33\hsize}{!}{\includegraphics{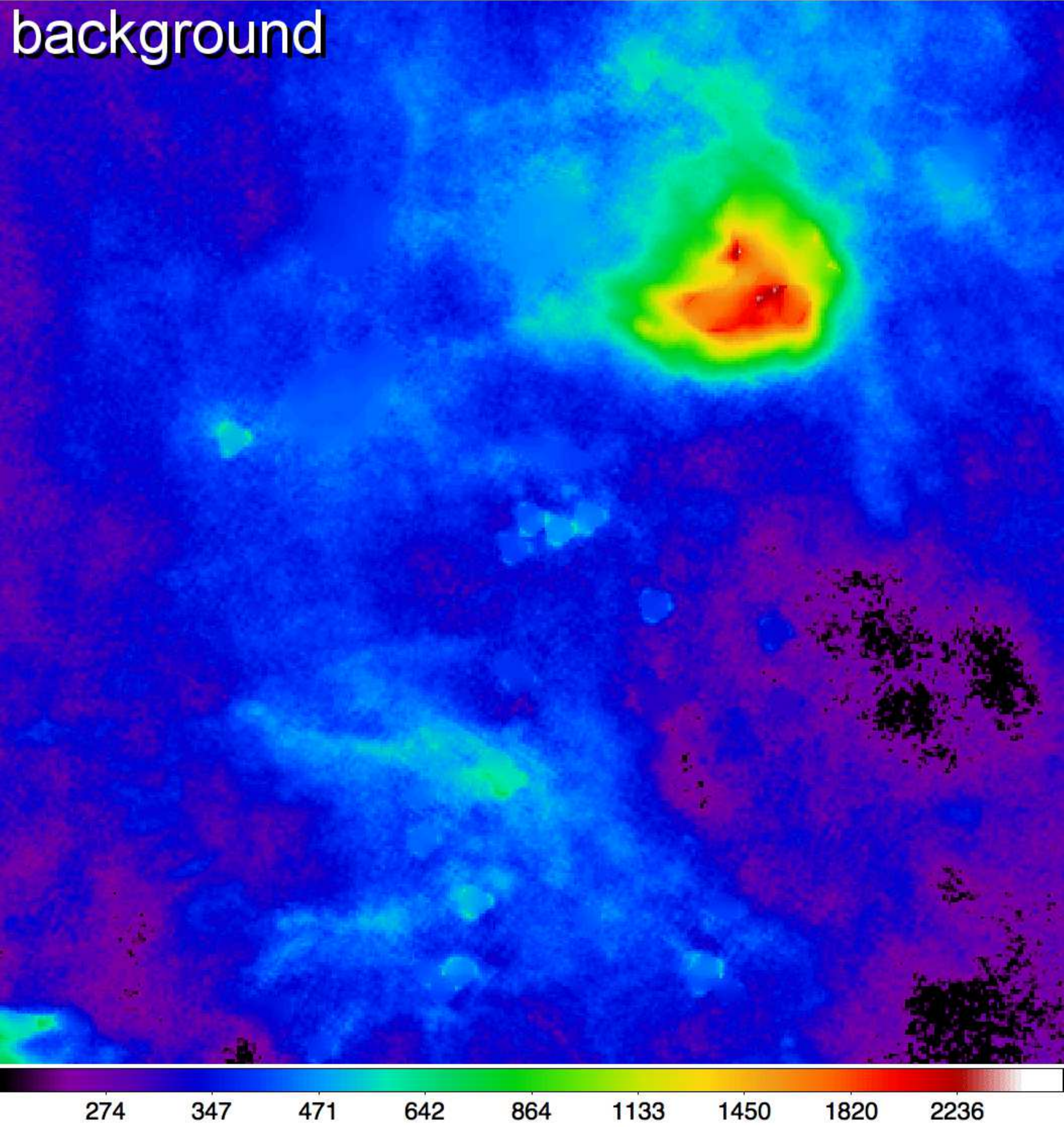}}}
\caption{
The \object{Rosette} sub-field is shown as the accumulated clean combined detection image containing spatial scales of up to 12{\arcsec}
(\emph{left}), obtained by summing up the single scales $\mathcal{I}_{{\rm D}{j}{\,\rm C}}$ in that range, with the
160\,{${\mu}$m} ellipses of all detected sources overplotted. Also shown are the background-subtracted image
$\mathcal{I}_{{\!\lambda}{\rm O}{\,\rm BS}}$ (\emph{middle}) and clean background $\mathcal{I}_{{\!\lambda}{\rm O}{\,\rm CB}}$
(\emph{right}) at 70\,{${\mu}$m}; when added together, they make up the original 70\,{${\mu}$m} image in
Fig.~\ref{rosette.ellipses}. For better visibility, the values displayed in the panels are somewhat limited in range; the color
coding is a function of the square root of intensity in MJy/sr.
} 
\label{rosette.visuals}
\end{figure*}

The observations, data reduction, and first results for the \object{Rosette} star-forming region (part of the HOBYS survey, adopted
distance $D{\,=\,}$1.6\,kpc) have been described by
\cite{Motte_etal2010,Schneider_etal2010,Hennemann_etal2010,diFrancesco_etal2010}. The \object{Rosette} sub-field (395$\times$395
1{\farcs}4 pixels), shows a central part of the \object{Rosette} field with an extended bright area at 70\,{${\mu}$m} and a number
of unresolved isolated and clustered protostars in the PACS wavebands. This sub-field of \object{Rosette} was chosen to illustrate
the performance of \textsl{getsources} for studying faint unresolved protostars in distant star-forming regions
(Figs.~\ref{rosette.ellipses}, \ref{rosette.visuals}).

Similarly to Fig.~\ref{aquila.ellipses}, the lower-right panel of Fig.~\ref{rosette.ellipses} shows a 3-color composite image of the
extracted sources in the \object{Rosette} sub-field, represented by their elliptical deblending shapes with the measured peak
intensities, sizes, and orientations. Note, however, that the color image is strongly affected by the difference in spatial
resolutions in the wavebands (by factors of $\sim$3 and 2) than the color image shown in Fig.~\ref{aquila.ellipses}, as most of the
sources in the distant \object{Rosette} sub-field are unresolved even at 70\,{${\mu}$m}. Most of the sources remain in all of the
\emph{Herschel} images displayed in the other panels of Fig.~\ref{rosette.ellipses}; they are deblended and remain measurable all
the way to the lowest resolution of the 500\,{${\mu}$m} band. Figures \ref{rosette.ellipses}, \ref{rosette.visuals} also highlight
severe problems encountered by the usual ``monochromatic'' algorithms when they extract sources independently at individual
wavelengths and then \emph{associate} sources based on their positions in the images with such greatly varying resolutions.

The redder (colder) sources are mostly found in the lower half of the image, whereas the bluer sources are scattered over the
\object{Rosette} sub-field. Here we will focus on the compact unresolved sources visible quite clearly in the 70\,{${\mu}$m}
waveband in relatively low-background conditions, which can be used to judge whether \textsl{getsources} is able to separate faint
peaks that are close to each other. One can count 35 such sources that are mostly clustered in groups of two, three, or four; we
will limit our discussion to the six groups labeled A--F in Fig.~\ref{rosette.ellipses}. Aiming to clarify why \textsl{getsources}
found the sources at those locations, the left panel of Fig.~\ref{rosette.visuals} shows the extraction ellipses of all sources on
top of the combined detection image (at scales $S_{\!j}{\,\le\,}12${\arcsec}). The middle panel of Fig.~\ref{rosette.visuals}
displays the background-subtracted image giving the cleanest view of the sources and its right panel shows the clean background
image $\mathcal{I}_{{\!\lambda}{\rm O}{\,\rm CB}}$ at 70\,{${\mu}$m} demonstrating, that no significant sources in the
\object{Rosette} sub-field were left undetected by \textsl{getsources}.

Group A consists of 4 protostars of different brightness but very similar separations between the neighbors, clearly distinct at
70\,{${\mu}$m} and extracted by \textsl{getsources} across all wavelengths. Group B presents a more difficult case of 4 sources with
decreasing brightess and distance between the members; the faintest protostar on top of group B is almost merged with its neighbor,
but still has been detected and measured in 4 bands. A similar case is displayed by group C, where one faint source is situated
between two other brighter sources; the faintest source is measurable only in the highest-resolution image at 70\,{${\mu}$m}. In
group D we have another similar cluster of 4 sources with an extremely faint source surrounded by the brighter ones; most members of
group D are measurable across all wavelengths. Group E is correctly extracted as two close companions, with only the brighter source
being measurable in all wavebands. Group F is two relatively bright protostars at a very small separation. Although they are
practically merged together, with a saddle point just a few percent below the peaks, the binary is detected as such and both
components are measurable up to 250\,{${\mu}$m}.

The results presented in Figs.~\ref{rosette.ellipses}, \ref{rosette.visuals} show that \textsl{getsources} handles very well the
multi-wavelength \emph{Herschel} observations of unresolved protostars, the main ingredient of the sub-field. It fully preserves the
highest-resolution information from the 70\,{${\mu}$m} waveband and uses it to correctly identify and measure close companions in
all groups of protostars at all wavelengths.

%||||||||||||||||||||||||||||||||||||||||||||||||||||||||||||||||||||||||||||||||||||||||||||||||||||||||||||||||||||||||||||||||||

\section{Conclusions}
\label{conclusions}

The multi-scale, multi-wavelength source extraction method \textsl{getsources} presented in this paper, was designed primarily for
use in large far-infrared and submillimeter surveys of star-forming regions with \emph{Herschel}. Instead of following the
traditional approaches of extracting sources directly in the observed images (Sect.~\ref{existing.methods}), the method analyzes
highly-filtered decompositions of original images over a wide range of spatial scales (Sect.~\ref{decomposing.detection.images}).
The algorithm separates the peaks of real sources from those produced by the noise and background fluctuations
(Sect.~\ref{removing.noise.background}) and constructs wavelength-independent sets of combined single-scale detection images
(Sect.~\ref{combining.clean.single.scales}) preserving spatial information from all wavebands. Sources are detected in the combined
detection images by tracking the evolution of their segmentation masks across all spatial scales (Sect.~\ref{detecting.sources}).
Source properties are measured in the original background-subtracted and deblended images at each wavelength in iterations
(Sect.~\ref{measuring.cataloging}). Additional catalogs and images are produced to aid in the analysis of the extraction results
(Sect.~\ref{visualizing.extractions}), complementing the main catalog of sources. Based on the results of the initial extraction,
detection images are flattened to produce much more uniform noise and background fluctuations in preparation for the second, final
extraction (Sect.~\ref{flattening.background.noise}). The performance of the new method on \emph{Herschel} images was illustrated by
extracting sources in small sub-fields of the \object{Aquila} and \object{Rosette} regions (Sect.~\ref{applications.herschel}).

There are several significant advantages of \textsl{getsources} over other existing methods of source extraction. (1) The fine
spatial decomposition filters out irrelevant spatial scales and improves detectability, especially in the crowded regions and for
extended sources. (2) The multi-wavelength design enables combining data over all wavebands, eliminating the need to match
independent extraction catalogs and enabling substantial super-resolution in the images with lower spatial resolution. (3) The
single-scale detection algorithm identifies sources and determines their characteristic sizes, avoiding spurious peaks on top of
large structures and filaments. (4) The background subtraction and deblending, based on the wavelength-dependent footprint of each
source, disentangle crowded regions with overlapping sources. (5) The extraction process is fully automatic and there are no free
parameters involved: the default configuration works best in all cases that have been tested.

A disadvantage of the algorithm is that it may not be very fast and it may require considerable storage space, depending on the
numbers of pixels, spatial scales, wavelengths, iterations, and potential sources detected (Appendix~\ref{installing.using}); most
of the space can be freed, however, after the extraction has been completed.

The method has been thoroughly tested using many simulated benchmark images and real-life observations. In particular, the overall
benchmarking results (Men'shchikov et al., in prep.) have shown that \textsl{getsources} comes on top of the other source extraction
methods that we have tested (Sect.~\ref{existing.methods}) in both the completeness and reliability of source detection and the
accuracy of measurements. The source extraction code is automated, very flexible, and easy-to-use; the code and validation images
with a reference extraction catalog are freely available (see Appendix~\ref{installing.using}).

%||||||||||||||||||||||||||||||||||||||||||||||||||||||||||||||||||||||||||||||||||||||||||||||||||||||||||||||||||||||||||||||||||

\begin{acknowledgements}
The authors employed \textsl{SAOImage DS9} (by William Joye) and \textsl{WCSTools} (by Douglas Mink) developed at the Smithsonian
Astrophysical Observatory (USA), the \textsl{CFITSIO} library (by William D Pence) developed at HEASARC NASA (USA), the
\textsl{STILTS} library (by Mark Taylor) developed at Bristol University (UK), the \textsl{PSPLOT} library (by Kevin E. Kohler) at
Nova Southeastern University Oceanographic Center (USA), and \textsl{SWarp} (by Emmanuel Bertin) developed at Institut
d'Astrophysique de Paris (France). We appreciate the feedback received from Guillaume LeLeu, Mika Juvela, Isabelle Ristorcelli,
Ana{\"e}lle Maury, Quang Nguyen Luong, Doris Arzoumanian, Sacha Hony, Glenn White, Michael Reid, Alana Rivera-Ingraham, Arabindo
Roy, Elaine Winston, Nick Cox, and Jason Kirk that helped improve the code and its usability. We are also grateful for useful
comments on a draft of the manuscript received from Nicolas Peretto, Tracey Hill, Vera K{\"o}nyves, Pedro Palmeirim, and the
anonymous referee, that helped improve the clarity of this paper.
\end{acknowledgements}

%||||||||||||||||||||||||||||||||||||||||||||||||||||||||||||||||||||||||||||||||||||||||||||||||||||||||||||||||||||||||||||||||||

\begin{appendix}
\section{Astrophysical objects: dense cores}
\label{astrophysical.objects}

The primary goal of this work is to develop a source extraction method suitable for the systematic detection and measurement of
dense cores in molecular clouds with \emph{Herschel}: one of the main observational objectives of the Gould Belt survey
\citep{Andre_etal2010} is to obtain a complete census of such prestellar cores in nearby molecular clouds.

The structure of molecular clouds is often filamentary \citep[e.g.,][]{SchneiderElmegreen_1979}; it is also known to be highly
hierarchical and self-similar over a wide range of scales \citep[e.g.,][]{Falgarone_etal1991}. This structure can be attributed to
the role of supersonic interstellar turbulence \citep[e.g.,][]{Larson1981} and is reasonably well described by fractal models
\citep[e.g.,][]{ElmegreenFalgarone1996}. However, interstellar turbulence dissipates on small scales; coupled to the effects of
gravity in gravitationally-bound clouds, this breaks the self-similarity of cloud structure on scales below $\sim 0.1$\,pc
\citep[e.g.,][]{Williams_etal2000}. The latter is the typical scale below which prestellar cores, the self-gravitating condensations
of gas and dust giving birth to individual stars or systems, are observed in molecular clouds
\citep[e.g.,][]{Motte_etal1998,Motte_etal2001,Andre_etal2000}. Prestellar cores are observed at the bottom of the hierarchy of
interstellar cloud structures and depart from \cite{Larson1981}'s self-similar scaling relations. They correspond to ``coherent''
regions of nearly constant and thermal velocity dispersion which do not obey Larson's power-law linewidth vs. size relation
\citep[e.g.,][]{Myers1983,Goodman_etal1998}. The $18${\arcsec} angular resolution of \emph{Herschel} at 250\,{${\mu}$m}, equivalent
to $0.03$\,pc at a distance of 350 pc, is sufficient to resolve the typical Jeans length in the nearby clouds; this is also the
characteristic diameter expected for Bonnor-Ebert-like cores.

To first order, known prestellar cores have simple, convex, not very elongated shapes, and their density structure approaches that
of the Bonnor-Ebert isothermal spheroids bound by the external pressure exerted by the parent cloud
\citep[e.g.,][]{Johnstone_etal2000,Alves_etal2001}. Conceptually, a dense core may be defined as the immediate vicinity of a local
minimum in the gravitational potential of a molecular cloud, corresponding to the part of the cloud under a given local
gravitational influence. While, in general, the gravitational potential cannot be inferred from observations, it turns out to be
directly related to the observable column density distribution for the post-shock, filamentary cloud layers produced by supersonic
turbulence in numerical simulations of cloud evolution \citep{GongOstriker2011}. In practical terms, this means that one can define
a dense core (more precisely, its projection onto the plane of the sky) as the immediate vicinity of a local maximum in observed
column density maps, such as those derived from \emph{Herschel} imaging, where dust continuum emission is largely optically thin and
directly traces column densities. The source extraction method presented in this paper offers a new approach to the detection and
measurements of sources, making full use of the multi-scale, multi-wavelength nature of the source extraction problem in the case of
\emph{Herschel} data. Analyzing a wide range of spatial scales, our method is able to detect the hierarchical structures of
molecular clouds (cf. Sect.~\ref{detecting.sources}). Unlike such techniques as the dendrogram analysis \citep{Rosolowsky_etal2008},
however, \textsl{getsources} is not explicitly designed to characterize the hierarchy of structures; our main focus is on the
``compact'' sources, at the end of the hierarchy.
\end{appendix}

%||||||||||||||||||||||||||||||||||||||||||||||||||||||||||||||||||||||||||||||||||||||||||||||||||||||||||||||||||||||||||||||||||

\begin{appendix}
\section{List of symbols}
\label{list.of.symbols}

For convenience of the readers, we list and define all symbols introduced in Sect.~\ref{getsources.extraction.method} of this paper:
\begin{tabbing}
Symbol \,\, \= Definition \kill
$\mathcal{A}_{\lambda}$                            \> images of the annuli around all detected sources\\
$\mathcal{D}_{\!\lambda}$                          \> images of local standard deviations for flattening\\
$\mathcal{F}_{\rm D}$                              \> image of the initial footprints after source detection\\
$\mathcal{F}_{\lambda}$                            \> images of source footprints in measurement iterations\\
$\mathcal{F}^{*}_{\lambda}$                        \> images of source footprints expanded by convolution\\
$\mathcal{G}_{j}$                                  \> smoothing Gaussians in successive unsharp masking\\
$\mathcal{G}_{\!\lambda}$                          \> smoothing Gaussians used to create detection images\\
$\mathcal{G}_{{\!\lambda}{\it A}}$                 \> smoothing Gaussians used in the flattening procedure\\
$\mathcal{I}_{{\rm D}{j}{\,\rm C}}$                \> clean detection images combined over wavelengths\\
$\mathcal{I}_{{\rm D}{j}{\,\rm C}{\,l}}$           \> partial detection images above the sub-level $\omega_{j\,l}$\\
$\mathcal{I}^{\prime}_{{\rm D}{j}{\,\rm C}}$       \> clean detection images combined over wavelengths\\
$\mathcal{I}_{{\rm D}{j}{\,\rm S}}$                \> single-scale segmentation images of source masks\\	
$\mathcal{I}_{{\!i}{{\lambda}{\rm O}}{\,\rm BSD}}$ \> background-subtracted, deblended images of sources\\
$\mathcal{I}_{\!\lambda}$                          \> original observed images produced by a map-maker\\
$\mathcal{I}_{{\!\lambda}{\rm D}{\rm F}}$          \> flattened detection images for the final extraction\\
$\mathcal{I}_{{\!\lambda}{\rm D}}$                 \> detection images: either $\mathcal{I}_{{\!\lambda}{\rm O}}$ or transformed
                                                     $\mathcal{I}_{{\!\lambda}{\rm O}}$\\
$\mathcal{I}_{{\!\lambda}{\rm D}{j}}$              \> single-scale decompositions of the images
                                                     $\mathcal{I}_{{\!\lambda}{\rm D}}$\\
$\mathcal{I}_{{\!\lambda}{\rm D}{j}{\,\rm C}}$     \> single-scale images cleaned of noise and background\\
$\mathcal{I}_{{\!\lambda}{\rm{D\,C}}}$             \> full clean images reconstructed from 
                                                     $\mathcal{I}_{{\!\lambda}{\rm D}{j}{\,\rm C}}$\\
$\mathcal{I}_{{\!\lambda}{\rm D}{j}{\,\rm R}}$     \> single-scale images of the cleaning residuals\\
$\mathcal{I}_{{\!\lambda}{\rm{D\,R}}}$             \> full residuals images reconstructed from 
                                                     $\mathcal{I}_{{\!\lambda}{\rm D}{j}{\,\rm R}}$\\
$\mathcal{I}_{{\!\lambda}{\rm F}}$                 \> scaling image smoothed by convolution\\
$\mathcal{I}_{{\!\lambda}{\rm M}}$                 \> images of the deblending shapes of all sources\\
$\mathcal{I}_{{\!\lambda}{\rm O}}$                 \> measurement images: $\mathcal{I}_{\!\lambda}$ resampled to pixel $\Delta$\\
$\mathcal{I}_{{\!\lambda}{\rm O}{\,\rm BS}}$       \> background-subtracted images of all detected sources\\
$\mathcal{I}_{{\!\lambda}{\rm O}{\,\rm CB}}$       \> clean-background images with all sources removed\\
$\mathcal{M}_{{\rm D}{\rm C}}$                     \> source mask images accumulated over scales and $\lambda$\\
$\mathcal{M}_{\lambda}$                            \> observational mask images defining areas of interest\\
$\mathcal{T}_{{\!\lambda}{j}}$                     \> threshold images with all pixels set to $\varpi_{{\lambda}{j}}$\\
$a_{i}$                                            \> major length of a source segmentation mask at $\omega_{j\,l}$\\
$A_{i}$                                            \> major FWHM size of a source $i$\\
$A_{{i}{\,\rm F}}$                                 \> major axis of the footprint ellipse of a source $i$\\
$A_{{i}{\,\rm F}{\lambda}}$                        \> major axis of the footprint ellipse of a source $i$ at $\lambda$\\
$A_{{i}{\lambda}}$                                 \> major FWHM size of a source $i$ measured at $\lambda$\\
$A^{\rm max}_{\lambda}$                            \> maximum FWHM sizes of sources to be extracted\\
$b_{i}$                                            \> minor length of a source segmentation mask at $\omega_{j\,l}$\\
$B_{i}$                                            \> minor FWHM size of a source $i$\\
$B_{{i}{\,\rm F}}$                                 \> minor axis of the footprint ellipse of a source $i$\\
$B_{{i}{\,\rm F}{\lambda}}$                        \> minor axis of the footprint ellipse of a source $i$ at $\lambda$\\
$B_{{i}{\lambda}}$                                 \> minor FWHM size of a source $i$ measured at $\lambda$\\
$C_{i}$                                            \> contrast of a source $i$ above the sub-level $\omega_{j\,l}$\\
$C^{\,\rm min}_{i}$                                \> required minimum contrast $C_{i}$ for real sources\\
$C_{i\,\rm A}$                                     \> amplification factor in the detection of noise peaks\\
$C_{i\,\rm E}$                                     \> elongation factor in the detection of noise peaks\\
$C_{{i\,\rm E}{j_{\rm F}}}$                        \> elongation factor at the footprinting scale $j_{\rm F}$\\
$C_{{i}{\lambda}{j}}$                              \> contrast of a source $i$ above the threshold $\varpi_{{\lambda}{j}}$\\
$f_{i}$                                            \> global flag: information on source global properties\\
$f_{{\,i}{\lambda}}$                               \> monochromatic flag: information on source at each $\lambda$\\
$f_{\rm S}$                                        \> scale factor defining relative spacing between scales\\
$f_{{\lambda}{j}}$                                 \> turn-on factor for combining scales when $S_{\!j}{\,<\,}O_{\!\lambda}$\\
$F_{i\,\rm hi}$                                    \> flux integrated over the source mask above $\omega_{j\,l}$\\
$F_{i\,\rm lo}$                                    \> flux integrated over the source mask below $\omega_{j\,l}$\\
$F_{{i}{\lambda}{\,\rm P}}$                        \> background-subtracted and deblended peak intensity\\
$F_{{i}{\lambda}{\,\rm T}}$                        \> background-subtracted and deblended total flux\\
$G_{i}$                                            \> goodness of an extracted source $i$\\
$i$                                                \> running number of a source in the extraction catalog\\
$I_{ij}$                                           \> peak intensity of a source $i$ at scale $j$\\
$I_{{i}{\lambda}{j}}$                              \> peak intensity of a source $i$ in the image 
                                                     $\mathcal{I}_{{\!\lambda}{\rm D}{j}{\,\rm C}}$\\
$I_{{i}{\lambda}{j_{\rm F}}}$                      \> peak intensity of a source $i$ in 
                                                     $\mathcal{I}_{{\!\lambda}{\rm D}{j}{\,\rm C}}$ at scale $j_{\rm F}$\\
$I_{{\lambda}{j}}$                                 \> pixel intensity in a single-scale detection image\\
$I^{\,\rm max}_{{\!\lambda}{\rm D}{j}{\,\rm C}}$   \> maximum intensity over the clean image 
                                                     $\mathcal{I}_{{\!\lambda}{\rm D}{j}{\,\rm C}}$\\
$j$                                                \> running number of a decomposed spatial scale\\
$j_{\rm F}$                                        \> number of the footprinting scale of a source\\
$k_{{\lambda}{j}}$                                 \> kurtosis in the single-scale residuals 
                                                     $\mathcal{I}_{{\!\lambda}{\rm D}{j}{\,\rm R}}$\\
$k^{\,\rm max}_{\lambda}$                          \> maximum allowed value of $k_{{\lambda}{j}}$ during cleaning\\
$l$                                                \> running number of the intensity sub-level $\omega_{j\,l}$\\
$n_{\rm det}$                                      \> minimum number of $\lambda$'s a source must be detected at\\
$n_{{\lambda}{j}}$                                 \> variable number of standard deviations $\sigma_{{\lambda}{j}}$ in 
                                                     $\varpi_{{\lambda}{j}}$\\
$N$                                                \> number of wavelengths used in the source extraction\\
$N_{\rm S}$                                        \> number of spatial scales in the image decomposition\\
$N_{{\Pi}{j}}$                                     \> number of pixels in a partial detection image 
                                                     $\mathcal{I}_{{\rm D}{j}{\,\rm C}{\,l}}$\\
$N^{\,\rm min}_{{\Pi}{j}}$                         \> minimum value of $N_{{\Pi}{j}}$ for detecting sources in
                                                     $\mathcal{I}_{{\rm D}{j}{\,\rm C}{\,l}}$\\
$N_{{\Pi}{\lambda}}$                               \> number of pixels in a cluster of connected pixels\\
$N^{\,\rm min}_{{\Pi}{\lambda}}$                   \> minimum value of $N_{{\Pi}{\lambda}}$ in 
                                                     $\mathcal{I}_{{\!\lambda}{\rm D}{j}{\,\rm C}}$ for making 
                                                     $\mathcal{I}_{{\rm D}{j}{\,\rm C}}$\\
$O_{\!\lambda}$                                    \> observational angular resolution: FWHM beam size\\
$\bar{O}$                                          \> observational beam size averaged over wavelengths\\
$R_{i}$                                            \> reliability of an extracted source $i$\\
$s_{{\lambda}{j}}$                                 \> skewness in the single-scale residuals
                                                     $\mathcal{I}_{{\!\lambda}{\rm D}{j}{\,\rm R}}$\\
$s^{\,\rm max}_{\lambda}$                          \> maximum allowed value of $s_{{\lambda}{j}}$ during cleaning\\
$S_{\!j}$                                          \> spatial scale: FWHM of a smoothing Gaussian beam\\
$S_{\!j_{\,\rm F}}$                                \> characteristic footprinting scale of a source in 
                                                     $\mathcal{I}_{{\rm D}{j}{\,\rm C}}$\\
$S_{\!j_{\,\rm F}{\lambda}}$                       \> characteristic footprinting scale of a source in 
                                                     $\mathcal{I}_{{\!\lambda}{\rm D}{j}{\,\rm C}}$\\
$S_{\rm max}$                                      \> largest spatial scale in a single-scale decomposition\\
$w_{\lambda}$                                      \> weight enhancing contribution of high-res. images\\
$x_i$, $y_i$                                       \> source coordinates obtained in the detection process\\
$\alpha_{i}, \delta_{i}$                           \> source coordinates in the equatorial system\\
$\eta$                                             \> parameter in the detection of noise peaks\\
$\gamma$                                           \> weighting power-law exponent defining $w_{\lambda}$\\
$\Delta$                                           \> pixel size (the same for all images in an extraction)\\
$\lambda$                                          \> wavelength (central wavelength of a waveband)\\
$\mu_{3{\lambda}{j}}$                              \> third statistical moments about the mean value\\
$\mu_{4{\lambda}{j}}$                              \> fourth statistical moments about the mean value\\
$\varpi_{{\lambda}{j}}$                            \> iterated cleaning thresholds (cut-off levels)\\
$\sigma_{{\lambda}{j}}$                            \> standard deviation in a single-scale image\\
$\sigma_{\lambda}$                                 \> standard deviation in a full image\\
$\sigma_{{\lambda}{j_{\rm F}}}$                    \> standard deviation in a detection image at scale $j_{\rm F}$\\
$\sigma_{{i}{\lambda}{\,\rm P}}$                   \> uncertainty of the peak intensity $F_{{i}{\lambda}{\,\rm P}}$ of a source\\
$\sigma_{{i}{\lambda}{\,\rm T}}$                   \> uncertainty of the total flux $F_{{i}{\lambda}{\,\rm T}}$ of a source\\
$\Pi_l$, $\Pi_m$                                   \> pixels $l$, $m$ in the clean combined images 
                                                     $\mathcal{I}_{{\rm D}{j}{\,\rm C}}$\\
$\Theta_{{i}{\,\rm F}{\lambda}}$                   \> position angle of the elongation of a footprint\\
$\Theta_{{i}{\lambda}}$                            \> position angle of the elongation of a source\\
$\Xi_{i}$                                          \> global significance over all wavelengths $\lambda$\\
$\Xi_{{i}{\lambda}}$                               \> monochromatic significance of a source $i$ at $\lambda$\\
$\Xi_{\rm rel}$                                    \> reliable level of signficance for extracted sources\\
$\Xi_{\rm ten}$                                    \> tentative level of signficance for extracted sources\\
$\Omega_{i}$                                       \> global signal-to-noise ratio over all wavelengths\\
$\Omega_{{i}{\lambda}}$                            \> monochromatic signal-to-noise ratio of a source $i$\\
$\omega_{j\,l}$                                    \> sub-levels of intensities during source detection
\end{tabbing}
\end{appendix}

%||||||||||||||||||||||||||||||||||||||||||||||||||||||||||||||||||||||||||||||||||||||||||||||||||||||||||||||||||||||||||||||||||

\begin{appendix}
\section{Simulated star-forming region}
\label{simulated.images}

To illustrate the spatial decomposition and all other processing steps of \textsl{getsources} in this paper, we use a simulated
star-forming region that we constructed well before the launch of \emph{Herschel} in order to have a reasonably realistic model; it
is sufficient to give here the following brief description\footnote{Full description of this and other synthetic skies will be
given in another paper (Men'shchikov et al., in prep.) devoted to benchmarking of source extraction algorithms.}.

The simulated star-forming region, placed at 140\,pc, consists of a synthetic (scale-free) cirrus background fitted to a typical
100\,{${\mu}$m} intensity of the backgrounds in the Gould Belt survey and scaled to all other \emph{Herschel} wavebands assuming a
blackbody with a dust temperature of 17.5\,K \citep[cf.][]{Lagache_etal1999}. The background was populated with 360 starless cores
and 107 protostars with realistic intensity distributions, obtained from a large grid of 129 one-dimensional dust continuum
radiative transfer models. The density distribution of the cores followed the structure of critical Bonnor-Ebert spheres, whereas
the protostars had $\rho \propto r^{-2}$ density profiles; both types of objects were embedded in background spherical clouds with a
uniform density. The standard isotropic interstellar radiation field was shining on the outer edges of the clouds, making the
temperature profile inverted (lower in the center) in all objects. Protostars, however, restored usual temperature distributions
deeper towards their central parts, as they produced an accretion luminosity ($L_{\rm acc}{\,\propto\,}M$); they had accreted half
of the mass of the cores they formed from \citep[the conceptual borderline between Class 0 and Class I objects,][]{Andre_etal2000}.
Starless cores consisted of low-, medium-, and high-density sub-populations and were distributed according to the $M_{\rm BE}
\propto R_{\rm BE}$ relation for the isothermal Bonnor-Ebert spheres ($T_{\rm BE}{\,=\,}$7, 14, 28\,K) in the area of the
mass-radius diagram occupied by prestellar cores observed in the Orion and Ophiuchus star-forming regions
\citep{Motte_etal1998,Motte_etal2001}. All populations span wide ranges of masses (0.01--10\,{$M_{\sun}$}) and radii
(0.001--0.1\,pc). Random noise (at the expected levels of the instrumental noise) was added to all pixels of the simulated images
and the latter were convolved to the expected observational resolutions of 5, 7, 11, 17, 24, and 35{\arcsec} in all PACS and SPIRE
bands at 70, 100, 160, 250, 350, and 500\,{${\mu}$m}.
\end{appendix}

%||||||||||||||||||||||||||||||||||||||||||||||||||||||||||||||||||||||||||||||||||||||||||||||||||||||||||||||||||||||||||||||||||

\begin{appendix}
\section{A look in the Fourier domain}
\label{fourier.domain}

\begin{figure*}                                                               
\centering
\centerline{\resizebox{0.33\hsize}{!}{\includegraphics{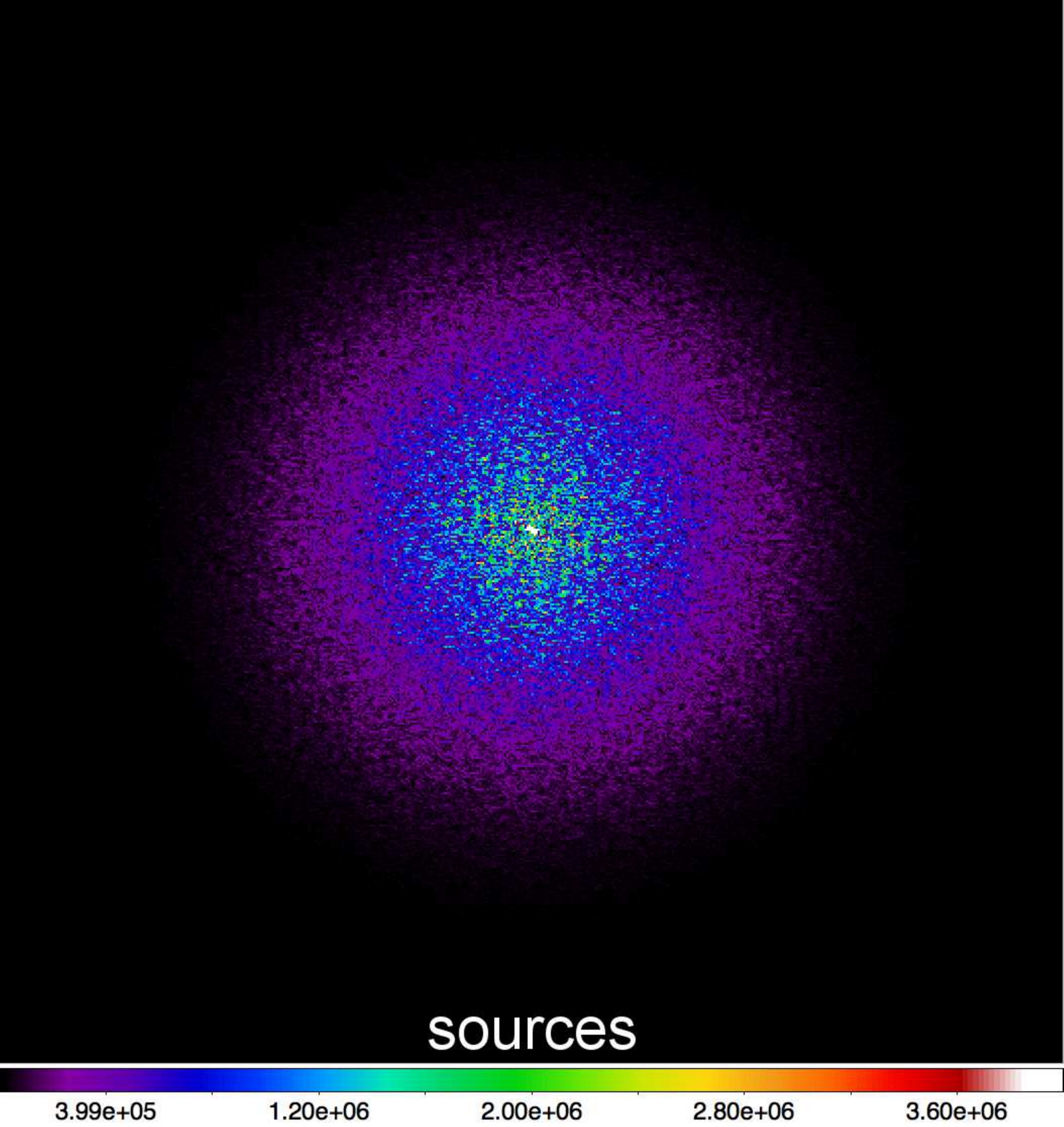}}
            \resizebox{0.33\hsize}{!}{\includegraphics{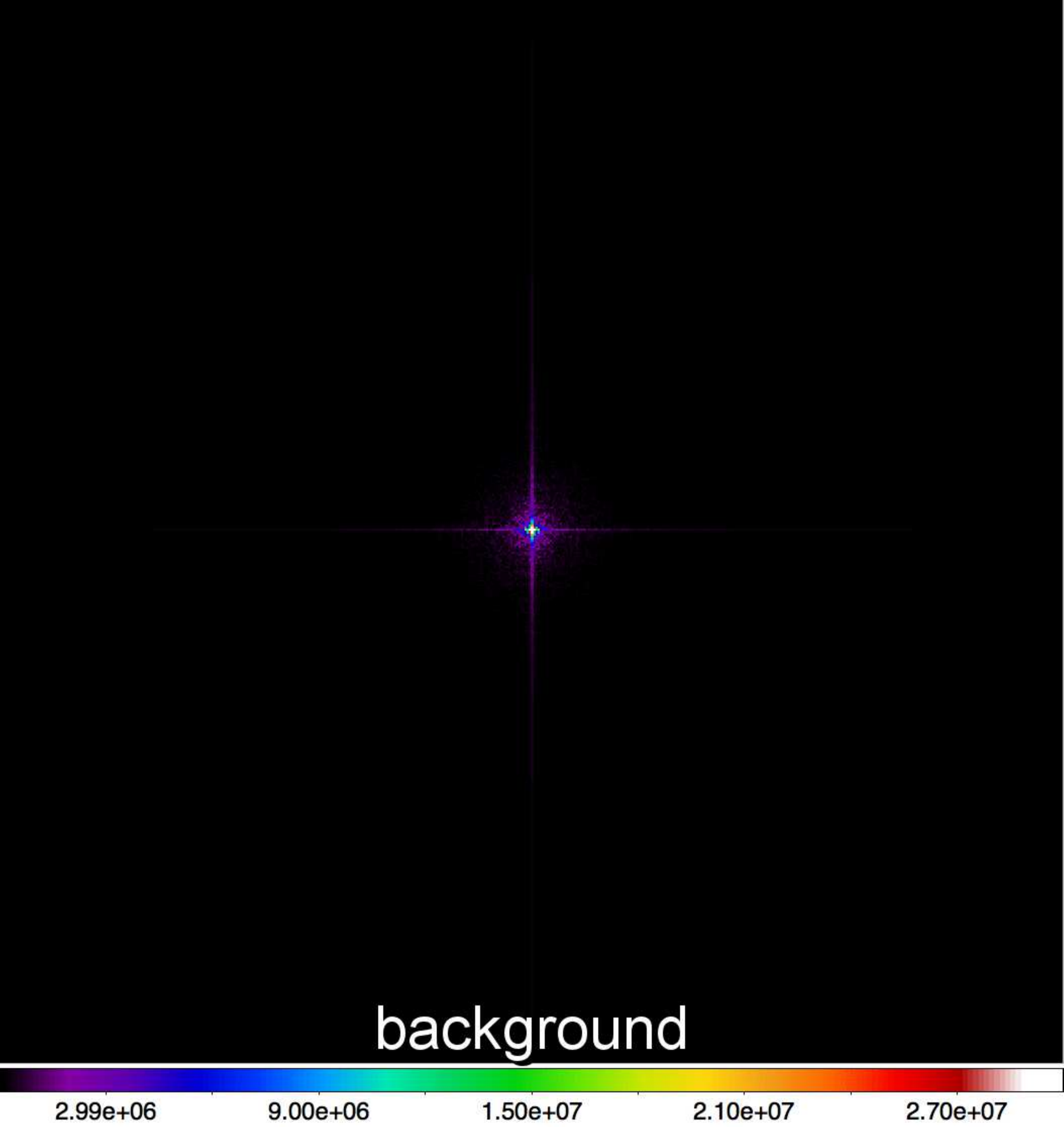}}
            \resizebox{0.33\hsize}{!}{\includegraphics{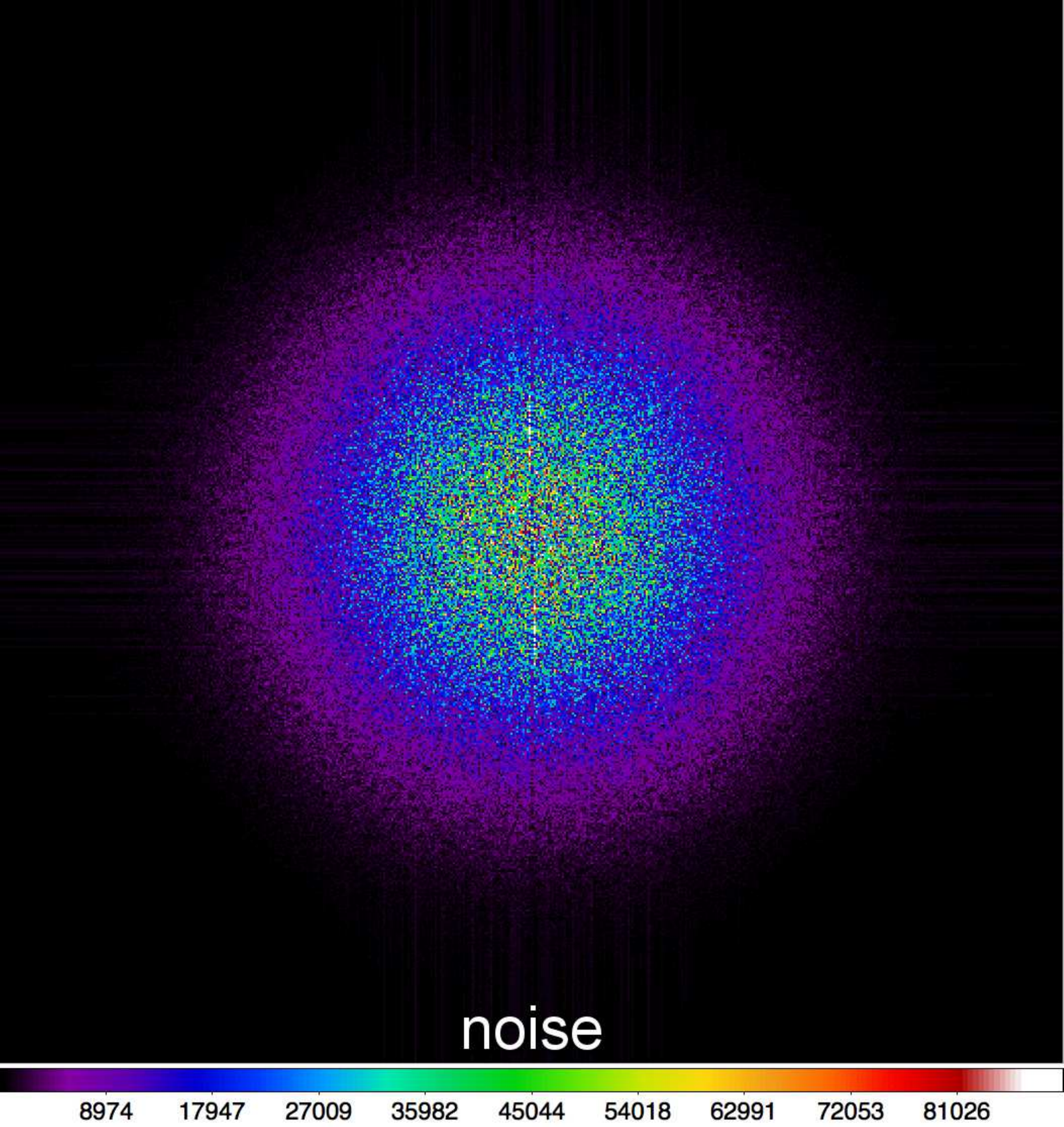}}}
\centerline{\resizebox{0.33\hsize}{!}{\includegraphics{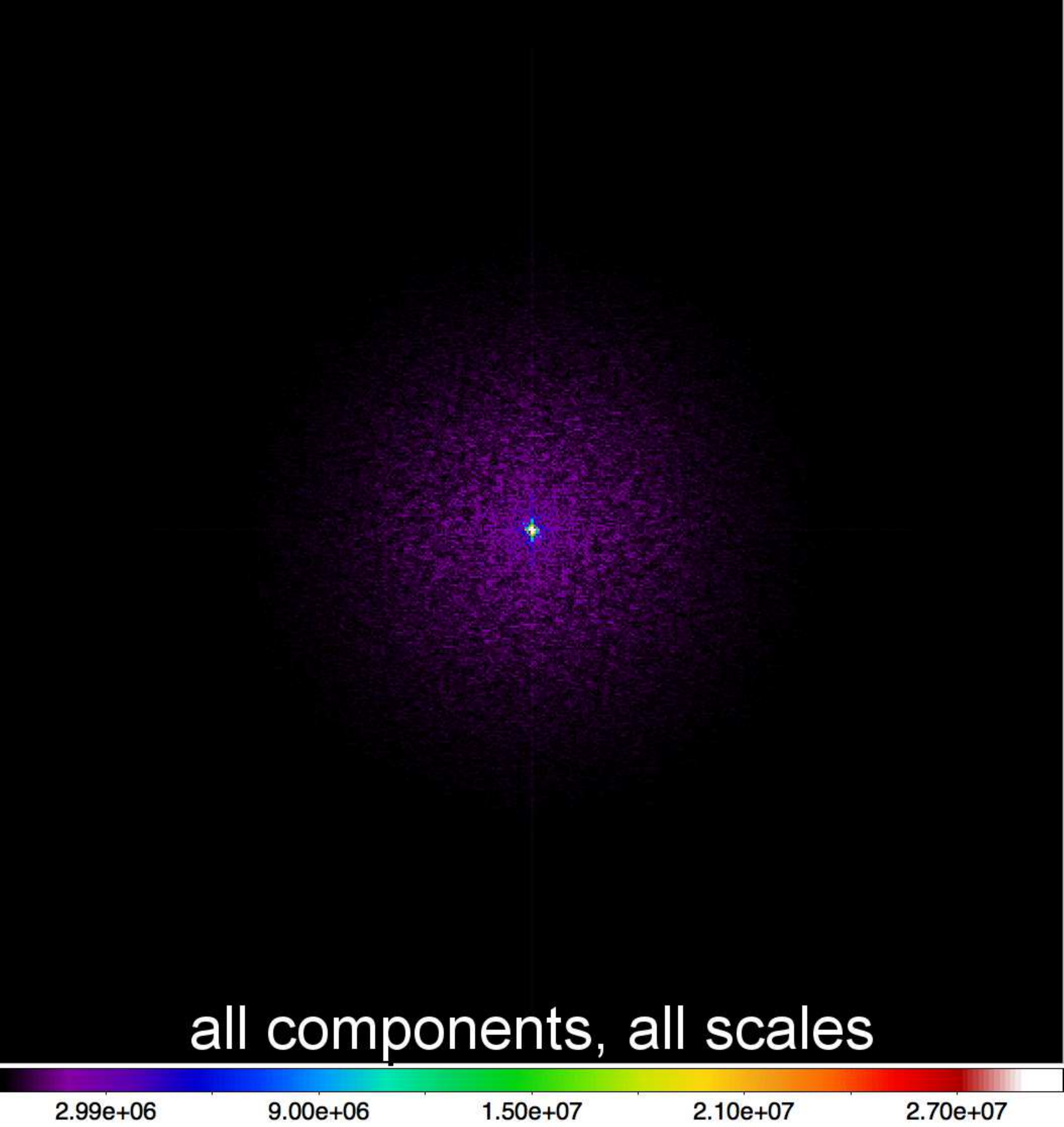}}
            \resizebox{0.33\hsize}{!}{\includegraphics{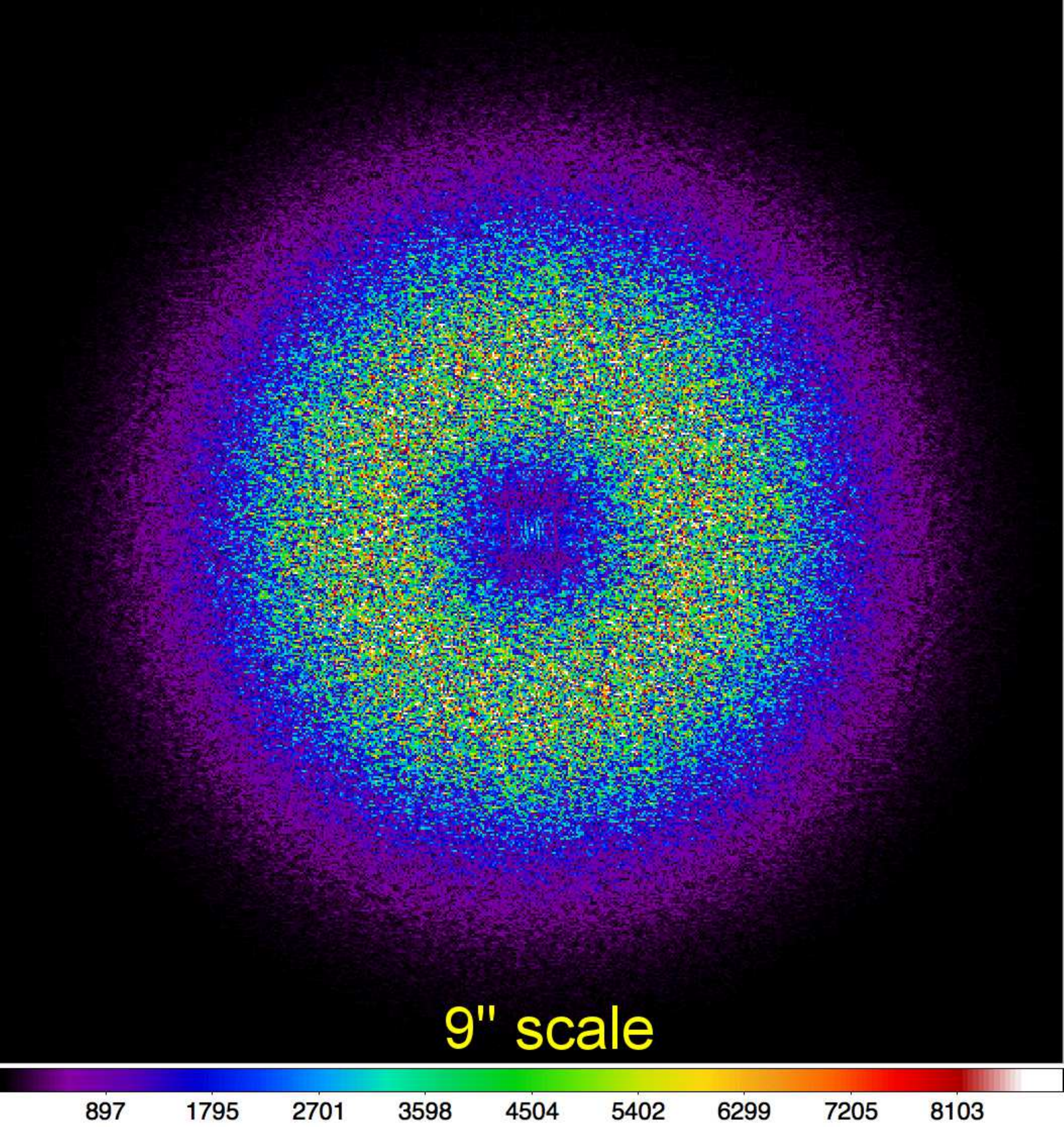}}
            \resizebox{0.33\hsize}{!}{\includegraphics{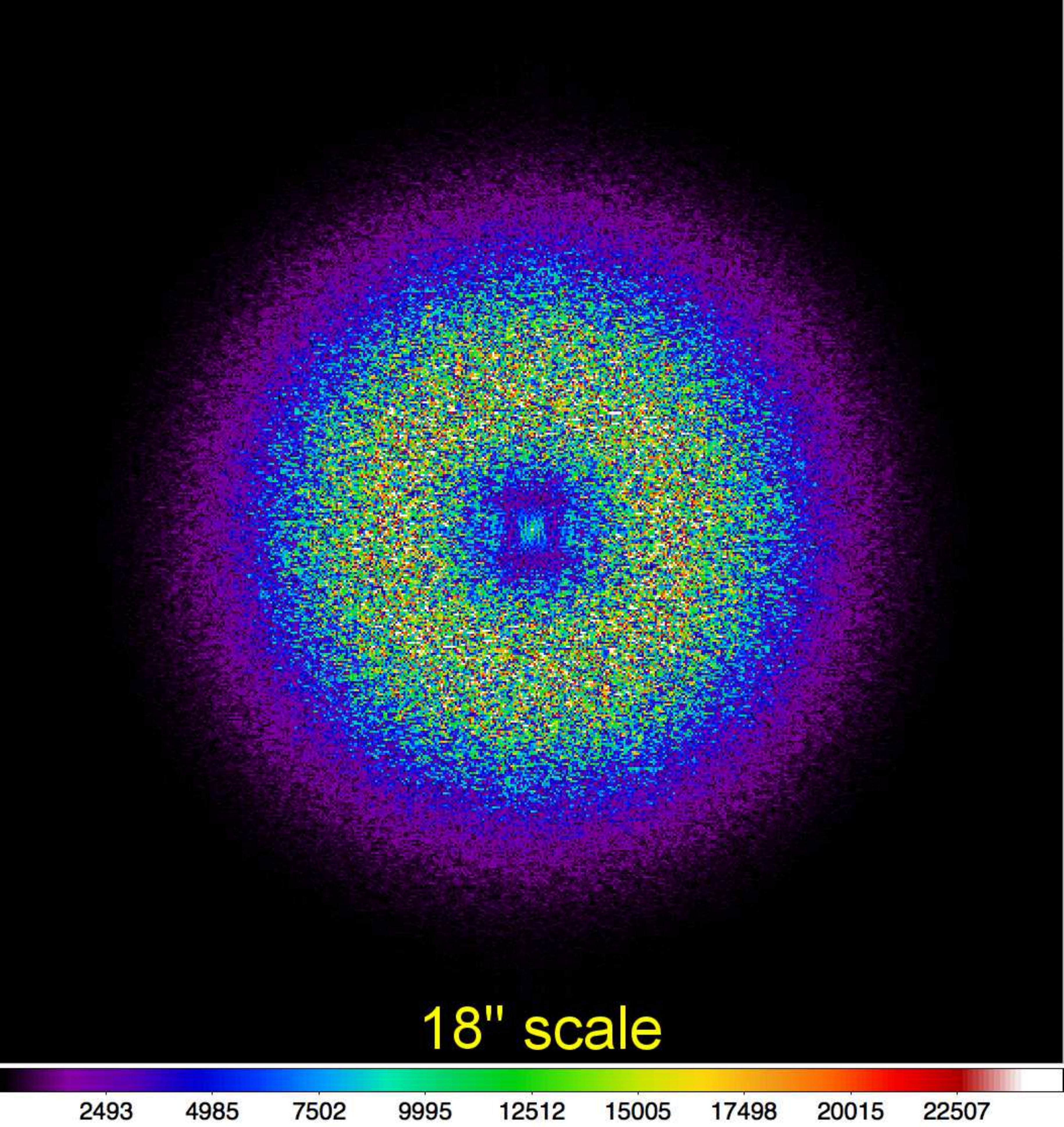}}}
\centerline{\resizebox{0.33\hsize}{!}{\includegraphics{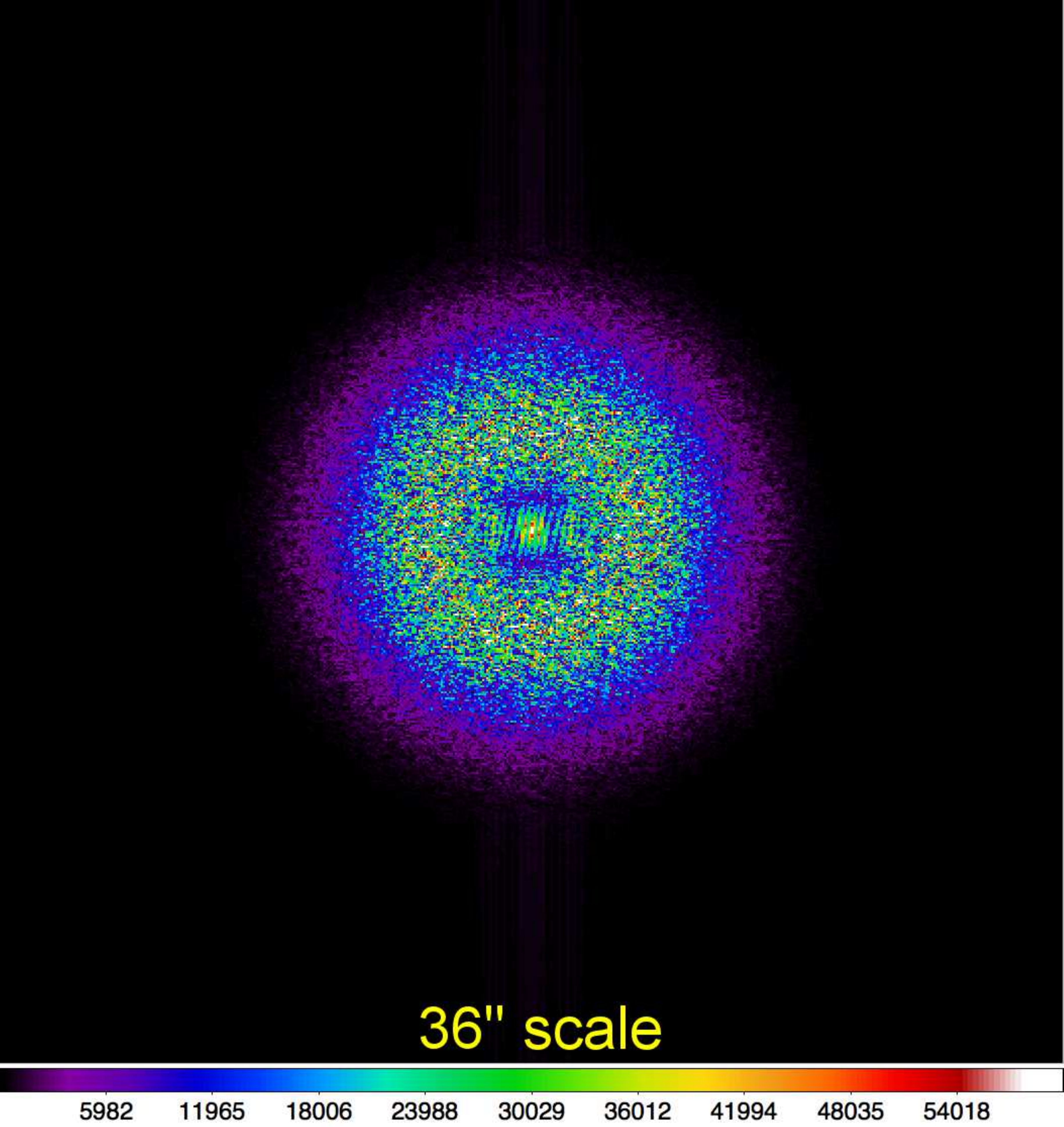}}
            \resizebox{0.33\hsize}{!}{\includegraphics{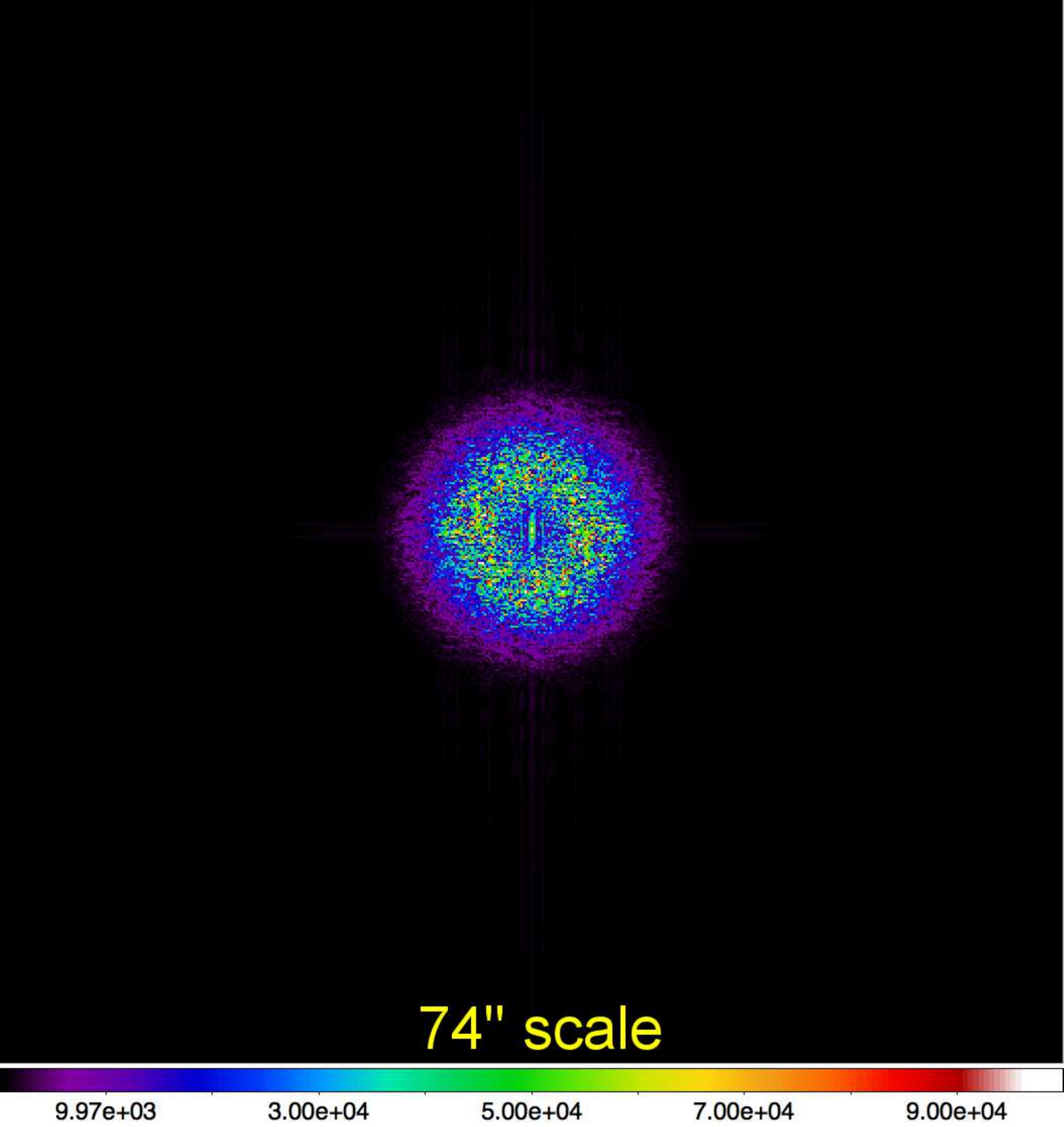}}
            \resizebox{0.33\hsize}{!}{\includegraphics{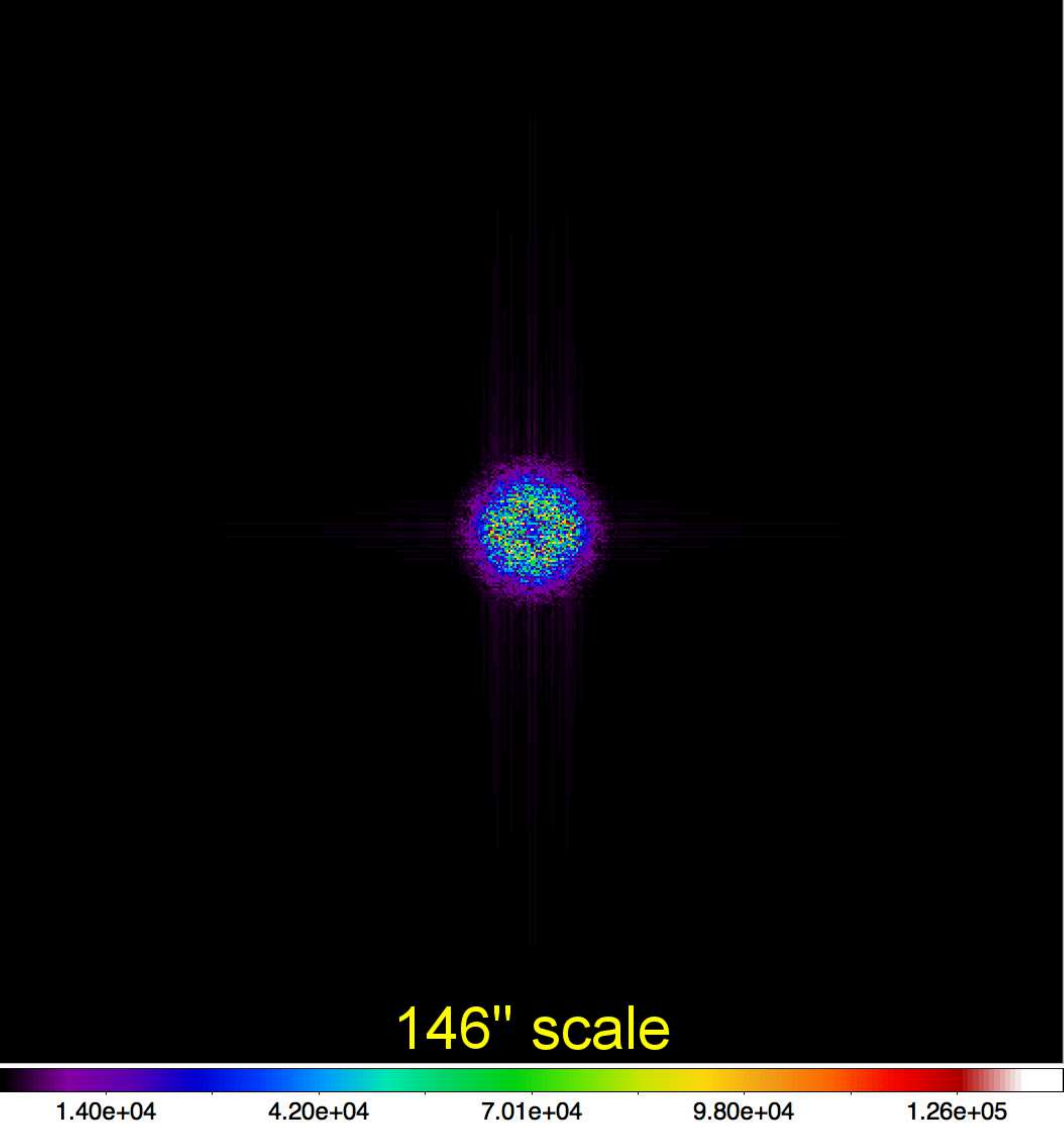}}}
\caption{
Fourier transform of the simulated star-forming region used in this paper (Appendix~\ref{simulated.images}) at 350\,{${\mu}$m} with
a 24{\arcsec} resolution (Fig.\,\ref{flattening.images}, \emph{upper-left}). The top panels show the Fourier amplitudes of separate
components: the model sources, synthetic background, and noise (\emph{top left to right}). The full image containing all the
ingredients (and all spatial scales) is displayed in the following (\emph{middle-left}) panel. The remaining panels (\emph{left to
right}, \emph{top to bottom}) show the images of the single-scale decompositions of the simulated sky (cf. Fig.~\ref{single.scales})
with the scale sizes $S_{\!j}$ annotated at the bottom of the panels. For the original images with 2048$\times$2048 \,2{\arcsec}
pixels, the panels present the Fourier amplitudes within the range of spatial frequencies from zero to one-fourth of the Nyquist
frequency (0.25 arcsec$^{-1}$). For better visibility, the pixel values are somewhat limited in range; the color coding is linear.
} 
\label{simulated.fourier}
\end{figure*}

\begin{figure*}                                                               
\centering
\centerline{\resizebox{0.33\hsize}{!}{\includegraphics{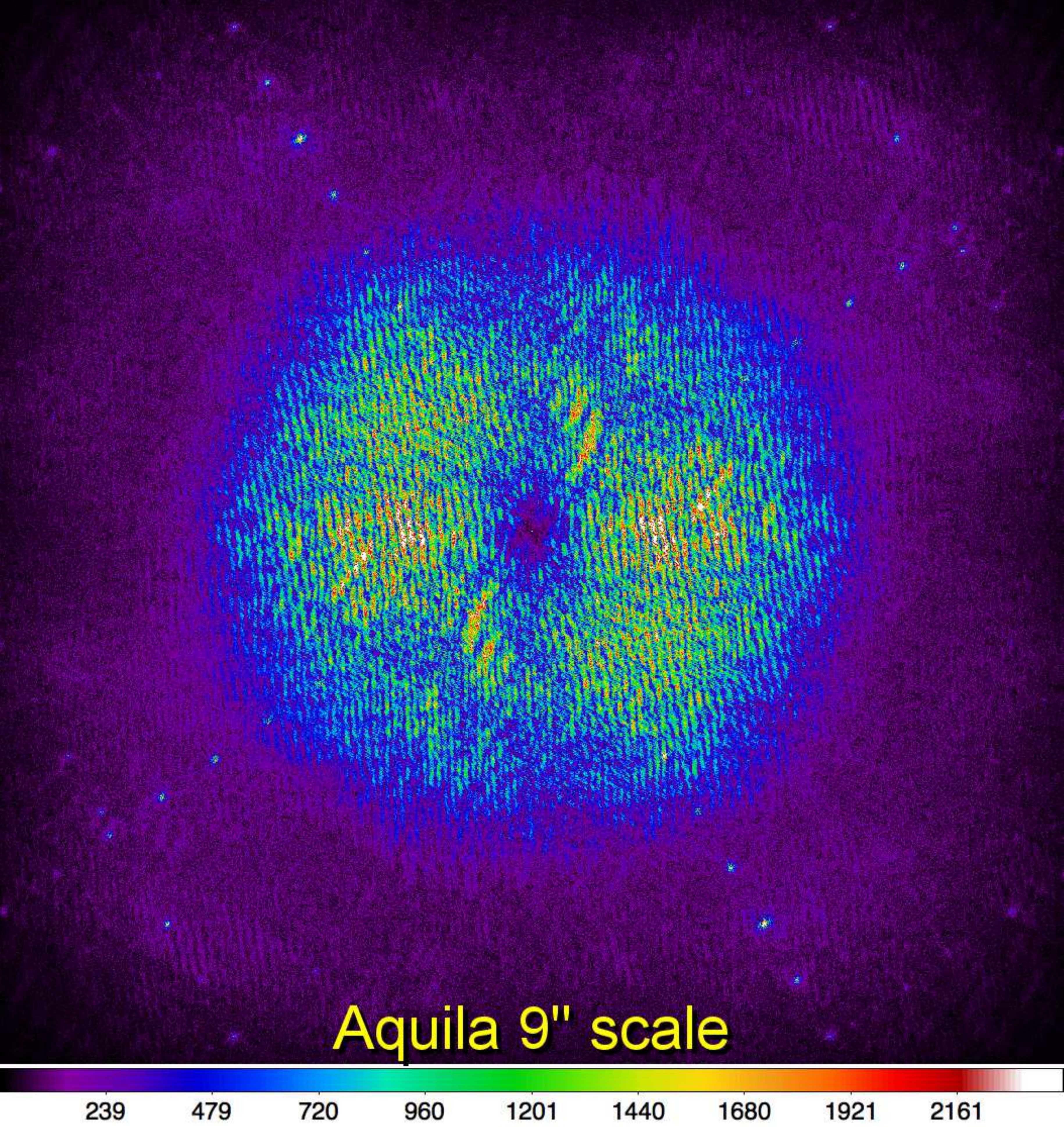}}
            \resizebox{0.33\hsize}{!}{\includegraphics{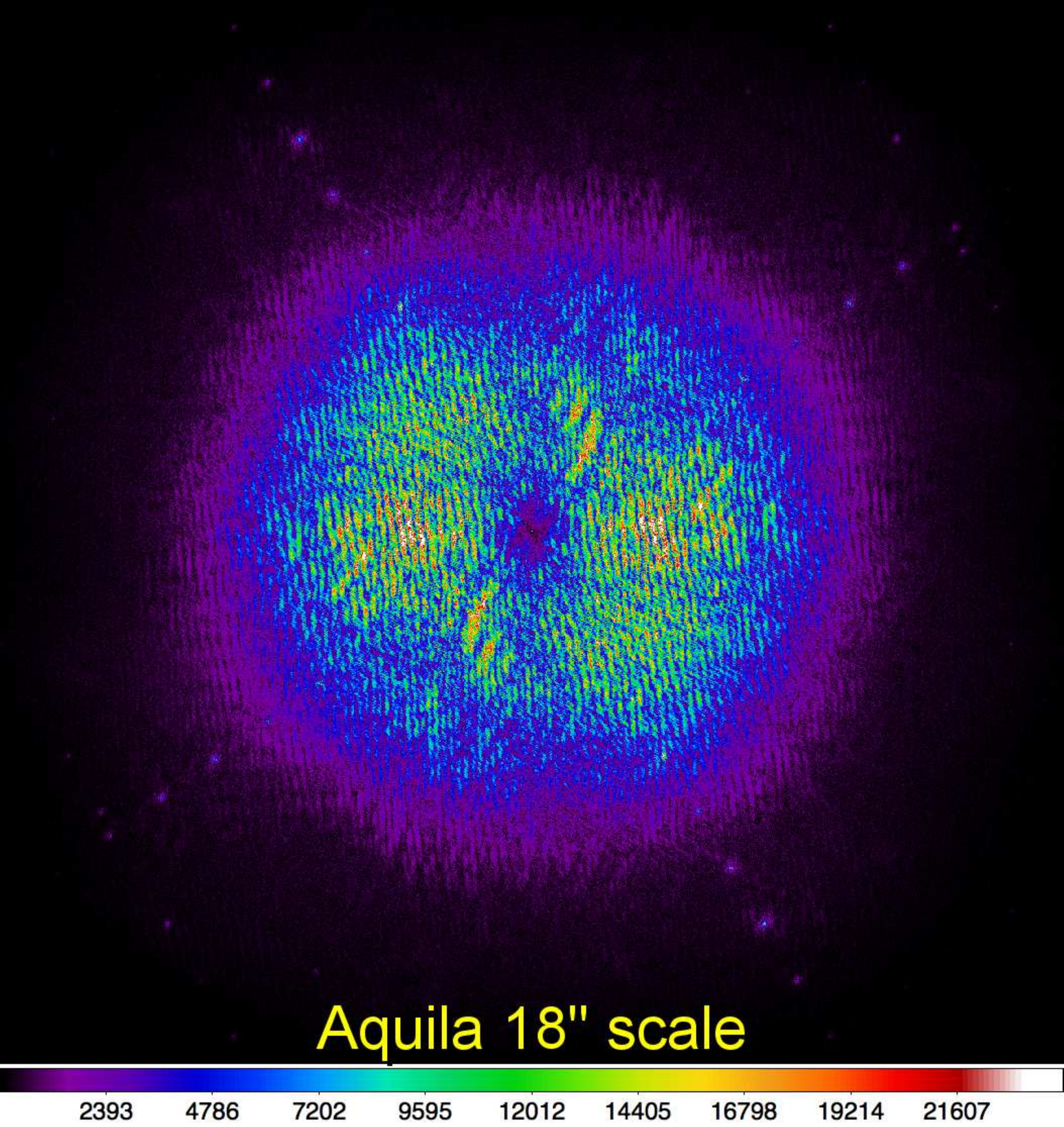}}
            \resizebox{0.33\hsize}{!}{\includegraphics{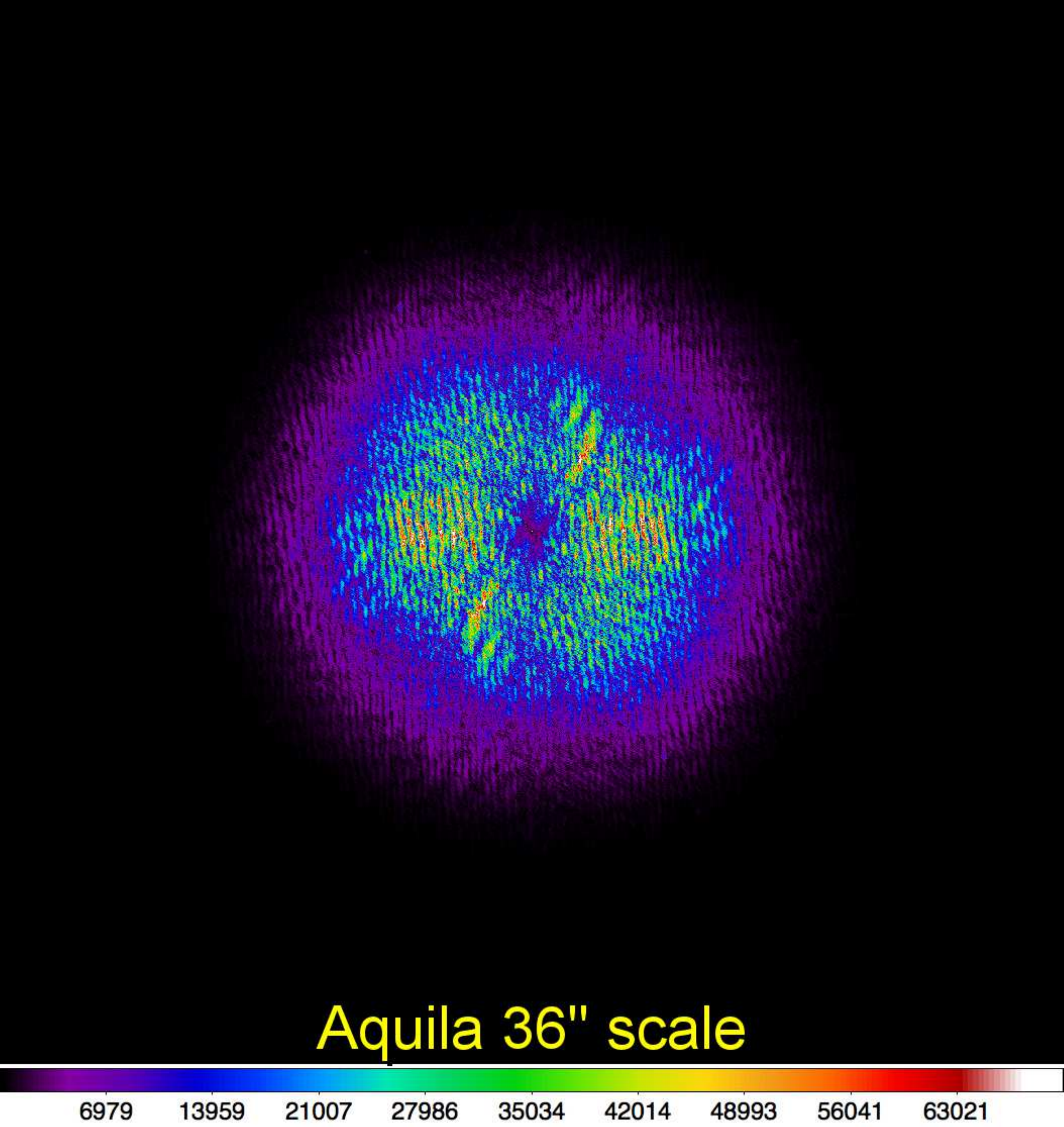}}}
\centerline{\resizebox{0.33\hsize}{!}{\includegraphics{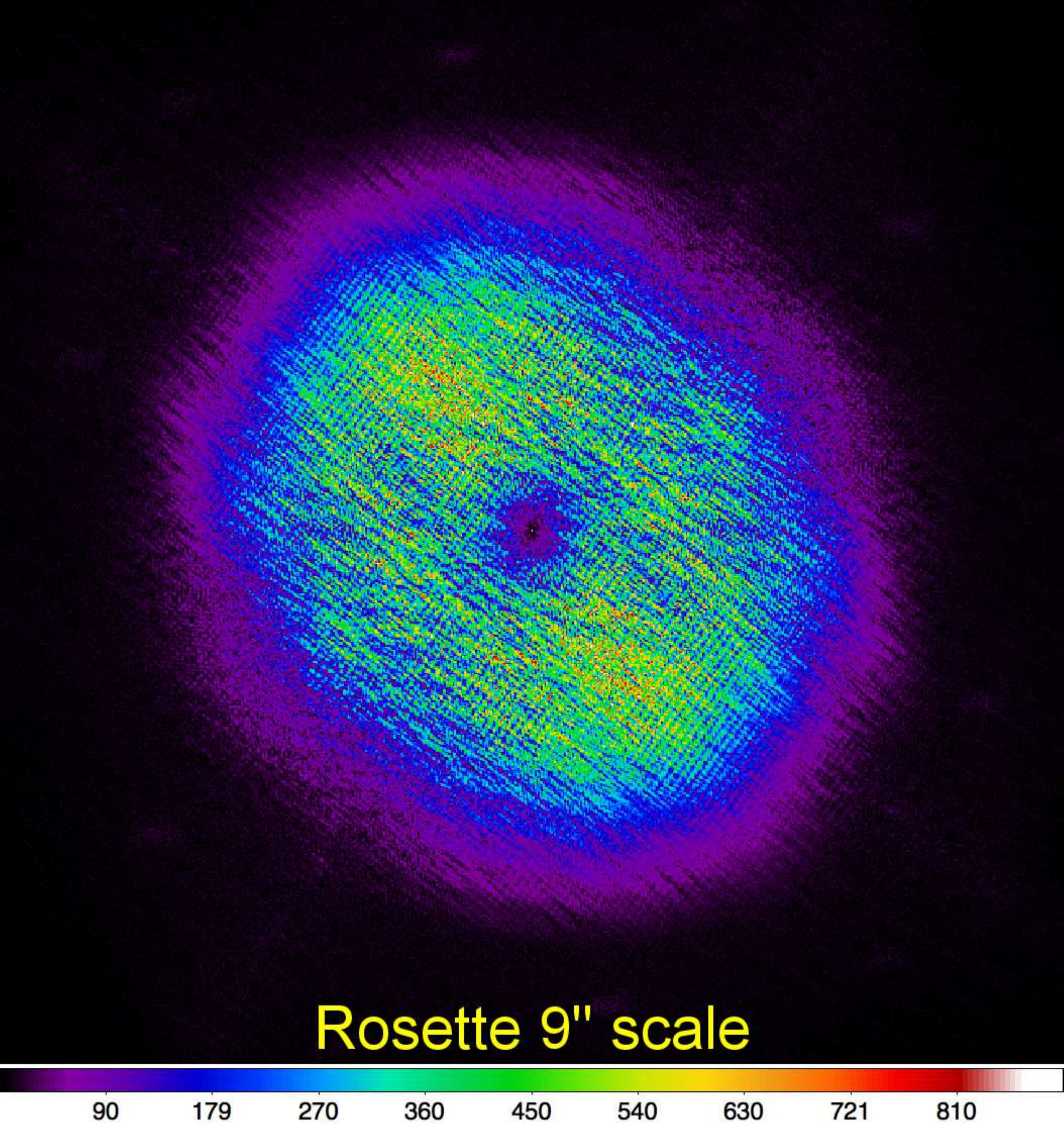}}
            \resizebox{0.33\hsize}{!}{\includegraphics{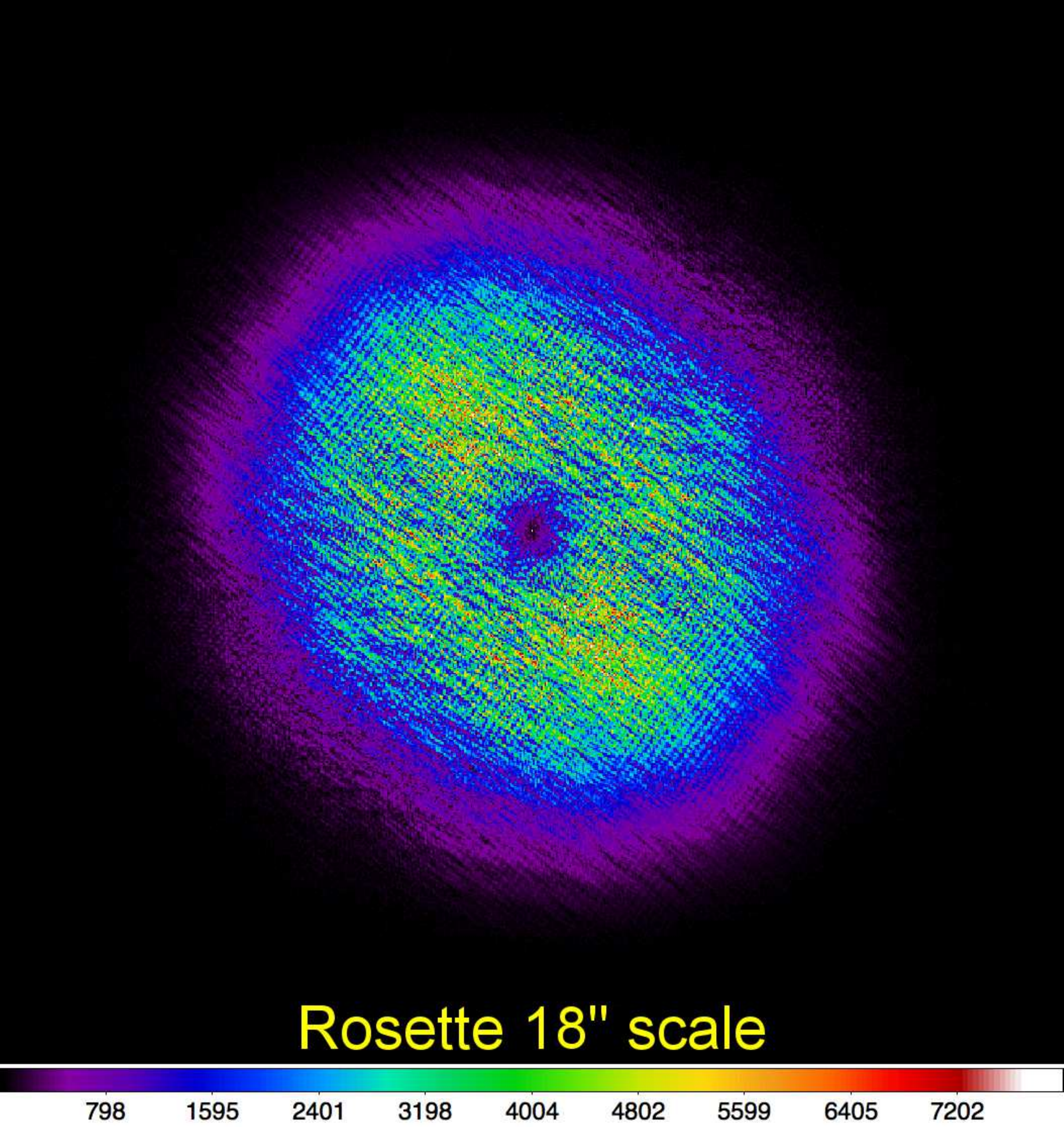}}
            \resizebox{0.33\hsize}{!}{\includegraphics{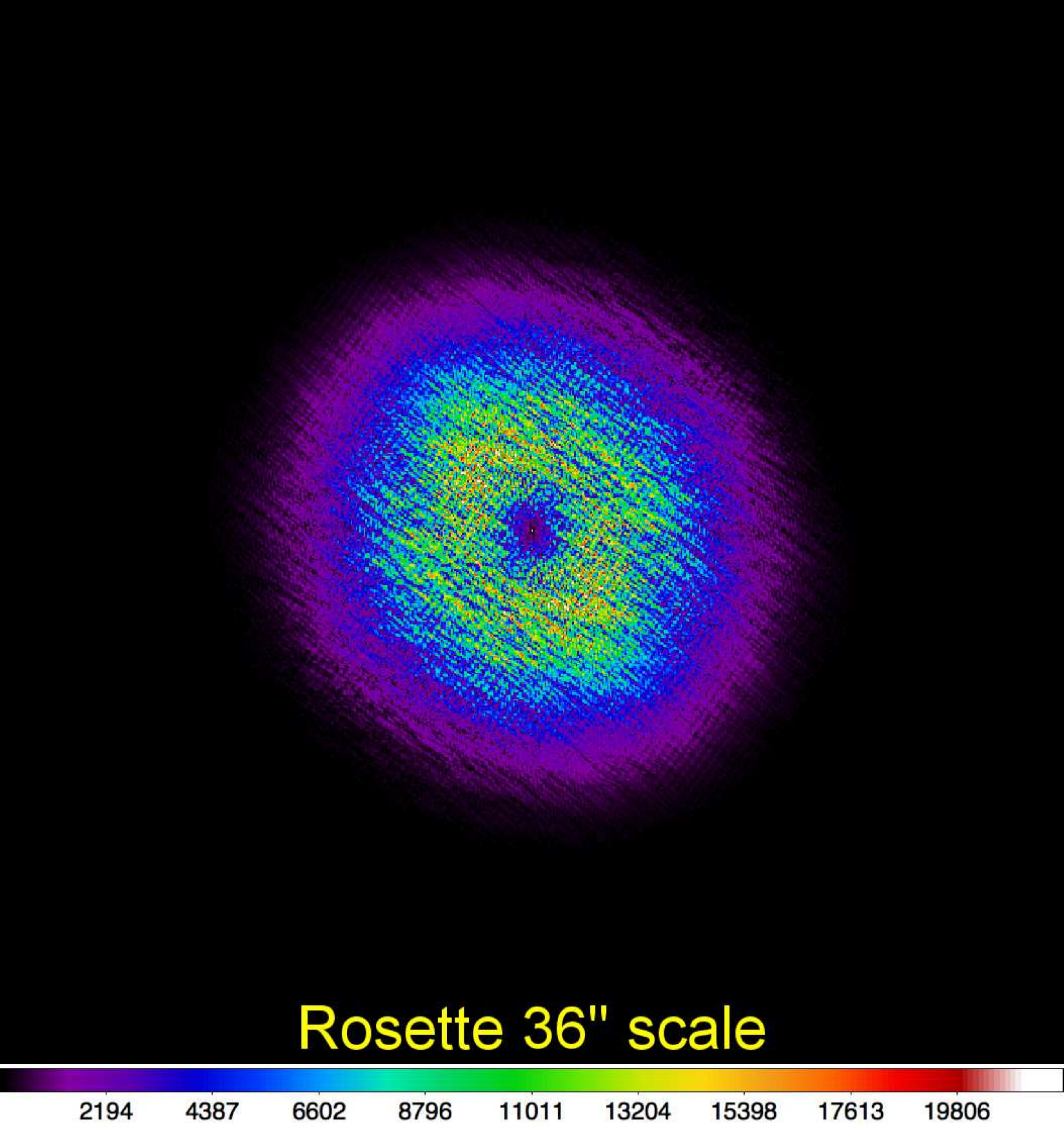}}}
\caption{
Fourier transform of the actual star-forming regions observed by \emph{Herschel} at 350\,{${\mu}$m} with a 25{\arcsec} resolution.
Shown are the amplitudes of the decomposed images of \object{Aquila} (\emph{top left to right}) and \object{Rosette} (\emph{bottom left to right});
the scale sizes $S_{\!j}$ are annotated at the bottom of the panels. For the original \object{Aquila} images with 4096$\times$4096
\,3{\arcsec} pixels and \object{Rosette} images with 2048$\times$2048 \,3{\arcsec} pixels, the panels present the Fourier amplitudes within
the range of spatial frequencies from zero to one-third of the Nyquist frequency (0.17 arcsec$^{-1}$). For better visibility, the
spatial frequencies and the pixel values are somewhat limited in range; the color coding is linear.
} 
\label{aquila.rosette.fourier}
\end{figure*}

The original images and their single-scale decompositions can be transformed into the Fourier domain. The successive unsharp masking
described by Eq.~\ref{successive.unsharp.masking} (Sect.~\ref{decomposing.detection.images}) can also be formulated in terms of the
Fourier transforms of the images. For those readers who are used to the tranformations between the image and Fourier domains, we
present some additional images that may be useful for better understanding of \textsl{getsources}.

In Fig.~\ref{simulated.fourier}, we show the amplitudes of the complex Fourier transform for the simulated star-forming region that
is used in this paper to illustrate \textsl{getsources} (cf. Appendix~\ref{simulated.images}). In the simulated images, one exactly
knows all individual components: the model sources, background, and noise; the amplitudes of the three components are displayed in
the top panels of Fig.~\ref{simulated.fourier}. The Fourier transforms of the sources and noise are clearly dominated by the fact
that the original images have a resolution of the 350\,{${\mu}$m} \emph{Herschel} band: higher spatial frequencies have been
suppressed by the convolution with the observational beam. Although the transform of the synthetic background is also shaped by the
limited angular resolution at high frequencies, the steep power spectrum $P(q){\,\propto\,}q^{-3}$ of the synthetic background
carries most of its power on large scales (cf. Appendix~\ref{power.spectra}) and hence the amplitude remains strongly peaked at zero
frequency. The full simulated image has all components added together and at each spatial frequency the Fourier amplitude becomes a
mixture of the noise, background, and sources that cannot be cleanly separated anymore (Fig.~\ref{simulated.fourier}, middle-left
panel).

Our source extraction method attempts to give a practical solution to the problem of separating sources from all other components
by decomposing the original images in a large number of fine spatial scales using a procedure that involves successive convolution
and subtraction of the images (cf. Sect.~\ref{decomposing.detection.images}, Eq.~\ref{successive.unsharp.masking}). In effect, such
a decomposition creates a set of the filtered ``single-scale'' images $\mathcal{I}_{{\!\lambda}{\rm D}{j}}$ each containing
considerable signals from only a limited range of spatial scales (or frequencies) that are determined by the size $S_{\!j}$ of the
smoothing Gaussian beam $\mathcal{G}_{j}$. Consequently, in the Fourier domain, the single-scale images look like toroidal
structures of variable widths heavily overlapping with each other over a substantial range of frequencies
(Fig.~\ref{simulated.fourier}). By construction, the simulated sky does not include any asymmetric or highly-elongated (filamentary)
structures, therefore the toroids look very regular and symmetric in the Fourier domain. The actual fields observed by
\emph{Herschel} are, however, more complicated, containing lots of filamentary structures, that make the toroids less symmetric. To
illustrate their appearance in the real observations, we present in Fig.~\ref{aquila.rosette.fourier} Fourier amplitudes of a few
single-scale images for the \object{Aquila} and \object{Rosette} star-forming regions.

The Fourier space representation is useful for understanding some aspects of our method, as well as for image convolution (for which
we apply a discrete fast Fourier transform algorithm). However, \textsl{getsources} is not entirely translatable into the Fourier
domain and remains fundamentally the image-oriented method of source extraction.
\end{appendix}

%||||||||||||||||||||||||||||||||||||||||||||||||||||||||||||||||||||||||||||||||||||||||||||||||||||||||||||||||||||||||||||||||||

\begin{appendix}
\section{Power spectra of image components}
\label{power.spectra}

\begin{figure}
\centering
\centerline{\resizebox{\hsize}{!}{\includegraphics{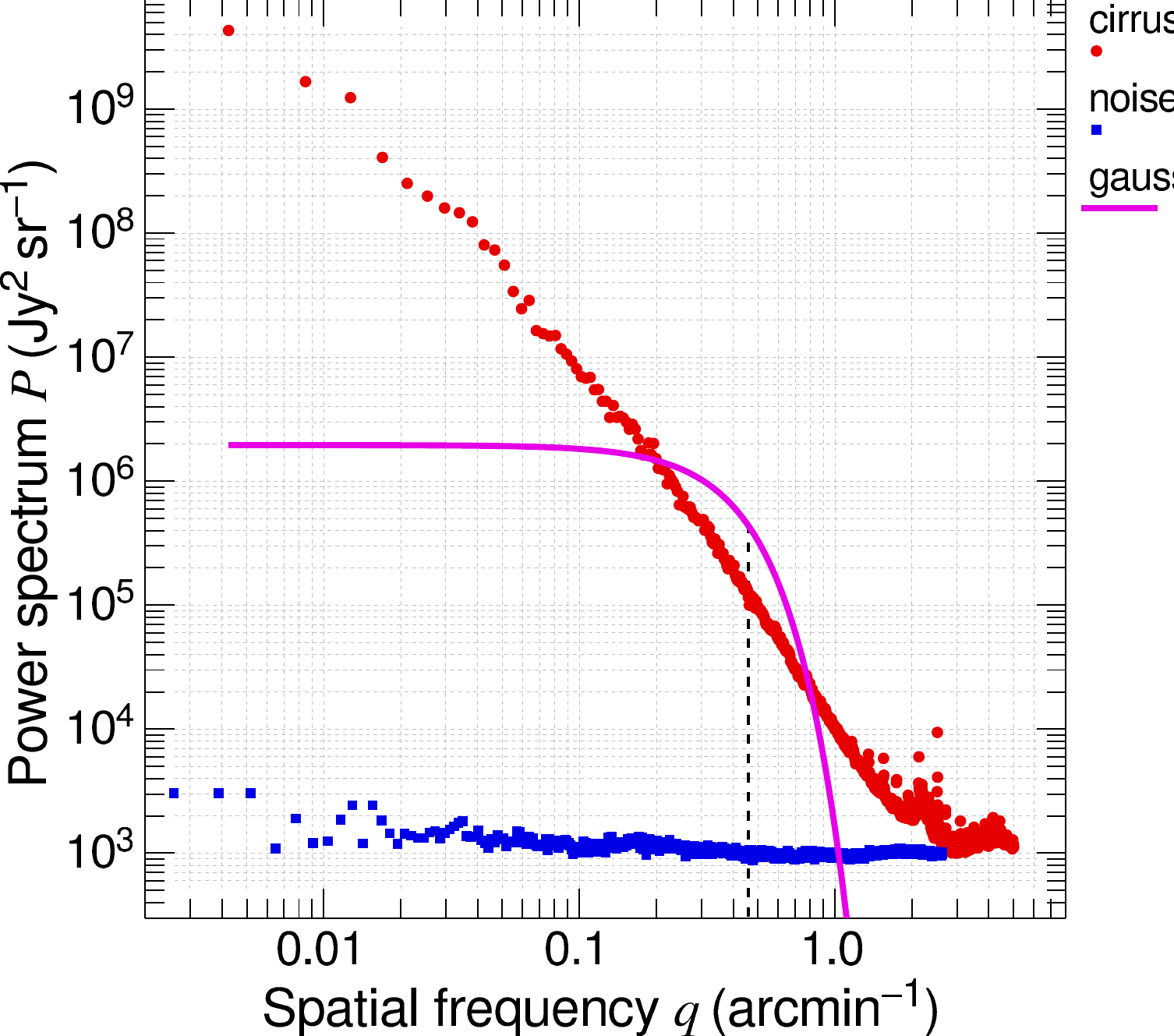}}} 
\caption{
Typical power spectra of several components contained in observed \emph{Herschel} images. Red \emph{circles} show the power 
spectrum of the SPIRE 250\,{${\mu}$m} image of the Polaris cirrus cloud taken as part of the \emph{Herschel} Gould Belt survey; no 
attempt was made here to correct the raw power spectrum for deviations of the SPIRE 250\,{${\mu}$m} beam from a Gaussian shape
\citep[cf.][]{Miville-Deschenes_etal2010}. This cirrus background is well described by a power law $P(q){\,\propto\,}q^{-3}$ in the
range of spatial frequencies $0.02{\,<\,}q{\,<\,}2$ arcmin$^{-1}$. Blue \emph{squares} show the estimated power spectrum of the
instrumental noise in a typical \emph{Herschel} image at 250\,{${\mu}$m}, which is flat over more than two orders of magnitude in
$q$. The \emph{solid} curve shows the power spectrum (scaled to an arbitrary level of power) of the intensity distribution of a 
Gaussian source with a size of 1 arcmin (FWHM). The vertical \emph{dashed} line marks the spatial frequency of $0.46$ arcmin$^{-1}$, 
at which the contrast of such a source is maximum over the background fluctuations (assumed to follow $P{\,\propto\,}q^{-3}$).
}
\label{plots.power.spectra}
\end{figure}

The ability to detect compact sources, such as dense cores, in the wide-field images obtained as part of the Gould Belt and HOBYS
surveys with \emph{Herschel} is primarily limited by the confusion arising from the small-scale cloud structure. In contrast, the
instrumental noise and mapping artifacts are often negligible. An important property of the background cloud fluctuations is that
they do not follow Gaussian statistics. In particular, it is well known that the power spectrum of interstellar cloud fluctuations
strongly depends on spatial scales \citep[$P(q){\,\propto\,}q^{-3}$, where $q$ is spatial frequency; e.g.,][and references
therein]{Roy_etal2010}. This is illustrated in Fig.~\ref{plots.power.spectra}, which shows the power spectrum of a SPIRE
250\,{${\mu}$m} image of the Polaris cirrus cloud \citep[see also Fig.\,3 of][]{Miville-Deschenes_etal2010}. In practice, this means
that the background fluctuations are much stronger on larger scales and, unlike Gaussian fluctuations, cannot be characterized by
a single value of the standard deviation. Indeed, they are better described by a wide range of standard deviations, as reflected in
their power spectrum (Fig.~\ref{plots.power.spectra}). Accordingly, the probability density function (PDF) of submillimeter
intensity (or column density) variations in the \emph{Herschel} images of Galactic fields is not Gaussian, but typically lognormal
in non-star-forming clouds, such as Polaris \citep[Schneider et al., in prep.; see also][]{Kainulainen_etal2009}.

In contrast to the astrophysical backgrounds, the instrumental noise fluctuations are approximately Gaussian, having a flat power
spectrum, essentially independent of spatial frequency (white noise). This is illustrated in Fig.~\ref{plots.power.spectra}, which
shows the estimated power spectrum of the instrumental noise in a typical \emph{Herschel} image at 250\,{${\mu}$m}, derived from the
power spectrum of the PACS 70\,{${\mu}$m} image of the Polaris field with no significant cloud emission
\citep[cf.][]{Men'shchikov_etal2010,Miville-Deschenes_etal2010,Andre_etal2010}. For a better comparison with the power-law cirrus
profile in Fig.~\ref{plots.power.spectra}, the spatial frequencies sampled in the original PACS 70\,{${\mu}$m} data were scaled down
by a factor $70/250$, to represent the typical range of spatial frequencies in the SPIRE 250\,{${\mu}$m} images of the
\emph{Herschel} Gould Belt survey. Likewise, the power spectrum of the intensity distribution of a Gaussian source with a
half-maximum width $D$ is itself Gaussian and thus flat approximately up to its half-maximum width ${2^{3/2}\ln 2}\,(\pi\,D)^{-1}$
in spatial frequencies.

Figure~\ref{plots.power.spectra} illustrates why decomposing the observed images into a finely-spaced set of filtered images is
advantageous for source extraction. Given the typical power spectra of the background fluctuations, instrumental noise, and
Gaussian-like sources, the maximum contrast of a source with a size $D$ over the background fluctuations is obtained at spatial
frequencies $0.46\,D^{-1}$ or, equivalently, for spatial scales close to $2.2\,D$. By performing source detection in a fine grid of
single-scale images, \textsl{getsources} can automatically identify and use the scale at which the contrast of each source over
the background is maximized, thereby improving source detectability.

Another great advantage of the fine spatial decomposition employed by \textsl{getsources} (Sect.~\ref{decomposing.detection.images})
is that the emission fluctuations in the \emph{decomposed} single-scale images of interstellar clouds follow Gaussian statistics
much more closely than the cloud fluctuations in the observed images containing all spatial scales (cf.
Fig.~\ref{single.scale.histograms}). This is because each single-scale image of a cirrus cloud can be characterized by a single
standard deviation value much better than the original image, since it selects only a narrow range of spatial frequencies from a
power-law spectrum of standard deviations, such as that shown in Fig.~\ref{plots.power.spectra}.
\end{appendix}

%||||||||||||||||||||||||||||||||||||||||||||||||||||||||||||||||||||||||||||||||||||||||||||||||||||||||||||||||||||||||||||||||||

\begin{appendix}
\section{Estimating shapes of sources}
\label{intensity.moments}

We derive the source coordinates, sizes, and orientations from the moments of the background-subtracted, deblended intensity 
distributions over the pixels of their footprints using the first and second moments
\begin{eqnarray}
\mathrm{E}(x)\!&=&\!\frac{\int{x\,I({\vec r})\,{\rm d}^2{\vec r}}}{\int{I(\vec r)\,{\rm d}^2{\vec r}}}, \\
\mathrm{E}(y)\!&=&\!\frac{\int{y\,I({\vec r})\,{\rm d}^2{\vec r}}}{\int{I(\vec r)\,{\rm d}^2{\vec r}}}, \\
\mathrm{E}(\tilde{x}^2)\!&=&\!\frac{\int{\tilde{x}^2\,I({\vec r})\,{\rm d}^2{\vec r}}}{\int{I(\vec r)\,{\rm d}^2{\vec r}}}, \\
\mathrm{E}(\tilde{y}^2)\!&=&\!\frac{\int{\tilde{y}^2\,I({\vec r})\,{\rm d}^2{\vec r}}}{\int{I(\vec r)\,{\rm d}^2{\vec r}}}, \\
\mathrm{E}(\tilde{x}\,\tilde{y})\!&=&\!\frac{\int{\tilde{x}\,\tilde{y}\,I({\vec r})\,{\rm d}^2{\vec r}}}{\int{I(\vec r)\,{\rm d}^2{\vec r}}},
\label{moments}	
\end{eqnarray}
where $x, y$ are the Cartesian coordinates of a vector ${\vec r}{\,=\,}(x, y)$ and the pixel coordinates relative to the source
barycenter are given by $\tilde{x}{\,=\,}x-\mathrm{E}(x)$ and $\tilde{y}{\,=\,}y-\mathrm{E}(y)$. Let us write the covariance 
matrix
\begin{equation}
M = \left[
\begin{array}{cc}
 \mathrm{E}(\tilde{x}^2) & \mathrm{E}(\tilde{x}\,\tilde{y}) \\
 \noalign{\medskip}
 \mathrm{E}(\tilde{x}\,\tilde{y}) & \mathrm{E}(\tilde{y}^2)
\end{array}
\right] = \left[
\begin{array}{cc}
 \sigma_x^2 & \sigma_{x y} \\
 \noalign{\medskip}
 \sigma_{y x} & \sigma_y^2
\end{array}
\right]
\label{covariance}
\end{equation}
where $\sigma_x^2$ and $\sigma_y^2$ are the variances and $\sigma_{x y}$ is the covariance of the intensity moments. For such a
symmetric matrix ($\sigma_{x y}{\,=\,}\sigma_{y x}$), one can find eigenvalues $\Lambda$ by solving the characteristic equation
\begin{equation}
\det\,\left(M-\Lambda I\right) = 0,
\label{characteristic.equation}
\end{equation}
where $I$ is the identity matrix. This leads to the quadratic equation
\begin{equation}
\Lambda^2 - \left(\sigma_x^2 + \sigma_y^2\right) \Lambda + \left(\sigma_x^2 \sigma_y^2 - \sigma_{x y}^2\right) = 0
\label{quadratic.equation}
\end{equation}
with the following two roots $\Lambda^+$ and $\Lambda^-$:
\begin{equation}
\Lambda^\pm = \frac{\sigma_x^2 + \sigma_y^2}{2} \pm \left(\left(\frac{\sigma_x^2 - \sigma_y^2}{2}\right)^2 \!+ 
\sigma_{x y}^2\right)^{1/2}
\label{two.roots}
\end{equation}
defining the major and minor axes of an ellipse. The major axis forms an angle $\theta$ with the $x$-axis, given by
\begin{equation}
\tan \left(2 \theta\right) = \frac{2\,\sigma_{x y}}{\sigma_x^2 - \sigma_y^2}.
\label{angle}
\end{equation}
The major and minor FWHM sizes, as well as the position angle (east of north) of a source $i$ at wavelength $\lambda$, are finally 
computed from
\begin{eqnarray}
A_{{i}{\lambda}}\!&=&\!\left(\Lambda^+\,8 \ln 2\right)^{1/2}, \\
B_{{i}{\lambda}}\!&=&\!\left(\Lambda^-\,8 \ln 2\right)^{1/2}, \\ 
\Theta_{{i}{\lambda}}\!&=&\!\pi - \arctan \theta.
\label{fwhm.sizes.and.position.angle}
\end{eqnarray}
\end{appendix}

%||||||||||||||||||||||||||||||||||||||||||||||||||||||||||||||||||||||||||||||||||||||||||||||||||||||||||||||||||||||||||||||||||

\begin{appendix}
\section{Installing and using \textsl{getsources}}
\label{installing.using}

The source extraction method described above has been developed by A.\,M. (since July 20, 2008) as a \textsl{Bash} script called
\textsl{getsources} and a suite of the FORTRAN utilities executed by the script that perform most of the work. This ensures a high
degree of portability and efficiency, as these two languages are the \emph{de facto} standards in the worlds of the UNIX-like
operating systems and numerical computations. Either Mac OS X or Linux and either \textsl{ifort 11.1} or \textsl{gfortran 4.5} can
be used to install \textsl{getsources}; other systems and compilers have not been tested. A preparation script called
\textsl{prepareobs} makes use of \textsl{SWarp} \citep{Bertin_etal2002} for image resampling and reprojection. To read and write
images in the FITS format, \textsl{getsources} uses the \textsl{CFITSIO} library \citep{Pence1999}; to convolve images, the fast
Fourier transform routine \textsl{rlft3} \citep{Press_etal1992} is used; the source coordinates ($\alpha$, $\delta$) are computed by
the utility \textsl{xy2sky} from \textsl{WCSTools} \citep{Mink2002}.

The total processing times are proportional to the numbers of pixels, spatial scales, wavelengths, iterations, and potential sources
detected, depending also on the computer CPUs available. The extraction time may vary between few minutes and a week, the latter
being the longest time we have experienced and it refers to a 6-wavelengths extraction for 5{\fdg}2$\times$5{\fdg}2 images with
6234$\times$6234 \,3{\arcsec} pixels, each image occupying $\sim$150 MB of disk space. On the other hand, a 6-wavelength extraction
for the simulated sky used for illustrations in this paper (1800$\times$1800 \,2{\arcsec} pixels, each image taking $\sim$13 MB of
disk space) was completed within one day. Although at first glance that may seem a long computational time, \textsl{getsources} is
not a real-time software; the completeness and reliability of the source extraction, not its speed, were the priorities in its
design. In the kind of astronomical research we are dealing with, even one week of processing is never a limiting factor, if the
work is properly planned. A researcher would spend much more time on the analysis of the information delivered by the code
(catalogs, images) and on the studies of the astrophysical reality of interest (in our case, the star formation).

Processing speed depends on many circumstances in a given computing system. Users are advised to always run \textsl{getsources} on
\emph{local} (internal) hard drives physically attached to the CPUs used for extractions. At some steps, the code performs lots of
read and write (I/O) operations on FITS files and the users would benefit from the fastest possible I/O throughput\footnote{To
optimize the processing times, it is a good idea to test available disk storage for speed. The script \textsl{iospeed} enables
comparisons of available hard drives by reading and writing a FITS image multiple times. One can perform tests of the local and
network disks, as well as of the virtual disks.}. Our algorithm is able to use \emph{virtual} random-access-memory (RAM) disks to
speed up the I/O for the images and to reduce the processing times. However, the gain compared to the speed of a \emph{local} hard
drive access may not be very significant. It is generally not optimal to use network-attached hard drives, as the disk access over
networks is \emph{very} slow compared to that of the local disks. If one \emph{must} use the network storage for running
\textsl{getsources} and has very large amounts of RAM, one might consider using the RAM disk facility of the code. This may lead up
to a considerable acceleration factor in processing times and thus compensate for the inefficient hard disk access over the
networks. For multi-wavelengths extractions, a natural way of cutting down the computational times is to run the first two
processing blocks (Fig.~\ref{algorithm}) in parallel, one wavelength per CPU, as they are decomposed and cleaned independently of
each other. If the computational time becomes an issue for enormously large images, the users may want to split them into several
sub-fields to obtain extractions much faster, running the extraction \emph{in parallel} on different CPUs. For example, one can get
an acceleration factor of $\sim$30 by splitting images in only 3 sub-fields of equal size (the factor assumes the same number of
sources in each sub-field and that the extractions are run in parallel).

Another consideration is the storage space necessary to keep available intermediate images for many spatial scales and wavelengths
until the end of the extraction process; the space needed scales between hundreds of MB to tens of GB, depending on the image size
and the numbers of the scales, wavelengths, and measurement iterations. In addition to the resulting catalogs and images produced
by \textsl{getsources}, a relatively large number of intermediate images and catalogs are also kept on the hard drive. Those are
useful in case the processing is interrupted for some reason or if the user needs to restart the extraction from some previous step
or to continue the measurement iterations until their convergence. Those restart images may also be very useful to inspect, in
order to better understand the extraction results; however, this time-saving feature works, of course, at the expense of the disk
space. It is up to the user to decide what is the biggest issue, the time or the space. Whenever the user becomes satisfied with the
extraction results or the time needed to re-run the extraction is not an issue, those extra files can be removed from the hard drive
by \textsl{getsources}.

The code is powerful, automated, flexible, easy-to-use, and very extensively tested; the algorithm is designed to be run on properly
prepared images twice (the initial and final extractions) and none of a few parameters of the code need to be changed, as their
default values have been carefully fine-tuned to work best in all cases. The multi-wavelength design of \textsl{getsources} is quite
flexible and it allows one to use it in some special ways. For example, one can detect sources using only selected wavelengths (even
a single image) but produce catalogs with measurements for all wavebands. It is also possible to add other non-\emph{Herschel}
images for either both detection and measurements or to only measurements, to use more information and get better results. One can
also use special mask images to exclude problematic (e.g., saturated) areas at some of the wavelengths to avoid using those areas in
combining images over wavelengths and in detecting sources. Users can re-run only selected steps of the extraction and also restart
the detection and measurements from any intermediate scale or iteration. There are also other possibilities; users are welcome to
request information on their specific needs from the author.

The source extraction code with an installation guide and a quick start guide are freely available upon request or they can be
downloaded from our web pages\footnote{http://gouldbelt-herschel.cea.fr/getsources,\\http://gouldbelt-herschel.cea.fr,
http://hobys-herschel.cea.fr}. Users installing \textsl{getsources} on their computers are advised to test it on a multi-wavelength
extraction using \emph{Herschel} images of the galaxy NGC4559 that was chosen as our validation field \citep[the galaxy was observed
as part of the KINGFISH project, see][]{Kennicutt_etal2011}. This relatively quick extraction performed by the author can also be
requested or downloaded by the users who wish to validate their installation and verify that they are able to reproduce the
reference extraction results.
\end{appendix}

%||||||||||||||||||||||||||||||||||||||||||||||||||||||||||||||||||||||||||||||||||||||||||||||||||||||||||||||||||||||||||||||||||

\bibliographystyle{aa}
\bibliography{aamnem99,getsources}

\begin{thebibliography}{45}
\expandafter\ifx\csname natexlab\endcsname\relax\def\natexlab#1{#1}\fi

\bibitem[{{Alves} {et~al.}(2001){Alves}, {Lada}, \& {Lada}}]{Alves_etal2001}
{Alves}, J.~F., {Lada}, C.~J., \& {Lada}, E.~A. 2001, \nat, 409, 159

\bibitem[{{Andr{\'e}} {et~al.}(2010){Andr{\'e}}, {Men'shchikov}, {Bontemps},
  {K{\"o}nyves}, {Motte}, {Schneider}, {Didelon}, {Minier}, {Saraceno},
  {Ward-Thompson}, {di Francesco}, {White}, {Molinari}, {Testi}, {Abergel},
  {Griffin}, {Henning}, {Royer}, {Mer{\'{\i}}n}, {Vavrek}, {Attard},
  {Arzoumanian}, {Wilson}, {Ade}, {Aussel}, {Baluteau}, {Benedettini},
  {Bernard}, {Blommaert}, {Cambr{\'e}sy}, {Cox}, {di Giorgio}, {Hargrave},
  {Hennemann}, {Huang}, {Kirk}, {Krause}, {Launhardt}, {Leeks}, {Le Pennec},
  {Li}, {Martin}, {Maury}, {Olofsson}, {Omont}, {Peretto}, {Pezzuto}, {Prusti},
  {Roussel}, {Russeil}, {Sauvage}, {Sibthorpe}, {Sicilia-Aguilar}, {Spinoglio},
  {Waelkens}, {Woodcraft}, \& {Zavagno}}]{Andre_etal2010}
{Andr{\'e}}, P., {Men'shchikov}, A., {Bontemps}, S., {et~al.} 2010, \aap, 518,
  L102+

\bibitem[{{Andr{\'e}} {et~al.}(2000){Andr{\'e}}, {Ward-Thompson}, \&
  {Barsony}}]{Andre_etal2000}
{Andr{\'e}}, P., {Ward-Thompson}, D., \& {Barsony}, M. 2000, Protostars and
  Planets IV, 59

\bibitem[{{Arzoumanian} {et~al.}(2011){Arzoumanian}, {Andr{\'e}}, {Didelon},
  {K{\"o}nyves}, {Schneider}, {Men'shchikov}, {Sousbie}, {Zavagno}, {Bontemps},
  {di Francesco}, {Griffin}, {Hennemann}, {Hill}, {Kirk}, {Martin}, {Minier},
  {Molinari}, {Motte}, {Peretto}, {Pezzuto}, {Spinoglio}, {Ward-Thompson},
  {White}, \& {Wilson}}]{Arzoumanian_etal2011}
{Arzoumanian}, D., {Andr{\'e}}, P., {Didelon}, P., {et~al.} 2011, \aap, 529,
  L6+

\bibitem[{{Bertin} \& {Arnouts}(1996)}]{BertinArnouts1996}
{Bertin}, E. \& {Arnouts}, S. 1996, \aaps, 117, 393

\bibitem[{{Bertin} {et~al.}(2002){Bertin}, {Mellier}, {Radovich}, {Missonnier},
  {Didelon}, \& {Morin}}]{Bertin_etal2002}
{Bertin}, E., {Mellier}, Y., {Radovich}, M., {et~al.} 2002, in Astronomical
  Society of the Pacific Conference Series, Vol. 281, Astronomical Data
  Analysis Software and Systems XI, ed. {D.~A.~Bohlender, D.~Durand, \&
  T.~H.~Handley}, 228

\bibitem[{{Bontemps} {et~al.}(2010){Bontemps}, {Andr{\'e}}, {K{\"o}nyves},
  {Men'shchikov}, {Schneider}, {Maury}, {Peretto}, {Arzoumanian}, {Attard},
  {Motte}, {Minier}, {Didelon}, {Saraceno}, {Abergel}, {Baluteau}, {Bernard},
  {Cambr{\'e}sy}, {Cox}, {di Francesco}, {di Giorgo}, {Griffin}, {Hargrave},
  {Huang}, {Kirk}, {Li}, {Martin}, {Mer{\'{\i}}n}, {Molinari}, {Olofsson},
  {Pezzuto}, {Prusti}, {Roussel}, {Russeil}, {Sauvage}, {Sibthorpe},
  {Spinoglio}, {Testi}, {Vavrek}, {Ward-Thompson}, {White}, {Wilson},
  {Woodcraft}, \& {Zavagno}}]{Bontemps_etal2010}
{Bontemps}, S., {Andr{\'e}}, P., {K{\"o}nyves}, V., {et~al.} 2010, \aap, 518,
  L85+

\bibitem[{{di Francesco} {et~al.}(2010){di Francesco}, {Sadavoy}, {Motte},
  {Schneider}, {Hennemann}, {Csengeri}, {Bontemps}, {Balog}, {Zavagno},
  {Andr{\'e}}, {Saraceno}, {Griffin}, {Men'shchikov}, {Abergel}, {Baluteau},
  {Bernard}, {Cox}, {Deharveng}, {Didelon}, {di Giorgio}, {Hargrave}, {Huang},
  {Kirk}, {Leeks}, {Li}, {Marston}, {Martin}, {Minier}, {Molinari}, {Olofsson},
  {Persi}, {Pezzuto}, {Russeil}, {Sauvage}, {Sibthorpe}, {Spinoglio}, {Testi},
  {Teyssier}, {Vavrek}, {Ward-Thompson}, {White}, {Wilson}, \&
  {Woodcraft}}]{diFrancesco_etal2010}
{di Francesco}, J., {Sadavoy}, S., {Motte}, F., {et~al.} 2010, \aap, 518, L91

\bibitem[{{Elmegreen} \& {Falgarone}(1996)}]{ElmegreenFalgarone1996}
{Elmegreen}, B.~G. \& {Falgarone}, E. 1996, \apj, 471, 816

\bibitem[{{Falgarone} {et~al.}(1991){Falgarone}, {Phillips}, \&
  {Walker}}]{Falgarone_etal1991}
{Falgarone}, E., {Phillips}, T.~G., \& {Walker}, C.~K. 1991, \apj, 378, 186

\bibitem[{{Gong} \& {Ostriker}(2011)}]{GongOstriker2011}
{Gong}, H. \& {Ostriker}, E.~C. 2011, \apj, 729, 120

\bibitem[{{Goodman} {et~al.}(1998){Goodman}, {Barranco}, {Wilner}, \&
  {Heyer}}]{Goodman_etal1998}
{Goodman}, A.~A., {Barranco}, J.~A., {Wilner}, D.~J., \& {Heyer}, M.~H. 1998,
  \apj, 504, 223

\bibitem[{{Griffin} {et~al.}(2010){Griffin}, {Abergel}, {Abreu}, {Ade},
  {Andr{\'e}}, {Augueres}, {Babbedge}, {Bae}, {Baillie}, {Baluteau}, {Barlow},
  {Bendo}, {Benielli}, {Bock}, {Bonhomme}, {Brisbin}, {Brockley-Blatt},
  {Caldwell}, {Cara}, {Castro-Rodriguez}, {Cerulli}, {Chanial}, {Chen},
  {Clark}, {Clements}, {Clerc}, {Coker}, {Communal}, {Conversi}, {Cox},
  {Crumb}, {Cunningham}, {Daly}, {Davis}, {de Antoni}, {Delderfield}, {Devin},
  {di Giorgio}, {Didschuns}, {Dohlen}, {Donati}, {Dowell}, {Dowell}, {Duband},
  {Dumaye}, {Emery}, {Ferlet}, {Ferrand}, {Fontignie}, {Fox}, {Franceschini},
  {Frerking}, {Fulton}, {Garcia}, {Gastaud}, {Gear}, {Glenn}, {Goizel},
  {Griffin}, {Grundy}, {Guest}, {Guillemet}, {Hargrave}, {Harwit}, {Hastings},
  {Hatziminaoglou}, {Herman}, {Hinde}, {Hristov}, {Huang}, {Imhof}, {Isaak},
  {Israelsson}, {Ivison}, {Jennings}, {Kiernan}, {King}, {Lange}, {Latter},
  {Laurent}, {Laurent}, {Leeks}, {Lellouch}, {Levenson}, {Li}, {Li},
  {Lilienthal}, {Lim}, {Liu}, {Lu}, {Madden}, {Mainetti}, {Marliani}, {McKay},
  {Mercier}, {Molinari}, {Morris}, {Moseley}, {Mulder}, {Mur}, {Naylor},
  {Nguyen}, {O'Halloran}, {Oliver}, {Olofsson}, {Olofsson}, {Orfei}, {Page},
  {Pain}, {Panuzzo}, {Papageorgiou}, {Parks}, {Parr-Burman}, {Pearce},
  {Pearson}, {P{\'e}rez-Fournon}, {Pinsard}, {Pisano}, {Podosek}, {Pohlen},
  {Polehampton}, {Pouliquen}, {Rigopoulou}, {Rizzo}, {Roseboom}, {Roussel},
  {Rowan-Robinson}, {Rownd}, {Saraceno}, {Sauvage}, {Savage}, {Savini},
  {Sawyer}, {Scharmberg}, {Schmitt}, {Schneider}, {Schulz}, {Schwartz},
  {Shafer}, {Shupe}, {Sibthorpe}, {Sidher}, {Smith}, {Smith}, {Smith},
  {Spencer}, {Stobie}, {Sudiwala}, {Sukhatme}, {Surace}, {Stevens}, {Swinyard},
  {Trichas}, {Tourette}, {Triou}, {Tseng}, {Tucker}, {Turner}, {Vaccari},
  {Valtchanov}, {Vigroux}, {Virique}, {Voellmer}, {Walker}, {Ward}, {Waskett},
  {Weilert}, {Wesson}, {White}, {Whitehouse}, {Wilson}, {Winter}, {Woodcraft},
  {Wright}, {Xu}, {Zavagno}, {Zemcov}, {Zhang}, \& {Zonca}}]{Griffin_etal2010}
{Griffin}, M.~J., {Abergel}, A., {Abreu}, A., {et~al.} 2010, \aap, 518, L3+

\bibitem[{{Hennemann} {et~al.}(2010){Hennemann}, {Motte}, {Bontemps},
  {Schneider}, {Csengeri}, {Balog}, {di Francesco}, {Zavagno}, {Andr{\'e}},
  {Men'shchikov}, {Abergel}, {Ali}, {Baluteau}, {Bernard}, {Cox}, {Didelon},
  {di Giorgio}, {Griffin}, {Hargrave}, {Hill}, {Horeau}, {Huang}, {Kirk},
  {Leeks}, {Li}, {Marston}, {Martin}, {Molinari}, {Nguyen Luong}, {Olofsson},
  {Persi}, {Pezzuto}, {Russeil}, {Saraceno}, {Sauvage}, {Sibthorpe},
  {Spinoglio}, {Testi}, {Ward-Thompson}, {White}, {Wilson}, \&
  {Woodcraft}}]{Hennemann_etal2010}
{Hennemann}, M., {Motte}, F., {Bontemps}, S., {et~al.} 2010, \aap, 518, L84+

\bibitem[{{Johnstone} {et~al.}(2000){Johnstone}, {Wilson}, {Moriarty-Schieven},
  {Joncas}, {Smith}, {Gregersen}, \& {Fich}}]{Johnstone_etal2000}
{Johnstone}, D., {Wilson}, C.~D., {Moriarty-Schieven}, G., {et~al.} 2000, \apj,
  545, 327

\bibitem[{{Kainulainen} {et~al.}(2009){Kainulainen}, {Beuther}, {Henning}, \&
  {Plume}}]{Kainulainen_etal2009}
{Kainulainen}, J., {Beuther}, H., {Henning}, T., \& {Plume}, R. 2009, \aap,
  508, L35

\bibitem[{{Kennicutt} {et~al.}(2011){Kennicutt}, {Calzetti}, {Aniano},
  {Appleton}, {Armus}, {Beir{\~a}o}, {Bolatto}, {Brandl}, {Crocker}, {Croxall},
  {Dale}, {Meyer}, {Draine}, {Engelbracht}, {Galametz}, {Gordon}, {Groves},
  {Hao}, {Helou}, {Hinz}, {Hunt}, {Johnson}, {Koda}, {Krause}, {Leroy}, {Li},
  {Meidt}, {Montiel}, {Murphy}, {Rahman}, {Rix}, {Roussel}, {Sandstrom},
  {Sauvage}, {Schinnerer}, {Skibba}, {Smith}, {Srinivasan}, {Vigroux},
  {Walter}, {Wilson}, {Wolfire}, \& {Zibetti}}]{Kennicutt_etal2011}
{Kennicutt}, R.~C., {Calzetti}, D., {Aniano}, G., {et~al.} 2011, \pasp, 123,
  1347

\bibitem[{{K{\"o}nyves} {et~al.}(2010){K{\"o}nyves}, {Andr{\'e}},
  {Men'shchikov}, {Schneider}, {Arzoumanian}, {Bontemps}, {Attard}, {Motte},
  {Didelon}, {Maury}, {Abergel}, {Ali}, {Baluteau}, {Bernard}, {Cambr{\'e}sy},
  {Cox}, {di Francesco}, {di Giorgio}, {Griffin}, {Hargrave}, {Huang}, {Kirk},
  {Li}, {Martin}, {Minier}, {Molinari}, {Olofsson}, {Pezzuto}, {Russeil},
  {Roussel}, {Saraceno}, {Sauvage}, {Sibthorpe}, {Spinoglio}, {Testi},
  {Ward-Thompson}, {White}, {Wilson}, {Woodcraft}, \&
  {Zavagno}}]{Ko"nyves_etal2010}
{K{\"o}nyves}, V., {Andr{\'e}}, P., {Men'shchikov}, A., {et~al.} 2010, \aap,
  518, L106+

\bibitem[{{Lagache} {et~al.}(1999){Lagache}, {Abergel}, {Boulanger},
  {D{\'e}sert}, \& {Puget}}]{Lagache_etal1999}
{Lagache}, G., {Abergel}, A., {Boulanger}, F., {D{\'e}sert}, F.~X., \& {Puget},
  J. 1999, \aap, 344, 322

\bibitem[{{Larson}(1981)}]{Larson1981}
{Larson}, R.~B. 1981, \mnras, 194, 809

\bibitem[{{Maury} {et~al.}(2011){Maury}, {Andr{\'e}}, {Men'shchikov},
  {K{\"o}nyves}, \& {Bontemps}}]{Maury_etal2011}
{Maury}, A.~J., {Andr{\'e}}, P., {Men'shchikov}, A., {K{\"o}nyves}, V., \&
  {Bontemps}, S. 2011, \aap, 535, A77

\bibitem[{{Men'shchikov} {et~al.}(2010){Men'shchikov}, {Andr{\'e}}, {Didelon},
  {K{\"o}nyves}, {Schneider}, {Motte}, {Bontemps}, {Arzoumanian}, {Attard},
  {Abergel}, {Baluteau}, {Bernard}, {Cambr{\'e}sy}, {Cox}, {di Francesco}, {di
  Giorgio}, {Griffin}, {Hargrave}, {Huang}, {Kirk}, {Li}, {Martin}, {Minier},
  {Miville-Desch{\^e}nes}, {Molinari}, {Olofsson}, {Pezzuto}, {Roussel},
  {Russeil}, {Saraceno}, {Sauvage}, {Sibthorpe}, {Spinoglio}, {Testi},
  {Ward-Thompson}, {White}, {Wilson}, {Woodcraft}, \&
  {Zavagno}}]{Men'shchikov_etal2010}
{Men'shchikov}, A., {Andr{\'e}}, P., {Didelon}, P., {et~al.} 2010, \aap, 518,
  L103+

\bibitem[{{Mink}(2002)}]{Mink2002}
{Mink}, D.~J. 2002, in Astronomical Society of the Pacific Conference Series,
  Vol. 281, Astronomical Data Analysis Software and Systems XI, ed.
  {D.~A.~Bohlender, D.~Durand, \& T.~H.~Handley}, 169--+

\bibitem[{{Miville-Desch{\^e}nes} {et~al.}(2010){Miville-Desch{\^e}nes},
  {Martin}, {Abergel}, {Bernard}, {Boulanger}, {Lagache}, {Anderson},
  {Andr{\'e}}, {Arab}, {Baluteau}, {Blagrave}, {Bontemps}, {Cohen},
  {Compiegne}, {Cox}, {Dartois}, {Davis}, {Emery}, {Fulton}, {Gry}, {Habart},
  {Huang}, {Joblin}, {Jones}, {Kirk}, {Lim}, {Madden}, {Makiwa}, {Menshchikov},
  {Molinari}, {Moseley}, {Motte}, {Naylor}, {Okumura}, {Pinheiro Gon{\c
  c}alves}, {Polehampton}, {Rod{\'o}n}, {Russeil}, {Saraceno}, {Schneider},
  {Sidher}, {Spencer}, {Swinyard}, {Ward-Thompson}, {White}, \&
  {Zavagno}}]{Miville-Deschenes_etal2010}
{Miville-Desch{\^e}nes}, M.-A., {Martin}, P.~G., {Abergel}, A., {et~al.} 2010,
  \aap, 518, L104

\bibitem[{{Moffat}(1969)}]{Moffat_1969}
{Moffat}, A.~F.~J. 1969, \aap, 3, 455

\bibitem[{{Molinari} {et~al.}(2011){Molinari}, {Schisano}, {Faustini},
  {Pestalozzi}, {di Giorgio}, \& {Liu}}]{Molinari_etal2011}
{Molinari}, S., {Schisano}, E., {Faustini}, F., {et~al.} 2011, \aap, 530, A133+

\bibitem[{{Motte} {et~al.}(1998){Motte}, {Andr{\'e}}, \&
  {Neri}}]{Motte_etal1998}
{Motte}, F., {Andr{\'e}}, P., \& {Neri}, R. 1998, \aap, 336, 150

\bibitem[{{Motte} {et~al.}(2001){Motte}, {Andr{\'e}}, {Ward-Thompson}, \&
  {Bontemps}}]{Motte_etal2001}
{Motte}, F., {Andr{\'e}}, P., {Ward-Thompson}, D., \& {Bontemps}, S. 2001,
  \aap, 372, L41

\bibitem[{{Motte} {et~al.}(2007){Motte}, {Bontemps}, {Schilke}, {Schneider},
  {Menten}, \& {Brogui{\`e}re}}]{Motte_etal2007}
{Motte}, F., {Bontemps}, S., {Schilke}, P., {et~al.} 2007, \aap, 476, 1243

\bibitem[{{Motte} {et~al.}(2003){Motte}, {Schilke}, \& {Lis}}]{Motte_etal2003}
{Motte}, F., {Schilke}, P., \& {Lis}, D.~C. 2003, \apj, 582, 277

\bibitem[{{Motte} {et~al.}(2010){Motte}, {Zavagno}, {Bontemps}, {Schneider},
  {Hennemann}, {di Francesco}, {Andr{\'e}}, {Saraceno}, {Griffin}, {Marston},
  {Ward-Thompson}, {White}, {Minier}, {Men'shchikov}, {Hill}, {Abergel},
  {Anderson}, {Aussel}, {Balog}, {Baluteau}, {Bernard}, {Cox}, {Csengeri},
  {Deharveng}, {Didelon}, {di Giorgio}, {Hargrave}, {Huang}, {Kirk}, {Leeks},
  {Li}, {Martin}, {Molinari}, {Nguyen-Luong}, {Olofsson}, {Persi}, {Peretto},
  {Pezzuto}, {Roussel}, {Russeil}, {Sadavoy}, {Sauvage}, {Sibthorpe},
  {Spinoglio}, {Testi}, {Teyssier}, {Vavrek}, {Wilson}, \&
  {Woodcraft}}]{Motte_etal2010}
{Motte}, F., {Zavagno}, A., {Bontemps}, S., {et~al.} 2010, \aap, 518, L77+

\bibitem[{{Myers}(1983)}]{Myers1983}
{Myers}, P.~C. 1983, \apj, 270, 105

\bibitem[{{Pence}(1999)}]{Pence1999}
{Pence}, W. 1999, in Astronomical Society of the Pacific Conference Series,
  Vol. 172, Astronomical Data Analysis Software and Systems VIII, ed.
  {D.~M.~Mehringer, R.~L.~Plante, \& D.~A.~Roberts}, 487--+

\bibitem[{{Pilbratt} {et~al.}(2010){Pilbratt}, {Riedinger}, {Passvogel},
  {Crone}, {Doyle}, {Gageur}, {Heras}, {Jewell}, {Metcalfe}, {Ott}, \&
  {Schmidt}}]{Pilbratt_etal2010}
{Pilbratt}, G.~L., {Riedinger}, J.~R., {Passvogel}, T., {et~al.} 2010, \aap,
  518, L1+

\bibitem[{{Poglitsch} {et~al.}(2010){Poglitsch}, {Waelkens}, {Geis},
  {Feuchtgruber}, {Vandenbussche}, {Rodriguez}, {Krause}, {Renotte}, {van
  Hoof}, {Saraceno}, {Cepa}, {Kerschbaum}, {Agn{\`e}se}, {Ali}, {Altieri},
  {Andreani}, {Augueres}, {Balog}, {Barl}, {Bauer}, {Belbachir}, {Benedettini},
  {Billot}, {Boulade}, {Bischof}, {Blommaert}, {Callut}, {Cara}, {Cerulli},
  {Cesarsky}, {Contursi}, {Creten}, {De Meester}, {Doublier}, {Doumayrou},
  {Duband}, {Exter}, {Genzel}, {Gillis}, {Gr{\"o}zinger}, {Henning},
  {Herreros}, {Huygen}, {Inguscio}, {Jakob}, {Jamar}, {Jean}, {de Jong},
  {Katterloher}, {Kiss}, {Klaas}, {Lemke}, {Lutz}, {Madden}, {Marquet},
  {Martignac}, {Mazy}, {Merken}, {Montfort}, {Morbidelli}, {M{\"u}ller},
  {Nielbock}, {Okumura}, {Orfei}, {Ottensamer}, {Pezzuto}, {Popesso},
  {Putzeys}, {Regibo}, {Reveret}, {Royer}, {Sauvage}, {Schreiber}, {Stegmaier},
  {Schmitt}, {Schubert}, {Sturm}, {Thiel}, {Tofani}, {Vavrek}, {Wetzstein},
  {Wieprecht}, \& {Wiezorrek}}]{Poglitsch_etal2010}
{Poglitsch}, A., {Waelkens}, C., {Geis}, N., {et~al.} 2010, \aap, 518, L2+

\bibitem[{{Press} {et~al.}(1992){Press}, {Teukolsky}, {Vetterling}, \&
  {Flannery}}]{Press_etal1992}
{Press}, W.~H., {Teukolsky}, S.~A., {Vetterling}, W.~T., \& {Flannery}, B.~P.
  1992, {Nu\-merical recipes in FORTRAN. The art of scientific computing}

\bibitem[{{Rosolowsky} {et~al.}(2008){Rosolowsky}, {Pineda}, {Kauffmann}, \&
  {Goodman}}]{Rosolowsky_etal2008}
{Rosolowsky}, E.~W., {Pineda}, J.~E., {Kauffmann}, J., \& {Goodman}, A.~A.
  2008, \apj, 679, 1338

\bibitem[{{Roy} {et~al.}(2010){Roy}, {Ade}, {Bock}, {Chapin}, {Devlin},
  {Dicker}, {Griffin}, {Gundersen}, {Halpern}, {Hargrave}, {Hughes}, {Klein},
  {Marsden}, {Martin}, {Mauskopf}, {Miville-Desch{\^e}nes}, {Netterfield},
  {Olmi}, {Patanchon}, {Rex}, {Scott}, {Semisch}, {Truch}, {Tucker}, {Tucker},
  {Viero}, \& {Wiebe}}]{Roy_etal2010}
{Roy}, A., {Ade}, P.~A.~R., {Bock}, J.~J., {et~al.} 2010, \apj, 708, 1611

\bibitem[{{Schneider} {et~al.}(2010){Schneider}, {Motte}, {Bontemps},
  {Hennemann}, {di Francesco}, {Andr{\'e}}, {Zavagno}, {Csengeri},
  {Men'shchikov}, {Abergel}, {Baluteau}, {Bernard}, {Cox}, {Didelon}, {di
  Giorgio}, {Gastaud}, {Griffin}, {Hargrave}, {Hill}, {Huang}, {Kirk},
  {K{\"o}nyves}, {Leeks}, {Li}, {Marston}, {Martin}, {Minier}, {Molinari},
  {Olofsson}, {Panuzzo}, {Persi}, {Pezzuto}, {Roussel}, {Russeil}, {Sadavoy},
  {Saraceno}, {Sauvage}, {Sibthorpe}, {Spinoglio}, {Testi}, {Teyssier},
  {Vavrek}, {Ward-Thompson}, {White}, {Wilson}, \&
  {Woodcraft}}]{Schneider_etal2010}
{Schneider}, N., {Motte}, F., {Bontemps}, S., {et~al.} 2010, \aap, 518, L83+

\bibitem[{{Schneider} \& {Elmegreen}(1979)}]{SchneiderElmegreen_1979}
{Schneider}, S. \& {Elmegreen}, B.~G. 1979, \apjs, 41, 87

\bibitem[{{Smith}(1979)}]{Smith_1979}
{Smith}, A.~R. 1979, in SIGGRAPH'79: Proc. of the 6th annual conference on
  Computer graphics and interactive techniques (New York: ACM), 276--283

\bibitem[{{Starck} \& {Murtagh}(2006)}]{StarckMurtagh2006}
{Starck}, J.-L. \& {Murtagh}, F. 2006, {Astronomical Image and Data Analysis},
  ed. {{Starck}, J.-L.~\& {Murtagh}, F.}

\bibitem[{{Stutzki} \& {Guesten}(1990)}]{StutzkiGuesten1990}
{Stutzki}, J. \& {Guesten}, R. 1990, \apj, 356, 513

\bibitem[{{Williams} {et~al.}(2000){Williams}, {Blitz}, \&
  {McKee}}]{Williams_etal2000}
{Williams}, J.~P., {Blitz}, L., \& {McKee}, C.~F. 2000, Protostars and Planets
  IV, 97

\bibitem[{{Williams} {et~al.}(1994){Williams}, {de Geus}, \&
  {Blitz}}]{WilliamsdeGeusBlitz1994}
{Williams}, J.~P., {de Geus}, E.~J., \& {Blitz}, L. 1994, \apj, 428, 693

\end{thebibliography}

\end{document}